\documentclass[a4paper,12pt,onecolumn,twoside]{book}
\pdfoutput=1

\usepackage{geometry}

\usepackage[english]{babel} 

\usepackage[square,numbers,sort&compress,merge]{natbib}
\usepackage{hypernat}
\usepackage{bibentry}
\nobibliography*

\usepackage{enumitem}
\makeatletter
\def\namedlabel#1#2{\begingroup
    #2%
    \def\@currentlabel{#2}%
    \phantomsection\label{#1}\endgroup
}
\makeatother

\usepackage[sfdefault,lf,t]{carlito}

\usepackage{setspace}
\usepackage{amsmath}
\usepackage{amssymb}
\usepackage{mathrsfs}
\usepackage{dsfont}
\allowdisplaybreaks

\usepackage[dvipsnames]{xcolor}

\usepackage{footnote}
\makesavenoteenv{tabular}
\makesavenoteenv{table}

\definecolor{EarthBlue}{HTML}{5e82b5}
\definecolor{EarthYellow}{HTML}{DBCA69}
\definecolor{EarthGreen}{HTML}{668D3C}
\definecolor{EarthRed}{HTML}{8F3B1B}
\definecolor{EarthGray}{HTML}{A3ADB8}

\definecolor{FlowChartBackground}{HTML}{D1D6DB}

\usepackage[
colorlinks=true, 
linkcolor=.,
citecolor=EarthRed,
urlcolor=EarthBlue,
pdfborder={0 0 0},
pagebackref,
plainpages=false,
hypertexnames=true
]{hyperref}
\renewcommand*{\backref}[1]
{
}
\renewcommand*{\backrefalt}[4]
{
   	\footnotesize
   	\color{gray}
   	\ifnum#1>0
   	\mbox{Cited on %
   	\ifnum#1=1 %
		   page~%
	   \else
   		the pages~%
   	\fi
   	#2.}
   	\fi
}

\usepackage{graphicx}
\usepackage{subcaption}
\newlength{\twosubht}
\newsavebox{\twosubbox}
\usepackage{tikz}
\usetikzlibrary{arrows,decorations.pathmorphing,backgrounds,positioning,fit,petri}

\usepackage{fancyhdr}
\renewcommand{\headrulewidth}{0.25pt}
\renewcommand{\footrulewidth}{0pt}

\fancypagestyle{plain}{%
\fancyhf{} 
\fancyfoot[LE,RO]{\bfseries\thepage}
\renewcommand{\headrulewidth}{0pt}
\renewcommand{\footrulewidth}{0pt}
}

\usepackage{titling}
\newcommand{\subtitle}[1]{%
  \posttitle{%
    \par\end{center}
    \begin{center}\large#1\end{center}
    \vskip0.5em}%
}
\numberwithin{equation}{section} 

\usepackage{afterpage}
\newcommand\blankpage{%
    \null
    \thispagestyle{empty}%
    \newpage}

\usepackage{siunitx}
\DeclareSIUnit \parsec {pc}

\usepackage[withpage,printonlyused]{acronym}

\newcommand{\dd}{\; \mathrm{d}}
\newcommand{\commutator}[2]{\left[ #1 , #2 \right]}

\newcommand{\isotope}[2]{${}^{\text{#2}}$#1}
\newcommand{\erf}[1]{{\rm erf}\left(#1\right)}

\title{Dark Matter in the Earth and the Sun - Simulating Underground Scatterings for the Direct Detection of Low-Mass Dark Matter}

\author{Timon Emken}

\date{14.02.2019}

\begin{document}
\pagenumbering{Roman} 
\setcounter{tocdepth}{1}
\setcounter{secnumdepth}{0}

\pagestyle{empty}

\begin{titlepage}
	\centering
	\includegraphics[width=0.33\textwidth]{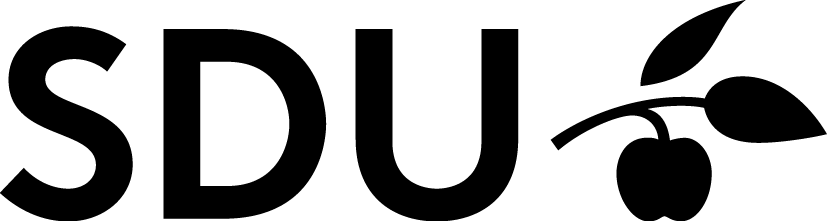}\par\vspace{1cm}
	{\scshape\LARGE University of Southern Denmark \par}
	\vspace{1.5cm}
	{\scshape\Large Doctoral Dissertation\par}
	\vspace{0.5cm}
	{\huge\bfseries Dark Matter in the Earth and the Sun\par}
	{\Large\bfseries  -- \par}
	{\Large\bfseries  Simulating Underground Scatterings for the Direct~Detection of Low-Mass Dark Matter \par}
	\vspace{0.8cm}
	{\LARGE\itshape Timon Emken\par}
	\vfill
	{\large
	\vspace{0.4cm}
	supervised by\par
	Chris Kouvaris
	\vspace{0.8cm}
	}
	\vfill
	\includegraphics[width=0.33\textwidth]{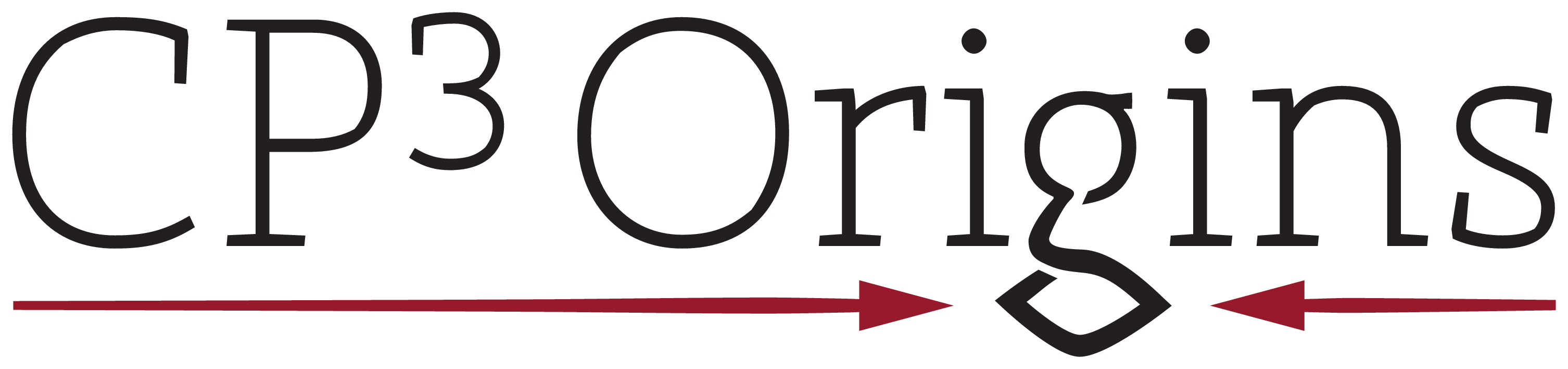}\par\vspace{0.5cm}
	{\large 
	Centre for Cosmology and Particle Physics Phenomenology\par
	Department of Physics, Chemistry, and Pharmacy\par
	\vspace{0.5cm}
	Submitted to the University of Southern Denmark\par 
	January  31, 2019\par}
\end{titlepage}

\newgeometry{
    top=3cm,
    bottom=3cm,
    outer=3cm,
    inner=4cm,
}
\blankpage

\vspace*{\fill} 
\hfill {\large \textit{Inveniam viam aut faciam.}}
\vspace*{\fill} 
\newpage
\blankpage

\pagenumbering{roman}
\pagestyle{plain}
\tableofcontents

\newpage
\section{Abstract}
Astrophysical and cosmological observations provide compelling evidence that the majority of matter in the Universe is \textit{dark}. Showing no interactions with electromagnetic radiation, this dark matter~(DM) eludes direct observations, and its nature and origin remains unknown to this day. Direct detection experiments search for interactions between halo~DM and nuclei inside a detector. So far, a variety of experiments were only able to set stringent limits on the~DM parameter space. These constraints weaken for sub-GeV~DM masses, as light particles are not energetic enough to trigger most detectors. New experimental efforts shift the focus towards lower masses, for example by looking for inelastic DM-electron scatterings.

If scatterings between~DM and ordinary matter are assumed to occur in a detector's target material, collisions will naturally take place inside the bulk of planets and stars as well. For sufficiently large cross sections, these scatterings might occur in the Earth or Sun even prior to the detection. In this thesis, we study the impact of these pre-detection scatterings on direct searches of light~DM with the use of Monte Carlo~(MC) simulations. By simulating the trajectories and scatterings of many individual DM~particles through the Earth or Sun, we determine the local distortions of the statistical properties of~DM at any detector caused by elastic DM-nucleus collisions.

Scatterings inside the Earth distort the underground DM~density and velocity distribution. Any detector moves periodically through these inhomogeneities due to the Earth's rotation, and the expected event rate will vary throughout a sidereal day. Using MC~simulations, we can determine the exact amplitude and phase of this diurnal modulation for any experiment. For even higher scattering probabilities, collisions in the overburden above the typically underground detectors start to attenuate the incoming~DM flux. The critical cross section above which an experiment loses sensitivity to~DM itself is determined for a variety of DM-nucleus and DM-electron scattering experiments and different types of interactions.

Furthermore, we develop the idea that sub-GeV DM~particles can enter the Sun, gain kinetic energy by colliding on hot nuclei and get reflected with great speeds. By deriving an analytic expressions for the particle flux from solar reflection via a single scattering, we demonstrate the prospects of future experiments to probe reflected~DM and extend their sensitivity to lower masses than accessible by halo~DM alone. We present first results for~MC simulations of solar reflections. Including reflection after multiple scatterings greatly amplifies the reflected DM~flux and thereby the potential of solar reflection for direct searches for light~DM.

\newpage
\section{Sammenfatning}
Astrofysiske observationer giver overbevisende tegn p\aa, at st\o rstedelen af stof i universet er \textit{m\o rkt}. Dette m\o rke stof (DM) viser ingen observationelle interaktioner med elektromagnetisk str\aa ling, og dets natur og oprindelse er stadig ukendt. Direkte detektionseksperimenter s\o ger efter interaktioner mellem DM fra galaksehaloen og atomkerner i en detektor. Hidtil har eksperimenter kun v\ae ret i stand til at s\ae tte strenge gr\ae nser i DM parameterrummet. Disse begr\ae nsninger l\o snes for sub-GeV DM-masser, da lette partikler ikke har nok energi til at udl\o se en m\aa ling i de fleste detektorer. Ny eksperimentel indsats skifter fokus mod lavere masser, for eksempel ved at lede efter uelastiske sammenst\o d mellem DM og elektroner.

Hvis sammenst\o d mellem DM og almindeligt stof antages at ske i detektoren, vil kollisioner ogs\aa~finde sted i hovedparten af planeter og stjerner. For tilstr\ae kkeligt store tv\ae rsnit kan disse sammenst\o d forekomme i Jorden eller Solen, endda f\o r de m\aa les i detektoren. I denne afhandling unders\o ger vi virkningen af disse pr\ae -detektor sammenst\o d i direkte s\o gninger af let DM ved brug af Monte Carlo (MC) simuleringer. Ved at simulere baner og sammenst\o d af mange individuelle DM-partikler gennem Jorden eller Solen bestemmer vi de lokale forvr\ae ngninger i de statistiske egenskaber af DM for\aa rsaget af elastiske DM-atomkernekollisioner i en given detektor.

Sammenst\o d inde i Jorden \ae ndrer den underjordiske DM-densitet og hastighedsfordeling. Enhver detektor bev\ae ger sig periodisk gennem disse inhomogeniteter mens planeten roterer, og den forventede h\ae ndelsesrate vil variere i l\o bet af en siderisk dag. Ved hj\ae lp af MC-simuleringer kan vi bestemme den n\o jagtige amplitude og fase af denne d\o gnmodulering i givent eksperiment. For endnu h\o jere sammenst\o dssandsynligheder begynder sammenst\o d i lagene ovenover det underjordiske eksperiment at d\ae mpe den indkommende DM-flux. Vi bestemmer det kritiske tv\ae rsnit, hvor et eksperiment mister f\o lsomheden overfor DM for en r\ae kke atomkerne- og elektronspredningsfors\o g samt forskellige typer interaktioner.

Desuden udvikler vi ideen om, at sub-GeV DM-partikler kan tr\ae nge ind i Solen, f\aa~kinetisk energi ved at kollidere med varme atomkerner og blive reflekteret med h\o je hastigheder. Vi udleder analytiske udtryk for partikelfluxen fra solreflektion via et enkelt sammenst\o d og demonstrerer udsigterne for fremtidige eksperimenter til at lede efter reflekteret DM og udvide f\o lsomheden over for lavere masser end hvad der er tilg\ae ngelig ved halo-DM alene. Vi pr\ae senterer de f\o rste MC-resultater. Vi inkluderer refleksion efter flere sammenst\o d, som forst\ae rker den reflekterede DM-flux og derved potentialet ved solreflektion til direkte s\o gninger efter let DM.

\newpage
\vspace*{\fill} 
\section{List of Publications}
This thesis contains results, which were published in the following papers.
\begin{description}
	\item[\namedlabel{paper5}{Paper V}] \label{paper5}\bibentry{Emken2019}
	\item[\namedlabel{paper4}{Paper IV}] \label{paper4}\bibentry{Emken:2018run}
	\item[\namedlabel{paper3}{Paper III}] \label{paper3}\bibentry{Emken:2017hnp}
	\item[\namedlabel{paper2}{Paper II}] \label{paper2}\bibentry{Emken:2017qmp}
	\item[\namedlabel{paper1}{Paper I}] \label{paper1}\bibentry{Emken:2017erx}
\end{description}
I will refer to these publications throughout this thesis as e.g.~\ref{paper2}. Results reported in these papers were partially generated with scientific codes I developed during the preparation of this thesis. They were made publicly available alongside the corresponding publications.
\begin{description}
	\item[\namedlabel{code2}{DaMaSCUS-CRUST}] \label{code2}\bibentry{Emken2018a}
	\item[\namedlabel{code1}{\textsc{DaMaSCUS}}] \label{code1}\bibentry{Emken2017a}
\end{description}
\vspace*{\fill} 

\cleardoublepage
\phantomsection
\addcontentsline{toc}{section}{\listfigurename}
\listoffigures

\cleardoublepage
\phantomsection
\addcontentsline{toc}{section}{\listtablename}
\listoftables

\newpage
\vspace*{\fill} 
\section{List of Acronyms}
\begin{acronym}[MACHOsxxxx]
	\acro{BBN}{Big Bang Nucleosynthesis}
	\acro{CL}{Confidence Level}
	\acro{CMB}{Cosmic Microwave Background}
	\acro{CMS}{Center of Mass System}
	\acro{CDF}{Cumulative Distribution Function}
	\acro{COBE}{Cosmic Background Explorer}
	\acro{DaMaSCUS}{Dark Matter Simulation Code for Underground Scatterings}
	\acro{DM}{Dark Matter}
	\acro{EFT}{Effective Field Theory}
	\acro{GIS}{Geometric Importance Splitting}
	\acro{GAST}{Greenwich Apparent Sidereal Time}
	\acro{GMST}{Greenwich Mean Sidereal Time}
	\acro{IS}{Importance Sampling}
	\acro{KDE}{Kernel Density Estimation}
	\acro{LAST}{Local Apparent Sidereal Time}
	\acro{LHC}{Large Hadron Collider}
	\acro{LIGO}{Laser Interferometer Gravitational-Wave Observatory}
	\acro{LNGS}{Laboratori Nazionali del Gran Sasso}
	\acro{LSM}{Laboratoire Souterrain de Modane}
	\acro{MACHO}{Massive Compact Halo Object}
	\acro{MC}{Monte Carlo}
	\acro{MET}{Missing Transverse Energy}
	\acro{MOND}{Modified Newtonian Dynamics}
	\acro{MSSM}{Minimal Supersymmetric Standard Model}
	\acro{NREFT}{Non-Relativistic Effective Field Theory}
	\acro{PBH}{Primordial Black Hole}
	\acro{PDF}{Probability Density Function}
	\acro{PE}{Photoelectron}
	\acro{PMF}{Probability Mass Function}
	\acro{PMT}{Photomultiplier Tube}
	\acro{PREM}{Preliminary Reference Earth Model}
	\acro{QCD}{Quantum Chromodynamics}
	\acro{RHF}{Roothaan-Hartree-Fock}
	\acro{SD}{Spin-Dependent}
	\acro{SHM}{Standard Halo Model}
	\acro{SI}{Spin-Independent}
	\acro{SIMP}{Strongly Interacting Massive Particle}
	\acro{SM}{Standard Model of Particle Physics}
	\acro{SSM}{Standard Solar Model}
	\acro{SUPL}{Stawell Underground Physics Laboratory}
	\acro{SURF}{Sanford Underground Research Facility}
	\acro{SUSY}{Supersymmetry}
	\acro{TT}{Terrestrial Time}
	\acro{UT}{Universal Time}
	\acro{WIMP}{Weakly Interacting Massive Particle}
	\acro{WMAP}{Wilkinson Microwave Anisotropy Probe}
\end{acronym}
\vspace*{\fill} 

\newpage
\section{Acknowledgments}
No dark matter detector exists in isolation, and neither do PhD students. This dissertation would not have been possible without a great number of people whose direct and indirect support I am deeply grateful for.

I am grateful to Chris Kouvaris for the opportunity to work in this exciting field of research and for a fruitful collaboration. His supervision allowed me to grow as an independent researcher. I would also like to thank the Faculty of Science at SDU and in particular the PhD School Secretariat for assistance especially during the final sprint.

More thanks go to my collaborators, namely Rouven Essig, Niklas G. Nielsen, Ian M. Shoemaker, and Mukul Sholapurkar for their contributions to the research in this thesis and countless valuable discussions. I learned a lot from them, and it is a pleasure to work with them.

During my PhD studies, I had the privilege to visit and work at the \href{http://insti.physics.sunysb.edu/itp/www/}{C.N. Yang Institute for Theoretical Physics} in Stony Brook, NY. For the hospitality and kindness that I encountered during my stay there, I am deeply grateful.

The computations of this thesis involved months of coding\footnote{I should also acknowledge the thousands and thousands of faceless heroes of~\href{https://stackoverflow.com}{Stack Overflow} who unknowingly and without reward solved a vast number of my everyday problems and `taught' me the art and ordeal that is programming.} and simulations running on the \textit{ABACUS~2.0} supercomputer. I would like to express my gratitude to the team of the \href{https://www.deic.dk/en/HPCCenter_SDU}{DeiC National HPC Center} at SDU for their always swift and helpful assistance.

At~$\text{CP}^{\text{3}}$, I was privileged not only to carry out my research, but also meet amazing colleagues, many of which have become dear friends. I especially appreciated the atmosphere and activities among the PhD students and postdocs. In particular, I would like to thank my collaborator, office-buddy, and friend Niklas for the time and work we shared in the party office.

A special thanks goes to my entire family, who I often missed during my time in Odense. I thank especially my parents for their unconditional love and support.

When I came to Denmark three years ago, little did I know who I would meet. I want to thank you, Majken, for your words of love, encouragement, and wisdom. Your humour and support never fail to brighten even the most stressful day.

There are without a doubt many names missing on this page, names of people who deserve to be mentioned here. You know who you are. You made my time here at~$\text{CP}^{\text{3}}$ worthwhile. Thank you. 

\cleardoublepage
\thispagestyle{empty}

\newpage

\pagestyle{fancy}
\fancyhf{} 
\fancyfoot[LE,RO]{\bfseries\thepage}
\fancyhead[LE]{\leftmark}
\fancyhead[RO]{\rightmark}
\renewcommand{\headrulewidth}{0.25pt}
\renewcommand{\footrulewidth}{0pt}
\setcounter{secnumdepth}{3}
\pagenumbering{arabic}

\clearpage
\chapter{Introduction}
\label{c:intro}

The nature of dark matter is one of the most exciting open questions of natural science in general and astro- and particle physics in particular. The field of high-energy physics finds itself in a peculiar situation. With the discovery of the Higgs particle at the Large Hadron Collider at CERN in 2012~\cite{Aad:2012tfa,Chatrchyan:2012xdj}, the~\ac{SM} was confirmed to describe the behavior of fundamental particles on all tested energy scales with remarkable precision. The physics of visible matter, fundamentally composed of leptons, quarks, and their interactions, seems very well understood. The success of the~SM clashes at the same time with a series of astrophysical observations, all of which substantiate the notion that the visible matter, the matter we observe in forms of stars, galaxies, gas, or planets, the matter we can describe so accurately, can only account for about 15\% of the total matter of our Universe. In order to make sense of various independent astrophysical measurements from galactic to cosmological scales, it seems vital to make the astounding assumption that~85\% of matter is~\textit{dark}. Showing no interactions with electromagnetic radiation, this~\ac{DM} eludes all direct observations, yet affects and dominates gravitational dynamics on astronomical scales. DM is the umbrella term to capture the unidentified explanation of these observations which can not be attributed to any particle of the~SM, as its ultimate origin is entirely unknown.

Ordinary matter is fundamentally composed of particles, and it is not farfetched to assume that this applies to the dark sector of matter as well. If so, the Earth would consequently get traversed by a continuous stream of a vast number of DM~particles at any moment. Should these particles interact with ordinary matter through some interaction besides gravity and occasionally scatter with atoms, it could be possible to observe these collisions inside a detector. Experiments on Earth should be able to discover~DM, provided that some ``portal'' between the light and the dark sector exists. Many of such \textit{direct detection} experiments have been conducted in the last three decades, thus far unable to discover~DM on Earth.

If we expect these scatterings to occur inside a detector at a non-vanishing rate, they should also happen without being detected inside the Earth's or Sun's bulk mass. For sufficiently strong interactions, this might even happen prior to detection. Underground scatterings before passing through a detector affect the expected outcome of the experiment. Elastic collisions on nuclei change the trajectory and speed of DM~particles on their way through the medium, with potentially strong implications for direct detection experiments.

Nuclear scatterings modify the underground spatial and energetic distribution of DM~particles inside the Earth through deflection and deceleration. For a significant scattering probability, the expected signal rate for any detector would depend on its exact location, because the average underground distance for~DM to reach the detector and therefore also its scattering probability vary periodically. Since the experiment does not stand still but rotates around the Earth axis, the signal rate will show a diurnal modulation. The phase and amplitude of this modulation, which we will predict over the course of this thesis, depends on the DM~model and the experiment's location on Earth. Diurnal modulations would not only be a clean signature distinguishing a signal from background, they could also tell us something about the interaction itself.

Direct detection experiments are usually set up deep underground in order to shield off background sources. However, if the DM-matter interactions are so strong that incoming DM~particles from the halo collide on nuclei of the experiment's overburden already, the shielding layers~(typically~$\sim$~1~km of rock) could weaken the DM~signal itself, up to the point where terrestrial experiments lose sensitivity to strongly interacting DM~particles entirely. Their scatterings on nuclei in the Earth crust or atmosphere would then attenuate the observable flux below detectability. This is a natural limitation of any direct DM~search on Earth and needs to be quantified.

It turns out that pre-detection scatterings can also \textit{extend} an experiment's sensitivity. Through collisions with highly energetic nuclei of the hot solar core, low-mass DM~particles could gain energy. These particles fall into the Sun's gravitational well, get further accelerated by elastic collisions and leave the star much faster than the initial speed. The solar reflection flux of~DM can extend an experiment's sensitivity to lower masses, since faster DM~particles can deposit more energy in a detector.

\begin{figure*}
	\centering
	\includegraphics[width=0.9\textwidth]{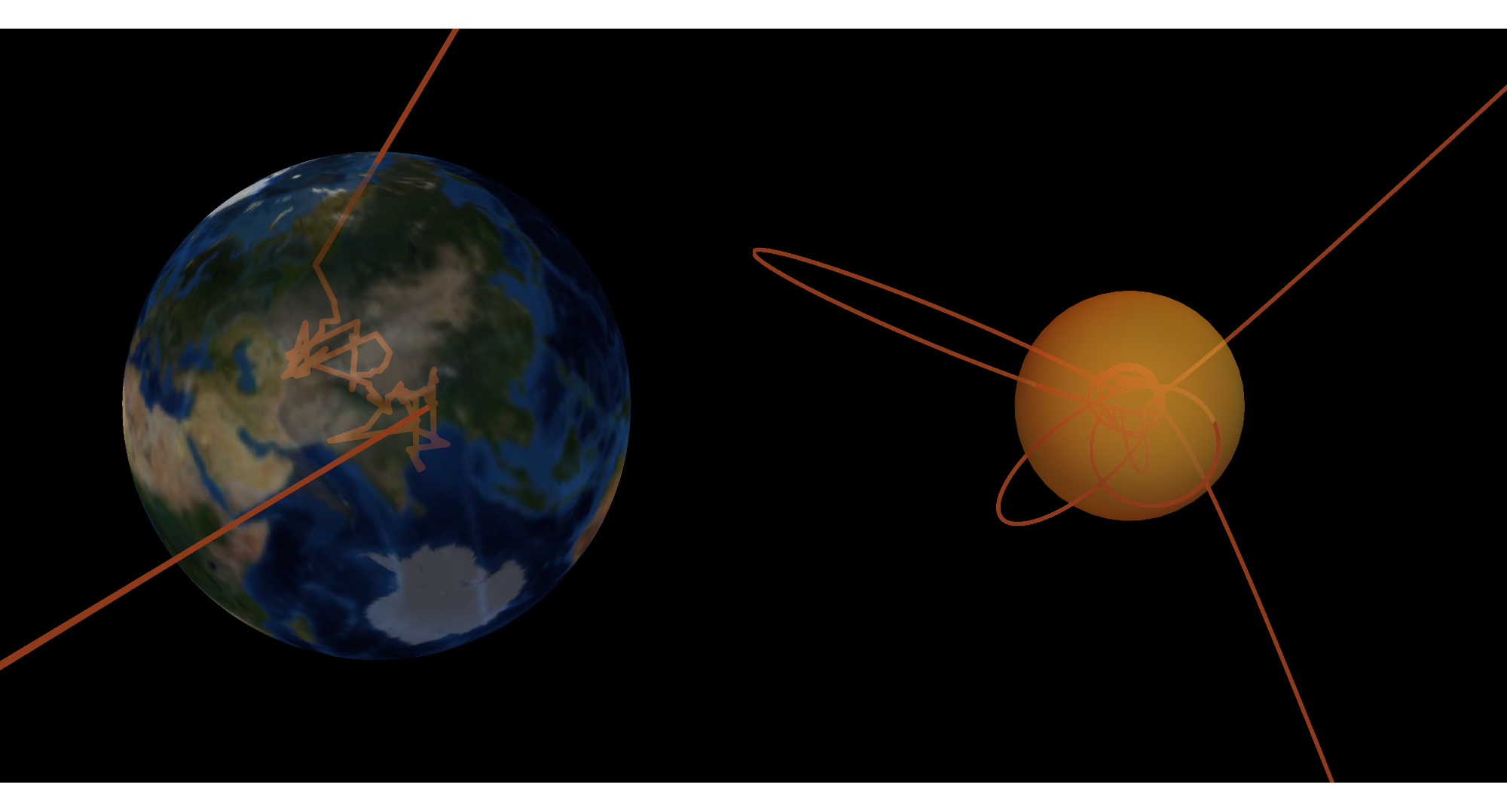}
	\caption{Simulated~DM trajectories in the Earth and Sun with multiple scatterings.}
	\label{fig: trajectories}
\end{figure*}

A powerful tool to investigate the effect of many underground scatterings are \ac{MC}~simulations. By simulating individual trajectories of particles passing through the Earth or Sun while colliding with terrestrial or solar nuclei, we can quantify the phenomenological impact by numerical and statistical methods. Examples of simulated trajectories in the Earth and the Sun are shown in figure~\ref{fig: trajectories}. Most of the results of this thesis have been obtained by setting up dedicated~\ac{MC} codes and running simulations on a supercomputer. The code used to generate the published results have been released together with the corresponding papers. In particular, the~\ac{DaMaSCUS}~\cite{Emken2017a} and~\ref{code2}~\cite{Emken2018a} are publicly available.\\[0.3cm]

Concerning the thesis' structure, the first two chapters introduce dark matter and the attempts to directly detect it. In chapter~\ref{c:DM}, we review the evidence for~DM and follow its history over the course of the~20th century. The evolution from~DM as a purely astronomical question to an active field of particle physics in the century's second half is emphasized. The direct detection of~DM is the main topic of chapter~\ref{c:directdetection}, where we will summarize previous detection attempts, prioritizing direct searches for low-mass~DM. This chapter also reviews the basics and essential computations of recoil spectra and signal rates for both conventional direct detection via nuclear recoils and electron-scattering experiments.

The next two chapters contain the main results of this thesis. The foundations and results of the simulations of~DM inside the Earth are compiled in chapter~\ref{c:earth}. Therein, we formulate the general algorithms for the \ac{MC}~simulations of underground trajectories. This is followed by the application of the algorithm to the entire Earth to quantify diurnal modulation of detection rates. The second application of the terrestrial simulations concerns trajectories through the overburden of a given experiment, e.g. the Earth crust or atmosphere. This allows to determine the exact constraints on strongly interacting~DM. In chapter~\ref{c:sun}, we focus the attention on DM~particles scattering and getting accelerated inside the Sun. The theoretical framework to describe DM~scatterings in a star is formulated and applied to study the detection prospects of solar reflection of~DM via a single scattering with analytic methods. Furthermore, the \ac{MC}~algorithms are extended for DM~trajectories inside the Sun by including its gravitational force and thermal targets. These simulations can shed light on the contribution of multiple scatterings to solar reflection.

Finally, we conclude in chapter~\ref{c:conclusion}. In addition, a number of appendices are included in this thesis containing details on the astronomical prerequisites of the simulations, on the experiments, various numerical methods, and more. These appendices are supposed to give a broad and extensive overview of the more technical, yet essential fundamentals and techniques applied throughout the thesis.

\clearpage
\chapter{Dark Matter}
\label{c:DM}
\begin{quote}
	``The weight of evidence for an extraordinary claim must be proportioned to its strangeness.''\\
	\hspace*{\fill}--Pierre-Simon Laplace~(1749-1827)~\cite{Laplace1814}.
\end{quote}
The claim that our Universe is dominated by a form of matter which we cannot see nor directly measure is indeed extraordinary and requires justification. Although the different pieces of evidence in favour of dark matter have been presented and reviewed in a plethora of publications, books, and presentations, we believe it is vital to keep in mind the compelling reasons why a lot of scientists spend a great amount of time and resources on the search for dark matter. This is why we will once more review the evidence for dark matter in the Universe and also shed some light on the rich history of dark matter research. While it started as a purely astronomical discipline, over time it evolved into a large interdisciplinary field of research bringing together astrophysicists, cosmologists, high energy physicists, and many more. 

The details of the evidence and its historic development are by no means complete. For further reading on this subject, we recommend a series of informative reviews~\cite{vandenBergh:1999sa,Bergstrom:2000pn,Dolgov:2002wy,Bertone:2004pz,Einasto:2009zd,Durrer:2015lza,Bertone:2016nfn}.

\clearpage
\section{History and Evidence of Dark Matter in the Universe}
\label{s:evidence}
When the term \textit{`dark matter'} first started to appear in the astronomical literature during the early 20th century, it had a very different meaning than it does today. `Dark matter' was used descriptively, simply to refer to ordinary matter which neither shines nor reflects light -- stars, which are too distant or too cool and faint to be observed, dim gas clouds, and other solid objects.  At this point, no one had any reason to entertain the idea of dark matter as some new, exotic form of matter, since little was known about the non-stellar mass in the Milky Way.

 In order to estimate the total mass of the galaxy, which could be compared to the observed amount, Henri Poincar\'e applied Lord Kelvin's idea to treat the galaxy as a thermodynamic gas of gravitating stars~\cite{Poincare1906}. Furthermore, two Dutch astronomers, Jacobus Kapteyn in 1922~\cite{Kapteyn:1922zz} and his student Jan Oort in 1932~\cite{Oort1932}, analysed stellar velocity in our galactic neighbourhood to estimate the local dark matter density. While these studies, among many others, showed no evidence for a large discrepancy between bright and dark matter\footnote{Oort did indeed find a discrepancy between the total and stellar density, but attributed this to neglecting faint stars close to the galactic plane.}, newer observations on intergalactic and galactic scales started to indicate otherwise.

\subsection{The `missing mass' problem of galaxies}
\label{ss: galactic evidence}
\paragraph{Galaxy clusters}
Following Lord Kelvin's and Poincar\'e's approach, the astronomer Fritz Zwicky applied the virial theorem to astronomical observations in 1933~\cite{Zwicky:1933gu}. The virial theorem relates the average total kinetic energy~$\langle T\rangle$ and the average potential energy~$\langle V\rangle$ of a stable system,
\begin{align}
	\langle T\rangle = -\frac{1}{2}\langle V\rangle\, . \label{virial theorem}
\end{align}
Zwicky used the virial theorem to the Coma galaxy cluster in order to estimate its mass. For this purpose, he measured the Doppler shifts of spectral lines to measure the galaxy's velocities in the line of sight. Furthermore, he estimated the total mass of the Coma cluster to be the sum of all stars times the solar mass,
\begin{align}
	M\approx \underbrace{800}_{\text{\footnotesize number of galaxies}} \times \underbrace{10^9}_{\text{\footnotesize stars per galaxy}}\times M_\odot \approx \SI{1.6e42}{\kg}\, .
\end{align}
Based on this estimate, the average velocity and velocity dispersion were estimated to be
\begin{align}
	\sqrt{\langle v^2\rangle} &=\left(\frac{3}{5}G_N \frac{M}{R}\right)^{1/2}\approx \SI{82}{\km\per\second}\, ,\\
	\sigma_v&=\sqrt{\frac{\langle v^2\rangle}{3}} \approx \SI{47}{\km\per\second}\, ,
\end{align}
where he estimated the cluster's radius~$R$ to be of order~$\text{10}^{\text{6}}$~ly. This estimate was however in direct conflict with Zwicky's observations. He measured the apparent velocities of eight galaxies and found a large velocity dispersion of~\SI{1100}{\km\per\second}. Zwicky concludes that, if we want to obtain a velocity dispersion of the same order from the virial theorem, we have to assume matter densities of at least~400 times larger than the stellar density. The Coma cluster could otherwise not be considered a bound system and would disperse over time. Only a small fraction of the mass was observable, most of it seemed missing. Four years later, Zwicky speculated that this `dark matter' should be made off cold stars, gases, and other solid bodies, which might also absorb background light and thereby reduce the observed luminosities further~\cite{Zwicky:1937zza}.

\begin{figure*}
\centering
\sbox\twosubbox{%
  \resizebox{\dimexpr.95\textwidth-1em}{!}{%
    \includegraphics[height=3.5cm]{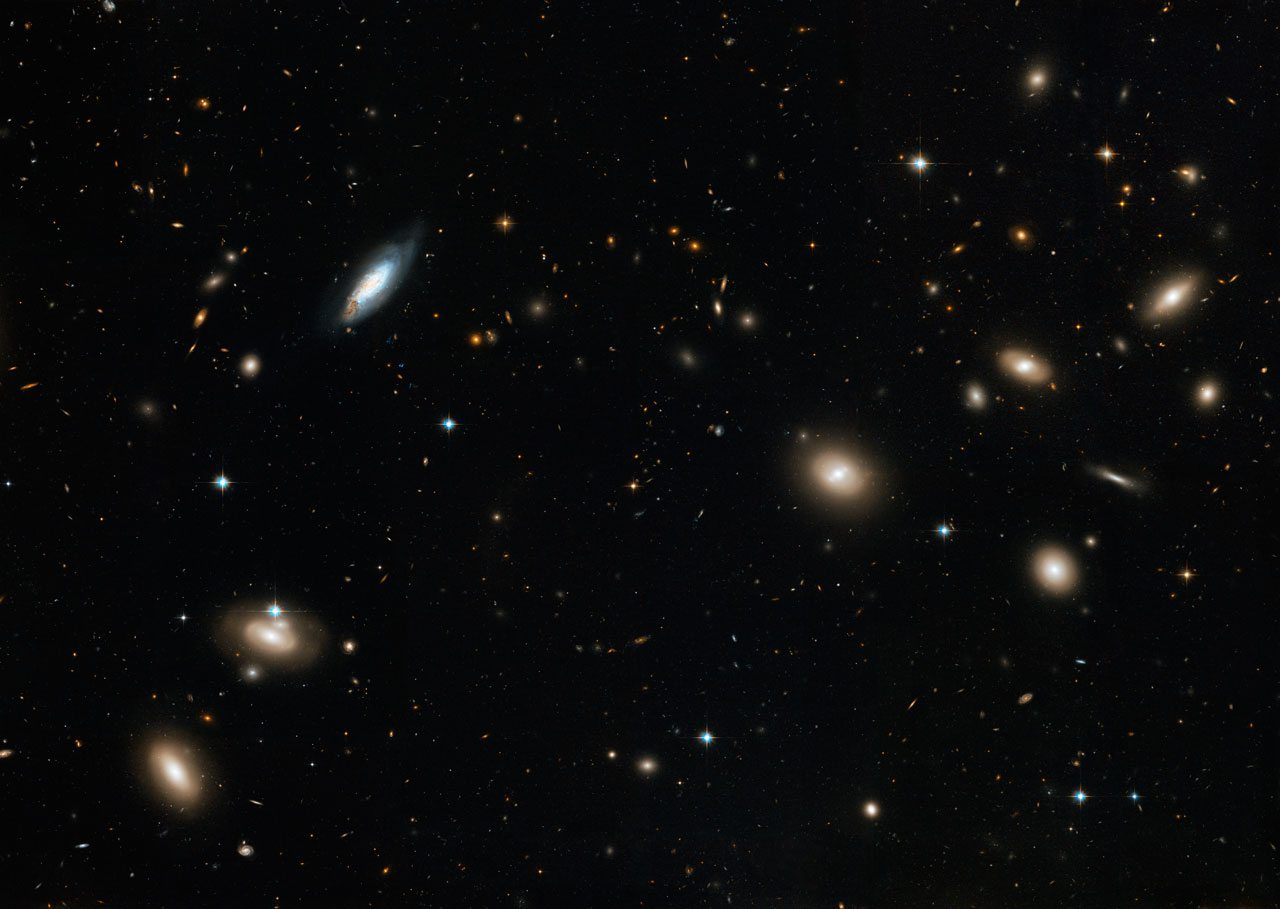}%
    \includegraphics[height=3.5cm]{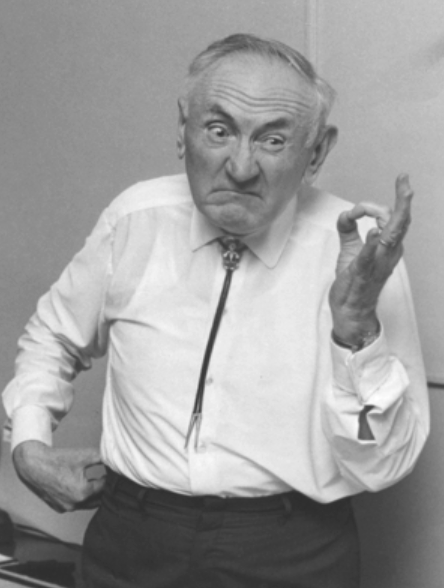}%
  }%
}
\setlength{\twosubht}{\ht\twosubbox}
   	\centering
   	\subcaptionbox{The Coma cluster~\cite{PhotoComa}.}{%
  	\includegraphics[height=\twosubht]{images/Coma_Cluster}%
	}
	\quad
	\subcaptionbox{Fritz Zwicky (1971)~\cite{PhotoZwicky}.}{%
	  \includegraphics[height=\twosubht]{images/Zwicky}%
	}

	\caption{Zwicky and the Coma cluster}
	\label{fig:zwicky}
\end{figure*}

The mass-to-light ratio refers to ratio between the total mass and its luminosity. Many observations of large mass-to-light ratios of galaxy clusters were published in the following years\footnote{The first measurements of large galactic mass-to-light ratios actually preceded even Zwicky's observations by three years and were made by the Swedish astronomer Knut Lundmark in 1930~\cite{Lundmark1930}.}. But many astronomers did not accept the hypothesis of large amounts of dark matter in galaxy clusters, questioning whether galaxy clusters are truly bound systems. This interpretation was disfavoured by the age of the universe, as unbound clusters should have disintegrated by now. Clusters were indeed gravitationally bound systems. Others tried to find the missing mass not in the galaxies, but in the intergalactic space, as proposed already in 1936 by the astronomer Sinclair Smith\footnote{Just as Zwicky for the Coma cluster, Smith found a high mass-to-light ratio for the Virgo cluster early on. Instead of the virial theorem, he used the circular orbits of the outermost galaxies to estimate the cluster's total mass.}~\cite{Smith:1936mlg}. They looked e.g. for hydrogen gas or ions outside galaxies. None of these observations showed a sufficient amount of ordinary matter to explain the large mass-to-light ratios, and the missing mass problem of galaxy clusters remained.

\paragraph{Galactic rotation curves}
Historically, the most important evidence in favour of large amounts of dark matter in the Universe came from galactic rotation curves, i.e. the speed~$v$ of visible matter orbiting the galactic center as a function of the galactocentric distance~$r$~\cite{Sofue:2000jx,Rubin2000}. Assuming circular orbits, Newtonian dynamics predicts the rotation curve,
\begin{align}
	v(r) = \sqrt{\frac{G_N M(r)}{r}}\, ,\label{eq: radial velocity}
\end{align}
where~$G_N$ is Newton's constant and~$M(r)$ denotes the mass found within a sphere of radius~$r$, which follows from the mass density distribution~$\rho(\mathbf{r})$ of the galaxy,
\begin{align}
	M(r)&= \iiint\dd^3\mathbf{r}^\prime\rho(\mathbf{r}^\prime)\Theta(r-r^\prime)\, .
\end{align}
For stars well outside the bulk mass,~$M(r)$ is approximately constant, and the predicted rotation curve should follow
\begin{align}
	v(r)\sim \frac{1}{\sqrt{r}}\, .
\end{align}
This Keplerian speed drop is most notably observed in the planetary orbits of the solar system.

Since it is difficult to infer the rotation curve of our own galaxy, the Milky Way, the first observations of galactic rotation curves were obtained for the Andromeda galaxy~(M31). The first astronomer to make spectrographic observations of its rotation curve and extract information about the galaxy's mass distribution, was the American Horace Babcock in~1939~\cite{Babcock1939}. He failed to observe the expected Keplerian behavior, instead Babcock found that the orbital velocities approach a constant value for the outer spiral arms and concluded that there must be much more mass at large radii than observed. Ultimately, he tried to explain the large mass-to-light ratio with light absorption with additional material in the outer regions of Andromeda or some new modification of the galactic dynamics. Even though Babcock's measurements turned out inconsistent with newer measurements, his description of the qualitative behavior of the rotation curve was correct. One year later, Oort reported a similar discrepancy between the light and mass distribution in the galaxy NGC 3115, where he found a mass-to-light ratio of~250 at large galactocentric distances~\cite{Oort1940}. Just as Babcock, Oort speculated on light absorption and diffusion by interstellar gas and dust, as well as the existence of faint dwarf stars, as the source of this puzzle, but also mentioned the idea of the galaxy being embedded in a larger dense mass.

\begin{figure*}
	\centering
	\includegraphics[width=0.8\textwidth]{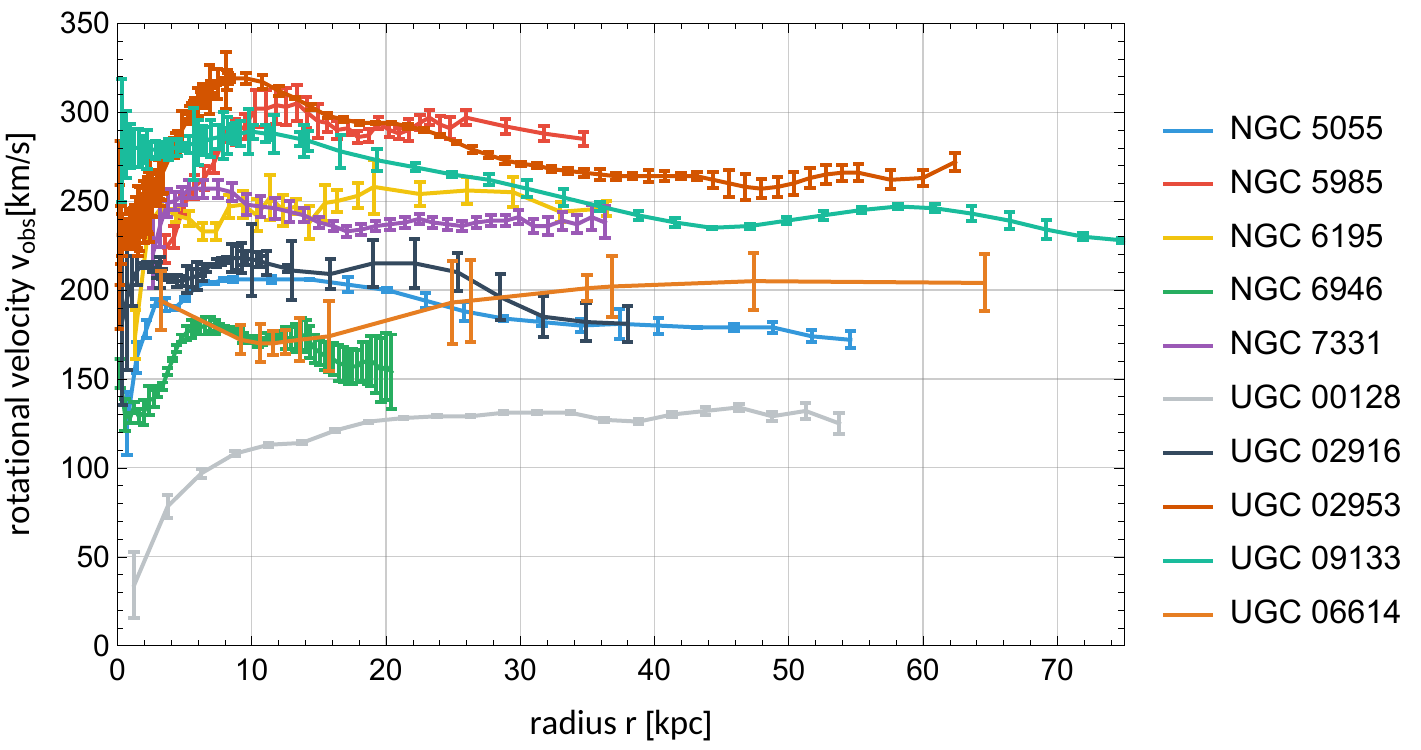}
	\caption{A random selection of ten galactic rotation curves of the SPARC sample.}
	\label{fig: rotation curves}
\end{figure*}

In the last sentences of his paper, Oort states the need for rotation speed observations at larger radii. These was made possible by the advent of radio astronomy. The 21cm spectral line of neutral hydrogen, predicted by Oort's student Hendrik van de Hulst in 1944~\cite{Hulst1944} and discovered in 1951 by Ewen and Purcell~\cite{Ewen1951}, allowed the measurements of the rotation curve to much higher radii. Van de Hulst himself, among many others, was involved in measuring both the Milky Way's~\cite{Hulst1954} and Andromeda's~\cite{Hulst1957} rotation curve by means of~21cm line observations. When Vera Rubin and Kent Ford revisited Andromeda in~1970 and measured the optical rotation curve with high accuracy~\cite{Rubin:1970zza}, they found a constant rotational speed at large radii far exceeding the galactic disk in agreement with radio observations. They concluded that the mass of the galaxy increases approximately linear with radius in the outer regions. We can see from eq.~\eqref{eq: radial velocity} that~$M(r)\sim r$ would indeed explain the flat rotation curve. Many more galaxies were analysed more systematically in the late '70s and '80s, on the basis of optical light and in particular using the 21cm spectral line. The puzzling conclusion was that not a single observed rotation curve showed the Keplerian speed drop, but instead had a flat rotation curve~\cite{Roberts1975}

Confronted with these new observations, the general interpretation started to shift~\cite{Faber1979}. Astronomers started to appreciate that the `missing mass' problem was indeed a real issue, that galaxies are bigger, and the outer galactic regions much more massive than they appear~\cite{Freeman:1970mx,Roberts1973,Ostriker:1974lna,Rubin:1980zd}. Others started to see a connection to Zwicky's `missing mass' problem on cluster scales~\cite{Einasto1974}. Up to today, thousands of galactic rotation curves have been measured, a small selection taken from the SPARC sample is shown in figure~\ref{fig: rotation curves}~\cite{Lelli:2016zqa}. Their flatness is one of the most convincing arguments that galaxies are embedded in a large DM~halo\footnote{We briefly mention a prominent alternative to~DM, which can reproduce the galactic rotation curves by modifying Newton's laws of motion, \ac{MOND}~\cite{Milgrom:1983ca}, and its relativistic realization~\cite{Bekenstein:2004ne}. A review can be found in~\cite{Famaey:2011kh}.}.

\begin{figure*}
\centering
\sbox\twosubbox{%
  \resizebox{\dimexpr.95\textwidth-1em}{!}{%
    \includegraphics[height=3.5cm]{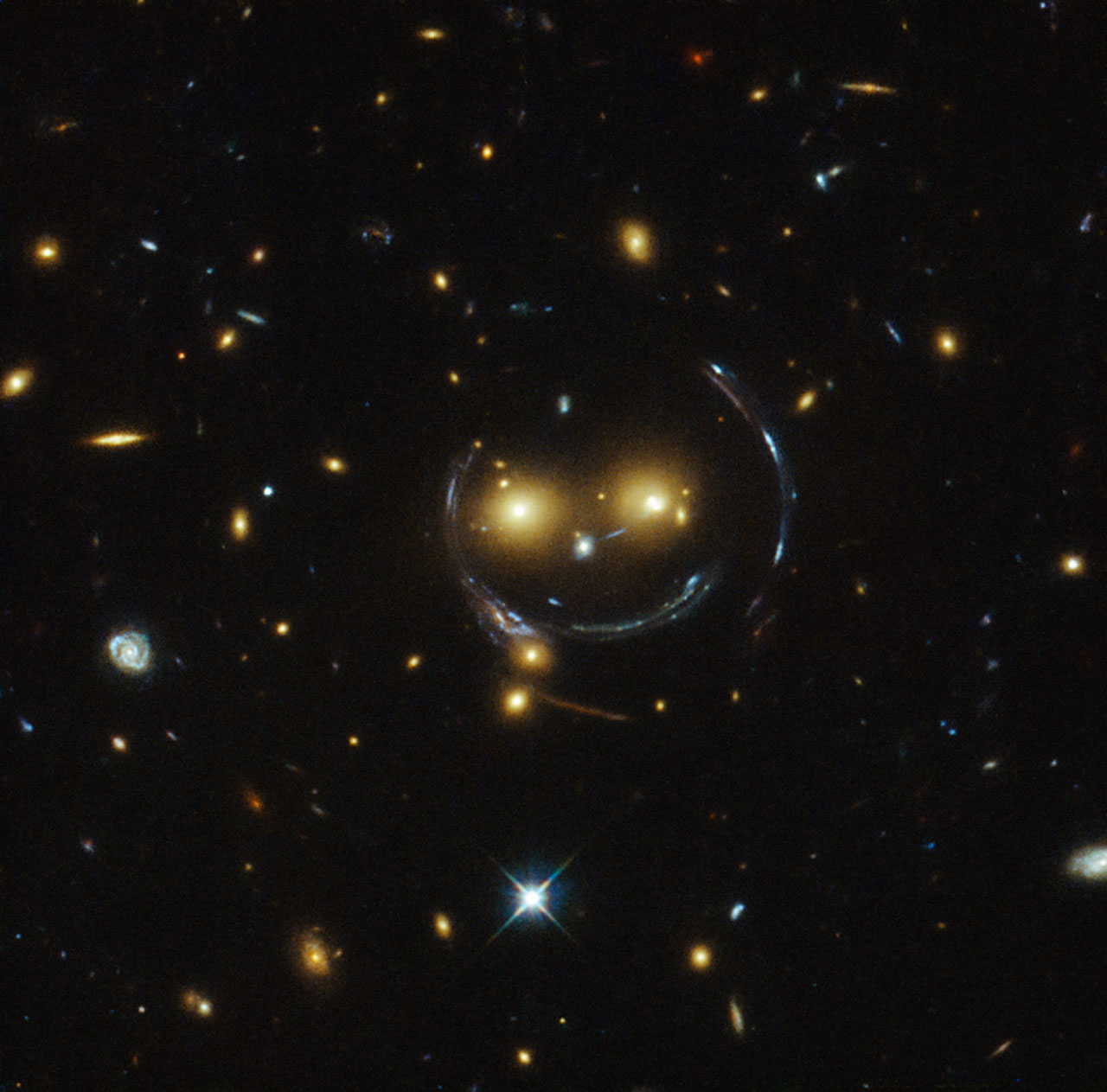}%
    \includegraphics[height=3.5cm]{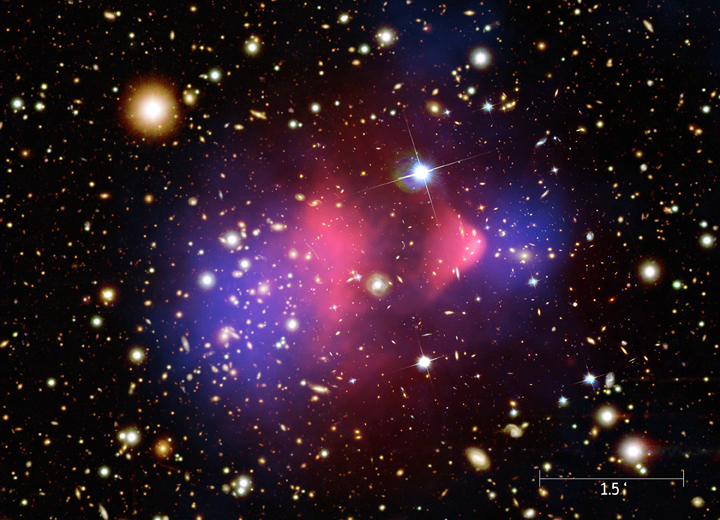}%
  }%
}
\setlength{\twosubht}{\ht\twosubbox}
   	\centering
   	\subcaptionbox{The `smiling cluster'~\cite{PhotoSmile}.\label{fig:lensing}}{%
  	\includegraphics[height=\twosubht]{images/Lensing}%
	}
	\quad
	\subcaptionbox{The bullet cluster~\cite{PhotoBulletCluster}.\label{fig:bulletcluster}}{%
	  \includegraphics[height=\twosubht]{images/Bullet_Cluster}%
	}
	\caption{Strong gravitational lensing of the galaxy cluster SDSS J1038+4849~(left) and an overlay of the optical, X-ray, and weak gravitational lensing observation of the bullet cluster~(1E 0657-558)~(right).}
	\label{fig:zwicky}
\end{figure*}

\paragraph{Gravitational lensing}
One of the predictions of Einstein's general theory of relativity was the effect that light gets deflected by large masses, called gravitational lensing, which was first observed during a solar eclipse in~1919~\cite{Dyson:1920cwa}. Zwicky proposed already in 1937 that galaxies and galaxy clusters would act as huge gravitational lenses with observable consequences~\cite{Zwicky:1937zzb}. But it took 42 years before strong gravitational lensing was first observed~\cite{Walsh1939}. A beautiful example of strong gravitational lensing due to a galaxy cluster is shown in figure~\ref{fig:lensing}. The mass of a heavy object is the crucial parameter determining the lensing effect and can be inferred this way. This was achieved for a galaxy cluster by e.g. Fischer and Tyson in 1997, who observed a mass-to-light ratio of around~200~\cite{Fischer:1997si}. During the '90s, it became more and more clear that the total masses of galaxy clusters obtained from gravitational lensing was consistent with independent measurements based on e.g. velocity dispersions~\cite{Taylor:1998uk,Wu:1998ju}. This consistency solidified the need for large amounts of undetected matter in clusters. Based on the idea by Kaiser and Squires in 1993~\cite{Kaiser:1992ps}, weak lensing observations allowed to directly map the spatial distribution of~DM in clusters in the following years, without any assumptions about its nature~\cite{Wittman:2000tc,Refregier:2003ct}\footnote{For more details on lensing evidence for dark matter, we recommend the review by Massey et al.~\cite{Massey:2010hh}.}.

\paragraph{The bullet cluster}
Another famous, more recent piece of evidence for~DM on cluster scales is the observation of the `bullet cluster'~\cite{Clowe:2003tk,Markevitch:2003at,Clowe:2006eq}, shown in figure~\ref{fig:bulletcluster}. The bullet cluster consists of two sub-clusters, which are drifting apart after having passed through each other. During this collision, the X-ray emitting gas, visible in red in the figure, was separated by the galaxies due to their electromagnetic interactions. The galaxies act like collision-less particles and simply passed by unaffected. Without the presence of~DM in this cluster, the predicted mass distribution of this system should follow the X-ray observations, as the gas makes up the majority of baryonic mass~\footnote{In astrophysics, `baryonic matter' or `baryonic mass' is often used rather loosely to refer to ordinary matter including the non-baryonic electrons, as the protons and neutrons contribute most to the mass.}. However, the simultaneous measurement of weak gravitational lensing allowed to directly map the gravitational potential of the bullet cluster (visible in blue in the figure), which traces not the gas, but the galaxies. It strongly suggests that the majority of mass is in the form of an undetected and collision-less matter. The spatial separation of gravitational and visible mass, which was now observed in multiple instances~\cite{Harvey:2015hha}, not only provides additional evidence for the existence of~DM, but also challenges alternative proposals of modified gravity such as~\ac{MOND}.\\[0.3cm]

Despite the strong evidence on the scales of galaxies and galaxy clusters, the most compelling evidence emerged on even larger scales.

\subsection{Cosmological evidence}
\paragraph{Cosmic Microwave Background}
During the early universe, all matter and radiation made up an almost homogeneous plasma. Baryons and electrons were undergoing Thomson scattering and were in thermal equilibrium. As such, the universe was opaque to photons. This changed abruptly during recombination, when the universe cooled down and electrons and baryons formed neutral atoms. The universe became transparent rapidly, and photons could propagate freely after their last scattering on a proton or electron. These photons form a radiation background present throughout the cosmos, which is present to this day. This cosmic background radiation was first predicted for the hot big bang model in 1948 by Alpher, Hermann and Gamow~\cite{Gamow1948,Alpher1948,Alpher1948-2}. In 1965, Penzias and Wilson accidentally detected an isotropic source of microwave radiation with a temperature of around~3.5~K~\cite{Penzias:1965wn}\footnote{Today's best measurement of the~\ac{CMB} temperature is~$T_{\text{CMB}}=$(2.72548~$\pm$~0.00057)K~\cite{Fixsen:2009ug}.}. In the same year, Robert Dicke and his collaborators, who were scooped by the discovery, identified this radiation as the cosmic microwave background radiation, dating back to the time of recombination, and predicted~17 years prior~\cite{Dicke:1965zz}. 

Since the photons were in thermal equilibrium prior to their last scattering, the background radiation should follow a Planckian black body spectrum~\cite{Gamow1948-2}. The discovery of the~\ac{CMB} and the confirmation of its thermal spectrum in the '70s established the radiation's cosmic source and therefore the hot big bang model of the Universe~\cite{Peebles:1991ch}. 

Small fluctuations of the~\ac{CMB}'s temperature were expected, because they should trace gravitational fluctuations necessary to grow and evolve into the cosmological structure of galaxy and clusters. The primary anisotropies originated in the baryon-photon plasma around the time of recombination. They result from the opposing processes of gravitational clustering, which forms regions of higher density, and radiation pressure of the photons, which erases baryon over-densities. The resulting acoustic oscillations leave a characteristic mark in the~\ac{CMB}.

The \ac{CMB} temperature fluctuations' directional dependence in the sky is usually expressed in terms of spherical harmonics,
\begin{align}
\delta T (\theta,\phi)&=\sum_{l=2}^\infty\sum_{m=-l}^l a_{lm}Y_{lm}(\theta,\phi)\, .
\intertext{The \ac{CMB}'s so-called power spectrum is nothing but the variance of the coefficients,}
 C_l&=\frac{1}{2l+1}\sum_{m=-l}^l|a_{lm}|^2\, ,
\end{align}
which are directly related to the two-point correlation function of the temperature fluctuations. The acoustic oscillations' signature is a number of peaks in the power spectrum, which encode the Universe's geometry and content. In particular, the acoustic peaks depend critically on the density of baryonic and dark matter, since only the baryonic matter experiences the photon's radiation pressure, and dark matter starts to cluster even before recombination.

For a long time, only the dipole~($l=2$) was observed, which is due to Earth's motion relative to the cosmic rest frame. In 1992, the satellite-borne~\ac{COBE} first reported the observation of tiny \ac{CMB} anisotropies of order~$\delta T/T\sim 10^{-5}$~\cite{Smoot:1992td}. Following~\ac{COBE}, a number of ground and balloon based experiments were performed to extend the measurements to smaller angular scales, i.e. higher multipoles~$l$~\cite{deBernardis:2000sbo,Stompor:2001xf,Kuo:2002ua,Grainge:2002da}.  Another great milestone was the~\ac{WMAP}, a spacecraft located at the Lagrange point~L2, which measured the first three peaks in~2003~\cite{Spergel:2003cb,Hinshaw:2012aka}.

\begin{figure*}
	\centering
	\includegraphics[width=0.8\textwidth]{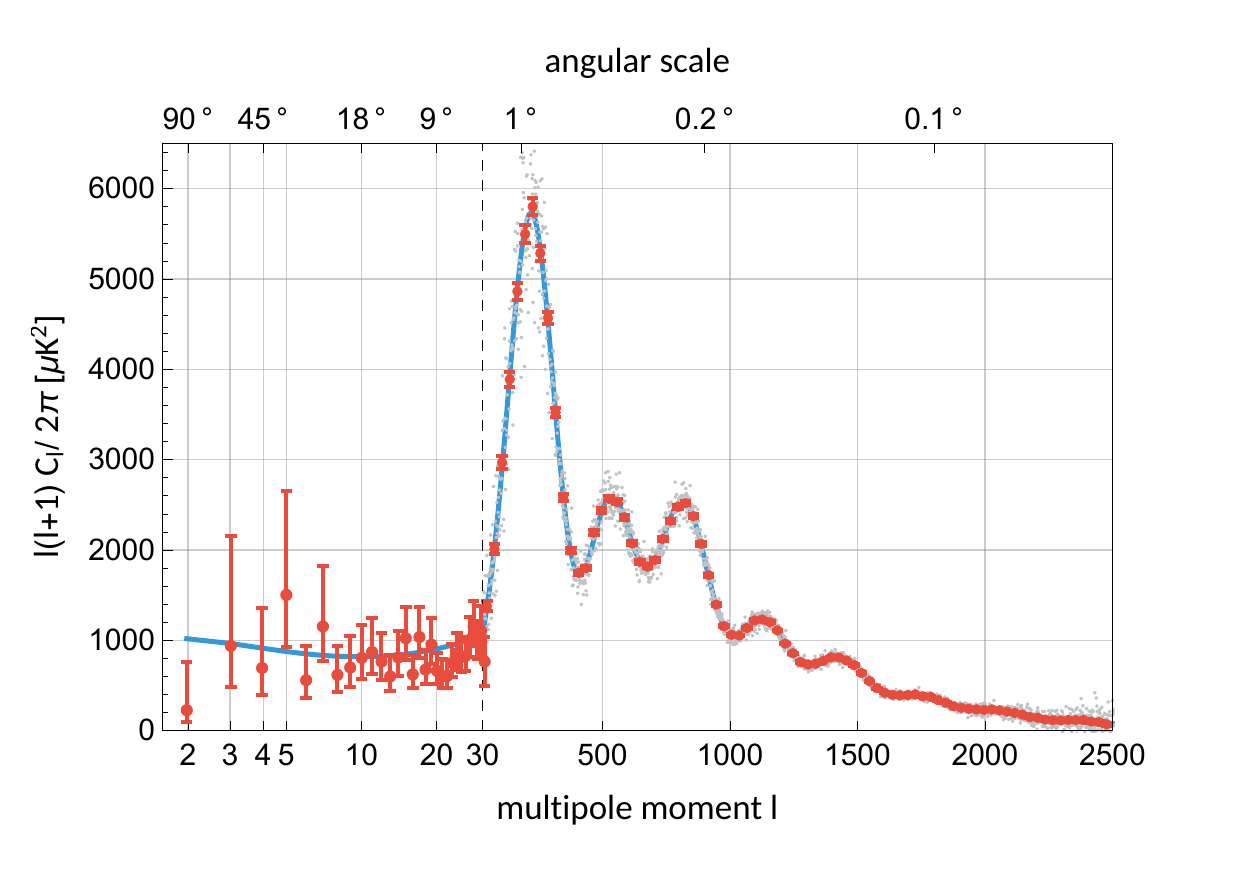}
	\caption{The power spectrum of the CMB, both the data and the best fit of the cosmological~$\Lambda$CDM model.}
	\label{fig: cmb power spectrum}
\end{figure*}

The most recent measurement of the \ac{CMB} power spectrum was obtained by \ac{WMAP}'s successor Planck in 2013~\cite{Ade:2013zuv} and is shown in figure~\ref{fig: cmb power spectrum}~\cite{Aghanim:2018eyx}\footnote{The data used for this plot was taken from the Planck Legacy Archive~\cite{PLA}. For~$l\leq 30$ the scale is logarithmic, otherwise the scale is linear. For~$l>30$ the data is binned with a bin width of 30. The unbinned data is visible in light gray.}. It also shows the best fit of the~$\Lambda$CDM model, the standard model of cosmology, which describes the Universe as a homogenous, isotropic, and flat spacetime, whose total energy density consists of ordinary matter, dark matter, and dark energy in the form of a cosmological constant~$\Lambda$. In terms of the parameters~$\Omega_i\equiv \rho_i/\rho_c$, hence the relative contribution of the different constituents to the critical density, the most recent fit of the Planck power spectrum yielded~\cite{Aghanim:2018eyx}
\begin{align}
	\Omega_M = 0.315\pm 0.007\, ,\quad \Omega_b =0.0493 \pm  0.0002\, , \label{eq: omega dm}
\end{align}
where~$\Omega_M$ is the relative amount of matter and~$\Omega_b$ accounts for the baryonic mass only. In order to explain the power spectrum, baryonic matter can make up~$\sim$15\% of the total matter in the Universe only, and the majority of matter must be dark. But as opposed to the interpretations during the first half of the 20th century,~`dark matter' does not just refer to unobserved matter, but is inherently different from baryonic matter, as it does not interact with photons at all.

The fact that baryons contribute to the cosmological density by only such a small amount was confirmed by the independent predictions of~\ac{BBN}, the production of light nuclei during the early Universe, first described by Alpher and Gamow in the infamous Alpher-Bethe-Gamow paper~\cite{Alpher:1948ve} and later refined e.g. in~\cite{Yang:1983gn,Walker:1991ap}. In order to account for the observed abundance of light nuclei, the baryonic density is determined to lie between 0.046 and 0.055~\cite{Fields:2014uja}, in perfect agreement with~\ac{WMAP}'s or Planck's finding. In combination with observations of high red shift type Ia supernovae, which also suggest a flat universe with~$\sum_i \Omega_i=1$~\cite{Schmidt:1998ys,Cai:2015pia}, we have a completely independent confirmation of the~\ac{CMB} results.\\[0.3cm]

This draws a consistent cosmological picture, which cannot work without large amounts of non-baryonic~DM, in perfect agreement with evidence from smaller scales. Yet, this is not the last cosmological argument in favour of~DM. It turned out that~DM also played a critical role in the evolution of the Universe's structure.

\paragraph{Structure formation}
During the evolution of the cosmos, the initial over-densities, whose seeds we can observe in the~\ac{CMB} grew under the influence of gravity into large-scale structures of galaxies and clusters. The main approach to study cosmological structure formation are numerical N-body simulations of many gravitating particles, which describe their clustering under various assumptions and compare this to observations of large structures. The pioneer of these simulations was arguably Erik Holmberg, a Swedish astronomer who studied the gravitational interaction of galaxies in 1941~\cite{Holmberg1941}. He set up an array of 37 light bulbs and used photocells to measure the ``gravitational force'', employing the~$\sim 1/r^2$ behaviour of the light intensity to mimic the gravitational force. The first numerical simulations for cosmological scales were performed in the '70s modelling galaxies as a gas of self-gravitating particles~\cite{Press:1973iz}.

Almost from the start, it was clear that, without~DM, galaxies would have formed much too late, as baryonic matter starts to cluster later due to its dissipative non-gravitational interactions. In addition, it turned out that the formation of smaller structures depends critically on the velocity of the~DM and that fast thermal motion of the~DM would wash out and suppress the formation of small structures~\cite{Schramm:1980xv,Peebles:1982ib}. Therefore, gravitational clustering of `hot' DM would initially create large structures. In contrast, non-relativistic or `cold'~DM can collapse into low mass halos early on. In this case, cosmological structure also builds up hierarchical, but bottom up, from stars to stellar clusters, from galaxies to clusters and super clusters. The observation of small sub-structures in the first 3D surveys of galaxies~\cite{Davis1982} confirmed this hierarchical structure evolution and lead to the cold DM~paradigm~\cite{Blumenthal:1984bp,Davis:1985rj}.

\begin{figure*}
\centering
\sbox\twosubbox{%
  \resizebox{\dimexpr.99\textwidth-1em}{!}{%
    \includegraphics[height=4cm]{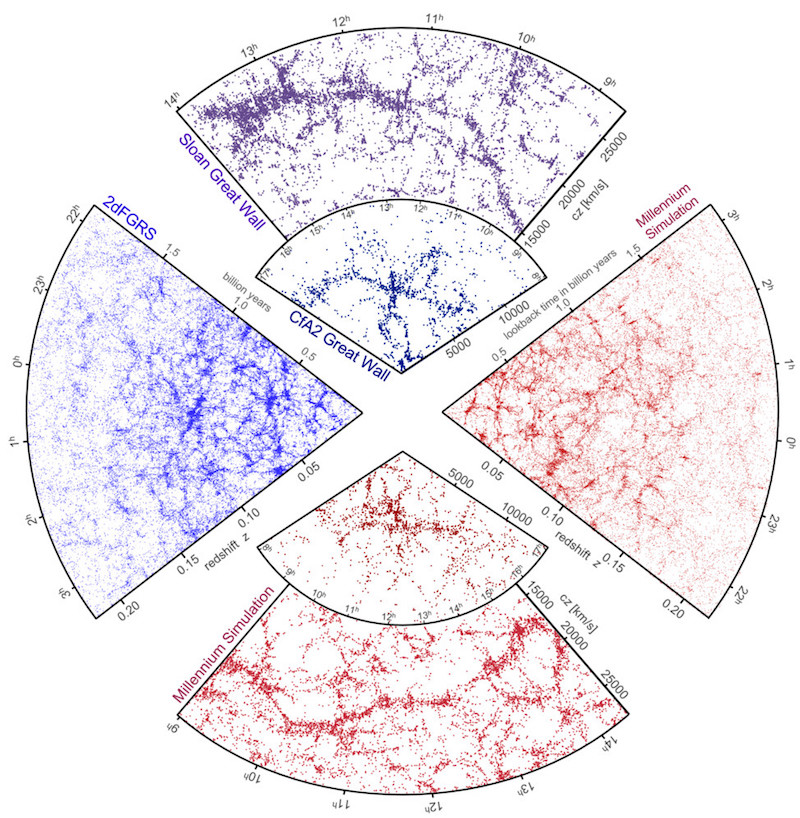}%
    \includegraphics[height=4cm]{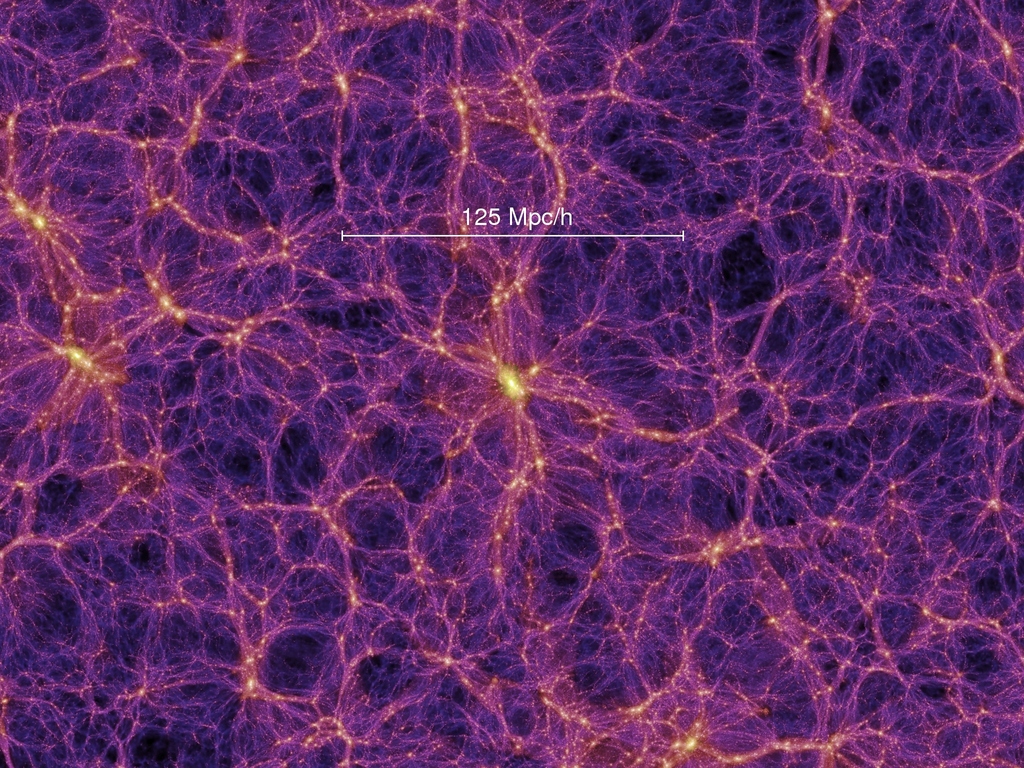}%
  }%
}
\setlength{\twosubht}{\ht\twosubbox}
   	\centering
   	\subcaptionbox{Observations vs. simulations of galaxy distributions~\cite{Springel:2006vs}.\label{fig:millenium1}}{%
  	\includegraphics[height=\twosubht]{images/Millenium_1}%
	}
	\quad
	\subcaptionbox{An example of the simulated  DM~distribution forming the cosmic web~\cite{PhotoMillenium}.\label{fig:millenium2}}{%
	  \includegraphics[height=\twosubht]{images/Millenium_2}%
	}
	\caption{N-body simulation of structure formation.}
	\label{fig: millennium}
\end{figure*}

Newer simulations with better resolutions beautifully show how the~DM and galaxies cluster and form the cosmic web, huge filaments surrounding enormous voids, most famously the Millennium simulations from 2005~\cite{Springel:2005nw,Springel:2006vs}. The comparison to galaxy surveys~\cite{York:2000gk,Cole:2005sx}, as shown in figure~\ref{fig:millenium1}, show excellent agreement with the observation on large scales.

While the `DM-only' type of simulations succeeded in explaining the largest scales, they fell short on galactic scales in some respects. The major small-scale problems were coined the `too-big-to-fail'~\cite{BoylanKolchin:2011de}, `cusp vs. core'~\cite{Moore:1994yx,Navarro:1996bv}, `diversity'~\cite{Oman:2015xda}, and `missing satellite' problem~\cite{Klypin:1999uc}. However, it is debated how severe these problems really are, as the inclusion of baryonic feedback e.g. from supernovae into the simulations and the observation of faint Milky Way satellites reduces the tension between observations and predictions~\cite{Pontzen:2011ty,Governato:2012fa,Simon:2007dq}. Especially the missing satellite problem seems to have disappeared over the last years.\\[0.3cm]

During this chapter, dark matter was treated as an almost purely astrophysical field. Yet the meaning of the term~`dark matter' shifted significantly in the second half of the 20th century. The term evolved slowly from referring to ordinary, but dim and unobserved matter, to something else entirely: an unknown form of matter, which seems to interact exclusively via gravity. Starting arguably with Gershtein and Zeldovich in 1966~\cite{Gershtein:1966gg}, the quest for dark matter would merge cosmology, astrophysics, and particle physics.

\section{Particle Dark Matter}
In the previous chapter, we reviewed the compelling set of astrophysical evidence for the existence of~DM in the Universe. Yet, very little is known about its nature. During the '70s, particle physicists naturally started to speculate about the identity of~DM, as discovering a new particle was not uncommon at this point. 

Before exotic new particles would be considered as the source of~DM, the possibility that the galactic halos consist of ordinary matter needed to be explored. This can be regarded as the continuation of earlier interpretations of the `missing mass' problem.  For a while, it seemed possible that~DM was comprised of~\acp{MACHO}\footnote{The term~MACHO was coined by Kim Griest, as contrast to the~WIMP, which we will discuss later.}, small, massive, and dark objects of baryonic matter, such as dim stellar remnants, black holes and neutron stars, rogue planets, rocks, and brown dwarfs, which drift unbound through the interstellar space and evade direct observation.

An early study by Hegyi and Olive in~1985 argued against baryonic halos~\cite{Hegyi:1985yu}, where they in particular pointed out the incompatibility of the hypothesis of large amount of baryonic MACHOs and the central result of~\ac{BBN}, which strongly suggested that baryonic matter constitutes only a small fraction of the cosmological energy density. This was also verified by~\ac{CMB} observations as discussed in the previous chapter. During the same year, a new technique of looking for~\ac{MACHO}s, regardless of their composition, was presented by the Polish astronomer Bohdan Paczy\'nski. He suggested to look for temporary amplification of the brightness of a large number of background stars, caused by gravitational lensing of a massive, transient object~\cite{Paczynski:1985jf}. Paczy\'nski called this process~\textit{microlensing}. This method does not require of the massive object to emit light itself, hence it is ideal to search for all kinds of galactic~\ac{MACHO}s, not just baryonic ones. Large microlensing surveys searched for dark stellar bodies, most notably the MACHO project~\cite{Alcock:2000ph} and EROS-2\cite{Lasserre:2000xw,Tisserand:2006zx}. While early results seemed to show an excess of microlensing events, eventually these surveys ruled out MACHOs as the primary source of~DM in the galaxy.

While the 1985 paper by Hegyi and Olive left the possibility of black hole~MACHOs open, it turned out that stellar remnant black holes did not have enough time to form and populate the halos. Yet, already 11 years earlier, the British theoreticians Bernard J. Carr and Stephen W. Hawking noted that density fluctuations in the early Universe, necessary for the formation of structure, could collapse gravitationally and form so-called~\acp{PBH}~\cite{Carr:1974nx}. These non-stellar black hole relics could have survived to this day while accreting more mass and possibly form the galactic halo, without the need to invoke new forms of matter. However, there are severe constraints on~\acp{PBH} as~DM. These constraints have been re-evaluated more recently, after the discovery of gravitational waves from black hole mergers by the~\ac{LIGO} in 2016~\cite{Abbott:2016blz}. While some conclude~\acp{PBH} to be excluded by microlensing and dynamical constraints, at least as the only contributor to~DM~\cite{Brandt:2016aco,Green:2016xgy}, others find that they remain a viable option~\cite{Carr:2016drx,Kuhnel:2017pwq}.\\[0.3cm]

In conclusion, the attempts to explain the observations with ordinary matter alone failed, and the non-baryonic nature of the galactic halo seemed unavoidable~\cite{Freese:1999ge}.

\subsection{When astronomy and particle physics merged}
With ordinary baryonic matter being excluded as the solution to the `missing mass' problem, the next question we should ask is, what conditions a particle would have to satisfy to act as~DM. It should be noted that there is no reason to assume that all observations of~DM can be attributed to a~\emph{single} particles. It could also be a dark sector with multiple particles and interactions. 

A DM~particle should
\begin{enumerate}
	\item obviously be non-luminous, i.e. not reflect, absorb or emit light,
	\item be non-relativistic and therefore able to drive structure formation,
	\item be stable, at least on the time scale of the age of the Universe,
	\item act collision-less and non-dissipative. This implies that the DM~particle should have no or only weak interactions with ordinary matter apart from gravity, and finally
	\item get produced in the right amount during the early Universe via some mechanism.	
\end{enumerate}
Naturally, the first approach is to check, if any of the known particles can satisfy these criteria. Indeed, the neutrinos, originally predicted by Wolfgang Pauli in 1930 to explain the continuous energy spectrum of~$\beta$-decay~\cite{Pauli:1930pc} and observed 26 years later by the Cowen-Reines neutrino experiment~\cite{Cowan:1992xc}, seemed to fit the bill. The Russian physicists Semyon~S. Gershtein and Yakov B. Zeldovich, two pioneers of what would become the field of astroparticle physics, discussed the cosmological implications of massive neutrinos in 1966~\cite{Gershtein:1966gg}. In analogy to the recently discovered~\ac{CMB}, they computed the thermal relic abundance of electron and muon neutrinos and derived upper limits on neutrino masses based on the expansion history. Even though they did not connect their findings to~DM, their work can be considered as essential groundwork for the idea of particle~DM and in particular \acs{WIMP}~DM. Soon, others drew the connection and started to consider neutrinos as potential~DM~\cite{Szalay:1976ef,Lee:1977ua,Gunn:1978gr}.

While the idea that~DM was nothing but neutrinos must have been very appealing, there are two problems. For one, the relic density of relic neutrinos was determined by known physics and turned out too low. The relative contribution to the cosmic energy density is given by~\cite{Lesgourgues:2006nd}
\begin{align}
	\Omega_\nu h^2 &= \frac{\sum_i m_i}{93.14\mathrm{eV}}\, ,
\intertext{where~$h$ is the Hubble parameter in units of~\SI{100}{\km\per\second\per\mega\parsec}. With an upper limit of around~$\sum_i m_i<\SI{0.11}{\eV}$ for the sum of neutrino masses~\cite{Tanabashi:2018oca}, we find that}
\Omega_\nu&<2.6\cdot 10^{-3}\, .
\end{align}
Comparing to eq.~\eqref{eq: omega dm}, neutrinos can only account for a small fraction of non-baryonic matter in the Universe. The second reason against neutrino~DM is that they would be relativistic during structure formation and behave as~\emph{hot} dark matter. As such, it cannot reproduce the observed hierarchical formation of cosmic structure~\cite{Schramm:1980xv,Peebles:1982ib,White:1984yj}. 

In the end, the neutrino did not satisfy two of our five conditions. But it can be regarded as the historic blueprint to many DM~particles proposed in the following years.

\subsection{DM models and particle candidates}
\label{ss: DM candidates}
When it became clear that the neutrino cannot account for the missing mass in the Universe, the full predictive power of particle theorists was unleashed to come up with well-motivated and ideally testable models of~DM. Often, new models proposed for independent reasons turned out to contain new particles, which could act like dark matter. In this chapter, we list a exemplary selection of the most common candidates for DM~particles in different contexts. Naturally, this list is not exhaustive.

\paragraph{Sterile Neutrinos}
The neutrinos of the~\ac{SM} are the only fermions, which appear exclusively with left handed chirality, yet nothing forbids adding neutrinos with right handed chirality. These leptons would be gauge singlets and not interact with the other fields, which is why they are also called~`sterile neutrinos'. Introducing heavy sterile neutrinos could also explain the small, non-vanishing neutrino masses via the seesaw mechanism~\cite{Minkowski:1977sc}. If these heavy fermions are of keV scale mass, they would get produced in the early Universe via a non-thermal mechanism and act as~DM~\cite{Dodelson:1993je,Shi:1998km}. For a recent review on the observational status and production mechanisms of sterile neutrino~DM, we refer to~\cite{Boyarsky:2018tvu}.

\paragraph{Supersymmetry}
In the beginning of the '70s, a fundamentally new kind of symmetry of quantum field theories was discovered, which is called~\ac{SUSY}\footnote{For a comprehensive review of \ac{SUSY} and the~MSSM we recommend the `Supersymmetric Primer' by Stephen P. Martin~\cite{Martin:1997ns}.}~\cite{Gervais:1971ji,Golfand:1971iw,Volkov:1972jx,Volkov:1973ix,Akulov:1974xz,Wess:1974tw}. \ac{SUSY} relates the fermionic and bosonic degrees of freedom and therefore predicts that the fermions (bosons) of a particle model are accompanied by a bosonic (fermionic) partner. Supersymmetric field theories remain theoretically and phenomenologically well-motivated and gain a great deal of attention to this day. In addition, supersymmetric models have arguably been the most generous providers of DM~particle candidates. The introduction of~SUSY roughly doubles the number of particles, and the lightest of the supersymmetric particles is typically regarded a candidate for~DM, provided that it is stabilized by R-parity~\cite{Jungman:1995df}. 

Due to its deep connection to the Poincar\'e group, supersymmetry as a local symmetry enforces that gravity has to be included. This is why local supersymmetry is usually called supergravity. In this context, the first supersymmetric DM~candidate was the gravitino, the spin-3/2 partner of the graviton~\cite{Hut:1977zn,Pagels:1981ke}. With the formulation of the~\ac{MSSM} in the '70s and '80s~\cite{Fayet:1974pd,Fayet:1976et,Fayet:1977yc,Dimopoulos:1981zb,Haber:1984rc}, the neutralino became one of the most studied potential DM~particles~\cite{Weinberg:1982tp,Goldberg:1983nd,Ellis:1983ew}. The neutralino is a mixture of the fermionic partners of the neutral bosons of the~\ac{SM}. It became the archetype of a~\ac{WIMP}, a much larger, more general class of DM~particle candidates with weak, but observable interactions with the~SM. 

Other, less popular supersymmetric candidates include the scalar partner of the neutrino, the sneutrino~\cite{Falk:1994es} and the axino~\cite{Rajagopal:1990yx}.

\paragraph{Extra dimensions}
One example for a non-supersymmetric WIMP arises from the consideration of compactified extra dimensions. The idea goes back to the German physicists Theodor Kaluza and Oskar Klein. In 1921, Kaluza tried to unify electromagnetism with Einstein's theory of gravity by postulating a fifth spatial dimension~\cite{Kaluza:1921tu}, whereas Klein provided an interpretation of this extra dimension as being microscopic and periodic~\cite{Klein:1926tv}. Around 60 years later, Edward W. Kolb and Richard Slansky realized that compact extra dimensions could be associated with additional stable, heavy particles~\cite{Kolb:1983fm}. These particles arise due to the conservation of the quantized momentum along the compact extra dimension. The high-momentum states build up a tower of so-called Kaluza-Klein states, which appear in the four large dimensions as particles with increasing mass of scale~$M_{\rm KK}\sim R^{-1}$, where~$R$ is the compactification scale.

In models of Universal Extra Dimensions~(UED) all~SM fields are allowed to propagate through the extra dimension~\cite{Appelquist:2000nn}.  If the lightest of the Kaluza-Klein states is stabilized by some symmetry and cannot decay into the ground state, which corresponds to the usual~SM particle, this particle would be a qualified DM~particle. As such it could get produced thermally and be observed e.g. via direct detection~\cite{Cheng:2002ej,Servant:2002aq,Servant:2002hb}.

\paragraph{Axions}
In 1977, Roberto D. Peccei and Helen R. Quinn proposed a solution to the strong CP problem of~\ac{QCD}~\cite{Peccei:1977hh,Peccei:1977ur}, which predicted a new particle, as was quickly noted by the American theorists Stephen Weinberg and Frank Wilczek~\cite{Weinberg:1977ma,Wilczek:1977pj}. 

The symmetries of~QCD allow for the CP violating term
\begin{align}
	\mathcal{L}_{\rm QCD}\supset \frac{\theta}{32\pi^2}\mathrm{Tr}\left[ G^{\mu\nu}\tilde{G}_{\mu\nu}\right]\, ,
\end{align}
where $G_{\mu\nu}$ ($\tilde{G}_{\mu\nu}$) is the (dual) field strength tensor of~QCD and the trace runs over the SU(3) color indices. This term introduces CP violating interactions into QCD. However, from upper limits of the neutron's electric dipole moment~\cite{Hertzberg:2008wr}, we know that
\begin{align}
	\theta \lesssim 10^{-10}\, .
\end{align}
The seemingly fine-tuned suppression of an otherwise allowed term is the strong CP problem. In their solution, Peccei and Quinn explain the parameter's suppression by introducing a new global, anomalous~$U(1)$ symmetry, which is spontaneously broken by a complex scalar field. In their model, the effective $\theta$ parameter depends on this field and vanishes dynamically once the field attains its vacuum expectation value. Furthermore, an additional light particle arises naturally as the pseudo-Nambu-Goldstone boson.  This particle was called the~\emph{axion}. The standard Weinberg-Wilczek axion was excluded by prompt experimental searches, and more general realizations such as the Invisible Axion were developed~\cite{Abbott:1982af}, culminating in a large class of Axion-Like Particles~(ALPs).

The axion might not just be the by-product of the Peccei-Quinn mechanism, it could also be another candidate for~DM, solving two problems at once~\cite{Preskill:1982cy,Turner:1985si}. These light scalar particles could get produced non-thermally and non-relativistically during the early universe and act as a collision-less fluid satisfying all five DM~conditions. Consequently, a large number of experimental searches were performed aiming at the discovery of the axion~\cite{Rosenberg:2000wb,Graham:2015ouw}.

\subsection{Origin of~DM in the early Universe}
An essential aspect of DM~phenomenology is to explain how it was produced in the right amount during the early Universe. A lot of the proposed DM~particles share a common production mechanism similar to the neutrinos' and get produced thermally. They make up the large class of the~\ac{WIMP}s~\cite{Steigman:1984ac}. A~WIMP is characterized by
\begin{itemize}
	\item a mass of MeV-TeV scale,
	\item its non-gravitational interactions with ordinary matter. This interaction should not be stronger than the weak interaction of the~SM.
	\item its thermal production via `freeze-out'.
\end{itemize}
\acp{WIMP} are in thermal equilibrium with the~SM bath during the early Universe, continuously created by and annihilating into lighter particles. As the Universe kept expanding and cooling, the light particles lost energy and with it the ability to generate new DM~particles. At this point, the DM~population got depleted due to their ongoing annihilations. But even the annihilations stopped being effective, when the number density has dropped so low that the annihilation rate dropped below the expansion rate, and particles and anti-particles no longer came into contact. Afterwards, a constant number of DM~particles remained in the Universe as a thermal relic, not unlike the photons of the~\ac{CMB} or the neutrino background. This production mechanism is called~`thermal freeze-out'.

\begin{figure*}
	\centering
	\includegraphics[width=0.67\textwidth]{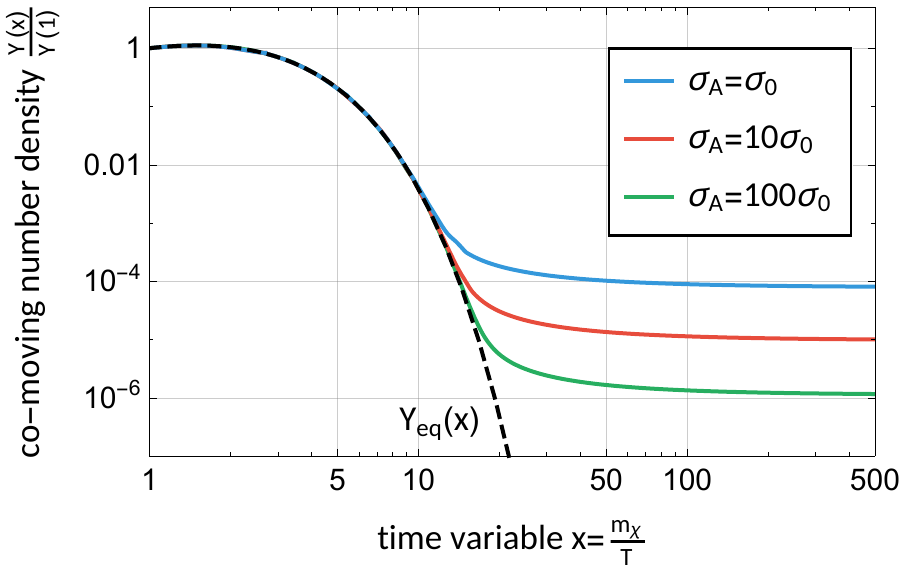}
	\caption{Thermal freeze-out of WIMPs.}
	\label{fig: freeze out}
\end{figure*}

The relic abundance of a particle can be computed using the Boltzmann equation~\cite{Kolb:1988aj,Jungman:1995df}, which determines the evolution of the DM~number density~$n_\chi$ in an expanding Universe,
\begin{align}
	\frac{\dd n_\chi}{\dd t} +\underbrace{3 H(t) n_\chi}_{\text{\footnotesize Hubble dilution}} = -\underbrace{\langle \sigma_A v\rangle \left( n_\chi^2 -(n_\chi^{\rm eq})^2\right)}_{\text{\footnotesize creation and annihilation}}\, .
\end{align}
Here,~$\sigma_A$ is the total annihilation cross section,~$v$ is the relative velocity, and~$\langle\cdot\rangle$ denotes the thermal average. In order to scale out the dilution due to expansion, we can use the conservation of the entropy density~$s$, i.e. $s R^3 = \mathrm{const}$. We define the DM~density in a co-moving volume~$Y\equiv\frac{n_\chi}{s}$, such that~$\frac{\dd Y}{\dd t}=\frac{\dd n_\chi}{\dd t} +3 H(t) n_\chi$. Usually the time parameter is replaced by~$x\equiv \frac{m_\chi}{T}$. Hence, the Boltzmann equation becomes
\begin{align}
	\frac{\dd Y}{\dd x} = -\frac{x s \langle \sigma_A v\rangle}{H(T=m_\chi)}\left(Y^2-Y_{\rm eq}^2\right)\, .
\end{align}
The freeze-out occurs during the radiation dominated epoch of the Universe, which fixes the Hubble parameter $H(x)\sim x^{-2}$. The entropy density per co-moving volume stays constant,~$s(x)\sim x^{-3}$. Some example solutions for~$Y(x)$ with different annihilation cross sections are shown in figure~\ref{fig: freeze out} and illustrate the thermal production of a non-relativistic relic. While the equilibrium number density falls exponentially~$Y_{\rm eq}(x)\sim e^{-x}$, at some point the~DM decouples and freezes out to a constant co-moving volume number density. The higher the annihilation cross section the longer the~DM keeps annihilating in thermal equilibrium and the lower the final density. An approximate expression for the present \ac{WIMP} density in units of the critical density is given by
\begin{align}
	\Omega_\chi h^2 \simeq  \left\langle\frac{\sigma_A}{\mathrm{pb}}\frac{v}{0.1c}\right\rangle^{-1}\, .  
\end{align}
 This procedure to compute the thermal relic of WIMPs gives a good estimate, but is not very precise. The calculation has been greatly refined to yield more precise predictions~\cite{Srednicki:1988ce,Gondolo:1990dk,Griest:1990kh,Steigman:2012nb}. However, it was quickly noted that a weak scale cross section would lead to~$\Omega_\chi = \mathcal{O}(1)$ , naturally explaining the origin of the right amount of~DM. This rough agreement between the WIMP relic density and the observed DM~density was called the~`WIMP miracle'. It motivated a great number of strategies and experiments in the following decades, aiming to detect \ac{WIMP}~DM. Unfortunately, these efforts have not been successful so far, and the~WIMP paradigm is getting constrained more and more~\cite{Bertone:2010at}. Nonetheless, it has not been excluded altogether~\cite{Leane:2018kjk}.
 
 The DM~particles in the Universe do not need to be a thermal relic, many other production mechanism have been proposed which do not rely on~DM being in thermal equilibrium. Examples include the `freeze-in' mechanism of very weakly interacting~DM~\cite{McDonald:2001vt,Hall:2009bx} or DM~production from heavy particle decays~\cite{Gherghetta:1999sw,Merle:2013wta}. For more information on non-thermal DM~production we refer to~\cite{Gelmini:2010zh,Baer:2014eja}.

\subsection{Detection strategies}
\begin{figure*}
	\centering
	\includegraphics[width=0.67\textwidth]{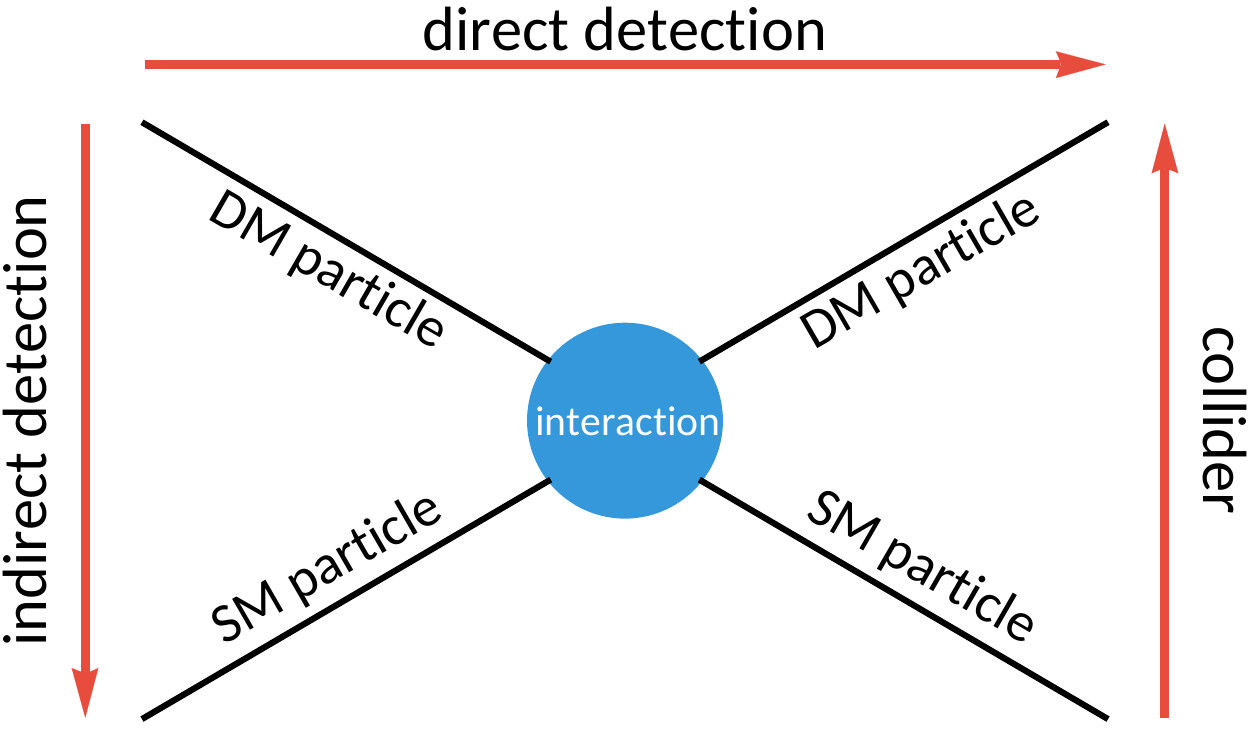}
	\caption{The three main search strategies for non-gravitational DM~interactions.}
	\label{fig: strategies}
\end{figure*}

Although the evidence for~DM is almost conclusive, it is unfortunately purely gravitational. There is little hope to directly observe or produce DM~particles, if the only DM-matter interaction is via gravity. However, there are good reasons that the dark and the bright sector share some other interaction. The~DM particle might not be a gauge singlet under the \ac{SM}~gauge groups and participate in some of the known forces. Alternatively, there might exist some interactive portal between the light and dark sector. The field acting as that portal might be known, such as the Z or the Higgs portal, or be a new field, e.g. a dark photon. There are three major strategies to search for non-gravitational effects of~DM, illustrated in figure~\ref{fig: strategies}.

\paragraph{Indirect detection}
If DM~particles could annihilate or decay into~SM model particles, these particles could be observed with cosmic ray and neutrino telescopes. The approach to look for observational excesses of~SM particle fluxes originating from regions of high DM~density is called \textit{indirect detection}~\cite{Cirelli:2012tf,Gaskins:2016cha}. These high-density regions could be the galactic center of the Milky Way, neighbouring dwarf galaxies, or galaxy clusters. In the case of the WIMP, the annihilation cross section determined the relic density. The indirect observation of such annihilations could thereby probe not just the existence of DM~particles, but also its origin.

Possible detection channels are DM~annihilations into gamma-rays~\cite{Gunn:1978gr,Stecker:1978du}, anti-protons and positrons~\cite{Silk:1984zy,Stecker:1985jc,Ellis:1988qp,Kamionkowski:1990ty}, or neutrinos~\cite{Bergstrom:1998xh}. The clearest and most conclusive indirect detection of DM~annihilation would be a monochromatic feature in the cosmic-ray spectrum.

 Another possible discovery channel of indirect detection are neutrinos from the Sun. DM~particles could get gravitationally captured by the Sun, aggregate in the solar core and annihilate resulting in an additional neutrino flux from the Sun~\cite{Silk:1985ax,Krauss:1985ks,Press:1985ug}. Due to their weak interactions with matter, neutrinos could escape even those dense regions.

The main challenge of indirect detection is the distinction between the flux of the annihilation products and background from astrophysical sources. Another challenge arises for the observations of charged particles which get reflected and diffused by galactic magnetic fields or scatterings, impeding the assignment of an observation to a particular source~\cite{MedinaTanco:1997rt,Dolag:2003ra}. The interpretation of an observed excess is difficult and drawing a definitive conclusion of a DM~discovery even more so. A number of experiments have indeed measured excesses. One example is an observed rise of the cosmic-ray positron fraction with energy,  measured by the satellite-borne cosmic-ray observatories PAMELA and AMS-02~\cite{Adriani:2008zr,Aguilar:2013qda}. The excess could in principle be interpreted as the product of DM~annihilations~\cite{Bergstrom:2008gr,Cholis:2013psa}.

\paragraph{Direct production}
If~DM couples to matter, it could be possible to produce DM~particles in high energy particle collisions such as the~\ac{LHC}~\cite{Beltran:2010ww,Fox:2011pm,ATLAS:2012ky,Chatrchyan:2012me}. After their creation, these particles would leave the detector without a trace. Hence, the main signature of~DM in colliders is \ac{MET}\footnote{A DM~particle would not be the first particle to be discovered through an \ac{MET} signature, as the W boson was discovered through this technique in 1983~\cite{Arnison:1983rp}.}, and neutrino production will be an important background. At hadron colliders, the typical signature event for the production of a pair of DM~particles is a single jets or photons from initial state radiation, plus \ac{MET}.

There are two draw-backs of collider searches of~DM. For one, the search for new physics with a collider is always to some degree model-dependent, since a model is necessary to make predictions. This problem can be eased by a more general and standardized formulation of DM-matter interactions using~\acsp{EFT} for contact interactions and simplified models in the presence of light mediators~\cite{Abercrombie:2015wmb,DeSimone:2016fbz}. More critically, colliders alone would not be able to confirm a newly discovered weakly interacting particle as the source of~DM. They could e.g. not test the stability of the particle and only find weak lower bounds on its lifetime. The discovery of the same particle by complementary astrophysical experiments is necessary to conclusively draw the connection between the galactic halo and the observation at a collider~\cite{Kane:2008gb}.

\paragraph{Direct detection}
The basic idea of direct detection is to search for recoiling nuclei or electrons in a terrestrial detector resulting from an elastic collision between an incoming DM~particle from the halo and a particle of the detector's target mass. Direct detection is the main topic of this thesis and will be discussed in detail in the next chapter.

\clearpage
\chapter{Direct Detection of Dark Matter}
\label{c:directdetection}
If the galactic halo is comprised of one or more new particles which interact not just gravitationally with ordinary matter, there is hope to detect these particles here on Earth. These particles would continuously pass through our planet in great numbers. The basic idea of the direct detection of~DM is to search for nuclear recoils resulting from an elastic collision between an incoming DM~particle from the halo and a nucleus inside a terrestrial detector.

In this chapter, we review the development of the field of direct DM~searches, as well as the basic relations and tools to make predictions for direct detection experiments. We will treat both the standard nuclear recoil experiments and newer techniques to search for sub-GeV~DM based on inelastic~DM-electron interactions. The necessary model of the galactic halo, the scattering kinematics, the description of~DM-matter interactions, as well as the computation of event rates for any detection experiment are presented in detail, before we conclude with a short summary to direct detection statistics and limits.

\clearpage
\section{The Historical Evolution of Direct DM Searches}
\label{s: direct detection development}
\subsection{Standard WIMP detection}
\label{ss: direct detection nuclear}
The fundamental technique of direct detection experiments was originally proposed by Andrzej Drukier and Leo Stodolsky in 1983 as a way to detect neutrinos via coherent elastic scatterings on nuclei~\cite{Drukier:1983gj}. Soon after, three groups, one including Drukier himself, independently pointed out that this method might also serve as a possible detection technique for DM~particles~\cite{Goodman:1984dc,Wasserman:1986hh,Drukier:1986tm}. Especially the WIMP scenario can be probed by this strategy since GeV scale DM~particle would cause observable rates of keV scale nuclear recoils, and it did not take long until the first direct detection experiments took place~\cite{Ahlen:1987mn,Caldwell:1988su}, setting the first direct detection constraints on DM-matter interactions.

In their 1986 paper, Drukier, Freese, and Spergel proposed to utilize the expected annual signal modulation to distinguish between a potential DM~signal and background from other sources~\cite{Drukier:1986tm}. This modulation occurs due to the orbital motion of the Earth around the Sun, which leads to a small variation of the DM~flux and event rate in a detector.  About twenty years ago, the DAMA experiment first reported the observation of such a modulation~\cite{Bernabei:1998fta}. The observed events in their sodium-iodide (NaI) target crystal showed not just an annual rate modulation, the rate also peaked at the expected time of the year. DAMA/NaI and the upgrade DAMA/LIBRA continued to confirm their observations over the years~\cite{Bernabei:2000qi,Bernabei:2008yi,Bernabei:2010mq}. In the latest DAMA/LIBRA phase 2 run, they finished the observation of the 20th annual cycle. Remarkably, they report a modulation period of~(0.999$\pm$0.001)yr and phase of (145$\pm$5)~days, just as expected for~DM\footnote{The theoretically predicted event rate under standard assumptions peaks around June 2nd, corresponding to a phase of 152 days.}.

The reason, why the DAMA claim is met with great scepticism, is the fact that no other experiment was able to confirm the observation. Even worse, the findings of many other DM~searches are in serious tension with the DAMA results and repeatedly exclude the parameter space favoured by~DAMA. Since these experiments use different target materials, it is not impossible that some unexpected type of interaction would show up in a NaI crystal, but not e.g. in liquid xenon experiments. The comparisons rely on standard assumptions, which could of course be inaccurate. Until an experiment of the same target material confirms or refutes the discovery claim, the DAMA signal's origin remains unsolved. A number of direct detection experiments with sodium-iodide crystals are being planned~\cite{Cherwinka:2011ij,Adhikari:2017esn,Shields:2015wka,Coarasa:2018qzs}. The first result by DM-Ice17~\cite{deSouza:2016fxg} and COSINE-100~\cite{Adhikari:2018ljm} did not show evidence for an annual modulation and seem to indicate that the DAMA/LIBRA modulation is indeed not due to particles from the DM~halo. Over the next few years, these experiments will find a definitive answer.\\[0.2cm]

Direct detection experiments can be classified in two broad categories of two-channel detectors, using either crystal or liquid noble targets. In both cases, the discrimination between background events and a potential DM~signal is realized by splitting the signal into two components, which are detected independently.

The historically first category are detectors with cryogenic solid-state targets. One two-channel technique of background rejection is to simultaneously measure both ionization and heat signals~\cite{Cabrera:1984rr,Krauss:1985aaa}. This method was applied to pure germanium targets by the EDELWEISS experiments~\cite{Benoit:2001zu,Armengaud:2009hc,Armengaud:2016cvl} and silicon/germanium semiconductors by the CDMS detectors located in the Soudan Mine~\cite{Akerib:2004fq,Ahmed:2008eu,Agnese:2014aze}. The newest generation of SuperCDMS will be taking data from SNOLAB~\cite{Agnese:2016cpb}.  Another similar method is to simultaneously detect scintillation photons and heat phonons, as done by the three CRESST experiments. Where CRESST-I used a sapphire target~\cite{Bravin:1999fc,Angloher:2002in}, the other two generations employed~$\text{CaWO}_{\text{4}}$ crystals~\cite{Angloher:2011uu,Angloher:2015ewa,Petricca:2017zdp}. Finally, a new generation of solid-state detection experiments looking for light~DM, namely DAMIC~\cite{Barreto:2011zu,Aguilar-Arevalo:2016ndq} and SENSEI~\cite{Crisler:2018gci}, uses silicon CCDs as target, which read out ionized charges in each pixel, potentially caused by a DM-atom interaction in the silicon crystal. In these experiments, background events are rejected due to their characteristic spatial signal correlation in neighbouring pixels, as opposed to point-like energy depositions by a WIMP.

So far, the results of these experiments have been a series of null results with a few exceptions. The CoGENT experiment, a germanium detector in the Soudan Underground Laboratory~(SUL), has reported evidence for annual signal modulations~\cite{Aalseth:2010vx,Aalseth:2011wp,Aalseth:2014eft}, however a re-analysis of the data found an underestimation of background due to surface events~\cite{Davis:2014bla,Aalseth:2014jpa}. Furthermore, both CRESST-II~(phase 1) and CDMS~II reported an observed signal excess~\cite{Angloher:2011uu,Agnese:2013rvf}. For CRESST, these signals could also be attributed to background sources, and a DM~interpretation was eventually excluded~\cite{Angloher:2014myn}. The CDMS signal is in tension with constraints from other experiments, and a DM~origin seems disfavoured as well~\cite{Witte:2017qsy}. These anomalies and their interpretation, including the DAMA observation, have been discussed in greater detail in~\cite{Davis:2015vla}.
 
In order to increase the sensitivity of detectors to weaker interactions, larger targets were needed. However, the upscaling of solid state detectors to sizes beyond the kg scale is costly. In the year 2000, a new detection technology was proposed, the use of a two-phase noble target~\cite{Arneodo:2000vc,Cline:2000wt}. Xenon has been the most common noble target, as it is easily purified, radio-pure, chemically inactive, has large ionization and scintillation yield, combined with a heavy nucleus ideal to probe coherent spin-independent interactions. In addition, a xenon target can be realized even in ton scales. Just as for crystal targets, the signal is split in two parts to reject background. An incoming DM~particle would scatter on a xenon nucleus in the liquid phase, and the nuclear recoil causes the first scintillation signal~(S1) followed by a time-retarded second scintillation signal~(S2) in the detector's gas phase. Both signals are observed with~\acp{PMT}. The S2-signal is caused by electrons which get ionized in the original scattering and drift towards the gas phase due to an external electric field. This ingenious idea to use the time-separated combination of scintillation and ionization became the blueprint for a series of experiments with increasing target sizes all around the world. 

While single-phase liquid xenon detectors have already been proposed and performed in the~'90s~\cite{Belli:1990bx,Davies:1994bz,Bernabei:1998ad} and early '00s~\cite{Aprile:2002ef,Alner:2005pa}, the first two-phase xenon detector was the ZEPLIN-II experiment at the Boulby Underground Laboratory in England, which reported the first results in~2007~\cite{Alner:2007ja}. ZEPLIN-II was followed by a third generation in 2009~\cite{Lebedenko:2008gb}. Within Europe, they competed with the XENON experiments located at the~\ac{LNGS}, with their three generations, XENON10~\cite{Angle:2007uj,Angle:2011th}, XENON100~\cite{Aprile:2010um,Aprile:2012nq}, and XENON1T~\cite{Aprile:2017iyp,Aprile:2018dbl}. The similar LUX experiment, installed at the~\ac{SURF} in the Homestake Mine, set leading constraints on \ac{WIMP}~DM~\cite{Akerib:2013tjd,Akerib:2016vxi}. In the last few years, two more direct DM~searches were performed by the PandaX collaboration with xenon detectors at the China Jinping Underground Laboratory~(CJPL)~\cite{Xiao:2014xyn,Xiao:2015psa,Tan:2016zwf,Cui:2017nnn}. The Japanese XMASS-I experiment started to use single phase liquid noble targets again~\cite{Abe:2015eos}, while others switched from xenon to an argon target, namely the DarkSide-50 dual-phase detector at Gran~Sasso~\cite{Agnes:2014bvk,Agnes:2018ves} and the single-phase detector DEAP-3600 at SNOLAB~\cite{Amaudruz:2017ekt}. Planned future experiments include LUX's successor, the next-generation LUX-ZEPLIN~(LZ) experiment, which is expected to take data with a 7~ton dual phase xenon target by~2020~\cite{McKinsey:2016xhn}, as well as proposals for XENONnT in Gran Sasso~\cite{Aprile:2015uzo,Aprile2017}, and a multi-ton dual phase xenon experiment DARWIN~\cite{Aalbers:2016jon}.

The continuing null results of these enormous experimental efforts were certainly a disappointment for conventional direct DM~searches. While larger and larger detectors continue to constrain and rule out weaker and weaker \ac{WIMP} interactions, no conclusive evidence for non-gravitational DM~interactions in terrestrial detectors has ever been found. This caused a shift away from the standard~WIMP paradigm. One way, to loosen the usual assumptions is to search for low-mass DM~particles.

\subsection{Low-mass~DM searches}
\label{ss: low mass DM detection}
The results presented in this thesis do not consider a particular particle physics model which contains a DM~candidate. Instead it focuses on light, mostly sub-GeV, DM~particles and their phenomenology in direct detection experiments, while staying mostly agnostic about the particle's origin. Nonetheless, it should be noted that light~DM arises in a number of well-motivated particle models. The standard WIMP picture can be modified in order to circumvent the Lee-Weinberg limit~\cite{Lee:1977ua} such that DM of masses below~$\sim$2~GeV can be produced as a thermal relic. This was shown to work for scalar~DM~\cite{Boehm:2003hm}, in supersymmetric models~\cite{Hooper:2008im,Feng:2008ya,Kumar:2009bw}, e.g. as light axinos~\cite{Covi:1999ty,Choi:2011yf}, for~\acp{SIMP}, meaning~DM with strong self interactions~\cite{Hochberg:2014dra,Hochberg:2014kqa}\footnote{The abbreviation~\ac{SIMP} is not used consistently in the literature, where some refer to strong self-interactions and others to strong couplings to ordinary matter. This is why we refrain from using it.}, for elastically decoupling~DM~\cite{Kuflik:2015isi}, secluded sector~DM~\cite{Pospelov:2007mp}, or light DM~particles which annihilate into heavier particles during the early Universe~\cite{DAgnolo:2015ujb}. Sub-GeV~DM can also be produced non-thermally as asymmetric~DM~\cite{Nussinov:1985xr,Kaplan:1991ah,Kaplan:2009ag,Falkowski:2011xh,Lin:2011gj} or via freeze-in~\cite{Hall:2009bx}.

Direct detection experiments only probe~particles down to some minimal mass. The energy deposits of even lighter~DM fall below the experiment's threshold and are insufficient to trigger the detector. Conventional nuclear recoil experiments typically have a recoil threshold of the order of~$\sim$keV and therefore probe DM~particles with masses above a few~GeV. It is a great challenge to probe masses below the GeV scale using nuclear recoils. The experiments of the CRESST collaboration are leading in this field and have pushed the limits of low-mass DM~searches. They realized recoil thresholds of the order of~$\mathcal{O}$(10)eV for a gram scale detector, setting constraints on~DM of masses down to~$\sim$140~MeV~\cite{Angloher:2017sxg}. The next generation~CRESST-III experiment is expected to reduce the threshold further~\cite{Petricca:2017zdp}. 

There are a number of ideas, which do not require a new generation of experiments. One possibility is to consider signatures of non-standard interactions at conventional large-scale detectors. A sub-GeV DM~particle could cause an otherwise unobservable nuclear recoil, where the recoiling nucleus emits Bremsstrahlung. The emitted photons are able to trigger the detector~\cite{Kouvaris:2016afs,McCabe:2017rln}. Another idea is to employ the Migdal effect~\cite{Migdal1941} and observe electrons dissociated from the atom through the nuclear scattering. As the nucleus recoils, the atomic electrons do not immediately follow and might therefore get excited or ionized, leading to a signal~\cite{Vergados:2004bm,Moustakidis:2005gx,Bernabei:2007jz,Ibe:2017yqa,Dolan:2017xbu}. Both techniques of extending a detector's sensitivity to lower masses were recently applied by the LUX collaboration to derive constraints on masses down to 400~MeV~\cite{Akerib:2018hck}. Another idea is to look for processes which could accelerate DM~particles as faster particles are able to deposit larger recoil energies in a detector. If sub-GeV DM~particles scatter on hot constituents of the Sun, they could gain energy and make up a highly energetic solar  DM~flux. This was shown to increase detectors' sensitivity for DM-electron interactions~\cite{An:2017ojc} and independently by us for DM-nucleus scatterings in~\ref{paper3}. A similar idea is to apply the same argument to relativistic cosmic rays, which could also transfer energy to halo particles, which could then be observed in terrestrial detector, even though they might have been undetectable before the scattering due to their low mass~\cite{Bringmann:2018cvk}. The solar reflection of sub-GeV~DM is the topic of chapter~\ref{c:sun}.

Arguable the most promising idea to sub-GeV DM~searches is to consider~DM-electron scatterings~\cite{Bernabei:2007gr,Kopp:2009et,Essig:2011nj}. MeV scale DM~particles can transfer almost their entire kinetic energy to bound electrons and are hence able to ionize atoms. In particular, the scattering on electrons in xenon experiments such as XENON10~\cite{Angle:2011th} and XENON100~\cite{Aprile:2016wwo} have been investigated. These experiments set strong constraints on DM-electron scatterings for DM~masses as low as a few MeV~\cite{Essig:2012yx,Essig:2017kqs}. With DarkSide-50, the first argon target detector has probed DM-electron scatterings in 2018~\cite{Agnes:2018oej}. Despite their smaller exposures, semiconductor targets are even more promising due to their small band gaps of order~$\mathcal{O}$(1)~eV~\cite{Graham:2012su,Lee:2015qva,Essig:2015cda,Hochberg:2016sqx}. The first experiments to test electron scatterings were SENSEI~(2018)~\cite{Tiffenberg:2017aac,Crisler:2018gci} and SuperCDMS~(2018)~\cite{Agnese:2018col}, both using a silicon semiconductor targets. In the near future, the DAMIC-M collaboration is planning to install a detector with a remarkably large semiconductor target mass of~kg scale~\cite{Settimo:2018qcm}.
 
These are not the only new proposals for experimental techniques and search strategies aimed at light~DM over the last few years. Others have suggested the use of scintillating materials~\cite{Derenzo:2016fse}, two-dimensional targets such as graphene for directional detection~\cite{Hochberg:2016ntt}, superfluid helium targets~\cite{Guo:2013dt,Schutz:2016tid,Knapen:2016cue,Hertel:2018aal}, molecule dissociation~\cite{Essig:2016crl}, super conductors~\cite{Hochberg:2015fth,Hochberg:2015pha,Hochberg:2016ajh}, and many other effects and techniques~\cite{Cavoto:2016lqo,Hochberg:2017wce,Bunting:2017net,Budnik:2017sbu,Cavoto:2017otc,Fichet:2017bng}. Many of these new ideas have been reviewed in~\cite{Battaglieri:2017aum}.

\section{The Galactic Halo}
\label{s: halo model}
To interpret the outcome of a direct detection experiment, it is crucial to know how many and how energetic DM~particles are expected to pass through the detector. It is necessary to estimate the local DM density and velocity distribution, which are related through the gravitational potential of the halo. The differential number density of DM~particles close to the Sun's orbit within the galactic halo with velocity within~$(\mathbf{v}_\chi,\mathbf{v}_\chi+\dd^3 \mathbf{v})$ can be written as
\begin{align}
	&\dd n =n_{\chi}f_{\rm halo} (\mathbf{v})\dd^3\mathbf{v} = \frac{\rho_\chi}{m_\chi} f_{\rm halo} (\mathbf{v})\dd^3\mathbf{v} \, , \label{eq: differential density}
\end{align}
where $f_{\rm halo}(\mathbf{v})$ is the normalized velocity distribution and $n_\chi$ and $\rho_\chi$ are the local DM number and energy density respectively. The DM density is assumed to remain constant along the Sun's orbit around the galactic centrum throughout this thesis. However, this does not need to be true, which can have critical consequences for direct detection~\cite{Freese:2012xd}. Throughout this thesis, we use the canonical value of $\rho_\chi=\SI{0.3}{\GeV\per\cm\cubed}$~\cite{Bovy:2012tw}, even though newer evidence suggests a value closer to~$\sim\SI{0.4}{\GeV\per\cm\cubed}$~\cite{Catena:2009mf}. The reason for this is the fact, that the former value is widely used in the literature and serves as a fiducial value.

The velocity distribution depends on the DM halo model. The conventional choice is the \ac{SHM}, which models the DM of a galaxy as a self-gravitating gas of collision-less particles in equilibrium, which form a spherical and isothermal halo~\cite{Binney2008}. As such, the density profile scales as $\sim r^{-2}$ and the mass function as~$M(r)\sim r$, which explains the observed flatness of galactic rotation curves we discussed in section~\ref{ss: galactic evidence}. The velocities follow an isotropic Maxwell-Boltzmann distribution,
\begin{align}
	f_{\text{halo}}(\mathbf{v}) \sim \exp\left(-\frac{\mathbf{v}^2}{2\sigma_v^2}\right)\, .
\end{align}
Even though this distribution does not fit results from N-body simulations very well~\cite{Fairbairn:2008gz,Vogelsberger:2008qb,Kuhlen:2009vh}, and a newer analysis of the stellar distribution using the SDSS-Gaia sample shows that the halo shows more substructure~\cite{Necib:2018iwb}, the \ac{SHM} is by far the conventional choice in DM~detection, since it does not introduces a critical error and simplifies the comparison between different results and experiments, similarly to the choice of the DM density\footnote{We also mention that an update to the~SHM was presented recently~\cite{Evans:2018bqy}, which has only a small impact on non-directional direct detection experiments.}.

The halo particles' speed will not exceed the galaxy's escape velocity $v_{\rm esc}$, faster particles would have left the galaxy a long time ago. Consequently, it is reasonable to truncate the distribution,
\begin{subequations}
\label{eq:fhalo}
\begin{align}
f_{\rm halo}(\mathbf{v}) &=\frac{1}{N_{\rm esc}} \frac{1}{\left(2 \pi \sigma_v^2\right)^{3/2}}\exp\left(-\frac{\mathbf{v}^2}{2\sigma_v^2}\right)\Theta\left(v_{\rm esc}-v\right)\, ,
\intertext{where $\Theta(x)$ is the Heaviside step function and the normalization constant $N_{\rm esc}$ is given by}
N_{\rm esc} &={\rm erf}\left( \frac{v_{\rm esc}}{\sqrt{2\sigma_v^2}}\right)-\sqrt{\frac{2}{\pi}}\frac{v_{\rm esc}}{\sigma_v}\exp \left(-\frac{v_{\rm esc}^2}{2\sigma_v^2} \right)\, .
\end{align}
\end{subequations}
The galactic escape velocity typically chosen in the DM~detection literature is given by $v_{\rm esc}\approx\SI{544}{\km\per\second}$, as obtained by the RAVE survey, $\text{544}^{\text{+54}}_{\text{-41}}$ \SI{}{\km\per\second} (90\% confidence)~\cite{Smith:2006ym}. This value is still the conventional choice, even though it was updated to~$\text{533}^{\text{+64}}_{\text{-46}}$ \SI{}{\km\per\second} (90\% confidence) more recently~\cite{Piffl:2013mla}. The velocity dispersion $\sigma_v$ is set to  $\sigma_v=v_{0}/ \sqrt{2}$, where $v_0 \approx \SI{220}{\km\per\second}$ is the IAU value for the Sun's circular velocity~\cite{Kerr:1986hz}.

\begin{figure*}
	\centering
	\includegraphics[width=0.66\textwidth]{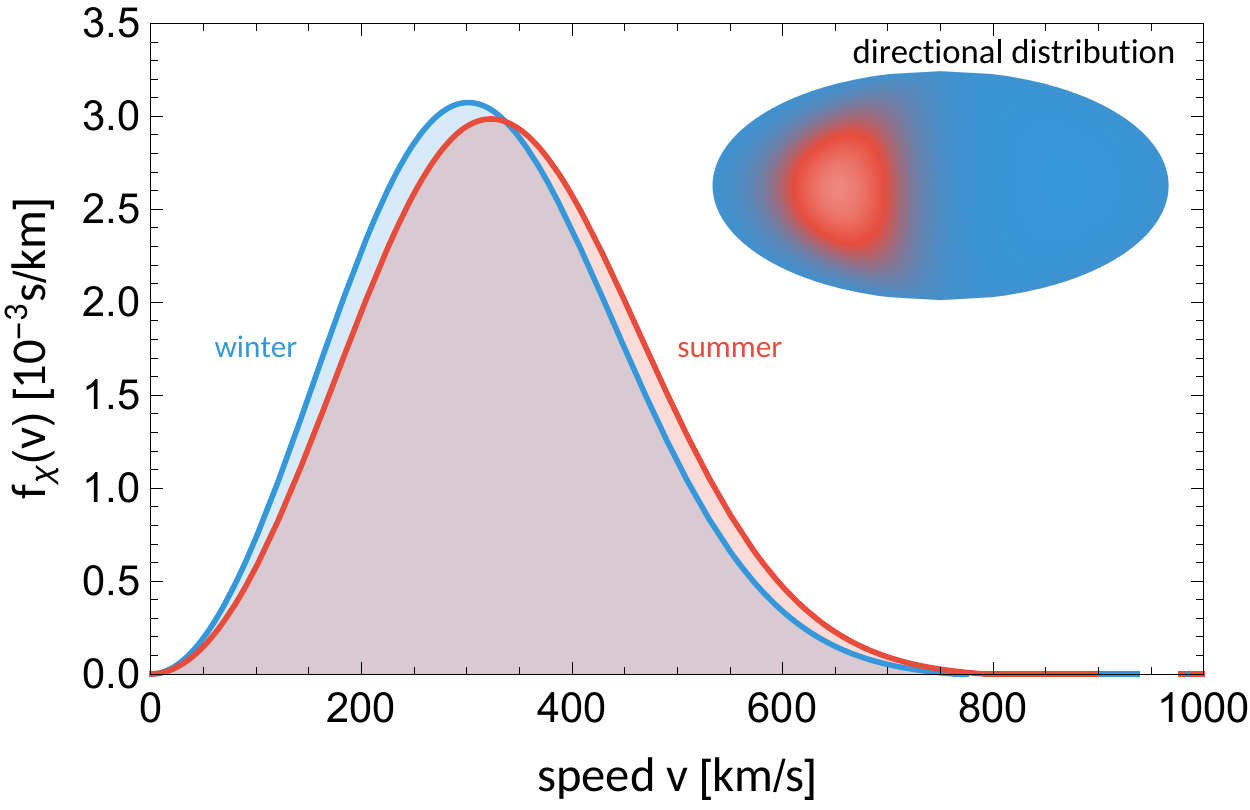}
	\caption{The local speed distribution of~DM and its annual modulation. A Mollweide projection of the directional distribution with the characteristic dipole of the DM~wind is shown as well.}
	\label{fig: halo pdf}
\end{figure*}

Finally, a Galilean transformation into the Earth's rest frame is necessary to obtain the local DM phase space $f_\chi(\mathbf{v})$,
\begin{subequations}
\label{eq:fearth}
\begin{align}
	f_{\chi}(\mathbf{v}) &\equiv f_{\rm halo}(\mathbf{v} +\mathbf{v}_{\oplus})\\
	&= \frac{1}{N_{\rm esc}}\frac{1}{\pi^{3/2}v_0^3} \exp\left(-\frac{(\mathbf{v}+\mathbf{v}_{\oplus})^2}{v_0^2}\right)\Theta(v_{\rm esc}-|\mathbf{v}+\mathbf{v}_\oplus|)\, .
\end{align}
\end{subequations}
Even though the original velocity distribution was isotropic, this transformation into the Earth's rest frame breaks this isotropy and is the cause for the `DM wind'. More and faster particles are expected to hit the Earth from its direction of travel, which can be seen in the inset directional distribution in figure~\ref{fig: halo pdf}. The Earth's velocity $\mathbf{v}_{\oplus}$ is given in app.~\ref{a:velocity}. As the Earth orbits the Sun during a year, the Earth's speed relative to the DM halo varies, which in turn causes a modulation of the DM phase space and an annual signal modulation of a direct detection experiment~\cite{Drukier:1986tm}\footnote{A similar, smaller \emph{diurnal} signal modulation can be included, if, instead of $\mathbf{v}_\oplus$, we use the laboratory velocity $\mathbf{v}_{\rm lab}$, which additionally varies daily as the Earth rotates.}. 

In many contexts, the directional information of eq.~\eqref{eq:fearth} is irrelevant, and the marginal speed distribution suffices. We obtain it by integrating out the velocity angles,
\begin{subequations}
\label{eq: DM speed PDF}
\begin{align}
	f_\chi(v) &\equiv \int \dd \Omega\; v^2 f_\oplus(\mathbf{v})\\
	&=\frac{1}{N_{\rm esc}}\frac{v}{\sqrt{\pi}v_0 v_\oplus}\times\bigg[2\exp\left(-\frac{v^2+v_\oplus^2}{v_0^2}\right) \sinh\left(2\frac{v v_\oplus}{v_0^2}\right)\nonumber\\
	&\;+\left(\exp\left(-\frac{(v+v_\oplus)^2}{v_0^2}\right)-\exp\left(-\frac{v_{\rm esc}^2}{v_0^2}\right)\right)\Theta\left(|v+v_\oplus|-v_{\rm esc}\right)\nonumber\\
	&\;-\left( \exp\left(-\frac{(v-v_\oplus)^2}{v_0^2}\right)-\exp\left(-\frac{v_{\rm esc}^2}{v_0^2}\right)\right)\Theta\left(|v-v_\oplus|-v_{\rm esc}\right) \bigg]\, .
\end{align}
\end{subequations}
The speed distribution and its annual modulation are plotted in figure~\ref{fig: halo pdf}.

Provided with a local DM density and phase space distribution, it is possible to compute the particle flux of DM through a detector, the first necessary ingredient to predict scattering rates in a detector.

\clearpage
\section{Scattering Kinematics}
\label{s:kinematics}

\subsection{Elastic DM-nucleus collisions}
\label{ss: nuclear scattering kinematics}
In most \ac{WIMP} searches, the fundamental processes which experiments look for are elastic coherent collisions between DM~particles and target nuclei in a detector. In this section, we summarize the kinematic relations required to describe direct detection experiments~\cite{Landau1976}.

\begin{figure*}
\sbox\twosubbox{%
  \resizebox{\dimexpr.95\textwidth-1em}{!}{%
    \includegraphics[height=3.5cm]{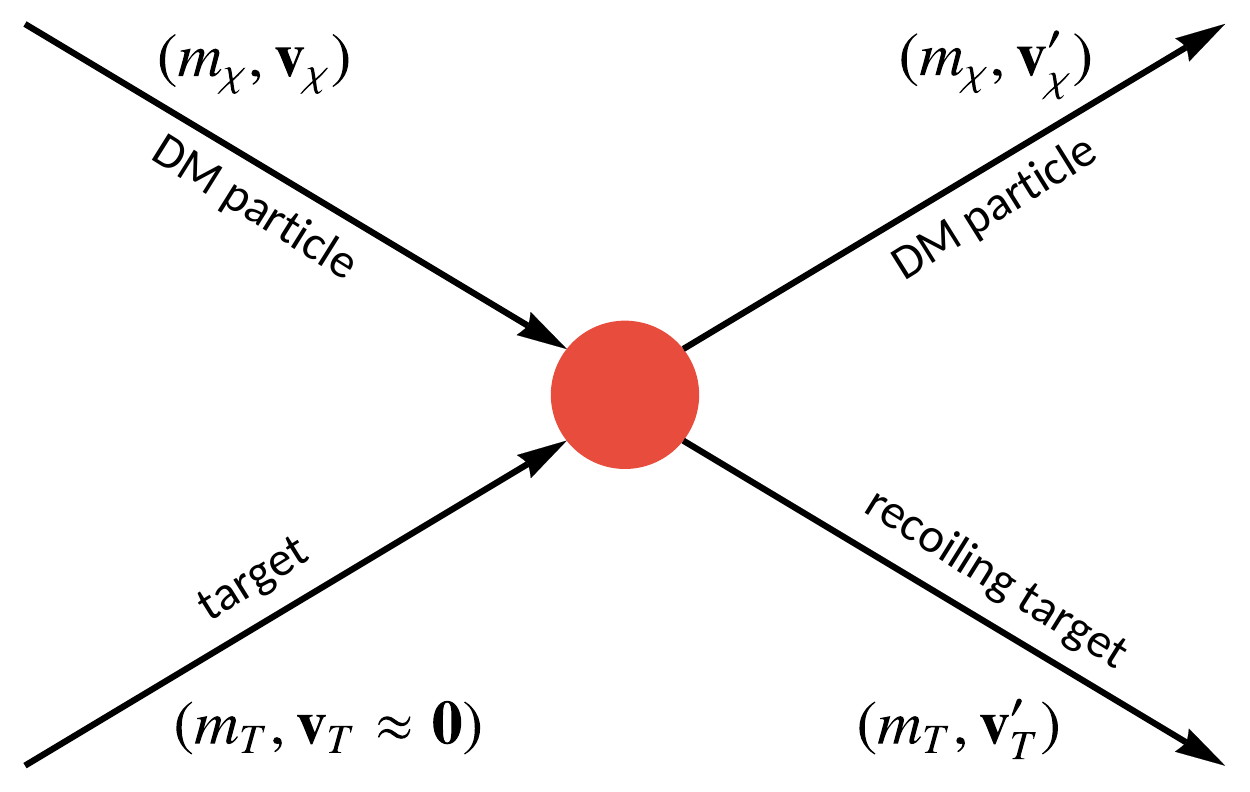}%
    \includegraphics[height=3.5cm]{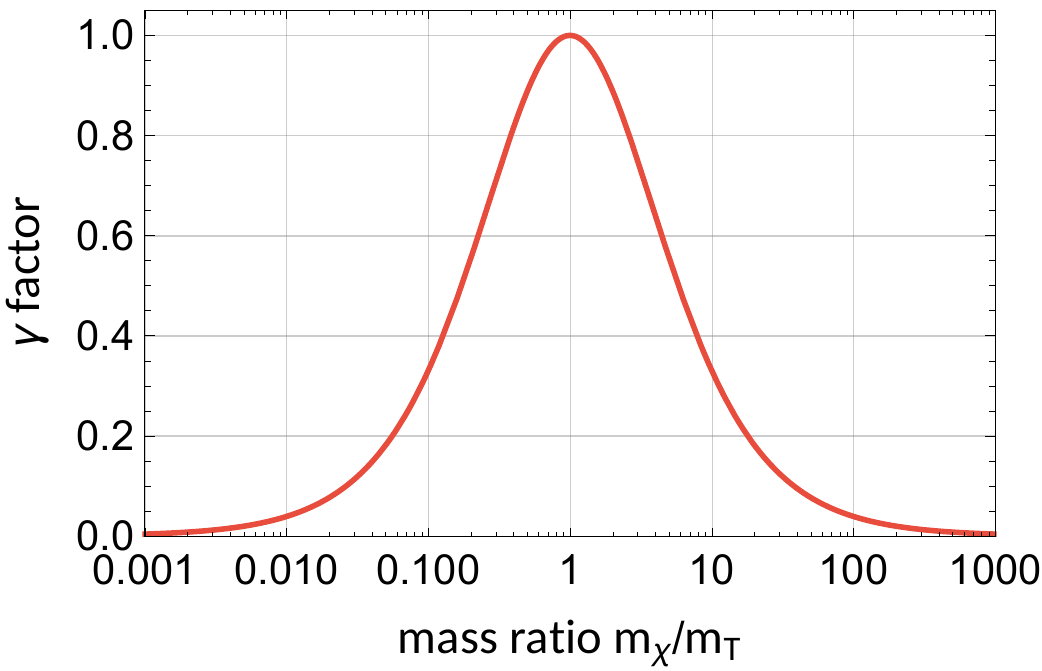}%
  }%
}
\setlength{\twosubht}{\ht\twosubbox}
   	\centering
   	\subcaptionbox{Sketch of the scattering.\label{fig:kinematics-a}}{%
  	\includegraphics[height=\twosubht]{plots/Kinematics_Sketch.pdf}%
	}
	\quad
	\subcaptionbox{The $\gamma$ factor.\label{fig:kinematics-b}}{%
	  \includegraphics[height=\twosubht]{plots/Kinematics_Gamma.pdf}%
	}
	\caption{Elastic scattering kinematics between a DM and a target particle.}
	\label{fig:kinematics}
\end{figure*}

A DM~particle of mass~$m_\chi$ and initial velocity~$\mathbf{v}_\chi$ scatters on a target nucleus of mass~$m_T$, as illustrated in figure~\ref{fig:kinematics-a}. Energy and momentum conservation alone determine the final velocities of both particles after the elastic scattering,
\begin{subequations}
\begin{align}
	\mathbf{v}_\chi^\prime &= \frac{m_T \left|\mathbf{v}_\chi-\mathbf{v}_T\right|}{m_T+m_\chi} \mathbf{n}+\underbrace{\frac{m_\chi\mathbf{v}_\chi+m_T\mathbf{v}_T}{m_T+m_\chi}}_{\text{\footnotesize velocity of the~\acs{CMS}}}\, ,\label{eq:vDMfinal sun}\\
\mathbf{v}_T^\prime &=-\frac{m_\chi \left|\mathbf{v}_\chi-\mathbf{v}_T\right|}{m_T+m_\chi} \mathbf{n}+\frac{m_\chi\mathbf{v}_\chi+m_T\mathbf{v}_T}{m_T+m_\chi}\, ,
\end{align}
\end{subequations}
For terrestrial nuclei, the relative velocity is completely dominated by the DM speed, $\left|\mathbf{v}_\chi-\mathbf{v}_T\right|\approx v_\chi\sim 10^{-3}$. For example, in the atmosphere the thermal velocities of oxygen/nitrogen molecules are well below 1~km/s. The two momenta in the second term's numerator can become comparable for very light DM. However, then the second term is negligible altogether. This leaves us with the approximation of
\begin{subequations}
\begin{align}
\mathbf{v}_\chi^\prime &\approx\frac{m_T v_\chi \mathbf{n}+m_\chi\mathbf{v}_\chi}{m_T+m_\chi}\qquad (\mathbf{v}_T\approx\mathbf{0})\label{eq:vDMfinal}\, ,\\
\mathbf{v}_T^\prime &\approx\frac{-m_\chi v_\chi \mathbf{n}+m_\chi\mathbf{v}_\chi}{m_T+m_\chi}\qquad (\mathbf{v}_T\approx\mathbf{0})\, .\label{eq:vTfinal}
\end{align}
\end{subequations}
We continue under the assumption of resting targets, keeping in mind that it will not hold for hot nuclei in the core of the Sun.

The only unknown part is the unit vector~$\mathbf{n}$, which points into the direction of the DM~particle's final velocity in the \ac{CMS}. This vector is determined by the \emph{scattering angle}~$\theta$, defined as the angle between the initial and final velocity of the DM~particle in the \ac{CMS},
\begin{align}
	\theta = \sphericalangle(\mathbf{v}_\chi,\mathbf{n})\in [0,\pi]\, .
\end{align}
The distribution of the scattering angle critically depends on the specific interaction between the DM and the nuclei and will be discussed in chapter~\ref{ss:scattering angle}. Once a scattering angle is specified, the \emph{momentum transfer}~$\mathbf{q}$ to the nucleus is given by
\begin{subequations}
\begin{align}
	\mathbf{q} &= m_T \left(\mathbf{v}_T^\prime-\mathbf{v}_T\right)\approx m_T \mathbf{v}_T^\prime\qquad (\mathbf{v}_T\approx\mathbf{0})\, ,\\
	\Rightarrow q^2 &= 2\mu_{\chi T}^2v_\chi^2\left(1-\cos\theta\right)\, ,\label{eq: momentum transfer}
\end{align}
\end{subequations}
where $\mu_{ij}\equiv \frac{m_i m_j}{m_i+m_j}$ is the reduced mass of a two body system. For an incoming speed the momentum transfer is bounded by
\begin{align}
	q^2_{\rm max} = 4\mu_{\chi T}^2v_\chi^2\, .
\end{align}
The kinetic quantity closer to the measured observable is the \emph{recoil energy}~$E_R$, i.e. the kinetic energy transferred to the nucleus in a collision,
\begin{align}
	E_R&\equiv E_T^\prime - E_T = \frac{q^2}{2m_T} \approx \frac{1}{2}m_T v_T^{\prime 2}\qquad (\mathbf{v}_T\approx\mathbf{0})\nonumber\\
	&=\frac{m_T m_\chi^2 v_\chi^2}{(m_T+m_\chi)^2}(1-\cos\theta) = \gamma E_\chi \frac{1-\cos\theta}{2}\, \quad \text{with }\gamma \equiv \frac{4\mu_{\chi T}^2}{m_\chi m_T}\, . \label{eq: nuclear recoil energy}
\end{align}
For an incoming speed the nuclear recoil energy is bounded by
\begin{align}
	E_R^{\rm max} = \gamma E_{\chi} = \frac{2\mu^2_{\chi T}v_{\chi}^2}{m_T} \label{eq:maximumrecoil}\, .
\end{align}
 This determines the minimum speed~$v_{\rm min}(E_R)$, for which a DM~particle is capable to cause a nuclear recoil of energy~$E_R$,
\begin{align}
	v_{\rm min}(E_R) =\sqrt{\frac{E_R m_T}{2 \mu_{\chi T}^2}}\, . \label{eq: vMin nuclear}
\end{align}
Furthermore, eq.~\eqref{eq:maximumrecoil} illustrates the relevance of the $\gamma$ factor. It is the maximum fraction of energy, the DM~particle can possibly transfer to the target. As shown in figure~\ref{fig:kinematics-b}, the factor $\gamma$ is close to one, if the two particles' masses are similar and exactly one if they are degenerate. In this case, the DM~particle may lose all its kinetic energy in a single scattering. This can also be seen from the deceleration of the DM~particle,
\begin{align}
	\frac{v_\chi^\prime}{v_\chi} = \sqrt{1-\gamma\frac{1-\cos\theta}{2}}<1\, .
\end{align}

In summary, the energy transfer is governed by the ratio of masses and the interaction type, which determines the distribution of the scattering angle. 

\subsection{Inelastic DM-electron scatterings}
\label{ss: kinematics electron}
For direct searches of sub-GeV DM~particles, it is possible to look for inelastic collisions between~DM and an electron of an atom. The kinematics are non-trivial since the bound electron's momentum is not uniquely defined, and there is no longer a direct relation between the final electron's recoil energy and the momentum transfer, as eq.~\eqref{eq: nuclear recoil energy} for elastic nuclear scatterings~\cite{Essig:2015cda}. The total energy transferred to the electron can be expressed as the energy lost by the~DM~particle,
\begin{subequations}
\begin{align}
	\Delta E_e &=-\Delta E_\chi = E_\chi- E_\chi^\prime\\
	&=\frac{m_\chi}{2}v_\chi^2-\frac{\left| m_\chi \mathbf{v}_\chi-\mathbf{q}\right|^2}{2m_\chi}\\
	&=\mathbf{v}\cdot\mathbf{q}-\frac{q^2}{2m_\chi}\, .\label{eq: electron energy}
\end{align}
\end{subequations}
Here, we neglected the fact that the atom also recoils as a whole. The maximum transferred energy~$\Delta E_e$ with respect to~$q$ is therefore
\begin{align}
	\Delta E_e^{\rm max} = \frac{1}{2}m_\chi v_\chi^2\, .
\end{align}
As opposed to elastic nuclear scatterings, where the~$\gamma$~factor of eq.~\eqref{eq:maximumrecoil} yields the maximum kinematically allowed relative energy transfer, a sub-GeV DM~particle is kinematically allowed to lose its entire energy in a DM-electron scattering, and the kinetic energy transfer is much more efficient for electron than for nuclear targets.

In analogy to eq.~\eqref{eq: vMin nuclear}, the minimum DM~speed kinematically required for a collision with momentum transfer~$q$ and transferred energy~$\Delta E_e$ is
\begin{align}
	v_{\rm min}(\Delta E_e,q) = \frac{\Delta E_e}{q}+\frac{q}{2m_\chi}\, .\label{eq: vMin electron}
\end{align} 
In sub-GeV DM-electron scatterings, the electron is not just the lighter particle. Noting that~$v_e\sim \alpha\sim \text{10}^{\text{-2}}$, the electrons are also faster by one order of magnitude and hence dominate the relative velocity.  Therefore, typical momentum transfers are of order~$q\sim \mu_{\chi e}|\mathbf{v}_\chi-\mathbf{v}_e|\simeq m_e \alpha \approx$~3.7~keV. The typical transferred energy is of the order of~eV, sufficient to ionize and excite atoms. 

\section{Describing DM-Matter Interactions}
\label{s:interactions}
At this point, all evidence in favour of the existence of DM is rooted in its gravitational influence on visible matter on astronomical scales, as discussed in chapter~\ref{s:evidence}. In contrast, we know next to nothing about possible non-gravitational interactions, which are probed by direct detection experiments. In the face of this ignorance, the question arises, how we should model and describe these interactions. One attempt is to look for UV~complete extensions of the~\ac{SM}, which are somehow well motivated and contain a DM~candidate particle. The most prominent examples have been discussed briefly in chapter~\ref{ss: DM candidates}. Due to the large number of proposed models, these model driven and dependent descriptions of DM-matter interactions hamper comparisons of different experiments. To avoid this, it is necessary to model the probed interactions in a more general framework. The use of an \ac{EFT} is a more suitable approach, since it is agnostic to particular proposals for extending the~\ac{SM}~\cite{Kurylov:2003ra}. It can be used to universally describe the low energy behavior of many proposed extensions of the~\ac{SM} and gives a general framework to model e.g. DM-quark interactions, independent on the underlying high energy degrees of freedom. Alternatively, it can make sense to formulate a simple prototype model with additional degrees of freedom and symmetries, so called~\textit{simplified models}, which can also be connected to more complete particle models.

\clearpage
\subsection{EFTs and DM-nucleus scatterings}
The Lagrangian of the~\ac{SM} can be extended by a new, stable, massive field~$\chi$, e.g. a massive Dirac fermion, which is the DM~particle. Its interaction with the~\ac{SM} quarks to leading order is written as Lorentz-invariant effective four-fermion operators,
\begin{align}
	\mathscr{L}_{\rm int}\supset \sum_q \alpha_q\, \left(\overline{\chi}\Gamma_\chi\chi\right)\, \left(\overline{q}\Gamma_q q\right)\, ,
\end{align}
where $\alpha_q$ are the effective~DM-quark couplings and~$\Gamma_i\in\left\{\mathds{I}, \gamma^\mu,\gamma^5\gamma^5\gamma^\mu,\sigma^{\mu\nu},\sigma^{\mu\nu}\gamma^5\right\}$ are the possible operators~\cite{Jungman:1995df,Cerdeno:2010jj,Freese:2012xd,Lisanti:2016jxe}. The mediator for this interaction is assumed to be heavier than the momentum transfers of the scatterings and was integrated out yielding a contact interaction. This operator needs to be mapped to a nucleon operator as described in~\cite{Jungman:1995df} in order to compute the relativistic matrix element for DM-nucleus scatterings. After taking the non-relativistic limit\footnote{Alternatively, the scattering cross sections can be described directly using~\ac{NREFT}~\cite{Fan:2010gt,Fitzpatrick:2012ix}.}, the resulting matrix element~$\mathcal{M}$ enters the differential cross section,
\begin{align}
	\frac{\dd \sigma_{N}}{\dd E_R} = \frac{1}{32\pi m_N m_\chi^2v_\chi^2}\overline{\left| \mathcal{M}\right|^2} \label{eq: differential cs general}\, .
\end{align}
Only the matrix element~$\mathcal{M}$ depends on the specifics of the particle physics model. The total scattering cross section is obtained by integrating over all recoil energies,
\begin{align}
	\sigma_N = \int_0^{E_R^{\rm max}}\dd E_R\; \frac{\dd \sigma_{N}}{\dd E_R}\, ,\quad\text{with }E_R^{\rm max} = \frac{2\mu_{\chi N}^2v_\chi^2}{m_N}\, . \label{eq: total cs general}
\end{align}

Typically, the leading order operators are divided into two categories, depending on the cross section's dependence on the nuclear spin. 

\paragraph{\ac{SI} interactions}
Spin independent interactions are mediated by scalar or vector fields, with~$\Gamma_{\chi / q}=\mathds{I}$ or~$\Gamma_{\chi / q}=\gamma^\mu$ respectively,
\begin{align}
	\mathscr{L}^{\rm SI}_{\rm int}= \sum_q \left[\underbrace{\alpha^S_q\, \left(\overline{\chi}\chi\right)\, \left(\overline{q} q\right)}_{\text{scalar}}+\underbrace{\alpha^V_q\, \left(\overline{\chi}\gamma_\mu\chi\right)\, \left(\overline{q}\gamma^\mu q\right)}_{\text{vector}}\right]\, ,
\end{align}
The matrix element for the scattering between a DM~particle and a nucleus~$N$ in the non-relativistic limit turns out as\footnote{For a more detailed derivation, we refer to~\cite{Jungman:1995df}.}
\begin{align}
	\mathcal{M}=4m_\chi m_N \left(f_p Z+f_n(A-Z)\right)F^{\rm SI}_N(q)\, ,
\end{align}
and the differential cross section can be evaluated to be
\begin{align}
	\frac{\dd\sigma^{\rm SI}_{N}}{\dd E_R} &=\frac{m_N}{2\pi v_\chi^2}\left[f_p Z+f_n(A-Z) \right]^2 \left|F^{\rm SI}_N\left(q\right)\right|^2_{q=\sqrt{2m_N E_R}}\, .
	\intertext{We can use the total DM-proton scattering cross section as a reference,}
	\sigma_p^{\rm SI}&=\frac{f_p^2\mu_{\chi p}^2}{\pi}\, .\label{eq: reference cross section SI}
\end{align}
Results of direct detection experiments are usually presented in terms of this reference cross section such that results from experiments with different target nuclei can be compared directly. The differential cross section in terms of~$\sigma_p^{\rm SI}$ is then
\begin{align}
	\frac{\dd\sigma^{\rm SI}_{N}}{\dd E_R} &=\frac{m_N\,\sigma_p^{\rm SI}}{2\mu_{\chi p}^2 v_\chi^2}\left[Z+\frac{f_n}{f_p}(A-Z) \right]^2 \left|F_N\left(q\right)\right|^2_{q=\sqrt{2m_N E_R}}\, .
\end{align}
For isospin conserving interactions, the coupling to protons and neutrons is identical~$(f_p=f_n)$ such that the differential cross section simplifies to
\begin{align*}
\left[Z+\frac{f_n}{f_p}(A-Z) \right]\rightarrow A\, ,	
\end{align*}
and the DM~particle essentially couples to the nuclear mass with a coherence enhancement~$\sim A^2$. Isospin invariance is a standard assumption in the context of direct detection. However, it is not a requirement, see e.g.~\cite{Feng:2011vu}. 

In most parts of this thesis, we will indeed assume that ~$f_p=f_n$. The exception is the dark photon model, a simplified model which we will consider in the context of DM-electron scattering experiments, where~$f_n=0$ such that the cross section is proportional to~$Z^2$.

\begin{figure*}
	\centering
	\includegraphics[width=0.66\textwidth]{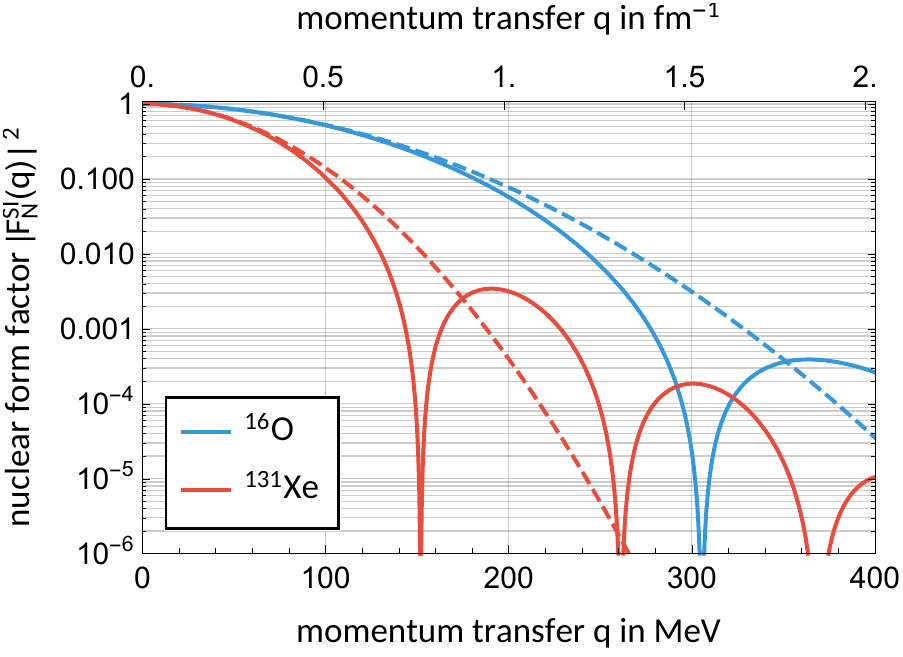}
	\caption{The nuclear Helm form factor (solid line) and its exponential approximation~(dashed line) for oxygen and xenon.}
	\label{fig: nuclear form factor}
\end{figure*}

The nuclear form factor~$F^{\rm SI}_N(q)$ describes the finite size of the nucleus and is defined as the Fourier transformed charge density. For large nuclei and large momentum transfers the~DM~particle does not scatter coherently on the nucleus and starts to resolve the nuclear structure. The loss of coherence for~\ac{SI} interactions can be described universally for all nuclei using the Helm form factor~\cite{Helm:1956zz},
\begin{subequations}
\label{eq: helm}
\begin{align}
	F^{\rm SI}_N(q)&=3\left(\frac{\sin (q r_n)}{(q r_n)^3}-\frac{\cos(q r_n)}{(q r_n)^2} \right)\exp\left(-\frac{q^2 s^2}{2}\right)\, ,\\
	\intertext{with}
	r_n &=\sqrt{c^2+\frac{7}{3}\pi^2a^2-5s^2}\, ,\\ c&=\left(1.23 A^{1/3}-0.6\right)\SI{}{\femto\meter}\, ,\\
	a&=\SI{0.52}{\femto\meter}\, ,\quad  s=\SI{0.9}{\femto\meter}\, .
\end{align}
\end{subequations}
While there are more involved and accurate ways to determine the nuclear form factor, the Helm form factor agrees with these newer results to high accuracy for the range of momentum transfers relevant for direct detections~\cite{Vietze:2014vsa}. In cases, where the momentum transfer is small, we can also use an exponential approximation of the Helm form factor,
\begin{align}
	F^{\rm SI}_N(q)\approx \exp\left[-\left(\frac{r_n^2}{10}+\frac{s^2}{2}\right)q^2\right]\, .\label{eq: helm approximation}
\end{align}
The Helm form factor and its approximation are shown for two examples in figure~\ref{fig: nuclear form factor}.

For light~DM with~$m_\chi\ll m_N$, the maximum momentum transfer in a collision on a nucleus is~$q_{\rm max}\approx 5 \left(\frac{m_\chi}{1\SI{}{\GeV}}\right)\SI{}{MeV}\approx \SI{0.03}{\per\femto\meter}$. It is therefore a good approximation, to neglect the nuclear form factor and set~$F_N(q)\approx 1$. In this limit, we can evaluate eq.~\eqref{eq: total cs general} and obtain the total scattering cross section,
\begin{align}
	\sigma^{\rm SI}_{N} = \sigma_p^{\rm SI}\left(\frac{\mu_{\chi N}}{\mu_{\chi p}}\right)^2 \left[Z+\frac{f_n}{f_p}(A-Z) \right]^2\, .
\end{align}
We can use the total cross section and rephrase the differential cross section for light~DM as
\begin{align}
	\frac{\dd\sigma^{\rm SI}_{N}}{\dd E_R} = \frac{\sigma_N^{\rm SI}}{E_R^{\rm max}}\, , \label{eq: diff cs SI light DM}
\end{align}
which shows that the recoil energies follow a uniform distribution. Since~$E_R$ is connected to the scattering angle via eq.~\eqref{eq: nuclear recoil energy}, we can already deduct that light~DM scatters on nuclei isotropically in the~\ac{CMS}. We will discuss this in more detail in chapter~\ref{ss:scattering angle}.

\paragraph{\ac{SD} interactions} Even though~\ac{SD} interactions play no major role in this thesis, we briefly introduce them at this point for completeness. The \ac{SD}~interaction arises from an axial-vector coupling between~DM and quarks,
\begin{align}
	\mathscr{L}^{\rm SD}_{\rm int}= \sum_q \alpha^A_q\, \left(\overline{\chi}\gamma^5\gamma^\mu\chi\right)\, \left(\overline{q} \gamma^5\gamma_\mu q\right)\, ,
\end{align}
with effective axial-vector DM-quark couplings~$\alpha_q^A$. The differential cross section for~\ac{SD}~DM-nucleus scatterings derived from this Lagrangian is~\cite{Engel:1992bf},
\begin{align}
	\frac{\dd \sigma_N^{\rm SD}}{\dd E_R} &= \frac{2m_N}{\pi v_\chi^2}\frac{J+1}{J}\left(f_p \langle S_p\rangle +f_n \langle S_N\rangle\right)^2 \left.F_N^{\rm SD}(q)^2\right|_{q=\sqrt{2m_N E_R}}\, .
	\intertext{ The total DM-proton cross section,~$\sigma_p^{\rm SD}=\frac{3\mu_{\chi p}^2f_p^2}{\pi}$ can serve as a reference cross section, similar to the~\ac{SI} case,}
	\frac{\dd \sigma_N^{\rm SD}}{\dd E_R}&=\frac{2m_N\sigma_p^{\rm SD}}{3\mu_{\chi p}^2v_\chi^2}\frac{J+1}{J}\left[\langle S_p\rangle +\frac{f_n}{f_p} \langle S_N\rangle\right]^2 \left.F_N^{\rm SD}(q)^2\right|_{q=\sqrt{2m_N E_R}}\, .
\end{align}
Here, $J$ is the nuclear spin, $f_n,\,f_p$ are again effective couplings to the nuclear spins, and $\langle S_p\rangle$,($\langle S_N\rangle$) is the isotope-specific average spin contributions of protons~(neutrons). Just as in the case of~\ac{SI} interactions, a nuclear form factor~$F_N^{\rm SD}(q)$ describes the nuclear structure. In this case, the spin structure depends on the specific isotope, and the nuclear form factor cannot be expressed universally~\cite{Bednyakov:2004xq,Bednyakov:2006ux}.

Compared to~\ac{SI} case, the~\ac{SD} interaction plays a subdominant role for direct detection. They lack the enhancement factor~$\sim A^2$, and the signal rate in a detector is consequently suppressed. Furthermore, most target isotopes have spin zero, as they include no unpaired nucleon, and the factors $J$, $\langle S_p\rangle$, and $\langle S_N\rangle$ often vanish.

\subsection{Simplified models and DM-electron scatterings}
\label{ss: dm electron interactions}
The idea that~DM consists of a single particle, and there exists nothing else than the~\ac{SM} and that particle, is certainly the minimal assumption. But it could be argued that the existence of a ``dark sector'' with more particles and interactions mediators is more likely, as it would mirror the diversity of the~SM sector. In that case, DM~could simply consist of the lightest stable of these dark particles or even consist of different dark particle species. The~\ac{EFT} approach used in the previous chapter applies best to the case, where the DM field is lighter than all other dark sector fields, including mediators, which can be integrated out. Then, the heavy fields would be kinematically inaccessible in experiments, and their indirect effect on the lighter degrees of freedom can be absorbed into effective operators. However, if there are fields lighter than the~DM~particle which mediate the interactions to~SM particles,~\textit{simplified models} are typically a more suitable description of the new interactions~\cite{Abercrombie:2015wmb,Abdallah:2015ter,DeSimone:2016fbz}.

Simplified models are prototype models which explicitly contain a small number of new degrees of freedom, in particular the mediators. As such, the simplified model approach lies conceptually between a UV-complete extensions of the~\ac{SM} and~\acp{EFT}. A good simplified model captures all the physical phenomena at the relevant energies and can be matched to more UV-complete models. The Lagrangian of such a simplified model of~DM contains one or more stable DM~candidates, a mediator coupling the visible to the invisible sector, and all renormalizable interactions consistent with the symmetries of the~SM and dark sector.

In the context of DM-electron scatterings, we are particularly interested in the case, where the interactions are mediated by an ultralight field. We set up a simplified model, where the~\ac{SM} is extended by a dark sector of a spin-1/2 DM~particle~$\chi$ of mass~$m_\chi$ and a new gauge group~$U(1)_D$. The new gauge boson, the so-called dark photon~$A^\prime$, can mix kinetically with the $U(1)$ gauge bosons of the~\ac{SM}, which would act as a portal between the two sectors~\cite{Galison:1983pa,Holdom:1985ag}. The dark sector's Lagrangian can be written as
\begin{subequations}
\begin{align}
	\mathscr{L}_D &= \bar{\chi} (i\gamma^{\mu}D_\mu-m_{\chi}) \chi  + \frac{1}{4}F'_{\mu \nu}F'^{\mu \nu} + m_{A^\prime}^{2} A'_{\mu}A'^{\mu}+ \varepsilon F_{\mu \nu}F'^{\mu \nu}\, ,
	\intertext{with the covariant derivative involving the new gauge coupling~$g_D$,}
	D_\mu&=\partial_\mu - i g_D A'_\mu\, .
\end{align}
\end{subequations}
The dark photon is assumed to have acquired a mass~$m_{A^\prime}$, the new symmetry might e.g. be spontaneously broken by a Higgs mechanism. Furthermore, its kinetic mixing is parametrized by~$\epsilon$. 

\paragraph{DM-nucleus scatterings}
Similarly to the previous case, the mediator's interaction to protons and neutrons can be expressed in terms of effective couplings~$f_i$~\cite{Kaplinghat:2013yxa}
\begin{align}
	\mathscr{L}_{\rm int} = e A^\prime_\mu \left(f_p \bar{p}\gamma^\mu p+f_n \bar{n}\gamma^\mu n\right)\, ,
\end{align}
which in turn can be related to the mixing parameters to the photon or~Z~boson,
\begin{align}
	f_p &=\epsilon_\gamma+\frac{1-4\sin^2\theta_W}{4\cos\theta_W \sin\theta_W}\epsilon_Z\, ,\quad f_n = -\frac{1}{4\sin\theta_W \cos\theta_W}\epsilon_Z\, .
\end{align}
This is also where the weak mixing angle~$\theta_W$ enters. The differential cross section for elastic DM-nucleus scatterings in terms of the momentum transfer~$q$ reads
\begin{align}
	\frac{\dd \sigma_{N}}{\dd q^2} &= \frac{4\pi \alpha\alpha_D}{(q^2+m_{A^\prime}^2)^2}\frac{1}{v_\chi^2} \left(f_p Z+f_n(A-Z)\right)^2\left|F_N(q)\right|^2\, .\label{eq: diff cs nucleus dark photon}
\end{align}
Here,~$\alpha\equiv\frac{e^2}{4\pi}$ and~$\alpha_D\equiv\frac{g_D^2}{4\pi}$ are the two fine structure constants, and~$F_N(q)$ is the same nuclear form factor as for~\ac{SI} interactions. Under the assumption that the dark photon mixes exclusively with the photon, i.e.~$\epsilon\equiv \epsilon_\gamma\neq 0$ and~$\epsilon_Z=0$, we find that the mediator naturally couples to charged particles only~($f_n=0$). In analogy to eq.~\eqref{eq: reference cross section SI}, we define a reference cross section,
\begin{align}
	\sigma_p &\equiv \frac{16\pi \alpha\alpha_D\epsilon^2 \mu_{\chi p}^2}{(q_{\rm ref}^2+m_{A^\prime}^2)^2}\, .\label{eq: reference cross section dark photon proton}
\end{align}
This cross section depends on a reference momentum transfer~$q_{\rm ref}$ which can be chosen arbitrarily. For contact interactions, this dependence vanishes if~$m_{A^\prime}^2\gg q^2$. With this reference cross section, the differential cross section of eq.~\eqref{eq: diff cs nucleus dark photon} takes the form

\begin{align}
	\frac{\dd \sigma_{N}}{\dd q^2} &= \frac{\sigma_p}{4\mu_{\chi p}^2v_\chi^2}F_{\rm DM}(q)^2\,F_N(q)^2\, Z^2\, ,\label{eq:dsdq2 nucleus}
		\intertext{where the~$q$~dependence is absorbed into the DM form factor}
	F_{\rm DM}(q) &\equiv \frac{q_{\rm ref}^2+m_{A^\prime}^2}{q^2+m_{A^\prime}^2}\, ,\label{eq: DM form factor}
\end{align}
We will use the dark photon model to study sub-GeV DM~particles, therefore the nuclear form factor can be neglected via~$F_N(q)\approx 1$. In addition, we focus on the two limits of ultraheavy and ultralight mediators, which can now be conveniently expressed in terms of the DM~form factor,
\begin{align}
	F_{\rm DM}(q) =
	\begin{cases}
		1\, ,\;&\text{for }m_{A^\prime}^2\gg q^2_{\rm max}\, ,\\
		\left(\frac{q_{\rm ref}}{q}\right)^2\, ,\;&\text{for }m_{A^\prime}^2\ll q^2_{\rm max}\, .
	\end{cases}
\end{align}
These are the two form factors of contact and long-range interactions respectively. We will also consider electric dipole interactions, characterized by
\begin{align}
	F_{\rm DM}(q) =\frac{q_{\rm ref}}{q}\, .
\end{align}
This type of interaction does not arise in the dark photon model, instead it originates from the operator~$\bar{\chi}\sigma_{\mu\nu}\gamma^5\chi\,F^{\mu\nu}$~with~$\sigma_{\mu\nu}=\frac{i}{2}\commutator{\gamma_\mu}{\gamma_\nu}$~\cite{Sigurdson:2004zp}. We will also consider results for electric dipole interactions in chapter~\ref{ss: constraints electron}.

\paragraph{DM-electron scatterings}
The differential cross section for DM-electron collisions is given by
\begin{align}
	\frac{\dd \sigma_{e}}{\dd q^2} &= \frac{4\pi \alpha\alpha_D\epsilon_\gamma^2}{(q^2+m_\phi^2)^2}\frac{1}{v_\chi^2}\, .
	\intertext{We introduce the reference DM-electron scattering cross section,}
	\sigma_e &\equiv \frac{16\pi \alpha\alpha_D\epsilon^2 \mu_{\chi e}^2}{(q_{\rm ref}^2+m_{A^\prime}^2)^2} \\
	\Rightarrow \frac{\dd \sigma_{e}}{\dd q^2} &=\frac{\sigma_e}{4\mu_{\chi e}^2v^2}F_{\rm DM}(q)^2\, , \label{eq: dsdq2 electron}
\end{align}
The reference cross section is conventionally set to the typical momentum transfer of DM-electron scatterings, hence~$q_{\rm ref}=\alpha m_e$. Despite this arbitrariness, the reference momentum transfer cancels in the ratio of the two cross sections,
\begin{align}
\frac{\sigma_p}{\sigma_e} = \left(\frac{\mu_{\chi p}}{\mu_{\chi e}}\right)^2\, .\label{eq:sigma ratio}
\end{align}
This ratio shows a hierarchy of cross sections for DM~masses above a few~MeV. In the dark photon model with kinetic mixing, DM-electron scatterings go hand in hand with much stronger DM-nucleus interactions as~$\sigma_p\gg \sigma_e$. Terrestrial effects due to elastic scatterings on underground nuclei can have strong implications for direct detection experiments, even if the nuclear recoils are undetectable and the experiment is probing DM-electron interactions. This has been studied e.g. in~\cite{Lee:2015qva}, as well as in~\ref{paper1} and~\ref{paper5}.

\paragraph{Charge screening}
\begin{figure*}
	\centering
	\includegraphics[width=0.66\textwidth]{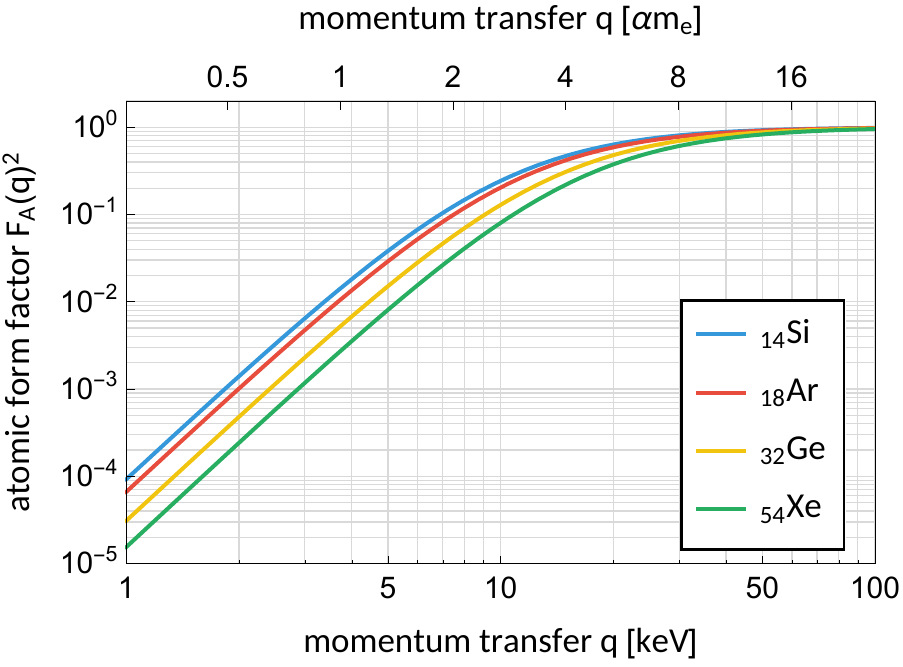}
	\caption{The atomic form factor, given in eq.~\eqref{eq: atomic form factor}, which describes the screening of the nuclear electric charge for silicon, argon, germanium, and xenon nuclei.}
	\label{fig: atomic form factor}
\end{figure*}
The cross sections in eq.~\eqref{eq:dsdq2 nucleus} and~\eqref{eq: dsdq2 electron} apply to free nuclei and electrons with their charges in isolation. In solids however, the electric charges are screened by the surrounding  on larger distances. In the end, a solid is electrically neutral. For DM-nucleus scatterings, we can re-scale the nuclear charge to the effective charge using an atomic form factor,
\begin{align}
	Z \rightarrow Z_{\rm eff} &= F_A(q) \times Z\, ,
\end{align}
with $\lim\limits_{q\rightarrow \infty}F_A(q)=1$ and $F_A(0)=0$. The atomic form factor decreases the effective nuclear charge on longer distances, hence for low momentum transfers. One particularly compact and simple form factor, which approximates the more elaborate Thomas-Fermi elastic form factor, is given in~\cite{Schiff:1951zza,Tsai:1973py} and reads
\begin{align}
	F_A(q) &= \frac{a^2 q^2}{1+a^2q^2}\, , \label{eq: atomic form factor}
	\intertext{where~$a$ is the Thomas-Fermi radius,}
	a&= \frac{1}{4}\left(\frac{9\pi^2}{2Z}\right)^{1/3}a_0\approx \frac{0.89}{Z^{1/3}}a_0\, ,
\end{align}
written in terms of the Bohr radius~$a_0\equiv\frac{1}{m_e\alpha}\approx\SI{5.29e-11}{\meter}$. This corresponds to a screened Coulomb potential~$\frac{Ze}{r}e^{-r/a}$. In turn, the electron charge is screened by the nuclei. Following~\cite{Tsai:1973py}, the corresponding atomic form factor takes the same form as eq.~\eqref{eq: atomic form factor}, but we have to use~$a^\prime\approx\frac{5.28}{Z^{2/3}}a_0$ instead of the Thomas-Fermi radius~$a$.

\paragraph{Total cross section}
For ultralight mediators and a DM~form factor~$F_{\rm DM}\sim 1/q^2$, the Rutherford type differential cross section diverges in the~IR, and the total cross section can not simply be obtained by integrating eq.~\eqref{eq:dsdq2 nucleus}. However, the charge screening removes these divergences, and the atomic form factor drives the differential cross section to zero for~$q\rightarrow 0$ as shown for some examples in figure~\ref{fig: atomic form factor}. Indeed, the total cross section
\begin{align}
	\sigma_{N} &= \int\limits_0^{q^2_{\rm max}}\dd q^2\;\frac{\dd \sigma_{N}}{\dd q^2}\left|F_A(q)\right|^2\, , \label{eq:totalCS}
\end{align}
is finite and can be evaluated analytically,
\begin{align}
	\sigma_{N} &=\sigma_p\left(\frac{\mu_{\chi N}}{\mu_{\chi p}}\right)^2Z^2\nonumber\\
	&\;\times
	\begin{cases}
		\left(1+\frac{1}{1+a^2 q^2_{\rm max}}-\frac{2}{a^2q^2_{\rm max}}\log(1+a^2 q^2_{\rm max})\right)\, , &\text{for }F_{\rm DM}(q)=1\, ,\\[2ex]
		\frac{q_{\rm ref}^2}{q^2_{\rm max}}\left(\log(1+a^2q^2_{\rm max})-\frac{a^2 q^2_{\rm max}}{1+a^2 q^2_{\rm max}}\right)\, , &\text{for }F_{\rm DM}(q)\sim\frac{1}{q}\, ,\\[2ex]
		\frac{a^4 q_{\rm ref}^4}{(1+a^2q^2_{\rm max})}\, , &\text{for }F_{\rm DM}(q)\sim \frac{1}{q^2}\, .
	\end{cases}
\end{align}
The first line is the usual result for contact interactions of low-mass~DM without charge screening. The case-specific factors depend on the screening length~$a$. For contact interaction it is of order one for~$m_{\chi}>$~100~MeV, in other words charge screening is an irrelevant effect for heavier~DM. Note that the arbitrary reference momentum transfer, appearing in the other two cases, cancels out due to the arbitrary definition of the reference cross section in eq.~\eqref{eq: reference cross section dark photon proton}.

\section{Recoil Spectra and Signal Rates}
\label{s: recoil spectra}

\subsection{Nuclear recoil experiments}
Once the DM flux and DM-matter interaction is specified, we can make predictions for direct detection experiments. The differential scattering rate between halo DM~particles and target nuclei of mass~$m_N$ per unit time and mass is given by~\cite{Lewin:1995rx}
\begin{subequations}
\begin{align}
	\dd R &= \underbrace{\frac{1}{m_N}}_{\text{number of targets per unit mass}} \times \underbrace{\sigma_N}_{\text{scattering cross section}} \times \underbrace{v \dd n_\chi}_{\text{differential DM flux}}\, ,
	\intertext{substituting eq.~\eqref{eq: differential density} and~\eqref{eq:fearth} results in}
		&=\frac{1}{m_N}\frac{\rho_\chi}{m_\chi} \sigma_N v f_\chi (\mathbf{v})\dd^3\mathbf{v}\, .
\end{align}
\end{subequations}
The recoil spectrum, i.e. the distribution of events over the recoil energies~$E_R$, requires the differential cross section and integration over velocities for which a DM~particle is kinematically able to cause that recoil,
\begin{align}
	\frac{\dd R}{\dd E_R} &= \frac{1}{m_N}\frac{\rho_\chi}{m_\chi}\iiint\dd^3\mathbf{v}\;  v f_\chi (\mathbf{v})\frac{\dd\sigma_N}{\dd E_R} \Theta(v-v_{\rm min}(E_R))\, . \label{eq: nuclear recoil spectrum}
\end{align}
For a given recoil energy, the step function limits the integral to the kinematically allowed velocities. The minimum speed~$v_{\rm min}$ is given by the kinematic relation of~\eqref{eq: vMin nuclear}.

If we are not interested in direction detection, we can integrate out the directional information of the distribution and compute the recoil spectrum with the marginal speed distribution,
\begin{align}
 \frac{\dd R}{\dd E_R}&= \frac{1}{m_N}\frac{\rho_\chi}{m_\chi}\int_{v>v_{\rm min}(E_R)}\dd v\;  v f_\chi (v)\frac{\dd\sigma_N}{\dd E_R}\, . \label{eq: nuclear recoil spectrum speed}
\end{align}
If the cross section $\sigma_N$ does not explicitly depend on the DM~speed $v$, then the differential cross section scales as $\frac{\dd\sigma_N}{\dd E_R}\sim \frac{1}{v^2}$ and
\begin{subequations}
 \label{eq: eta function}
\begin{align}
	\frac{\dd R}{\dd E_R} \sim \eta(v_{\rm min})&\equiv\iiint\dd^3\mathbf{v} \frac{f_\chi(\mathbf{v})}{v} \Theta(v-v_{\rm min})\\
	&=\int \dd v\; \frac{f_\chi(v)}{v} \Theta(v-v_{\rm min})\, .
\end{align}
\end{subequations}
\begin{figure*}
	\centering
	\includegraphics[width=.66\textwidth]{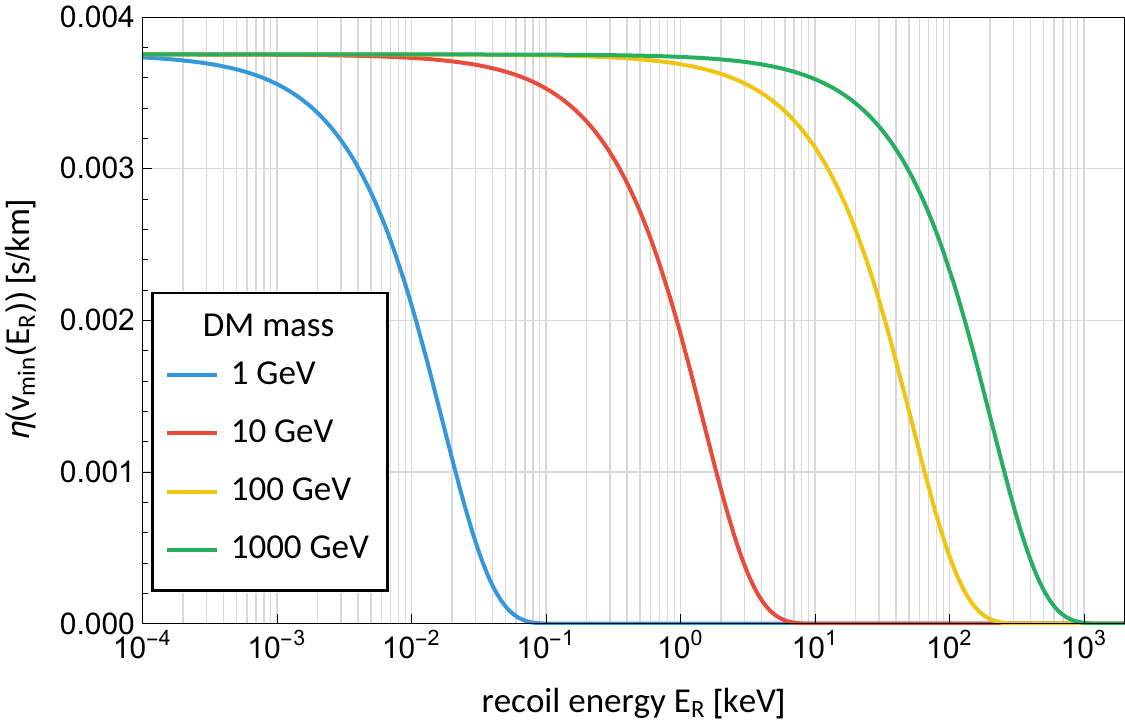}
	\caption{Examples of $\eta(v_{\rm min}(E_R))$ with varying DM~masses for a \isotope{Xe}{131} target.}
	\label{fig:eta}
\end{figure*}
Taking the distribution of the \ac{SHM} from~\eqref{eq:fearth}, the function $\eta(v_{\rm min})$ can be integrated analytically~\cite{Savage:2006qr},
\begin{subequations}
\label{eq:etav}
\begin{align}
	\eta(v_{\rm min}) &=
	\begin{cases}
 		\frac{1}{v_0 y} & \text{for }z<y\, ,\; x<|y-z|\, ,\\
 		\frac{1}{2N_{\rm esc}v_0 y}&\bigg[\erf{x+y}-\erf{x-y}-\frac{4}{\sqrt{\pi}}y e^{-z^2} \bigg]\\
 		& \text{for }z>y\, ,\; x<|y-z|\, ,\\
 		\frac{1}{2N_{\rm esc}v_0 y}&\bigg[\erf{z}-\erf{x-y}-\frac{2}{\sqrt{\pi}}\left(y+z-x\right) e^{-z^2} \bigg]\\
 		& \text{for }|y-z|<x<y+z\, ,\\
 		0 &\text{for } x>y+z\, ,
 	\end{cases}
 	\intertext{with}
 	&x\equiv \frac{v_{\rm min}}{v_0}\, ,\quad y\equiv \frac{v_{\oplus}}{v_0}\, ,\quad z\equiv \frac{v_{\rm esc}}{v_0}\, .
 \end{align}
\end{subequations}
Some examples for a xenon target and different DM masses are shown in figure~\ref{fig:eta}. The $\eta$ function also indicates the fact that a direct detection experiment cannot probe arbitrarily low DM masses. Every experiment comes with a recoil threshold $E_R^{\rm thr}$, below which recoil energies are too low to be detected. If even the fastest halo particles are kinetically incapable to cause a nuclear recoil above threshold, the experiment loses sensitivity to this mass. Using eq.~\eqref{eq:maximumrecoil}, the minimal probed DM mass is given by
\begin{align}
	m_\chi^{\rm min} = \frac{m_N}{\sqrt{\frac{2m_N}{E_R^{\rm thr}}}v_{\rm max}-1}\, ,\label{eq:minimal DM mass}
\end{align}
where the maximal speed of the galactic halo is the sum of the galactic escape velocity and the observer's relative speed, $v_{\rm max} = v_{\rm esc} + v_\oplus$. To extend the experiments sensitivity towards lighter DM, one needs to either lower the threshold or use lighter targets. In chapter~\ref{c:sun}, we will come back to this equation and present a third option concerning $v_{\rm max}$, which involves the Sun.

For an experiment with multiple target species of mass fractions $f_i$ we can simply add the spectra. The total scattering event rate $R$ for a given threshold is then
\begin{align}
	R &= \sum_i f_i\int\limits_{E_{\text{thr}}}^{E_{R,i}^{\rm max}}\dd E_R\;\frac{\dd R_i}{\dd E_R}\, .\label{eq:idealtotalrate}
\end{align}
Lastly, we have to multiply the total signal rate by the \emph{exposure}~$\mathcal{E}$ in order to obtain the expected total number of events~$N$.
The exposure can be understood as the product of the target mass and the run time of the experiment, since the signal rate is given in events per unit mass and time.
\begin{align}
	N=\mathcal{E} \times R\, .
\end{align}
Experimental observations can be interpreted statistically by comparison to the spectra and number of events for a given DM mass and cross section, as we will see in chapter~\ref{s:statistics}. 
 
\paragraph{Detector effects}The previous expressions are the theoretical spectra and rates. Real detectors however have a finite energy resolution, which might depend on the recoil energy, finite detection efficiencies, and do not directly measure recoil energies but e.g. the number of photons in~\acp{PMT}. The next step is to relate a theoretical spectrum to an observable spectrum.

Underlying eq.~\eqref{eq:idealtotalrate} is the assumption that the deposited energy corresponds \emph{exactly} to the nuclear recoil energy. Instead a fraction of the recoil energy will be lost into phonons, and the deposited energy $E^\prime$ is less than the true recoil energy $E_R$, which can be expressed by the quenching factor~$Q\equiv E^\prime/E_R$. In practice, this factor can often be taken as a constant over the relevant energy ranges. Furthermore, obtain a more realistic spectrum in terms of the actually detected energy~$E_D$ by including the detector's energy resolution~$\sigma_E(E_D)$ through convolution with a Gaussian and detection efficiency~$\epsilon(E_D)$,
\begin{subequations}
\begin{align}
\frac{\dd R}{\dd E_D} &= \int\limits_{0}^{\infty} \dd E^\prime \; \mathcal{K}(E_D,E^\prime)\sum_i f_i \left.\frac{\dd R_i}{\dd E_R}\right|_{E_R = E^\prime/Q_i}	\, . \label{eq: dRdE}
\intertext{with the detector response function}
\mathcal{K}(E_D,E^\prime)&\equiv\epsilon(E_D)\,\mathrm{Gauss}(E_D|E^\prime,\sigma_E(E_D))\, .
\end{align}
\end{subequations}
Alternatively, the detector might use \acp{PMT}, which measure small discrete photon counts $n$, which are Poisson distributed. The spectrum in this case can be related to the theoretical recoil spectrum via
\begin{align}
	\frac{\dd R}{\dd n} &=\sum_i f_i \int\limits_{E_{\rm thr}}^{E_{R\text{max}}}\dd E_R\; \mathrm{Poiss}\left(n|\nu(E_R)\right)\frac{\dd R_i}{\dd E_R}\, ,\label{eq:dRdn}
\intertext{where $\nu(E_R)$ is the expected number of signals for a given recoil energy. In the end, we can sum over all $n$ to get the total signal rate,}	
R&=\sum_{n=1}^\infty\frac{\dd R}{\dd n}\, .
\end{align}

\subsection{Detection of sub-GeV~DM with electron scatterings}
\label{ss: direct detection electron}
In chapter~\ref{ss: low mass DM detection}, we discussed DM-electron scatterings as a promising detection channel for sub-GeV~DM. An incoming DM~particle could ionize or excite an electron bound in an atom causing a detection signal. In this section, we will review how to compute spectra and signal rates for these events for liquid noble gas~\cite{Essig:2012yx,Essig:2017kqs,Agnes:2018oej} and semiconductor targets~\cite{Essig:2015cda,Lee:2015qva}.

Compared to nuclear recoils, the computation of the DM-electron scattering cross section is complicated by the quantum nature of the electrons bound in atoms. The electron's momentum is indeterminate, and there is no one-to-one relation between the momentum transfer and the deposited energy. As we showed in chapter~\ref{ss: kinematics electron}, the kinematics of this scattering allows the DM~particle to deposit its entire kinetic energy in a single interaction. With~$\Delta E_e\sim\mathcal{O}$(10-100)eV for~MeV scale~DM, there is an overlap of energy scales with the atomic scales. The electronic structure details of the initial bound state and the final outgoing states can be absorbed into an ionization/excitation form factor, which then enters the signal rate. The computation of these form factors is typically involved. Fortunately, they do not depend on the specific DM~model and have been tabulated and published for xenon and semiconductor targets~\cite{FormfactorLink}.

\paragraph{Liquid noble targets}
We discussed two-phase noble target detectors in chapter~\ref{ss: direct detection nuclear}, where the two time separation of the two scintillation signals (`S1' and~`S2') can distinguish DM-nucleus collisions from background. However, it turns out that experiments such as XENON or DarkSide-50 can also probe DM-electron interactions. After being ionized by a sub-GeV DM~particle, the primary electron drifts towards the gas-phase of the target. It may scatter on other atoms and cause the ionization of secondary electrons. Upon reaching the gas-phase, these electrons cause a scintillation signal (S2). The signature of DM-electron scatterings in two-phase xenon or argon detectors are therefore `S2 only' events, where no initial S1 scintillation photons were detected.

The differential ionization rate for atoms of mass~$m_N$ with an ionized electron of final kinetic energy~$E_e=k^{\prime 2}/(2m_e)$ is given~\cite{Kopp:2009et,Essig:2011nj,Essig:2012yx,Essig:2015cda} by
\begin{subequations}
\label{noble}
\begin{align}
	\frac{\dd R_{\rm ion}}{\dd E_e} & = \frac{1}{m_N} \frac{\rho_{\chi}}{m_{\chi}}\sum_{nl} \frac{\dd \langle \sigma_{\rm ion}^{nl}v\rangle}{\dd E_e}\, ,
\end{align}
with the differential thermally averaged ionization cross section,
\begin{align}
	\frac{\dd \langle \sigma_{\rm ion}^{nl}v\rangle}{\dd E_e}&=\frac{\sigma_e}{8\mu_{\chi e}^2E_e}\int\dd q\,q\left|F_{\rm DM}(q)\right|^2\left|f_{\rm ion}^{nl}(k^\prime,q)\right|^2\eta\left(v_{\rm min}(\Delta E_e,q)\right)\, ,
\end{align}
\end{subequations}
where~$\Delta E_e = E_e+|E_B^{nl}|$. The minimum speed~$v_{\rm min}(\Delta E_e,q)$ is given by eq.~\eqref{eq: vMin electron}. The sum runs over the quantum numbers~$(n,l)$ identifying the atomic shells with binding energy~$E_B^{nl}$. The ionization form factor~$f_{\rm ion}^{nl}(k^\prime,q)$ is essentially the overlap of the initial and final electron wave functions,
\begin{align}
	|f_{\rm ion}^{nl}(k^\prime,q)|^2=\frac{2k^{\prime 2}}{(2\pi)^3}\sum_{\text{occupied states}}\sum_{l^\prime m^\prime}\left|\int\dd^3\mathbf{x}\,\tilde{\psi}^{*}_{k^\prime l^\prime m^\prime}(\mathbf{x})\psi_i(\mathbf{x})e^{i\mathbf{q}\cdot \mathbf{x}}\right|^2\, .
\end{align}
 The initial state of the target electron is approximated as a single particle state of an isolated atom, the wave function~$\psi_i(\mathbf{x})$ can be evaluated numerically via tabulated~\ac{RHF} bound wave functions~\cite{Bunge:1993jsz}. The first sum runs over all degenerate occupied initial states. The final state wave function~$\tilde{\psi}_{k^\prime l^\prime m^\prime}(\mathbf{x})$ is obtained by solving the Schr\"odinger equation for the hydrogen-type potential of the new ion. The second sum runs over the final state angular momentum quantum numbers~$l^\prime$ and~$m^\prime$. The details of the ionization form factors' derivation are beyond the scope of this thesis and can be found in~\cite{Essig:2011nj,Essig:2012yx,Agnes:2018oej}.

The next step is to compute the number of electrons, which reach the gas phase and cause the S2 signal. The initial electron can ionize more atoms and cause the release of additional `primary quanta',
\begin{align}
	n^{(1)} = \left\lfloor \frac{E_e}{W}\right\rfloor\, ,
\end{align}
where~$W=\SI{13.8}{\eV}$ is the average energy to produce a single quanta for xenon~\cite{Doke:2002oab}. Furthermore, the ionization of an inner shell electron brings about the emission of a photon and even more `secondary quanta'. For an quantum leap between two atomic shells with binding energies~$E_i$ and~$E_j$, the number of secondary quanta from de-excitation is
\begin{align}
	n^{(2)}=\left\lfloor \frac{E_i-E_j}{W}\right\rfloor\, .
\end{align}
The number of secondary quanta depends on the atomic shell. In table~\ref{tab: xenon shells}, we list the number of secondary quanta for the different shells of xenon.

Assuming that the ionized xenon atom never recombines after ionization, the electron fraction of the primary quanta is~$f_e=0.83$~\cite{Essig:2012yx}. Under the same assumption, the number of electrons is~$n_e=n^\prime+n^{\prime\prime}$, where~$n^\prime=1$ is the primary electron, and~$n^{\prime\prime}$ follows a binomial distribution with~$n^{(1)}+n^{(2)}$ trials and success probability~$f_e$. Therefore the spectrum in terms of electrons is
\begin{align}
	\frac{\dd R}{\dd n_e} = \int\dd E_e\; \frac{\dd R}{\dd E_e}\;\mathrm{Binomial}\left(n^{(1)}+n^{(2)},f_e|n_e-1\right)\, .
\end{align}
The electrons are not detected directly, instead their scintillation light is observed by~\acp{PMT}, which count the \acp{PE}. We have to perform a last conversion into the \ac{PE}~spectrum of the actual S2 signal. One electron causes a normal distributed number of~PEs. The Gaussian is determined by its mean~$n_e\mu_{\rm PE}$ and width~$\sqrt{n_e}\sigma_{\rm PE}$. Hence,the final \ac{PE}~spectrum reads
\begin{align}
	\frac{\dd R}{\dd n_{\rm PE}} = \sum_{n_e=1}^{\infty} \frac{\dd R}{\dd n_e}\mathrm{Gauss}(\mu_{\rm PE}n_e,\sqrt{n_e}\sigma_{\rm PE}|n_{\rm PE})\, . \label{eq: PE spectrum}
\end{align}
The values for~$\mu_{\rm PE}$ and~$\sigma_{\rm PE}$ for the different experiments a listed in table~\ref{tab: XENON10&100 data} of app.~\ref{ss: XENON10 and 100}. Next, the number of events for a given number of~PEs is nothing but
\begin{align}
	\frac{\dd N}{\dd n_{\rm PE}} =\epsilon(n_{\rm PE})\times \mathcal{E} \times \frac{\dd R}{\dd n_{\rm PE}} \, ,
\end{align}
where~$\epsilon(n_{\rm PE})$ is the overall detection efficiency, and~$\mathcal{E}$ is again the exposure. This expression can be used to make predictions at two-phase noble target experiments and derive constraints from experimental data. For the conversions from electron number~$n_e$ to PEs, we followed~\cite{Essig:2012yx,Essig:2017kqs}.

The signal rates derived in this chapter might be a slight underestimate, as we neglected the band structure of liquid noble targets, which could lower the ionization energy gap~\cite{Sorensen:2011bd}. A very recent work pointed out the importance of atomic many-body physics, which are also neglected here~\cite{Pandey:2018esq}.

\paragraph{Semiconductor targets}
Although semiconductor targets can not be scaled up as easily as liquid noble experiments, experiments with silicon or germanium crystals can probe even lower DM~masses thanks to their small band gaps. While the binding energy of atoms is of order~$\mathcal{O}$(10)eV, the band gap to excite an electron to the conducting band of a semiconductor is as low as~$\mathcal{O}$(1)eV. This enables these experiments to search for DM~particles with masses~$\lessapprox$~1~MeV~\cite{Graham:2012su,Lee:2015qva,Essig:2015cda}.

A semiconductor is a multi-body system, and the computation of the ionization form factors of periodic crystal lattices requires methods from condensed matter physics. The~\textit{QEdark}-module of the QuantumEspresso quantum simulation code is a publicly available tool to compute these form factors~\cite{Giannozzi2009}, which have also been tabulated for silicon and germanium~\cite{Essig:2015cda,FormfactorLink}. The event rate was derived in app.~A of~\cite{Essig:2015cda} and is given in~$\text{kg}^{-1}\text{s}^{-1}\text{keV}^{-1}$ by
\begin{subequations}
\begin{align}
	\frac{\dd R_{\rm crystal}}{\dd E_e} &=\frac{\rho_\chi}{m_\chi} \frac{1}{M_{\rm cell}}\frac{\sigma_e\alpha m_e^2}{\mu_{\chi e}^2}\nonumber\\
	&\quad \times\int \dd q\;\frac{1}{q^2}\eta(v_{\rm min}(E_e,q))\left|F_{\rm DM}(q)\right|^2 \left| f_{\rm crystal}(E_e,q)\right|^2\, .\label{eq: differential event rate semiconductors}
\end{align}
Here,~$E_e$ is the~\textit{total} deposited energy and~$\left| f_{\rm crystal}(q,E_e)\right|$ is the crystal form factor encapsulating the electronic band structure of the target lattice. The numbers of unit cells per unit mass is~$1/M_{\rm cell}$ with
\begin{align}
	M_{\rm cell} &= 
	\begin{cases}
 	2\times m_{\rm Si} &\approx \SI{52}{\GeV}\, ,\quad \text{for silicon,}\\
 	2\times m_{\rm Ge} &\approx \SI{135}{\GeV}\, ,\quad \text{for germanium.}
 	\end{cases}
\end{align}
\end{subequations}
Similarly to the previous case, we have to convert the spectrum to the observable quantity. Here, this signal is the ionization~$Q$, the number of electron-hole pairs created per event. The conversion from deposited energy to~$Q$ can be approximated linearly via
\begin{align}
	Q(E_e)&=1+\left\lfloor\frac{E_e-E_{\rm gap}}{\epsilon}\right\rfloor\, ,
	\intertext{where the average energy per electron-hole pair~$\epsilon$ and energy gap~$E_{\rm gap}$ are}
	\epsilon&=\begin{cases}\SI{3.6}{\eV}\, ,\\ \SI{2.9}{\eV}\, ,\end{cases}\quad E_{\rm gap}=\begin{cases}\SI{1.11}{\eV}\, ,\quad\text{for silicon,}\\\SI{0.67}{\eV}\, ,\quad\text{for germanium.}\end{cases}
\end{align}
Therefore, an energy deposition~$E_e\in \left[E_{\rm gap}+(Q-1)\epsilon,E_{\rm gap}+Q\epsilon\right)$ is assumed to generate~$Q$ electron-hole pairs. Furthermore, any semiconductor experiment has a ionization threshold~$Q_{\rm thr}$, corresponding to the energy threshold
\begin{align}
	E_{e}^{\rm thr} = \epsilon(Q_{\rm thr}-1)+E_{\rm gap}\, .
\end{align}
Obviously, the optimal threshold is~$Q_{\rm thr}=1$, such that the band gap energy is also the energy threshold. Assuming that~DM can deposit all its kinetic energy, the lowest theoretically observable DM~mass is obtained by solving~$\frac{1}{2}m_{\chi}^{\rm min}(v_{\rm esc}+v_\oplus)^2=E_e^{\rm thr}$. Hence,
\begin{align}
	m_{\chi}^{\rm min} = \frac{2\left( \epsilon(Q_{\rm thr}-1)+E_{\rm gap}\right)}{(v_{\rm esc}+v_\oplus)^2} \approx 
	\begin{cases}
		\text{0.3~(1.4)~MeV}\, ,\quad \text{for silicon,}\\
		\text{0.2~(1.0)~MeV}\, ,\quad \text{for germanium,}
	\end{cases}
\end{align}
where we substituted the ionization threshold~$Q_{\rm thr}=1$~($Q_{\rm thr}=2$). Remarkably, even sub-MeV DM~masses are within reach of semiconductor experiments with single electron-hole pair thresholds.

\section{Statistics and Exclusion Limits}
\label{s:statistics}

Apart from the controversial claims by the~DAMA collaboration, no direct detection experiments succeeded to observe a clear signal of halo~DM. In this situation, the question is how to interpret the null results and how to derive exclusion limits on physical parameters. One of the largest experimental challenges is the understanding of the background. There are a number of known processes which can trigger the detector, and a great deal of effort goes into the distinction between background and a potential signal. Given a reliable background model, one can attribute a number of observed signals to the background, a process called~\emph{background subtraction}. 

Let us assume some experimental run of a generic direct detection experiment, where either the background subtraction has already been performed or was conservatively omitted. Furthermore, we assume that this experiment observed~$N$ signals with energies~$\{E_1,\ldots,E_N\}$. Then the null result can be translated into exclusion limits on the physical parameters, here the DM~mass~$m_\chi$ and the interaction cross section~$\sigma$. Roughly speaking, if a point in parameter space predicts more events than observed, the point is excluded. We will discuss this in greater detail, particularly the most straight forward approach using Poisson statistics~\cite{Tanabashi:2018oca} and Yellin's maximum gap method for cases with unknown backgrounds~\cite{Yellin:2002xd}.

\paragraph{Poisson statistics}
The detector essentially counts events and measures their deposited energies. The probability to observe~$n$ events for an expected value of~$\mu$ is given by the~\ac{PMF} of the Poisson distribution,
\begin{align}
	P(n|\mu) = \frac{\mu^n}{n!}e^{-\mu}\, .
\end{align}
If a set of parameters predicts that an experiments should have observed more events than it actually did with a certain~\ac{CL}, then this point is excluded by that~CL.
\begin{align}
	\text{CL}=P(n>N|\mu) = \sum_{n=N+1}^{\infty}\frac{\mu^n}{n!}e^{-\mu} = 1-\text{CDF}(N|\mu)\label{eq: poisson CL}
\end{align}
For a given DM~mass, we find the cross section~$\sigma$ corresponding to the value of~$\mu_{\text{CL}}$ determined by eq.~\eqref{eq: poisson CL}. The easiest way is to choose an arbitrary reference cross section~$\sigma_{\rm ref}$, compute the number of events~$N_{\rm ref}$ and find the upper bound at confidence level~CL via
\begin{align}
	\sigma<\frac{\mu_{\text{CL}}}{N_{\rm ref}}\sigma_{\rm ref}\, .
\end{align}
This means that we need to solve~$\text{CDF}(N|\mu_{\text{CL}})=(1-\text{CL})$ for~$\mu_{\text{CL}}$ at a given~\ac{CL} and number of observed events~$N$. With the exception of~$N=0$, the Poisson~\ac{CDF} can not simply be inverted. Instead we can exploit its close connection to incomplete gamma functions,
\begin{align}
	\text{CDF}(N|\mu)&=e^{-\mu}\sum_{n=0}^N\frac{\mu^n}{n!} = \frac{\Gamma(N+1,\mu)}{N!} \equiv Q(N+1,\mu)\, ,
	\intertext{where we introduced the upper incomplete gamma function,}
	\Gamma(s,x)&\equiv \int_{x}^{\infty}\dd t\, t^{s-1}e^{-t}\, ,
	\intertext{such that~$\Gamma(s,0)=\Gamma(s)$. The corresponding regularized gamma function is defined as}
	Q(s,x)&\equiv\frac{\Gamma(s,x)}{\Gamma(s)}\, .
\end{align}
The regularized upper incomplete gamma function can be numerically inverted with methods described e.g. in~\cite{Press2007}. Some relevant solutions of eq.~\ref{eq: poisson CL} are listed in table~\ref{tab: poisson}.

\begin{table*}
\centering
\begin{tabular}{llll}
\hline
N		&	$\mu_{90\%}$	&	$\mu_{95\%}$	&	$\mu_{99\%}$\\
\hline
0	&	2.30	&	2.99	&	4.61	\\
1	&	3.89	&	4.74	&	6.64	\\
2	&	5.32	&	6.30	&	8.41	\\
3	&	6.68	&	7.75	&	10.05	\\
10	&	15.41	&	16.96	&	20.14	\\
100	&	114.08	&	118.08	&	125.84	\\
\hline
\end{tabular}
\caption{Some solutions of eq.~\eqref{eq: poisson CL} for different confidence levels and event numbers.}
\label{tab: poisson}	
\end{table*}

\begin{figure*}
	\centering
	\includegraphics[width=0.67\textwidth]{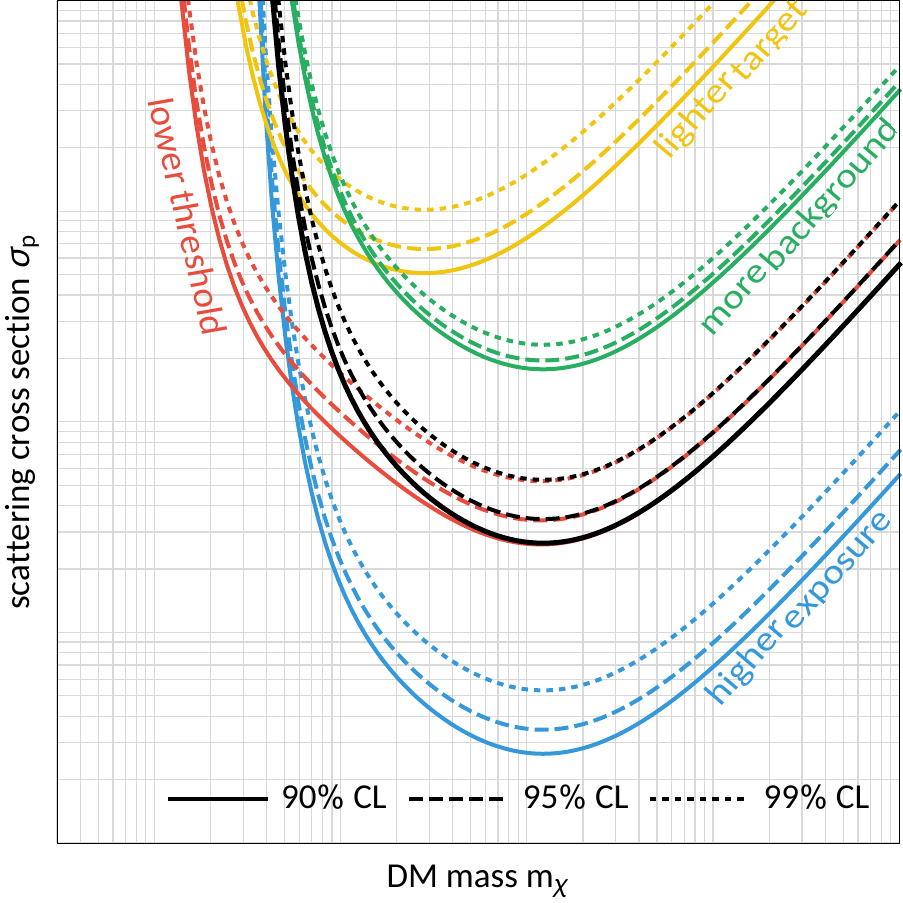}
	\caption{Generic exclusion limits from direct detection and their scaling.}
	\label{fig: generic DM limits}
\end{figure*}

A set of generic direct detection constraints is depicted in figure~\ref{fig: generic DM limits}. The limit curve is characterized by a sharp loss of sensitivity for masses below the minimum value, see eq.~\eqref{eq:minimal DM mass}, caused by the truncated tail of the DM~speed distribution. For larger masses the upper limit increases, because the DM~energy density is fixed and heavier DM~particles are more diluted. The figure also illustrates the effect of adjusting experimental parameters and how the constraints scale under change of threshold, target, background, and exposure.

These steps to find an upper confidence limit on~$\sigma$ can be applied to either the total number of observed events or to a number of energy bins, independently for each bin. In the latter case, the bin with the lowest value for the upper cross section bound sets the overall limit. It should be noted at this point that, strictly speaking, this procedure actually yields an exclusion at a lower~\ac{CL}~\cite{Green:2001xy}.

This method assumes no knowledge about the expected spectrum or the background and interprets all observed events as DM~signal of equal significance. As such, it is the most conservative method to obtain limits. It assigns an observed energy of low energy the same importance as a high energy signal, even though the expected spectrum is exponentially decreasing. Yellin proposed several alternative criterions for deciding if the expected signal is too high compared to observations, which take the expected energy spectrum into account.

\paragraph{Maximum Gap Method}
The Maximum Gap method is Yellin's simplest method to obtain exclusion limits, if the data is contaminated with background events of an unknown source, which cannot be subtracted~\cite{Yellin:2002xd}. It avoids event binning and takes the functional shape of the spectrum into account. 

For an observation of~$N$ events of energies~$\{E_1,\ldots,E_N\}$ and including the threshold~$E_{\rm thr}$ and the maximum energy~$E_{\rm max}$, there are~$N+1$ gaps without observed events. Using the expected event spectrum~$\frac{\dd N}{\dd E}$, we can compute the expected number of events in each gap. This number is called the gap size,
\begin{align}
	x_i = \int\limits_{E_i}^{E_{i+1}}\dd E\,\frac{\dd N}{\dd E}\, .
\end{align}
\begin{figure*}
	\centering
	\includegraphics[width=0.67\textwidth]{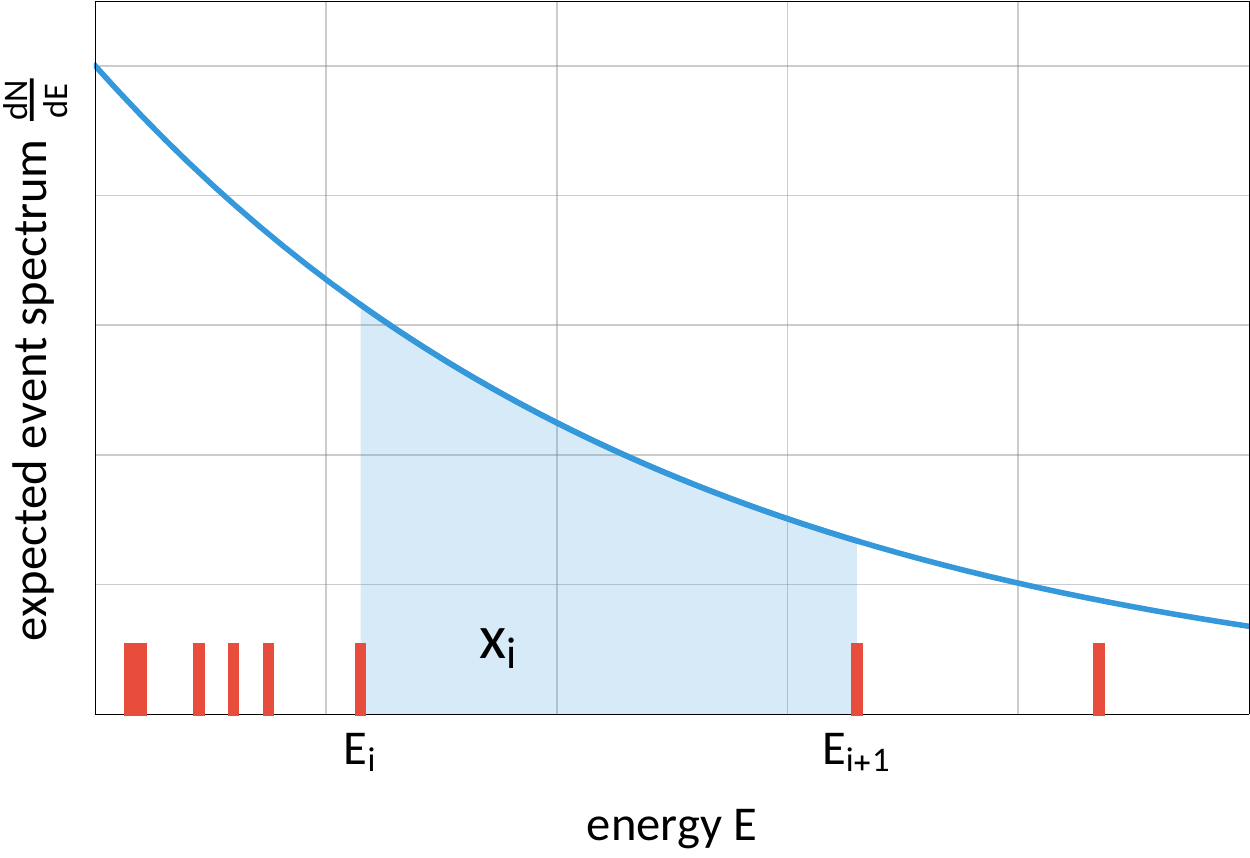}
	\caption{Yellin's Maximum Gap method for finding upper limits.}
	\label{fig: maximum gap}
\end{figure*}
The size of the largest gap is denoted with~$x_{\rm max}$, which can be regarded as a random variable on its own. For a given expectation value of the total number of events~$\mu$, the probability that the maximum gap is smaller than~$x$ is
\begin{align}
	C_0(x,\mu)\equiv  P(x_{\rm max}<x|\mu)= \sum_{n=0}^{\left\lfloor \frac{\mu}{x}\right\rfloor}\frac{(nx-\mu)^n}{n!}e^{-n x}\left(1+\frac{n}{\mu-kx}\right)\, .
\end{align}
This CDF remarkably depends only on~$\mu$ and not on the specific shape of the spectrum. The more signals are expected, the smaller is the expected size of the maximum gap, as we expect more and denser signals with smaller gaps. If a point in parameter space predicts a value for~$\mu$ such that the corresponding maximum gap should most likely be smaller than the observed~$x_{\rm obs}$, the point can be excluded with confidence level of that probability. Hence, we need to solve $C_0(x_{\rm obs},\mu)= \text{CL}$ for~$\mu$ and find the bound on the cross section~$\sigma$ on this way.

The Maximum Gap Method can be generalized in several ways. For one, it can be extended to the Maximum Patch Method for directional detection experiments~\cite{Henderson:2008bn}. Furthermore, it can be modified to not only consider gaps of zero events, but also intervals containing~1,2,\ldots~events. This is called the Optimum Interval Method~\cite{Yellin:2002xd,Yellin:2008da}. A drawback of this method is the additional complication that the interval's~\acp{CDF} no longer have a closed analytic form and require to be tabulated using \ac{MC}~simulations.\\[0.3cm]

All these constraints rely on a set of standard assumptions, especially concerning the DM~halo model. Despite its shortcomings, the use of the~\ac{SHM} simplifies the comparison of results. However, it is also possible to set halo-independent exclusion limits, see e.g.~\cite{Fox:2010bz,Cremonesi:2013bma,Feldstein:2014gza,Ferrer:2015bta}.

\clearpage
\chapter{Terrestrial Effects on Dark Matter Detection}
\label{c:earth}
The conventional detection strategy for direct DM~searches is to look for recoils of nuclei which collided with an incoming DM~particle of the halo. Given the basic assumption that this fundamental process can occur in a detector, nothing prevents the DM~particle to also scatter undetected on other terrestrial nuclei. For a moderately high cross section, it is then not unlikely that the particle could scatter twice, once underground and subsequently inside a detector. If the underground scattering rate is significant, pre-detection scatterings will alter the halo particles' density and distribution locally at the experiment's site and therefore also the expected DM~signal. This is particularly relevant for searches for low-mass~DM, where new detection channels aside from nuclear recoils have been proposed as discussed in section~\ref{ss: low mass DM detection}. Elastic~DM-nucleus collisions of light~DM might not be observable, but they still happen and affect experiments indirectly by deforming the DM~phase space. They can amplify or reduce the local DM~particle flux through the detector. Over the course of this chapter, we will investigate two phenomenological consequences of these scatterings using~\ac{MC}~simulations, namely diurnal modulations of the detection signal rate and experiments becoming insensitive to~DM itself due to a flux attenuation by the overburden. 
\begin{enumerate}
	\item Diurnal Modulations: Assuming a significant probability for a DM~particle to scatter on a nucleus of the Earth's core or mantle, the underground distribution of DM~particles, both spatial and energetic, will be distorted by the decelerating and deflecting collisions. Since the incoming particles arrive predominantly from a specific direction due to the Earth's velocity in the galactic frame, the underground distance a halo particle has to travel to reach the detector changes throughout a sidereal day due to the Earth's rotation. Therefore, the probability to scatter before being detected varies with the same frequency, and the signal rate in the detector should show a diurnal modulation~\cite{Collar:1992qc,Collar:1993ss}.
	\item Loss of sensitivity to strongly interacting~DM: With some exceptions from the more recent past, most DM~detectors are placed deep underground for the purpose of background reduction. But above a certain critical cross section, the typical~$\sim$~1~km of rock overburden would start to shield off even DM~particles. Therefore, underground detectors would be severely limited in their ability to probe strongly interacting~DM~\cite{Goodman:1984dc}. The constraints on the~DM-proton cross section extend up to this critical value only, and the parameter space above opens up and might be viable~\cite{Starkman:1990nj}.
\end{enumerate}
Each individual scattering decelerates and deflects the DM~particle on its path through the planet's interior. While the effect of a single scattering can be described analytically~\cite{Kavanagh:2016pyr}, for multiple scatterings the effect of a series of deflections is best treated with numerical simulations of particle trajectories. We developed and applied two scientific \ac{MC}~simulation codes, the~\acf{DaMaSCUS}~\cite{Emken2017a} and~\ref{code2}~\cite{Emken2018a}, to quantify diurnal modulations and constraints on strongly interacting~DM respectively. Both codes are publicly available.\\[0.3cm]

This chapter is structured as follows. In the first part, we introduce the fundamentals of the \ac{MC}~simulations of underground trajectories of DM~particles. The central outcome is the general formulation of the simulation algorithm. In chapter~\ref{s: diurnal modulation}, we apply this algorithm to the whole Earth to perform a \ac{MC}~study of diurnal signal modulations. The results have been published in~\ref{paper2}. In the third section, we determine the constraints on strongly interacting~DM by simulating particle trajectories inside the experiments' overburden. The results are divided into two parts, where chapter~\ref{ss: constraints nuclear} contains the constraints on low-mass~DM based on nuclear recoil experiments originally published in~\ref{paper4}, and chapter~\ref{ss: constraints electron} extends these results to light mediators and DM~searches based on inelastic electron interactions. The treatment of DM-electron scattering experiments was published in~\ref{paper1} and~\ref{paper5}.

\section{Monte Carlo Simulations of Underground~DM Trajectories}
\label{s: MC simulations}
There are three fundamental random variables at the core of the DM~particle transport simulation, which are required to get sampled repeatedly for every trajectory~\cite{Kalos2008,Haghighat2016}.
\begin{enumerate}
	\item The free path length: How far does a DM~particle propagate freely through the medium before it interacts with one of the constituent nuclei?
	\item The target nucleus: Once a scattering takes place, of all the different nuclei present, what isotope is involved in the collision?
	\item The scattering angle: What is the scattering angle in the \ac{CMS} of the DM-target system?
\end{enumerate}
After a brief review of different \ac{MC}~sampling methods, we will describe the distribution and sampling of each of these quantities. In the last part of this chapter, we combine them into a general \ac{MC}~simulation algorithm for underground trajectories of DM~particles.

\subsection{\acl{MC} sampling methods}
\label{ss:MCsampling}
It is usually not a problem to generate uniformly distributed (pseudo) random numbers. However, \ac{MC}~simulations always include the generation of some non-uniform random numbers, where the underlying distribution is known and determined by the simulated physical process. This generation is also called \emph{sampling}. In this section, we will describe how to sample random numbers for any underlying distribution~\cite{Tanabashi:2018oca}.

Assuming a continuous random variable~$X$ defined on the domain~$[a,b]$, the \acf{PDF}~$f_X(x)\dd x$ is defined as the probability of~$X$ to take a value between~$x$ and~$x+\dd x$,
\begin{align}
	f_X(x)\dd x = P(x<X<x+\dd x)\, .
\end{align}
As such it is a positive function, normalized on its domain,
\begin{align}
	\int\limits_a^b\dd x \; f_X(x) = 1\, .
\end{align}
Next, the \acf{CDF}~$F_X(x)$ is defined as the probability that~$X$ assumes a value below~$x$,
\begin{align}
	F_X(x)=P(X<x)=\int\limits_{a}^{x}\dd x^\prime\; f_X(x^\prime)\, .
\end{align}
The \ac{CDF} is a non-decreasing positive function~$[a,b]\rightarrow[0,1]$ satisfying~$F(a)=0$ and~$F(b)=1$.

\begin{figure*}
	\centering
	\includegraphics[width=0.49\textwidth]{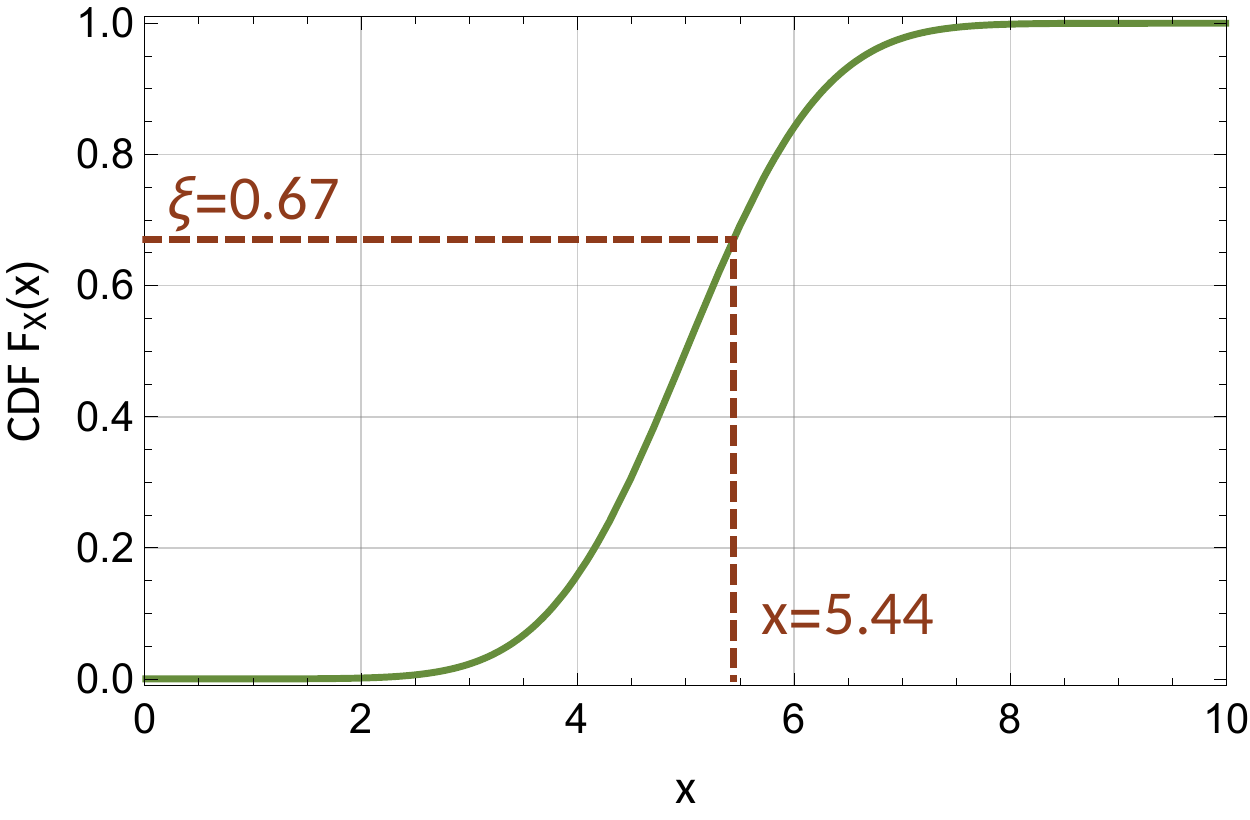}
	\includegraphics[width=0.49\textwidth]{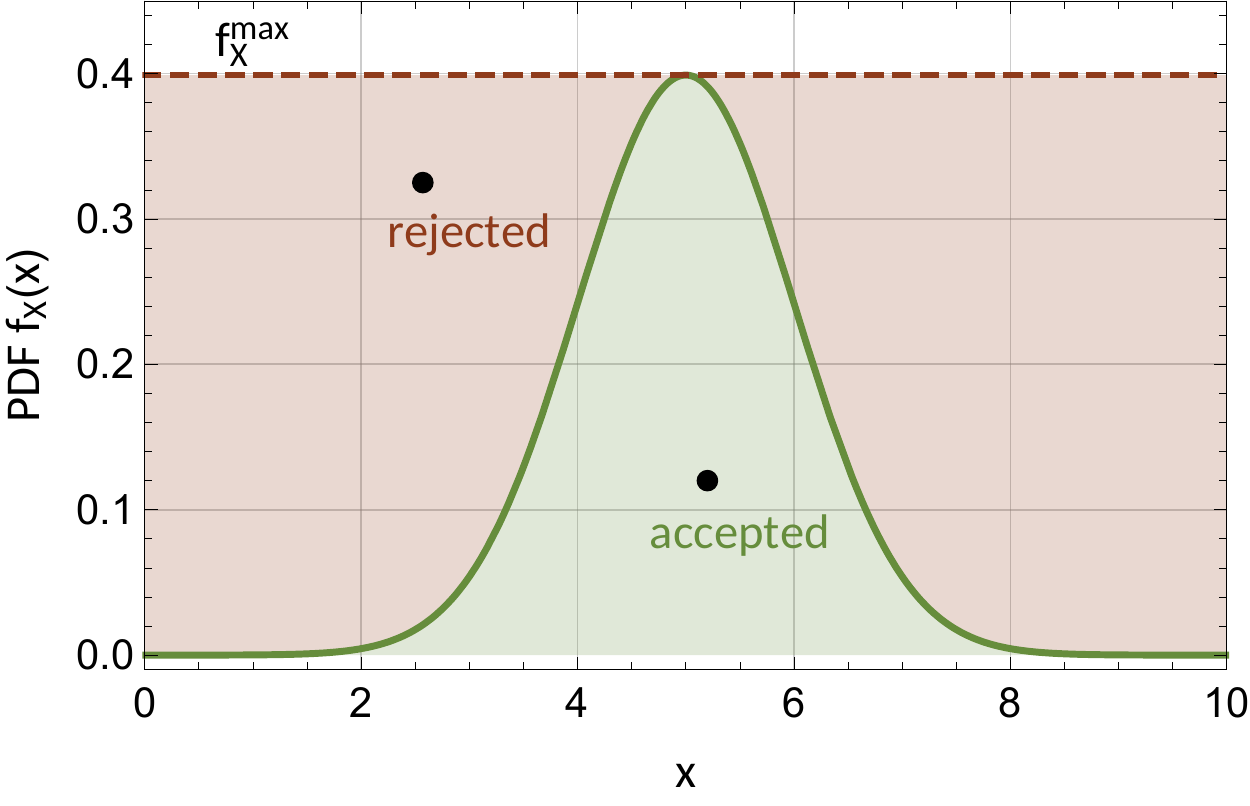}
	\caption{Examples of inverse transform sampling (left) and rejection sampling (right) for a normal distribution with mean~$\mu=5$ and standard deviation~$\sigma=1$.}
	\label{fig:sampling}
\end{figure*}

\paragraph{Inverse transform sampling} The \ac{CDF} is the central function for transforming the sample of a uniform random number into a sample of another, non-trivial but known \ac{PDF}~$f_X(x)$. Interpreting the~\ac{CDF} as a random variable on its own,~$Y=F_X(X)\in[0,1]$, we can compute its~\ac{CDF},
\begin{align}
	F_Y(y) &= P(Y<y) \nonumber\\
	&= P(F_X(X)<y) \nonumber\\
	&= P(X<F^{-1}(y))\nonumber\\
	&=F_X(F_X^{-1}(y)) = y\, .
\end{align}
This is nothing but the \ac{CDF} of a uniform random variable of domain~$[0,1]$, and the~\ac{CDF} of \textit{any} random variable is itself uniform, $Y=\mathcal{U}_{[0,1]}$. We can show that the reverse is true as well. If~$Y=\mathcal{U}_{[0,1]}$, then the random variable~$Z\equiv F_X^{-1}(Y)$ is identical to~$X$ itself, since their \acp{CDF} are identical,
\begin{align}
	F_Z(z) &= P(Z<z)\nonumber\\
	&=P(F_X^{-1}(Y)<z)\nonumber\\
	&=P(Y<F_X(z))\nonumber\\
	&=F_Y(F_X(z))\nonumber\\
	&= F_X(z)\, .
\end{align}
In conclusion, we can use this fact to transform a sample~$\xi$ of~$\mathcal{U}_{[0,1]}$ into a sample~$x$ of any random variable~$X$ by solving
\begin{align}
	F_X(x) = \xi\, . \label{eq: inverse transform sampling}
\end{align}
This is called \emph{inverse transform sampling} and is a very efficient way of sampling if the \ac{CDF} can be inverted explicitly, such that~$x = F^{-1}_X(\xi)$ can be computed directly. It is illustrated in the left panel of figure~\ref{fig:sampling} for the example of a normal distribution. If such an inversion is not possible, we can also solve eq.~\eqref{eq: integral} numerically with a root finding algorithm. However, in most cases there is a better alternative.

\paragraph{Rejection sampling} In many cases, the \ac{PDF} is a complicated expression, and the \ac{CDF} cannot be inverted analytically. Then, the \emph{acceptance-rejection method}, or simply \emph{rejection sampling}, provides an alternative efficient procedure to transform uniformly distributed random numbers into random numbers of some more complicated distribution~\cite{Robert2005}. The only condition for this method is that the \ac{PDF} can be evaluated efficiently. We already established that the \ac{CDF} or in other words, the area under the \ac{PDF} can be used to generate a sample of any distribution from uniform random numbers. Rejection sampling uses this fact and is closely related to \ac{MC}~integration.

Again, we assume a random variable~$X$ with a given \ac{PDF}~$f_X(x)$, defined on its support~$[a,b]$. Furthermore, the distribution is bounded,~$f_X(x)\leq f_X^{\rm max}$ for all~$x\in[a,b]$. We sample two random numbers $\xi_1,\xi_2$ of~$\mathcal{U}_{[0,1]}$ and find a random position in the domain,~$x=a+\xi_1(b-a)$. We accept~$x$ as a value of $X$, if 
\begin{align}
	\xi_2f_X^{\rm max}\leq f_X(x)
\end{align}
turns out to be true. Otherwise we start over with two new random numbers~$\xi_1,\xi_2$. The procedure is visualized for the example of a normal distribution in the right panel of figure~\ref{fig:sampling}. By sampling uniformly distributed points~$(x,y)$ in a plane and only accepting points with $y<f_X(x)$, we obtain random numbers distributed according to~$f_X(x)$.

The more values for~$x$ get rejected before finding an acceptable sample, the less efficient the method is. The efficiency can be quantified as the ratio of the two areas,
\begin{align}
	\epsilon = \frac{\int\limits_a^b\dd x\;f_X(x)}{(b-a)f_X^{\rm max}} = \frac{1}{(b-a)f_X^{\rm max}}\, .
\end{align}
Hence, if the probability is concentrated in a small region of the support, rejection sampling will involve a large number of rejections and is therefore inefficient. Luckily, rejection sampling, as described above, can be generalized.

In fact, rejection sampling allows to generate samples of~$X$ by sampling \emph{any}~other random variable, not just $\mathcal{U}_{[a,b]}$. Instead, we can choose any random variable~$Y$, whose domain contains the domain of~$X$ and which we know how to sample e.g. by inverse transform sampling. Next we have to find a constant~$M$ large enough that
\begin{align}
	f_X(x)\leq M f_Y(x)\quad\text{for all }x\in[a,b]\, ,
\end{align}
and $Mf_Y(x)$ envelopes~$f_X(x)$. Given two generated random values~$y$ of~$Y$ and~$\xi$ of~$\mathcal{U}_{[0,1]}$,~$y$ is accepted as a sample of~$X$, if 
\begin{align}
	\xi M f_Y(y)\leq f_X(y)\, .
\end{align}
The efficiency in this case is simply~$\epsilon=\frac{1}{M}$, and it takes on average $M$ trials before a value is accepted. 

We can convince ourselves of the validity of this procedure by considering the probability to accept a value~$y$ below~$x$,
\begingroup
\allowdisplaybreaks
\begin{align}
	P(\text{$y<x$ is accepted}) &= P\left(y<x\middle|\xi\leq\frac{f_X(y)}{Mf_Y(y)}\right)\nonumber\\
	&=\frac{P\left(y<x,\;\xi\leq\frac{f_X(y)}{Mf_Y(y)}\right)}{P\left(\xi\leq\frac{f_X(y)}{Mf_Y(y)}\right)}\nonumber\\
	&=\frac{\int_{a}^{x}\dd y\;f_Y(y)\int_{0}^{f_X(y)/(Mf_Y(y))}\dd \xi\;f_{\mathcal{U}_{[0,1]}}(\xi)}{\int_{a}^{b}\dd y\;f_Y(y)\int_{0}^{f_X(y)/(Mf_Y(y))}\dd \xi\;f_{\mathcal{U}_{[0,1]}}(\xi)}\, ,\nonumber
	\intertext{and using~$f_{\mathcal{U}_{[0,1]}}(\xi)=1$, we obtain}
	&=\frac{\frac{1}{M}\int_a^x\dd y\; f_X(y)}{\frac{1}{M}\int_a^b\dd y\; f_X(y)}\nonumber\\
	&=F_X(x)\, .
\end{align}
\endgroup
The accepted values of this method share the \ac{CDF} with~$X$ and therefore provide an accurate sample of~$X$.

After this review of general \ac{MC}~sampling methods, we come back to the concrete random variables for the DM~trajectory simulations.

\subsection{Free path length}
\label{ss:free path length}
\paragraph{Distribution of the free path length} In order to determine the location of the next collision of a DM~particle on its path through a medium, it is necessary to know the statistical distribution of the free path length. The probability to scatter depends on the type and strength of the interaction, as well as the properties of the medium, such as the density and composition. In general, the infinitesimal probability~$\dd P_{\rm scat}$ to scatter within an infinitesimal distance between~$x$ and~$x+\dd x$ along the particle's path is simply proportional to the distance and can be written in terms of the local interaction probability per unit length~$\Sigma(x,v)$,
\begin{align}
	\dd P_{\rm scat}(x) \equiv \Sigma(x,v) \dd x\, .\label{eq:dPscat Earth}
\end{align}
Furthermore, the cumulative probability to scatter within a finite distance~$x$ is denoted as~$P(x)$. For two distances~$x_1<x_2$, we can relate the corresponding probabilities intuitively via
\begin{align}
	P(x_2) &= P(x_1) + (1-P(x_1)) \kern-0.5cm\underbrace{P_{\rm scat} (x_1<x\leq x_2)}_{\text{\footnotesize prob. to scatter between~$x_1$ and~$x_2$}}\kern-0.5cm\, .
\intertext{For a small distance~$\Delta x$, this can be rewritten using eq.~\eqref{eq:dPscat Earth},}
	\frac{P(x+\Delta x)-P(x)}{\Delta x} &= (1-P(x)) \Sigma(x,v) +\mathcal{O}(\Delta x)^2\, ,\\
	\Rightarrow \frac{\dd P(x)}{\dd x} &= (1-P(x))\Sigma(x,v)\, ,
\end{align}
and finally integrated to the resulting probability to scatter within the following distance~$L$ (using~$P(0)=0$),
\begin{align}
P(L) &= 1-\exp\left(-\int_0^L\dd x\; \Sigma(x,v)\right)\, . \label{eq:cdf path length}
\end{align}
It should be noted at this point that eq.~\eqref{eq:cdf path length} is the \ac{CDF} of the random variable~$L$. Hence, the derivative yields the corresponding \ac{PDF},
\begin{subequations}
\begin{align}
	p(x) \dd x &= \frac{\dd P(x)}{\dd x} \dd x \\
	&=\underbrace{\exp\left(-\int_0^x\dd x^\prime\; \Sigma(x^\prime,v)\right)}_{\footnotesize =1-P(x)} \times\underbrace{\Sigma(x,v)\dd x}_{\footnotesize=\dd P_{\rm scat}(x)} \, .
\end{align}
\end{subequations}
The first factor is the probability to reach position~$x$ freely without scattering, whereas the second part is nothing but the probability to scatter between~$x$ and~$x+\dd x$.

\paragraph{Mean free path} With the \ac{PDF} for the free path at hand, it is straight forward to compute the mean free path,
\begin{align}
	\lambda &\equiv \langle x\rangle = \int_{0}^\infty\dd x\; x\, p(x)\, .
\intertext{However, to evaluate the integral, we would need to know the evolution of~$\Sigma(x,v)$ along the entire path. For the case of an infinite and homogenous medium, the expression simplifies thanks to~$\Sigma(x,v)=\Sigma(v)$,}
	\lambda &= \Sigma(v) \int_{0}^\infty\dd x\; x \exp\left(- \Sigma(v) x\right)=\Sigma(v)^{-1}\, .
\end{align}
This motivates to define the \emph{local mean free path}, also for inhomogeneous media, as
\begin{align}
	\lambda(x,v) &= \Sigma(x,v)^{-1}\, ,
	\intertext{such that}
	p(x) &= \frac{1}{\lambda(x,v)}\exp\left(-\int_0^x\frac{\dd x^\prime}{\lambda(x^\prime,v)}\right)\, ,\\
	P(L) &= 1-\exp\left(-\int_0^L\frac{\dd x}{\lambda(x,v)}\right)\, . \label{eq: MFP CDF}
\end{align}
The local mean free path at a given location depends on the density and composition of the medium at that location, as well as the interaction strength as quantified by the total scattering cross section,
\begin{align}
	\lambda(\mathbf{x},v)^{-1}=\sum_i \lambda_i(\mathbf{x},v)^{-1}\equiv \sum_i n_i(\mathbf{x}) \sigma_{\chi i}\, .\label{eq: MFP}
\end{align}
The index~$i$ runs over all targets present at~$\mathbf{x}$ with number density~$n_i$. The total cross section was introduced in chapter~\ref{s:interactions}. Furthermore, we treated the targets as essentially resting relative to the DM~particle. The medium is usually characterized by the mass density~$\rho(\mathbf{x})$ and the mass fractions~$f_i$ of the different target species of mass~$m_i$ such that the number densities are simply
\begin{align}
	n_i(\mathbf{x}) = \frac{f_i\rho(\mathbf{x})}{m_i}\, .
\end{align}

\paragraph{Sampling of free path lengths}
As described in section~\ref{ss:MCsampling}, we can determine the free path length between two scatterings for a particular trajectory via inverse transform sampling and solve
\begin{align}
	P(L) = \xi^\prime\, ,
\end{align}
for~$L$, where~$\xi^\prime$ is a sample of~$\mathcal{U}_{[0,1]}$. This is equivalent to finding the solution of
\begin{align}
	\Lambda(L)\equiv\int_0^L\frac{\dd x}{\lambda(x,v)} = -\log\xi \, , \quad\text{with }\xi = 1-\xi^\prime\, . \label{eq: sample L general}
\end{align}
This is especially easy for an infinite, homogenous medium, for which the mean free path does not depend on the position,
\begin{align}
	L= - \log\xi\, \lambda(v)\, . \label{eq: sample L simple}
\end{align}
\paragraph{Passing sharp region boundaries} Unfortunately, the situation is usually more complicated. The medium might change either gradually, as for example in the Earth's outer core, where the density increases continuously towards the center or have sharp boundaries, where the medium properties change abruptly. The latter occurs e.g. in transitions between the Earth core and the mantle or between the Earth crust and a lead shielding layer, where both composition and density, and therefore also the mean free path, change discontinuously along the trajectory. If a DM~particle passes one or several of such region boundaries, the left hand side of eq.~\eqref{eq: sample L general} takes the form 
\begin{align}
	\Lambda(L) = \underbrace{\sum_{i=1}^{l-1}\Lambda_i(L_i)}_{\text{\footnotesize layers passed freely}} + \underbrace{\Lambda_s(L_s)}_{\text{\footnotesize layer of scattering}}\, ,\quad \text{and }L=\sum_{i=1}^{l-1}L_i+L_s\, .\label{eq: boundary passing}
\end{align} 
Here, a particle passes~$(l-1)$ regions without interactions, where~$L_i$ is the distance to the next boundary inside region~$i$. In the last layer~$l$, the particle freely propagates a distance~$L_s$ before finally scattering. Hence, to find the solution~$L$ of~$\Lambda(L) = -\log\xi$, we have to determine the number of freely passed layers~$(l-1)$, as well as the distance~$L_s$ inside the final layer~$l$.

\begin{figure*}
	\centering
	\begin{tikzpicture}[scale=0.9, every node/.style={scale=0.9}]
		\node[draw,rectangle,fill=EarthGreen] (a0) at (0,0) {\color{white}START};
		\node[draw,rectangle,text width=3.2cm,fill=FlowChartBackground] (b0) at (0,-1.5) {\footnotesize Sample~$\xi\in(0,1)$.\\Start region:~$l=1$.};
		\draw[thick,->](a0.south)--(b0.north);
		\node[draw,rectangle,text width=4.1cm,fill=FlowChartBackground] (b1) at (4.5,-1.5) {\footnotesize Find the distance~$L_l$ to the next region boundary.};
		\draw[thick,->](b0.east)--(b1.west);
		\node[draw,rectangle,text width=4.1cm,fill=FlowChartBackground] (c1) at (4.5,-3.5) {\footnotesize Check if eq.~\eqref{eq: logxi inequality} holds.};
		\draw[thick,->](b1.south)--(c1.north);
		\node[draw,rectangle,text width=4.5cm,fill=FlowChartBackground] (c2) at (10,-3.5) {\footnotesize No interaction in layer~$l$.\\Move to next layer~$l\rightarrow l+1$.};
		\draw[thick,->](c1.east)--(c2.west)node[pos=0.5,above]{\small \color{EarthRed}No};
		\node[draw,rectangle,text width=4.5cm,fill=FlowChartBackground] (b2) at (10,-1.5) {\footnotesize Does the particle escape the simulation volume?};
		\draw[thick,->](c2.north)--(b2.south);
		\draw[thick,->](b2.west)--(b1.east)node[pos=0.5,above]{\small \color{EarthRed}No} ;
		\node[draw,rectangle,text width=4.1cm,fill=FlowChartBackground] (d1) at (4.5,-5.5) {\footnotesize Scattering in layer~$l$.\\Solve eq.~\eqref{eq: Ls} for~$L_s$.};
		\draw[thick,->](c1.south)--(d1.north)node[pos=0.5,right]{\small\color{EarthGreen}Yes};
		\node[draw,rectangle,text width=4.5cm,fill=FlowChartBackground] (d2) at (10,-5.5) {\footnotesize Final free path length:\\{\centering$L=\sum_i^{l-1}L_i+L_s$.}};
		\draw[thick,->](d1.east)--(d2.west);
		\node[draw,rectangle,fill=EarthRed] (e2) at (10,-7) {\color{white}STOP};
		\draw[thick,->](b2.east)--(13.25,-1.5) node[pos=0.5,above]{\small \color{EarthGreen}Yes}--(13.25,-7)--(e2.east) ;
		\draw[thick,->](d2.south)--(e2.north);
	\end{tikzpicture}
	\caption{Recursive algorithm to sample the free path length and therefore the location of the next scattering with different regions and layers of sharp boundaries.}
	\label{fig: L algorithm}
\end{figure*}
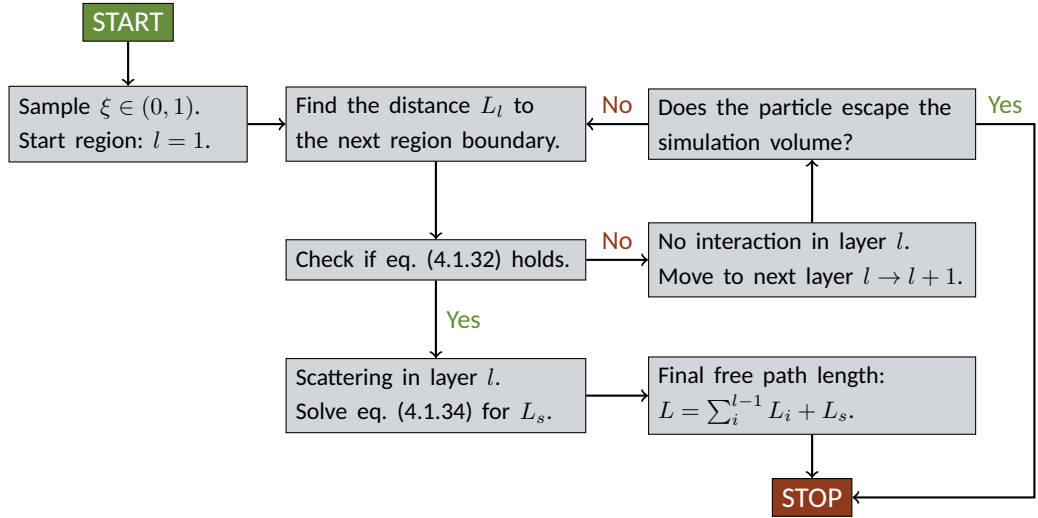

We start by sampling a uniformly distributed random number~$\xi\in(0,1)$ and determine the distance~$L_1$ to the next region boundary along the particle's path. The next step is to compare~$-\log\xi$ with~$\Lambda_1(L_1)$. If 
\begin{align}
	-\log\xi<\Lambda_1(L_1)\, ,
\end{align} 
the particle scatters within this layer. The exact location of this scattering is found by solving~$\Lambda_1(L)=-\log\xi$ for~$L$. On the other hand, if 
\begin{align}
-\log\xi>\Lambda_1(L_1)\, ,	
\end{align}
the particle passes through this region without interacting at all. We compute the distance~$L_2$ between the next two region boundaries along the trajectory and compare~$-\log\xi - \Lambda_1(L_1)$ with~$\Lambda_2(L_2)$ in the same way. If, at this point, 
\begin{align}
-\log\xi-\Lambda_1(L_1)<\Lambda_2(L_2)	
\end{align}
holds, then the particle scatters in region 2 after having travelled freely for~$L_1+L_s$, where~$L_s$ is the solution of~$-\log\xi-\Lambda_1(L_1)=\Lambda_2(L_s)$. Otherwise, we have to repeat these steps again for the next layer and continue these comparisons until we reach the layer~$l$ of scattering, such that
\begin{align}
	-\log\xi-\sum_{i=1}^{l-1}\Lambda_i(L_i)<\Lambda_l(L_l)\, . \label{eq: logxi inequality}
\end{align}
At last, the overall free path length is given by
\begin{align}
	L = \sum_{i=1}^{l-1}L_i + L_s\, ,\label{eq: free path length}
\end{align}
with~$L_s$ being the solution of
\begin{align}
	-\log\xi-\sum_{i=1}^{l-1}\Lambda_l(L_l) = \Lambda_l(L_s)\, .\label{eq: Ls}
\end{align}
If the DM~particle scatters before leaving the simulation volume, this procedure will determine where exactly this collision occurs. However, if the particle exits the simulation volume, as the DM~particle passes e.g. the Earth's surface towards space, then the trajectory's simulation is finished. These steps can be implemented as a recursive algorithm, depicted in figure~\ref{fig: L algorithm}.

\subsection{Target particle}
Once the location~$\mathbf{x}$ of the next collision is known, the identity of the target among the~$N$ different particle species present at~$\mathbf{x}$ needs to be determined. The probability is given by its relative `target size', which is proportional to its number density at~$\mathbf{x}$ and the total scattering cross section.
\begin{subequations}
\label{eq: probability target nucleus}
\begin{align}
	P\left(\text{scattering on target species~$T$}\right) &= \frac{n_T(\mathbf{x}) \sigma_{\chi T}(v_\chi)}{\sum_{i=1}^{N} n_i(\mathbf{x})\sigma_{\chi i}(v_\chi)}\, .
	\intertext{Using the definition of the mean free path in eq.~\eqref{eq: MFP}, this simplifies to}
	&=\frac{\lambda_T(\mathbf{x},v_\chi)^{-1}}{\lambda(\mathbf{x},v_\chi)^{-1}}\equiv P_j(\mathbf{x},v_\chi)\, .
\end{align}
\end{subequations}
With these probabilities, sampling the target species is straight forward. The DM~particle collides with a nucleus isotope~$T$ found via
\begin{align}
	T=\min\left\{ n\in \{1,\ldots,N\} \middle|\sum_{i=1}^{n}P_i(\mathbf{x},v_\chi) >\xi\right\} \, , \label{eq: sample target}
\end{align}
where a random number~$\xi$ of~$\mathcal{U}_{[0,1]}$ needs to be sampled.

\subsection{Scattering angle}
\label{ss:scattering angle}
The distribution of the scattering angle~$\theta$ follows from the differential cross section~$\frac{\dd\sigma_N}{\dd E_R}$ and is therefore a consequence of the chosen DM~interaction model. The recoil energy is related to the scattering angle via eq.~\eqref{eq: nuclear recoil energy}. Hence, the ~\ac{PDF} of the random variable~$\cos\theta$ can in general be obtained as
\begin{align}
	f_\theta(\cos\theta) = \frac{1}{\sigma_N}\frac{\dd\sigma_N}{\dd \cos\theta} =\frac{E_R^{\rm max}}{2 \sigma_N} \frac{\dd\sigma_N}{\dd E_R}\, .\label{eq: PDF scattering angle}
\end{align}
Together with the mass ratio of the DM~particle and the nuclei, this distribution determines the kinematics of an elastic scattering. As we will see, it will differ drastically between contact and long range interactions.

\paragraph{SI interactions} In the limit of light~DM the \ac{PDF} for \ac{SI} contact interactions is very simple. With eq.~\eqref{eq: diff cs SI light DM}, we find
\begin{align}
	f_\theta(\cos\theta) = \frac{1}{2} \label{eq: pdf scattering angle isotropic}\, .
\end{align}
Hence,~$\cos\theta$ is a uniform random variable, $\mathcal{U}_{[-1,1]}$, and the scattering is completely isotropic in the~\ac{CMS}. Inverse transform sampling for~$\cos\theta$ is trivial, we sample a random number~$\xi$ of~$\mathcal{U}_{[0,1]}$ as usual and set~$\cos\theta=2\xi-1$.

\begin{figure*}
	\centering
	\includegraphics[width=0.67\textwidth]{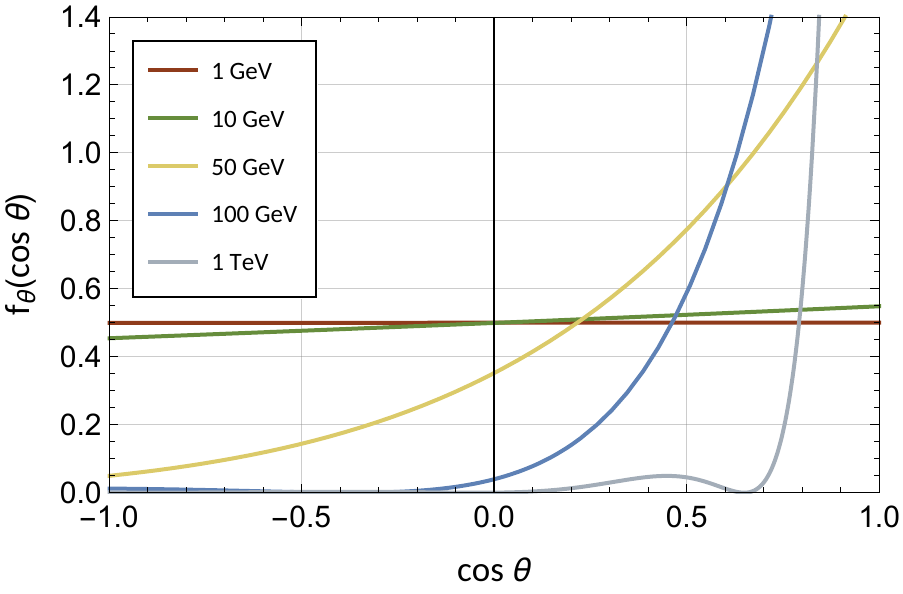}
	\caption{Distribution of the scattering angle~$\cos\theta$ for~SI interactions for different DM~masses.}
	\label{fig:pdf cosalpha SI}
\end{figure*}

When simulating heavier DM~particles, we cannot neglect the loss of coherence and have to include the nuclear form factor. As the loss of coherence suppresses large momentum transfers, this means that forward scattering will be favoured. The~\ac{PDF} of the scattering angle becomes
\begin{align}
	f_\theta(\cos\theta) = \frac{F^{\rm SI}_N\left(q(\cos\theta)\right)^2}{\int_{-1}^{+1}\dd \cos\theta\, F_N^{\rm SI}\left(q(\cos\theta)\right)^2}\, ,\label{eq: scattering angle pdf helm}
\end{align}
with $q(\cos\theta) = q_{\rm max}\sqrt{\frac{(1-\cos\theta)}{2}}$ and~$q_{\rm max} = 2\mu_{\chi N}v_\chi$. The distribution is shown for different masses in figure~\ref{fig:pdf cosalpha SI}.

The~\ac{CDF} cannot be inverted directly, and inverse transform sampling of the scattering angle is not an option whenever the full Helm form factor is taken into account. Instead, rejection sampling is a suitable way to generate scattering angles. It should be noted however that the sampling efficiency behaves as~$\epsilon\sim 1/f_\theta(1)$. For very heavy~DM and large nuclei, $f_\theta(1)$ increases, and rejection sampling could become inefficient. It could be a good idea to solve eq.~\eqref{eq: inverse transform sampling} numerically using a root-finding algorithm in that case.

\paragraph{Dark Photon}
In the dark photon model, the DM~particle couples to electric charge. Both the screening of charge and the presence of an ultralight mediator alter the scattering kinematics significantly, as they introduce a dependence on the momentum transfer~$q$ into the differential cross section. Starting from eq.~\eqref{eq:dsdq2 nucleus}, the PDF for the scattering angle can be obtained via
\begin{align}
	f_\theta(\cos \alpha) &=
	 \frac{1}{\sigma_N}\frac{\dd \sigma_N}{\dd \cos\alpha}\left|F_A(q)\right|^2=\frac{1}{2}\frac{q^2_{\rm max}}{\sigma_N}\frac{\dd \sigma_N}{\dd q^2}\left|F_A(q)\right|^2\, .
\end{align}
Here, we again express the momentum transfer in terms of the scattering angle using eq.~\eqref{eq: momentum transfer}. The~$q$ dependence, and therefore the~\ac{PDF}, is determined by the~DM and the atomic form factors,~$F_{\rm DM}(q)$ and~$F_A(q)$. The three~PDFs for contact, electric dipole, and long range interactions are
\begin{align}
	&f_\theta(\cos \theta) = \nonumber\\
	&\qquad\frac{1}{2}\times\begin{cases}
		\frac{x^3}{4}\frac{1+x}{x(2+x)-2(1+x)\log(1+x)}\frac{(1-\cos\theta)^2}{(1+\frac{x}{2}(1-\cos\theta))^2}\, ,\quad &\text{for }F_{\rm DM}(q)=1\, ,\\[2ex]
		\frac{x^2}{2}\frac{1+x}{(1+x)\log(1+x)-x}\frac{(1-\cos\theta)^2}{(1+\frac{x}{2}(1-\cos\theta))^2}\, ,\quad &\text{for }F_{\rm DM}(q)\sim\frac{1}{q}\, ,\\[2ex]
	 \frac{1+x}{(1+\frac{x}{2}(1-\cos\theta))^2}\, ,\quad &\text{for }F_{\rm DM}(q)\sim\frac{1}{q^2}\, ,
	\end{cases}
\end{align}
where~$x\equiv a^2q_{\rm max}^2\approx \text{2255}\times Z^{-2/3}\left(\frac{m_{\chi}}{\text{100 MeV}}\right)^2\left(\frac{v}{\text{10}^{\text{-3}}}\right)^2$ is an auxiliary parameter.

\begin{figure*}
	\centering
	\includegraphics[width=\textwidth]{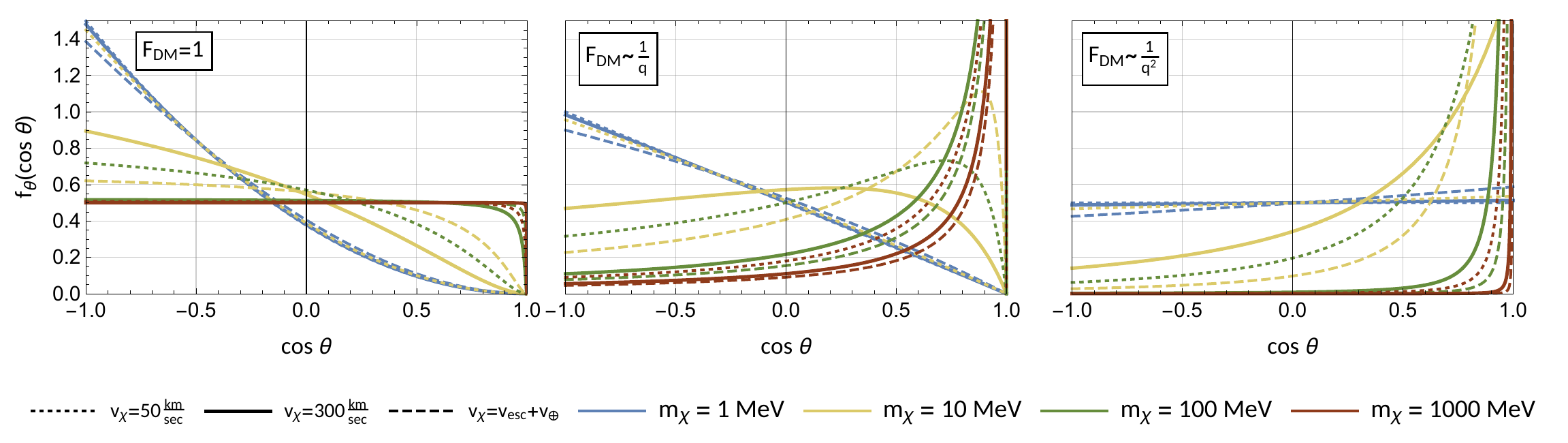}
	\caption{Distributions of the scattering angle for contact, electric dipole, and long range interactions.}
	\label{fig:pdf cosalpha dark photon}
\end{figure*}

The distributions are depicted in figure~\ref{fig:pdf cosalpha dark photon} and demonstrate the versatile kinematics of this model. For heavier DM~masses, the charge screening has little effect on a~DM-nucleus collision. For~GeV scale masses and contact interactions, the scattering is virtually isotropic with~$f_\theta\approx 1/2$, where the other two interaction types heavily favour forward scattering due to the suppression of large momentum transfers from the DM~form factor of the cross section. For lower masses, the typical momentum transfer decreases, and the effect of the charge screening on larger scales suppresses small momentum transfers. Hence, screened contact interaction favour backward scattering more and more, whereas long range interactions occur more and more isotropically. Here, the two effects of charge screening and the DM~form factor~$F_{\rm DM}\sim 1/q^2$ cancel. The electric dipole interaction behaves in an intermediate way, favouring forward or backward scattering depending on the DM~mass and speed. We can see that for~$m_\chi=$10~MeV, slow particles tend to scatter backwards, whereas faster halo particles scatter more into the forward direction. The behaviours of contact and long range interactions is entirely opposite. DM~particles of lower~(higher) mass and lower~(higher) speed scatter more isotropically if the interaction is mediated by a ultralight~(ultraheavy) mediator.

Simulating trajectories with the dark photon model, we sample scattering angles~$\cos\theta$ using inverse transform sampling, i.e. solving eq.~\eqref{eq: inverse transform sampling}. While we can evaluate the~\acp{CDF} directly, there are no closed form solutions to~$F_\theta(\cos\theta)=\xi$ for a~$\xi$~of~$\mathcal{U}_{[0,1]}$, and we have to find the solution numerically. Rejection sampling is extremely inefficient for these~\acp{PDF}.

\subsection{Trajectory simulation}
In this section, we will introduce the general algorithm to simulate the trajectory of a DM~particle, as it traverse through a medium. In the context of diurnal modulations in chapter~\ref{s: diurnal modulation}, this medium will be the whole Earth. For the purposes of section~\ref{s:simp constraints}, it will be the atmosphere, the Earth crust, and possibly some additional shielding layers made of e.g. copper or lead. Nonetheless, the fundamental algorithm will be the same, and the ground work is already done. The three central random variables introduced in the previous chapter will come together to form the \ac{MC}~simulation algorithm.

As the first step, we have to generate a set of initial conditions~$(t_0,\mathbf{x}_0,\mathbf{v}_0)$ for a DM~particle from the galactic halo in space. The details of the initial conditions are more specific and will be discussed in the respective context later on. The initial conditions should be chosen in a way that the simulated particle actually enters the simulation volume. It would be a waste of computational time, if e.g. half the particles miss the Earth.

The point of entry~$(t_1,\mathbf{x}_1,\mathbf{v}_1)$, where the particle enters the simulation volume, is determined by the distance~$d$ to its boundary along the particle's path. As long as a particle moves freely inside the Earth, we assume that it moves along a straight path, and we neglect the gravitational force of the Earth's mass acting on the DM~particle, which is a good approximation for this case. Hence,
\begin{align}
	t_1&= t_0+\frac{d}{v_0}\, ,\quad \mathbf{x}_1=\mathbf{x}_0+d\cdot\hat{\mathbf{v}}_0\, ,\quad\mathbf{v}_1 = \mathbf{v}_0\, ,
\end{align}
where we introduced the notation for the unit vector~$\hat{\mathbf{v}}\equiv\frac{\mathbf{v}}{v}$. Now that the particle is inside the simulation volume, the location of the first nuclear collision can be obtained by sampling the free path length~$L$ as discussed in chapter~\ref{ss:free path length}. The time and place of the scatterings are
\begin{align}
	t_2 = t_1 + \frac{L}{v_1}\, ,\quad\mathbf{x}_2 = \mathbf{x}_1 + L_1\cdot\hat{\mathbf{v}}_1\, .
\end{align}
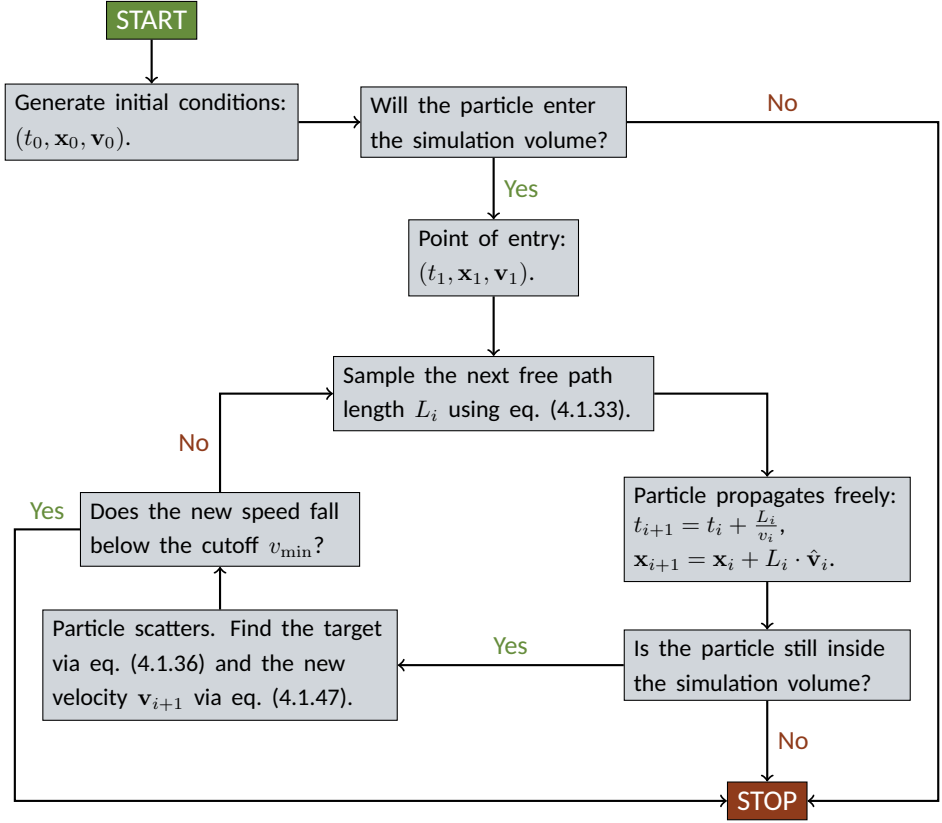
\begin{figure*}
	\centering
	\begin{tikzpicture}[scale=0.9, every node/.style={scale=0.9}]
		\node[draw,rectangle,fill=EarthGreen] (a0) at (0,0) {\color{white}START};
		\node[draw,rectangle,text width=4cm,fill=FlowChartBackground] (b0) at (0,-1.5) {\footnotesize Generate initial conditions:\\$(t_0,\mathbf{x}_0,\mathbf{v}_0)$.};
		\draw[thick,->](a0.south)--(b0.north);
		\node[draw,rectangle,text width=3.6cm,fill=FlowChartBackground] (b1) at (5,-1.5) {\footnotesize Will the particle enter the simulation volume?};
		\draw[thick,->](b0.east)--(b1.west);
		\node[draw,rectangle,text width=2.2cm,fill=FlowChartBackground] (c1) at (5,-3.5) {\footnotesize Point of entry:\\$(t_1,\mathbf{x}_1,\mathbf{v}_1)$.};
		\draw[thick,->](b1.south)--(c1.north)node[pos=0.5,right]{\small\color{EarthGreen}Yes};
		\node[draw,rectangle,text width=4.4cm,fill=FlowChartBackground] (d1) at (5,-5.5) {\footnotesize Sample the next free path length~$L_i$ using eq.~\eqref{eq: free path length}.};
		\draw[thick,->](c1.south)--(d1.north);
		\node[draw,rectangle,text width=3.9cm,fill=FlowChartBackground] (e2) at (9,-7.5) {\footnotesize Particle propagates freely:\\$t_{i+1} = t_i + \frac{L_i}{v_i}$,\\ $\mathbf{x}_{i+1} = \mathbf{x}_i + L_i\cdot \hat{\mathbf{v}}_i$.};
		\draw[thick,->](d1.east)--(9,-5.5)--(e2.north);
		\node[draw,rectangle,text width=3.9cm,fill=FlowChartBackground] (f2) at (9,-9.5) {\footnotesize Is the particle still inside the simulation volume?};
		\draw[thick,->](e2.south)--(f2.north);
		\node[draw,rectangle,text width=4.9cm,fill=FlowChartBackground] (f0) at (1,-9.5) {\footnotesize Particle scatters. Find the target via eq.~\eqref{eq: sample target} and the new velocity~$\mathbf{v}_{i+1}$ via eq.~\eqref{eq:new velocity}.};
		\draw[thick,->](f2.west)--(f0.east)node[pos=0.5,above]{\small\color{EarthGreen}Yes};
		\node[draw,rectangle,text width=3.8cm,fill=FlowChartBackground] (e0) at (1,-7.5) {\footnotesize Does the new speed fall below the cutoff~$v_{\rm min}$?};
		\draw[thick,->](f0.north)--(e0.south);
		\draw[thick,->](e0.north)--(1,-5.5) node[pos=0.5,left]{\small \color{EarthRed}No}--(d1.west) ;
		\node[draw,rectangle,fill=EarthRed] (g2) at (9,-11.5) {\color{white}STOP};
		\draw[thick,->](e0.west)--(-2,-7.5)node[pos=0.5,above]{\small \color{EarthGreen}Yes}--(-2,-11.5)--(g2.west);
		\draw[thick,->](f2.south)--(g2.north)node[pos=0.5,right]{\small\color{EarthRed}No};
		\draw[thick,->](b1.east)--(11.5,-1.5)node[pos=0.5,above]{\small \color{EarthRed}No}--(11.5,-11.5)--(g2.east);
	\end{tikzpicture}
	\caption{Flow chart for the trajectory simulation algorithm.}
	\label{fig: simulation algorithm}
\end{figure*} 
Assuming that $\mathbf{x}_2$ is still well inside the simulation volume, the particle will scatter on a nucleus of type $T$ with mass~$m_T$, sampled via eq.~\eqref{eq: sample target}. Since the two particles are assumed to scatter elastically, we can use the same kinematic relations as in the context of direct detection via nuclear recoils, discussed in chapter~\ref{ss: nuclear scattering kinematics}. In particular, we obtain the new velocity of the DM~particle using~eq.~\eqref{eq:vDMfinal},
\begin{align}
	\mathbf{v}_2=\frac{m_T v_1 \mathbf{n}+m_\chi\mathbf{v}_1}{m_T+m_\chi}\, .
\end{align}
The unit vector~$\mathbf{n}$ points towards the final velocity of the DM~particle in the~\ac{CMS}. For its determination, we sample the scattering angle~$\theta$ or rather its cosine as described in section~\ref{ss:scattering angle}. The vector~$\mathbf{n}$ is defined only up to the azimuth angle~$\phi$, which is distributed uniformly,~$\mathcal{U}_{[0,2\pi)}$. Having sampled a value for both angles, the vector reads
\begin{align}
		\mathbf{n}=\begin{pmatrix}
		\frac{\sin\theta}{\sqrt{1-e_3^2}}\left(e_2\sin\phi-e_3e_1\cos\phi\right)+e_1\cos\theta\\
		\frac{\sin\theta}{\sqrt{1-e_3^2}}\left(-e_1\sin\phi-e_3e_2\cos\phi\right)+e_2\cos\theta\\
		\sin\theta\sqrt{1-e_3^2}\cos\phi+e_3\cos\theta
		\end{pmatrix}\, , 
\end{align}
with $\hat{\mathbf{v}}_1\equiv(e_1,e_2,e_3)^T$ and a particular but arbitrary origin of the azimuth angle. It can easily be verified that~$\mathbf{n}\times\hat{\mathbf{v}}_1=\cos\alpha$ and $|\mathbf{n}|=1$. 

After having simulated the scattering, which deflects and decelerates the DM~particle, the procedure repeats: We sample the free path length $L_i$, find the location of the next scattering, 
\begin{align}
	t_{i+1} &= t_i + \frac{L_i}{v_i}\, ,\quad \mathbf{x}_{i+1} = \mathbf{x}_i + L_i\cdot \hat{\mathbf{v}}_i\, ,\label{eq:new time and location}
\end{align}
identify the target~$T$ of the collision, sample the scattering angles and obtain the new velocity,
\begin{align}
	\mathbf{v}_{i+1}&=\frac{m_T v_i \mathbf{n}+m_\chi\mathbf{v}_i}{m_T+m_\chi}\, , \label{eq:new velocity}
\end{align}
and so on. The repetition continues until either the particle exits the simulation volume or its speed falls below a minimal speed~$v_{\rm min}$, which we defined in order to avoid wasting time on particles too slow to contribute to our desired data sample. The simulation algorithm is illustrated as a flow chart in figure~\ref{fig: simulation algorithm}.

\clearpage
\section{Diurnal Modulations}
\label{s: diurnal modulation}

Due to the Earth's velocity in the galactic rest frame, there is a preferred direction from which more DM~particles hit the Earth with higher energies, for details see chapter~\ref{s: halo model}. Consequentially, the underground distance travelled by an average DM~particle from the halo before reaching a detector varies as the Earth rotates. This variation translates into a modulation of the probability to scatter prior to reaching a detector over a sidereal day. Assuming a moderately strong interaction between~DM and terrestrial nuclei, underground scatterings redistribute and decelerate the DM~particles inside the Earth, periodically modifying the local density and velocity distribution at any laboratory's location on Earth. If e.g. the average DM~particle has to travel through the bulk mass of the Earth crust and core before it can cause a detection signal, the chance of being deflected or decelerated below the energy threshold is much higher compared to when the DM~wind hits the laboratory from above. Especially experiments in the southern hemisphere are sensitive to this `Earth shadowing' effect. The phenomenological signature in a detector is a diurnal modulation of the signal rate, for which the amplitude depends on the latitude and the phase on the longitude of the laboratory. In this chapter, we will describe and study this modulation for light~DM.

Already in the 90s, diurnal modulations of direct detection rates due to underground scatterings have been investigated and quantified with early \ac{MC}~simulations in the context of the classic \ac{WIMP}~\cite{Collar:1992qc,Collar:1992jj,Collar:1993ss,Hasenbalg:1997hs}. Later on, similar modulations have been studied in the context of hidden sector~DM~\cite{Foot:2003iv,Foot:2011fh,Foot:2014uba,Foot:2014osa}, strong DM-nucleus interactions~\cite{Kouvaris:2014lpa}, and DM-electron scattering experiments~\cite{Lee:2015qva}. Multiple experiments have searched for diurnal modulations, including the COSME-II experiment in the early 90s~\cite{Collar:1992jj}, the DAMA/LIBRA experiment~\cite{Bernabei:2014jnz,Bernabei:2015nia}, and LUX~\cite{Akerib:2018zoq}. So far, no evidence for daily signal modulations has been reported. The future experiment SABRE, designed to test the discovery claim by the DAMA collaboration~\cite{Bernabei:2008yi}, is potentially relevant for diurnal modulations, since it is located at the Stawell Underground Physics Laboratory in Australia~\cite{Froborg:2016ova}. Being in the southern hemisphere increases the sensitivity to diurnal modulations.

The phenomenological signature of underground scatterings was studied with analytic methods in~\cite{Kavanagh:2016pyr}. The authors quantified the lab-frame DM~distribution modifications due to a \textit{single} Earth scattering and determined the diurnal modulations for different effective operators of the~\ac{NREFT} framework. As such, the procedure does not apply beyond the single scattering regime. For higher cross sections, where multiple scatterings before reaching the detector are expected, numerical methods such as \ac{MC}~simulations are necessary. As a tool to study the effect of multiple underground scatterings, we developed and published the Dark Matter Simulation Code for Underground Scatterings~(\ref{code1}), which complements and generalizes the~\textsc{EarthShadow} code by Kavanagh et al.~\cite{EarthShadow}. The analytic method will serve as a crucial consistency check for the new \ac{MC}~results. The simulation of billions of particles provides the statistical properties of~DM inside the Earth and allows to compute the expected distortion of the underground DM~density and velocity distribution taking into account the Earth's orientation in the galactic frame, its composition and internal structure, as well as its time dependent velocity. Based on the simulations, we can predict the diurnal modulation of the signal rate, both amplitude and phase, based on an experiment's setup, location and underground depth.
 
We already covered the basics for \ac{MC}~simulations of particle trajectory through a medium in the previous chapter. In this section of the thesis, we will apply them to simulate DM~particle trajectory propagating through the whole Earth undergoing multiple scatterings. After a brief description of the initial conditions, we present the procedure to sample the free path length inside the Earth's mantle and core, where the density is changing gradually. The next step will be to connect the trajectory simulations with predictions for signal rates in a DM~detector. Finally, we present and discuss results for a CRESST-II type experiment and different benchmark points.

\subsection{Initial conditions}
\label{ss: initial conditions earth}
The initial conditions consists of a starting time, location outside the Earth, and velocity. Naturally, the correct choice of initial conditions for the simulated DM~particles is crucial for the accuracy of the final results. 

\paragraph{Initial time}The initial time can be chosen arbitrarily, we can set~$t_0$ to a random value or just start at zero. 

\paragraph{Initial velocity} The initial velocity consists of two components, an isotropically distributed velocity sampled from the halo distribution in the galactic rest frame and the Galilean boost into the Earth's rest frame,
\begin{align}
	\mathbf{v}_0 = \mathbf{v}_{\text{halo}}-\mathbf{v}_\oplus(t)\, .\label{eq:vini}
\end{align}
The halo component is sampled from the velocity distribution in eq.~\eqref{eq:fhalo}. The Earth's velocity is given in eq.~\eqref{eq:vearth} of app.~\ref{a:velocity}. 

\paragraph{Initial position} To find the initial position of a DM~particle is slightly more delicate. The particles of the halo should effectively be distributed uniformly in space, however we would like the particle to be on a collision course with the Earth.

In previous works, the initial positions were chosen on top of the Earth's surface~\cite{Collar:1992jj,Collar:1993ss,Hasenbalg:1997hs}. This choice is arbitrary and does not correspond to uniformly distributed initial positions outside the planet. Starting all trajectories from the surface creates a finite volume bias at shallow underground depth. Too many particles are being sent into the Earth with narrow angle, creating an over-density of DM~particles close to the surface, i.e. exactly where direct detection experiments are located. This can be confirmed both analytically and with simulations for a transparent Earth.

\begin{figure*}
\centering
	\includegraphics[width=.45\textwidth]{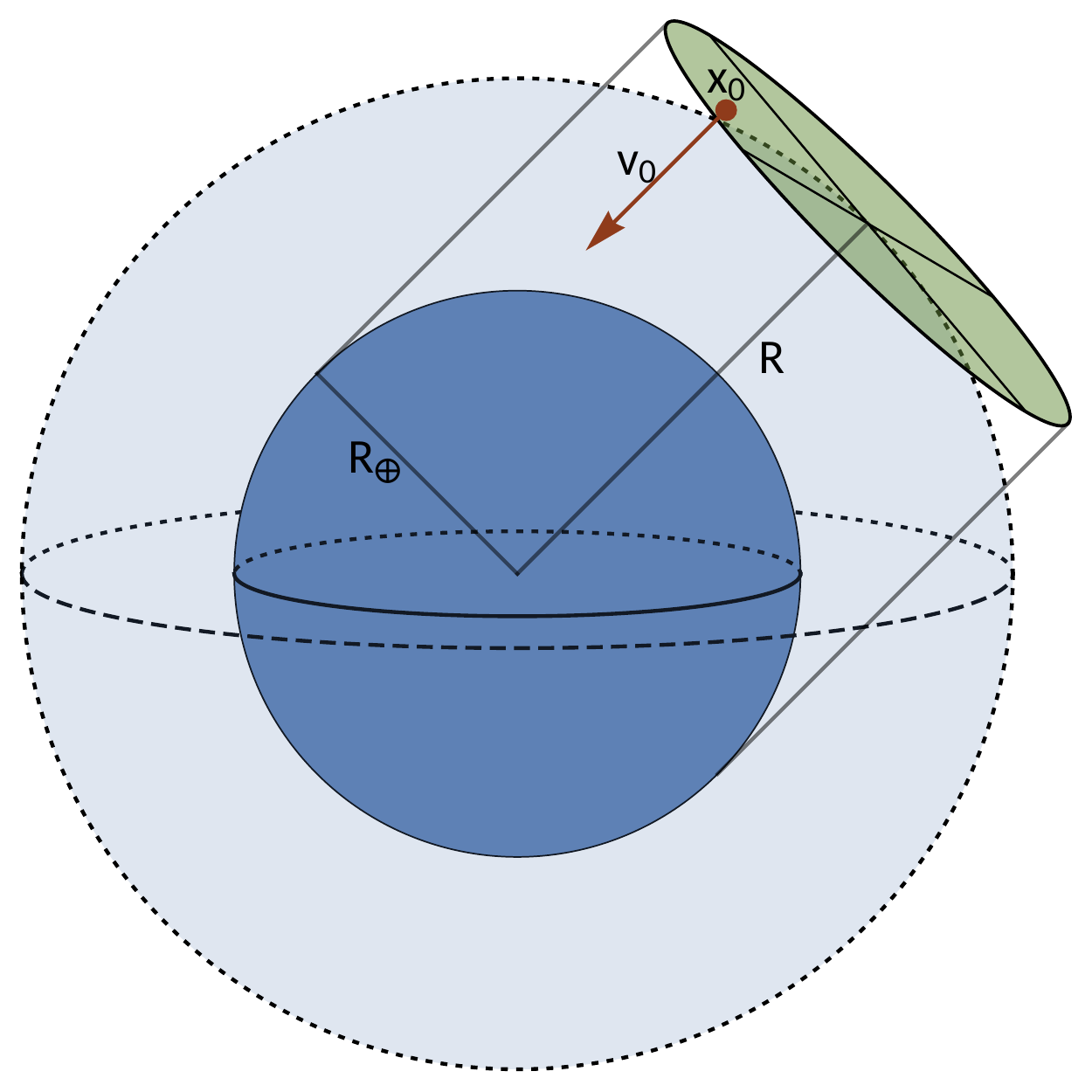}
	\caption{Sketch of the initial position.}
	\label{fig: IC}
\end{figure*}

Given an initial velocity~$\mathbf{v}_0$, all initial positions, which result in a collision with Earth, are located inside a cylinder oriented parallel to~$\mathbf{v}_0$. We can sample a random point within this cylinder, which is equivalent to picking a random point on a circular disk of radius~$R_\oplus$ at distance~$R>R_\oplus$ from the center of the Earth, as depicted in figure~\ref{fig: IC}, 
\begin{align}
	\mathbf{x}_0&=R \mathbf{e}_z+\sqrt{\xi}R_\oplus\left( \cos \phi\; \mathbf{e}_x+\sin \phi \;\mathbf{e}_y\right)\, ,
\end{align}
where~$\xi$ and~$\phi$ are sampled values of~$\mathcal{U}_{[0,1]}$ and~$\mathcal{U}_{[0,2\pi]}$ respectively, and the unit vectors $\mathbf{e}_x$ and $\mathbf{e}_y$ span the disk. This choice results in an effectively uniform distribution of particles, which we confirmed for the case of a transparent Earth using simulations.

\subsection{Free path length inside the Earth}
In chapter~\ref{ss:free path length}, we discussed how to sample the free path length of a DM~particle inside matter by solving eq.~\eqref{eq: sample L general}. We also presented how to handle sharp region boundaries. For the Earth, there is another complication in this context, as the mass density increases smoothly towards the planetary core within each layer, see eq.~\eqref{eq:densityprofile}. With this polynomial parametrization of the Earth's density profile, we can compute~$\Lambda_i(L_i)$ of eq.~\eqref{eq: boundary passing} analytically,
\begin{subequations}
	\label{eq:scatterprobabilityanalytic}
\begin{align}
	\Lambda_i(L_i)&=\int\limits_{0}^{L_i/v}\dd t\;v\lambda^{-1}(\mathbf{x}(t),v)\nonumber\\
	&=v g_l \int\limits_{0}^{L_i/v}\dd t\; \underbrace{\left[ a_l +b_l \frac{|\mathbf{x}(t)|}{r_{\oplus}} + c_l \left(\frac{|\mathbf{x}(t)|}{r_{\oplus}}\right)^2 + d_l \left(\frac{|\mathbf{x}(t)|}{r_{\oplus}}\right)^3\right]}_{=\rho_\oplus(|\mathbf{x}(t)|)}\, , \nonumber\\
	&=\frac{g_l}{r_{\oplus}^2}\left(C_1 a_l + C_2 b_l + C_3 c_l + C_4 d_l\right)\, ,
\end{align}
where we parametrized the particle's path through layer~$i$ as $\mathbf{x}(t)=\mathbf{x}_0+t\mathbf{v}$ and use $g_l \equiv \frac{\lambda^{-1}(\mathbf{x},v)}{\rho_\oplus(\mathbf{x})}$. These factors only depend on the layer's composition through the mass fractions~$f_i$, not on the location within that layer. Furthermore, if the interaction cross section does not depend explicitly on the DM~speed, which is the case if we neglect the nuclear form factor, they only have to be computed once for each mechanical layer. Otherwise they depend on the DM~speed and have to be re-computed after each scattering. The coefficients~$C_i$ are
\begin{align}
	C_1 &= L_i r_{\oplus}^2\, ,\\
	C_2 &=\frac{r_{\oplus}}{2}\Bigg[\tilde{L}(L_i+x_0 \cos \alpha)-x_0^2\cos \alpha\nonumber\\
	&\qquad\qquad+x_0^2\sin^2 \alpha  \log \left( \frac{L_i+\tilde{L}+x_0 \cos \alpha}{x_0 (1+\cos \alpha)}\right)\Bigg]\, ,\\
	C_3 &=L_i\left(x_0^2+x_0 L_i \cos \alpha +\frac{1}{3}L_i^2\right)\, ,\\
	C_4 &=\frac{1}{8r_{\oplus}}\Bigg[ (5-3\cos^2\alpha)(\tilde{L}- x_0)x_0^3\cos \alpha \nonumber\\
	&\qquad\qquad+ 2 L_i^2 \tilde{L}(L_i+3x_0\cos\alpha)+L_i\tilde{L}x_0^2(5+\cos^2\alpha)\nonumber\\
	&\qquad\qquad+3x_0^4\sin^4\alpha\log \left( \frac{L_i+\tilde{L}+x_0 \cos \alpha}{x_0 (1+\cos \alpha)}\right)\Bigg]\, .
\end{align}
Here, we used
\begin{align}
\cos \alpha &\equiv \frac{\mathbf{x}_0\cdot \mathbf{v}}{x_0v}\, ,\quad \tilde{L}\equiv \sqrt{L_i^2+x_0^2+2L_ix_0\cos\alpha}
\end{align}
\end{subequations}
With this closed form for~$\Lambda_i(L_i)$ inside a layer, the free path length inside the Earth can be found using the algorithm in figure~\ref{fig: L algorithm}. 

\subsection{Collecting data}
\label{ss: data collection earth}

\begin{figure*}
	\centering
	\begin{minipage}[b]{0.5\textwidth}
		\includegraphics[width=0.95\textwidth]{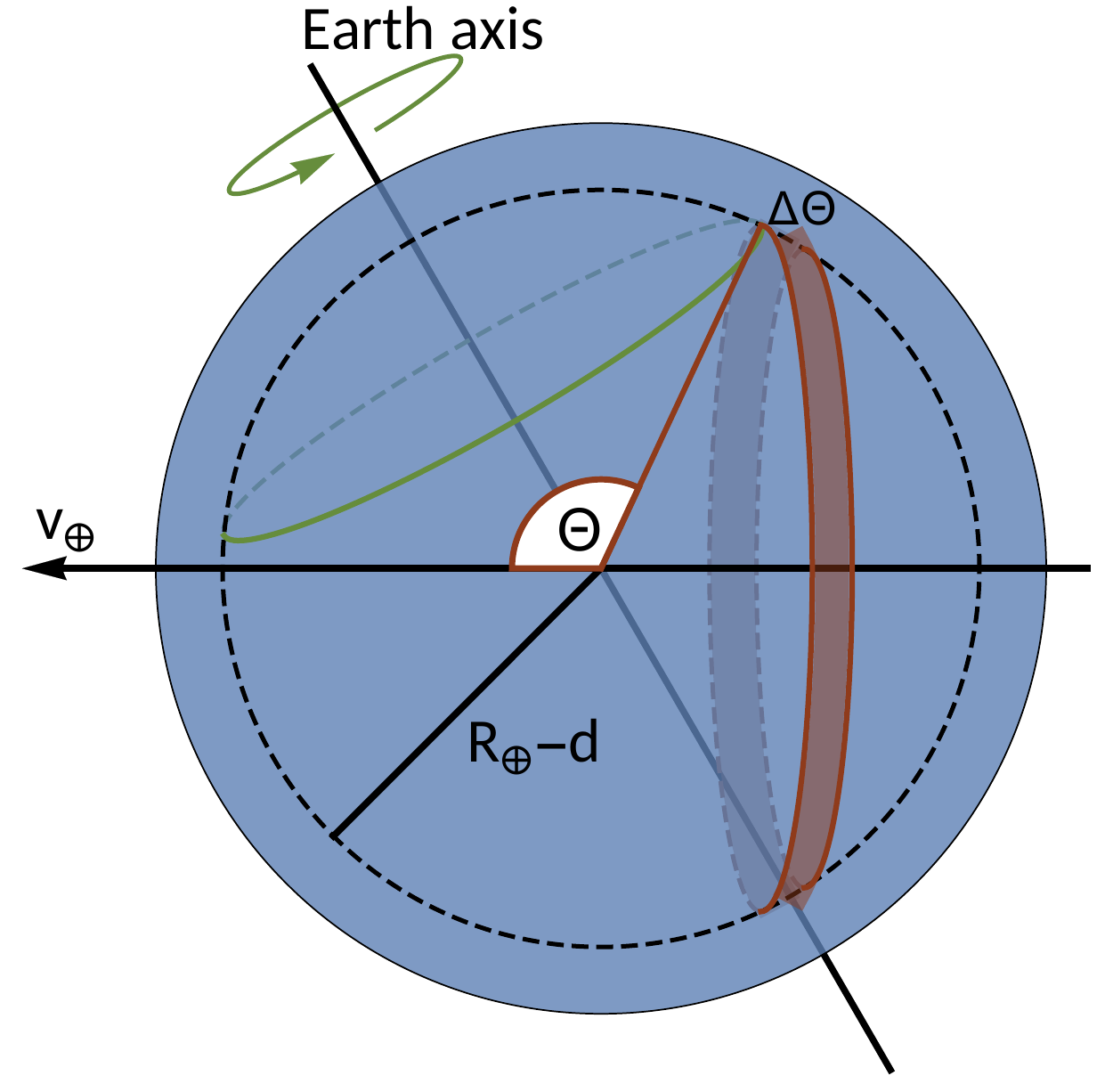}
	\end{minipage}
	\begin{minipage}[b]{0.45\textwidth}
		\includegraphics[width=0.95\textwidth]{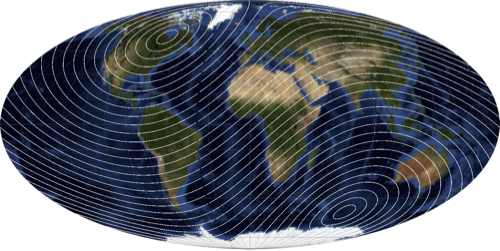}
		\includegraphics[width=0.95\textwidth]{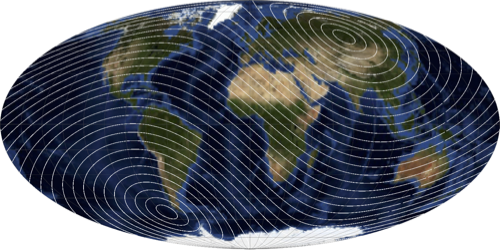}
	\end{minipage}
	\caption{Illustration of the isodetection angle and rings and their projection onto the Earth surface at 0:00 and 12:00 with $\Delta\Theta = \ang{5}$.}
	\label{fig: isodetection rings}
\end{figure*}

The main objective of the \ac{MC}~simulations is to obtain a good estimate of the DM distribution inside the Earth at a given underground depth. Even though the boost from the galactic to the Earth's rest frame breaks the isotropy of the halo distribution, the system preserves a residual rotational symmetry around the axis parallel to the Earth's velocity~$\mathbf{v}_\oplus$. The polar angle~$\Theta$ to this axis is called the \emph{isodetection angle}\footnote{An equivalent angle~$\gamma$ is being used in~\cite{Kavanagh:2016pyr} and the \textsc{EarthShadow} code, defined as~$180^{\circ}-\Theta$.}, as the DM~distribution and direct detection rates will always be constant along constant~$\Theta$. This symmetry is used in the simulation, to define small isodetection rings of equal finite angular size~$\Delta\Theta$, as proposed originally in~\cite{Collar:1992jj,Collar:1993ss}, placed at an underground depth~$d$ (e.g. \SI{1400}{\meter} for the \ac{LNGS}), which are shown in figure~\ref{fig: isodetection rings}. In this figure, we set~$\Delta \Theta = \ang{5}$ for illustrative purposes. In the actual simulations, we aim to have good resolution and choose~$\Delta \Theta = 1^{\circ}$. This should be sufficiently small that the local DM~distributions will not vary over a single ring.

\begin{figure*}
	\centering
	\includegraphics[width=0.67\textwidth]{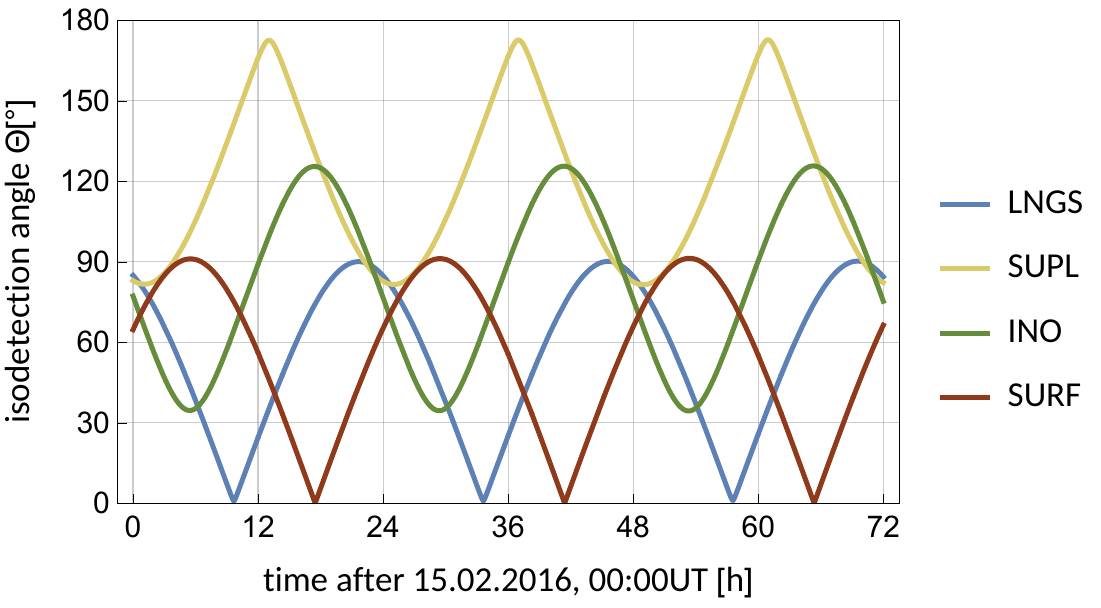}
	\caption{Time evolution of the isodetection angle for different laboratories around the globe over the duration of three days.}
	\label{fig: detector angle}
\end{figure*}

As the Earth rotates, a laboratory travels through the different isodetection rings, and its position in terms of the isodetection ring is
\begin{align}
	\Theta(t) = \arccos\left[\frac{\mathbf{v}_{\oplus}(t)\cdot \mathbf{x}^{\text{(gal)}}_{\text{lab}}(t)}{v_{\oplus}(t)(r_{\oplus}-d)}\right]\, ,\label{eq: isodetection angle}
\end{align} 
where we have to substitute eq.~\eqref{eq:vearth} for the velocity of the Earth and eq.~\eqref{eq:labpos} for the position of the experiment in the galactic frame. A given experiment will cover a certain range of~$\Theta$ during a sidereal day depending on its latitude, as shown in figure~\ref{fig: detector angle} for the \acs{LNGS}~(\ang{45.454} N, \ang{13.576} E), \acs{SUPL}~(\ang{37.07} S, \ang{142.81} E), INO~(\ang{9.967} N, \ang{77.267} E) and \acs{SURF}~(\ang{44.352} N, \ang{103.751} W). The plot shows how deeply a given experiment penetrates the Earth's `shadow' during its revolution around the Earth axis. Already at this point, we can conclude that experiments in the southern hemisphere are more sensitive to diurnal modulations due to underground scatterings.

The 180 rings are defined by their isodetection angle~$\Theta_k= k\Delta\Theta$ and labeled by~$k\in$~[0,179]. By choosing a constant~$\Delta \Theta$, their surface areas will differ, and the area of ring~$k$ is given by
\begin{align}
	A_k = 2\pi (r_{\oplus}-d)^2 \left[ \cos (\Theta_k) -\cos(\Theta_k+\Delta\Theta)\right]\, . \label{eq:ringarea}
\end{align}
For each of these rings, we have to find a \ac{MC}~estimate of the local DM~velocity distribution and density based on individually simulated particle trajectories.

Whenever a simulated particle crosses an isodetection ring, we record its velocity vector. By simulating a large number of trajectories, we gather a statistical velocity sample for each ring. The corresponding speed histograms will serve as a estimate of the~\ac{PDF} and contain all the information about how nuclear scatterings decelerate the halo particles depending on the distance of underground propagation. Since the rings vary in surface area, we have to ensure that the velocity sample sizes are sizeable even for the smaller rings close to~$\Theta=\ang{0}$ and~$\Theta=\ang{180}$. In order to extract the redistribution of particles due to deflections, we simply count the number of particles crossing a given isodetection ring. We will discuss the details of the data analysis in the next chapter.

We should point out a number of differences to previous simulations of this kind~\cite{Collar:1992jj,Collar:1993ss,Hasenbalg:1997hs}. Collar et al. focused on the classic WIMP scenario with masses above 50 GeV and their detection via nuclear recoil experiments, whereas we study sub-GeV~DM and have new detection strategies such as DM-electron scatterings in mind. Furthermore, there are substantial differences in the implementation of the simulations in the~\ref{code1} code, which include the corrected generation of initial conditions, which avoids to over-estimate the DM~density close to the surface, as well as the more refined~\ac{PREM}. By positioning the isodetection rings underground, the code is also able to simulate the shielding effect of the overburden for strongly interacting~DM. However, it turns out to be much more efficient to set up a dedicated simulation code for this problem, as we will discuss in chapter~\ref{s:simp constraints}.

\subsection{From MC simulations to local DM distributions and detection rates}

The DM~particle transport simulations in the Earth generate local velocity data for all isodetection rings. For each of these rings, an independent data analysis is performed to estimate the local DM~density, velocity distribution, and direct detection event rate. The technical details of this analysis are presented in this section.

\paragraph{Local DM Speed Distribution}

Since we do not study directional detection, it suffices to estimate the DM~speed distribution~$f_\chi^k(v)$ for a given isodetection ring~$k$. The simplest, non-parametric way to estimate the distribution underlying a data set is the histogram. However, the data must be weighted.

We mentioned earlier, that velocity data is recorded, once a DM~particle crosses an isodetection ring, i.e. a fictitious two-dimensional surface. If enough particles crossed a small surface, we can estimate the distribution in close proximity to the surface, but it is important to keep in mind that particles crossing a surface define a particle flux~$\Phi_\chi$, which is related to the~\ac{PDF} via
\begin{align}
	\Phi_\chi(\mathbf{v})\dd^3\mathbf{v} = n_{\chi}f_\chi(\mathbf{v})v\cos\gamma\;\dd^3\mathbf{v}\, , \label{eq:fluxvsf}
\end{align}
where~$\gamma$ is the angle between the velocity and the normal vector of the surface at the point of crossing~\cite{Reif1965}. Instead of continuously sending in a stream of DM~particles and time-tag each surface crossing, we effectively simulate a single burst of incoming particles and wait until even the slowest particle has left the Earth again. Hence, we do not actually track the flux~$\Phi_\chi$, but instead~$\Phi_\chi/v$. Consequently, given a particle crossing at~$\mathbf{x}$ with velocity~$\mathbf{v}$, we have to weigh this data point by the reciprocal cosine of~$\gamma$,
\begin{align}
	w = \frac{1}{|\cos \gamma|}\, ,\quad\text{where }\cos\gamma\equiv\frac{\mathbf{x}\cdot\mathbf{v}}{xv}\, ,
\end{align}
in order to really obtain an estimate of the distribution function in proximity of the isodetection ring's surface. An alternative approach to understand this weight factor is to assume a DM~particle passing through a small patch of area~$\dd A$. The effective area for the particle to pass the patch is~$\dd A\cos\gamma$, and a particle moving almost parallel to the surface is less likely to cross, but should contribute to the distribution just the same as a particle crossing the surface perpendicular to the surface.

Suppose we collected a MC~data sample of~$N_{\rm sample}$ weighted data points~$(v_i,w_i)$. The histogram's domain is $(v_{\rm min},v_{\rm esc}+v_\oplus)$, and the bin width~$\Delta v$ is set by Scott's normal reference rule~\cite{Scott1979},
\begin{align}
	\Delta v = \frac{3.5\sigma}{N^{1/3}_{\rm sample}}\, ,\quad \text{with }\sigma = \frac{v_0}{\sqrt{2}}\, ,\label{eq:scott}
\end{align}
such that the histogram consists of~$N_{\rm bins}=\lceil\frac{v_{\rm esc}+v_{\oplus}-v_{\rm min}}{\Delta v}\rceil$ bins,~$B_1=[v_{\rm min},v_{\rm min}+\Delta v)$,~$B_2=[v_{\rm min}+\Delta v,v_{\rm min}+2\Delta v)$,\ldots.  The bin height of bin~$i$ is then
\begin{align}
	W_i &= \sum\limits_{j=1}^{N_{\rm sample}}w_j\, \mathbf{1}_{B_i}(v_j)\, , \label{eq:W}
	\intertext{where we introduced the indicator function defined by}
	\mathbf{1}_A(x) &\equiv \begin{cases}
		1 \quad &\text{if }x\in A\, ,\\
		0\quad &\text{otherwise.}
	\end{cases}
\end{align}
Finally, the weighted histogram estimation of the speed distribution $f_\chi(v)$ is simply
\begin{align}
	\hat{f}_\chi(v) &= \frac{1}{N}\sum\limits_{i=1}^{N_{\rm bins}}W_i \,\mathbf{1}_{B_i}(v)\, ,\label{eq:histogram estimate of f}
\end{align}
where the denominator~$N=\Delta v\sum\limits_{i=1}^{N_{\rm bins}}W_i$ ensures that the histogram is normalized, and~$\hat{f}_\chi(v)$ can be regarded as a~\ac{PDF}. The variance of the bin height, and therefore the error of the~\ac{PDF} estimate, can be inferred from Poisson statistics,
\begin{align}
	\sigma_{W_i}^2\simeq \frac{1}{N^2}\sum\limits_{j=1}^{N_{\rm sample}}w_j^2\, \mathbf{1}_{B_i}(v_j)\, .\label{eq:histoerror2}
\end{align}
The average speed for some isodetection angle is just the weighted mean,
\begin{align}
	\langle v\rangle &= \frac{1}{w_{\rm tot}}\sum\limits_{i=1}^{N_{\rm sample}}w_i v_i\, ,\quad\text{with }w_{\rm tot}\equiv \sum\limits_{i=1}^{N_{\rm sample}}w_i \, ,
\end{align}
and for the associated standard error we can use the approximation by Cochran~\cite{Gatz1995},
\begin{align}
	(\text{SE})^2 \simeq& \frac{N_{\rm sample}}{(N_{\rm sample}-1)w_{\rm tot}^2}\sum\limits_{i=1}^{N_{\rm sample}}\Bigg[\left(w_iv_i-\langle w\rangle\langle v\rangle\right)^2\nonumber\\
	&\quad-2\langle v\rangle(w_i-\langle w\rangle)(w_iv_i-\langle w\rangle\langle v\rangle)+\langle v\rangle^2(w_i-\langle w\rangle)^2\Bigg]\, .
\end{align}
With an estimate of the DM~velocity distribution at hand, the next question is to determine the local density distortions.

\paragraph{Local DM Density}
In the absence of DM~interactions, the Earth becomes transparent, and the DM~density is constant throughout space with $\rho_\chi^{(0)}=\SI{0.3}{\GeV\per\cm\cubed}$. This can be utilized to estimate the local DM~density by comparing simulations with and without underground scatterings. Furthermore, the local number density in some given isodetection ring is proportional to the (weighted) number of passing particles. By initially running a number of free trajectories without nuclear collisions, we obtain a reference value~$w_{\rm tot}^{(0)}$ for particles passing the isodetection ring, which can be related to its local DM density~$\rho_\chi$ including Earth scatterings,
\begin{align}
	\frac{\hat{\rho}_{\chi}}{\rho^{(0)}_{\chi}}\sim \frac{w_{\rm tot}}{w^{(0)}_{\rm tot}}\, .
\end{align}
The number of simulated particles in the initial scatter-less simulation run and the main simulations, $N_{\rm tot}^{(0)}$ and $N_{\rm tot}$, generally differs, which can simply be taken into account resulting in the density estimate
\begin{align}
	\hat{\rho}_{\chi}= \frac{N^{(0)}_{\rm tot}}{N_{\rm tot}}\frac{w_{\rm tot}}{w^{(0)}_{\rm tot}}\; \rho^{(0)}_{\chi}\, .\label{eq:localrho}
\end{align}
The standard deviation of the density is obtained via error propagation,
\begin{align}
	\sigma_{\rho_{\chi}}^2 &= \left[\frac{\sigma_{w_{\rm tot}}^2}{w_{\rm tot}^2}+\frac{\sigma_{w_{\rm tot}^{(0)}}^2}{(w_{\rm tot}^{(0)})^2} \right]\rho_{\chi}^2\, ,\quad\text{where }\sigma^2_{w_{\rm tot}}=\sum\limits_{j=1}^{N_{\rm tot}}w_j^2\, .
\end{align}
All these steps are performed for each isodetection ring independently.

\paragraph{Direct Detection Rates}

The standard nuclear recoil spectrum for a direct detection experiment is given in eq.~\eqref{eq: nuclear recoil spectrum}. For \ac{SI}~interactions, it can also be written as
\begin{align}
	\frac{\dd R}{\dd E_R} &=\frac{\rho_{\chi}}{m_{\chi}} \frac{ \sigma^{\rm SI}_{N}}{2\mu_{\chi N}^2}\eta(v_{\rm min})\, ,\quad\text{with }v_{\rm min} =\sqrt{\frac{E_R m_N}{2 \mu_{\chi N}^2}}\, .\label{eq: dRdER with eta}
\end{align}
The DM density is already well estimated by eq.~\eqref{eq:localrho}. The~$\eta$ function has been defined in eq.~\eqref{eq: eta function}, the analytic expression for the Maxwell-Boltzmann distribution is given by eq.~\eqref{eq:etav}. If we can estimate this function on the basis of the \ac{MC}~data, we have a full \ac{MC}~estimate of recoil spectra, event rates, and signal numbers.

The integral of the speed distribution is particularly simple in our case as we use a histogram estimator. Therefore, we have to simply sum up the bins' areas in order to obtain a histogram estimator of the ~$\eta$ function. The bin height of bin~$i$ is
\begin{align}
	H_i &= \int\limits_{v>(i-1) \Delta v}\dd v\frac{\hat{f}_\chi(v)}{v} \nonumber\\
	&= \sum\limits_{j=i}^{N_{\rm bins}}\Delta v \frac{\hat{f}_\chi\left((j-1/2)\Delta v\right)}{(j-1/2)\Delta v}=  \frac{1}{N}\sum\limits_{j=i}^{N_{\rm bins}}\frac{W_j}{(j-1/2)}\, .
\end{align}
The bin height~$W_j$ was defined in~\eqref{eq:W}. Hence, the resulting histogram estimate is nothing but
\begin{align}
	\hat{\eta}(v_{\rm min}) &=  \sum\limits_{i=1}^{N_{\rm bins}}H_i\; \mathbf{1}_{B_i}(v_{\rm min})\label{eq:etahistogram}\, .
\end{align}

In the end, we can substitute the local DM~density~\eqref{eq:localrho} and the histogram estimate of~$\eta(v_{\rm min})$ into eq.~\eqref{eq: dRdER with eta} and obtain the \ac{MC}~estimate of the nuclear recoil spectrum for any given experiment. The residual steps to compute total signal numbers, likelihoods, to include detector resolutions, multiple targets, efficiencies, as described in chapter~\ref{s: recoil spectra}, do not differ from the usual analytic computations. The distortions due to underground scatterings enter only into the density and velocity distribution.

It has been verified as a consistency check that simulations of the transparent Earth, i.e. without underground scatterings, reproduce the correct Standard Halo DM distribution, as well as a spatially constant DM~density. The detection events agree perfectly to the standard analytic computation in this case.

\subsection{Results}
\label{ss: diurnal modulation result}
We investigate the effect of underground DM-nucleus scatterings for light~DM and large enough cross sections that DM~particles passing through the Earth will scatter on one or more terrestrial nuclei. The first part of our results focus on the single scattering regime, where the average scattering probability is around~10\%. In this case, the effect of multiple scatterings is of the order of~1\%, and \ac{MC}~simulations would not really be necessary, since this regime can be described analytically with the \textsc{EarthShadow} code~\cite{Kavanagh:2016pyr}. Nonetheless, the analytic results give us the great opportunity to compare and check some simulation results directly. Once we demonstrated that our \ac{MC}~results agree with the independent \textsc{EarthShadow} results, we can be confident in the accuracy of our simulation code and concentrate on its actual purpose, to study the effect of multiple Earth scatterings on direct detection experiments.

For every parameter set of DM~mass and DM-proton scattering cross section new simulations have to be performed, therefore it makes sense to present the characteristic results for a set of benchmark points. For all of these points, we choose a DM~mass of~\SI{500}{\MeV}, which marks the minimum mass the CRESST-II detector was able to probe, and four increasing values for the cross section. The lowest cross section is chosen to correspond to the single scattering regime. The three higher cross section are tuned to yield 1, 10, or 50 underground scatterings on average. The benchmark points are summarized in table~\ref{tab: benchmarks}. Here, $\langle N_{\rm sc}\rangle$ is the average number of scatterings and~$N_{\rm Sample}$ the sample size, i.e. the amount of velocity data points recorded per isodetection ring. The collection of such large data samples is computationally expensive. All simulations were performed on the~\textit{Abacus 2.0}, a 14.016 core supercomputer of the DeIC National HPC Center at the University of Southern Denmark. They involved simulations of up to $\text{10}^{\text{11}}$ individual DM~trajectories.

\begin{table*}
\centering
	\begin{tabular}{lcccc}
	\hline
		Simulation ID	&$m_\chi$~[MeV]&$\sigma^{\rm SI}_{p}$[pb]	&$\langle N_{\rm sc}\rangle$	&$N_{\rm Sample}$	\\
		\hline
		`SS'		&500	&0.5206							&0.12							&$10^7$		\\
		`MS1'		&500	&4.255								&1.0							&$10^7$		\\
		`MS10'		&500	&42.45								&10.0							&$2\times 10^6$\\
		`MS50'		&500	&297.5							&$\gtrsim$50.0						&$10^6$	\\
		\hline
	\end{tabular}
\caption{The four benchmark points to study the impact of DM-nucleus scatterings and diurnal modulations at direct detection experiments.}
\label{tab: benchmarks}
\end{table*}

\paragraph{The single scattering regime}

\begin{figure*}
\includegraphics[width=0.49\textwidth]{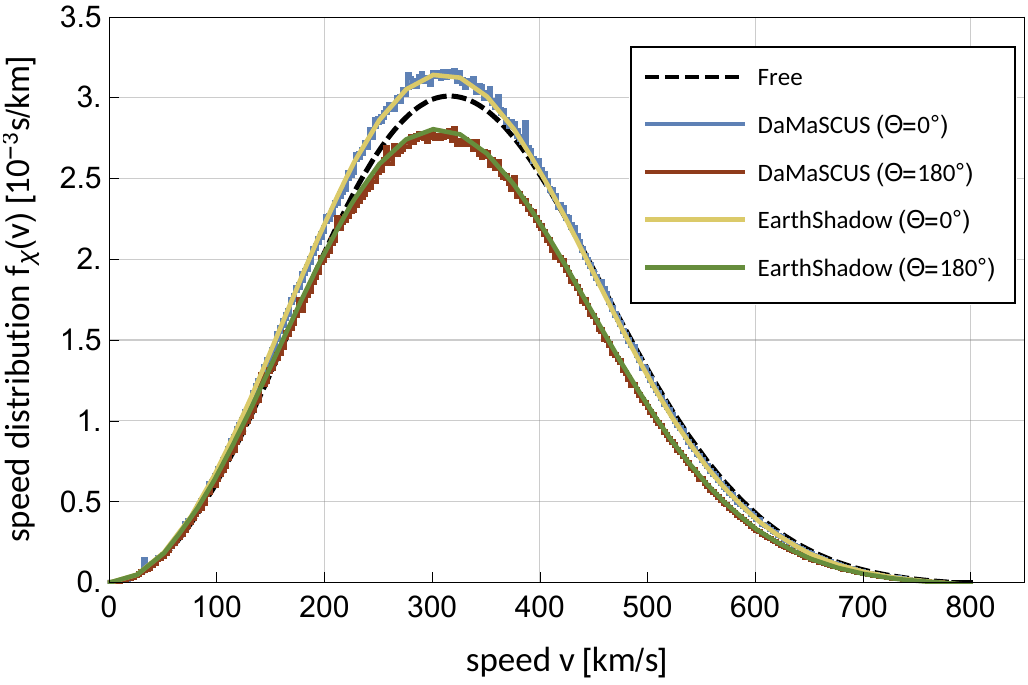}
\includegraphics[width=0.49\textwidth]{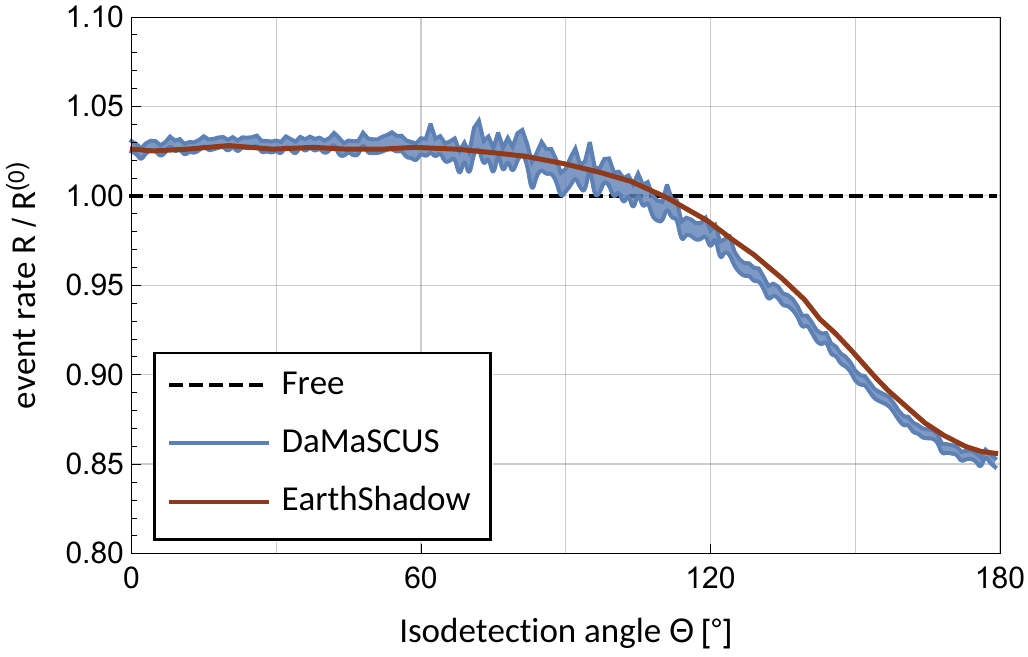}
	\caption{Comparison of the analytic~\textsc{EarthShadow} and the MC~results of~\textsc{DaMaSCUS}.}
	\label{fig:ss}
\end{figure*}

For the initial consistency check, we perform simulations with a cross section corresponding to an average scattering probability of~10\% and generate the equivalent result using the~\textsc{EarthShadow} tool. The DM~mass and DM-nucleon scattering cross section were set to be 500 MeV and~$\sim$0.5~pb respectively. This cross section was determined using one of \textsc{EarthShadow}'s analytic routines. The simulations yielded a relative amount of freely passing particles of~$\sim$~90\%, the remaining~10\% of the simulated particles scattered at least once, the average number of scatterings added up to 0.12. Therefore, the first comparison shows good agreement, as~$\sim$~1\% of the particles in the \ac{MC}~simulations are expected to scatter twice or more, especially if they pass the denser Earth core. This should result in small deviations between the analytic and the \ac{MC}~approach, which affect regions deeper in the Earth shadow most. There, the particles have to travel longer distances underground and the scattering probability is largest.

The central question is, how the elastic collisions on terrestrial nuclei distorts both the DM~speed distribution as well as the spatial distribution inside the planet. Both modifications can be seen from the speed distribution, if we stipulate that the normalized distribution corresponds to a DM~energy density of~$\rho^{(0)}_\chi=\SI{0.3}{\GeV\per\cm\cubed}$. Hence, the local speed distribution in the Earth's shadow will satisfy~$\int\dd v f(v)<1$, since~$\rho_\chi<\rho^{(0)}_\chi$. On the other hand, if an experiment faces the DM~wind, the local DM~density can also exceed the local halo density, such that~$\int\dd v f(v)>1$. In this case, many particles which were originally not moving towards this region get deflected. A portion of the underground DM~population gets redistributed from large to small values of~$\Theta$. The distributions are plotted in the left panel of~figure~\ref{fig:ss}. Even for a relatively low scattering probability, the distributions show this behaviour very well. Particles heading to regions in the Earth's shadow are likely to get scattered off their original path, and the local particle density there is decreased. The shape of the distributions of~\ref{code1} and~\textsc{EarthShadow} agree remarkably well. Both show that the major modification of the DM distribution is due to deflections and the resulting redistribution of the DM~particles. Deceleration plays only a minor role, because a single scattering is not capable to cause a significant energy loss for a relatively light particle.   

Given this excellent agreement, it is unsurprising that it propagates to the direct detection event rates. In the right panel of figure~\ref{fig:ss}, we show the event rate for a CRESST-II type detector\footnote{See app.~\ref{a:experiments} for a description of the CRESST-II experiment.} for the different isodetection rings and compare it to the equivalent calculations based on the analytic approach. Here, the analytic result was taken from figure~7 of~\cite{Kavanagh:2016pyr}. The Earth's shadow is clearly visible, the event rate drops significantly for~$\ang{120}\lesssim\Theta\leq\ang{180}$ down to~$\sim$ 85\% of the rate~$R^{(0)}$, which does not takes underground scatterings into account.

It is however important to note that an experiment in the northern hemisphere, e.g. located at the~\ac{LNGS} in Italy, is not sensitive to this effect. The event rate is slightly increased, but next to completely flat for~$\ang{0}\leq\Theta\lesssim\ang{90}$, and the isodetection angles covered by such an experiment during a sidereal day never reach higher values as we saw in figure~\ref{fig: detector angle}. In contrast, the same experiment located at the~\ac{SUPL} in Australia would observe a diurnal modulation of~$\sim$~15\%.

\paragraph{Multiple Scatterings}

\begin{figure*}
	\centering
	\includegraphics[width=\textwidth]{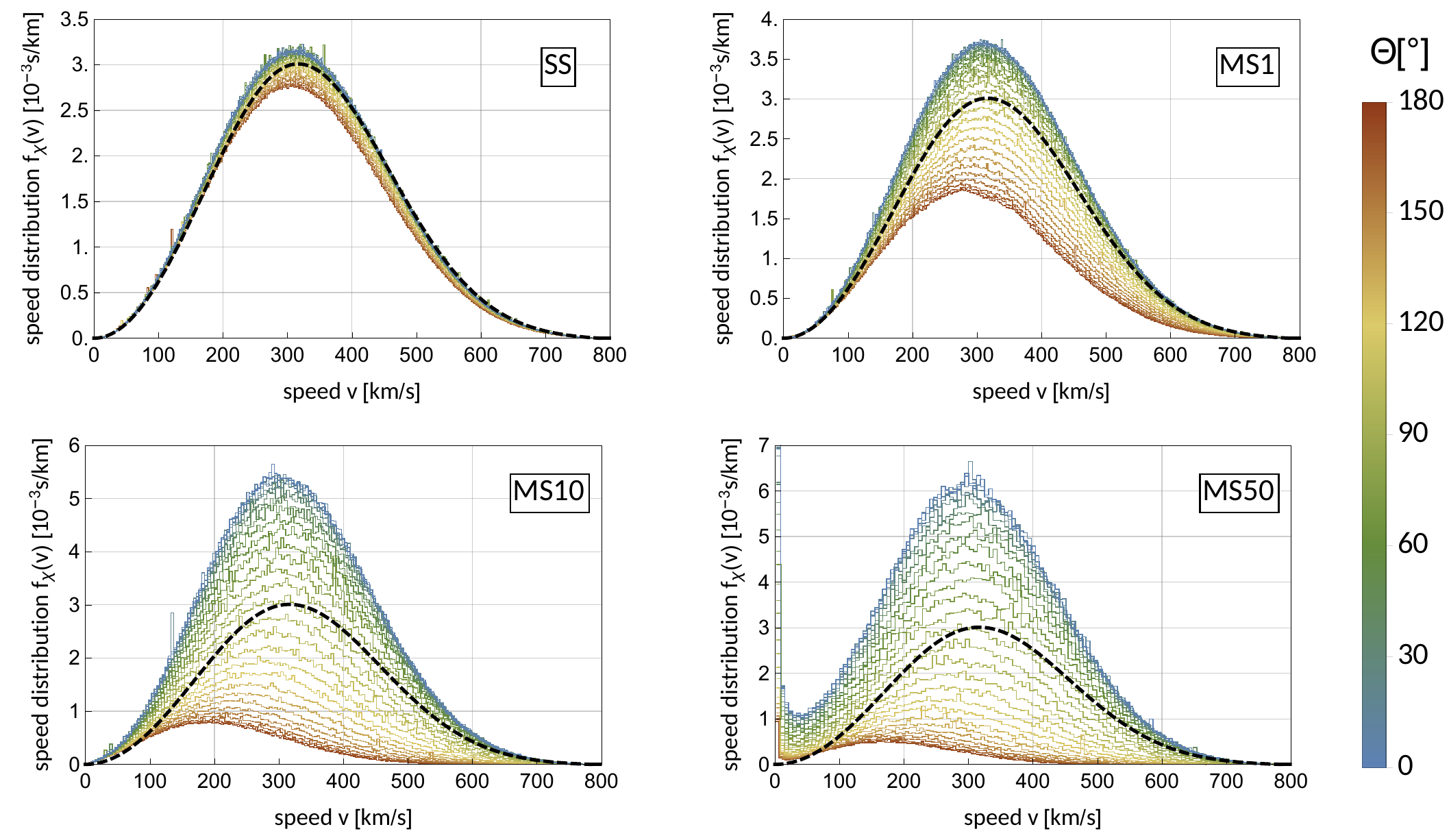}
	\caption{DM speed distributions across the globe for our four benchmark points. Note that they are normalized to \SI{0.3}{\GeV\per\cm\cubed}. The black dashed line shows the speed distribution of free DM.}
	\label{fig: MS1}
\end{figure*}

Having established the correspondence between analytic and simulation results for the single scattering regime, we can direct our focus on the actual purpose of our \ac{MC}~approach. We continue to explore the effect of multiple scatterings in the diffusion regime and increase the cross section, such that we can expect 1, 10 or more than 50 scatterings on average. The values of the cross section for the three benchmark points~`MS1',~`MS10' and~`MS50' have been fine-tuned to yield the respective result for~$\langle N_{\rm sc}\rangle$. 

The local distortions of the DM~speed distribution for all four benchmark points are shown in figure~\ref{fig: MS1}. As one might expect, the underground DM~particles get decelerated more severely for higher cross sections. As opposed to the single scattering regime, the peak of the speed distribution shifts significantly towards slower speeds as~$\Theta$ increases. Similarly, the depletion of the DM population at large isodetection angles occurs to a much higher degree, as the benchmark point~`MS50' demonstrates best. Locations on Earth facing the DM~wind show a vastly increased DM~density caused by multiple scatterings. The effect is essentially the same as in the single scattering regime, but is pronounced much stronger. For very high cross sections, the distributions show a new feature, a second peak close to zero speed, populated by very slow particles. Current experiments are not sensitive to such low-energetic~DM, but future experiments could potentially observe this peak as an increase in low recoil events. Furthermore, these particles could partially get captured gravitationally, leading to a similar effect~\cite{Catena:2016sfr,Catena:2016tlv}.

\begin{figure*}
	\centering
	\includegraphics[width=\textwidth]{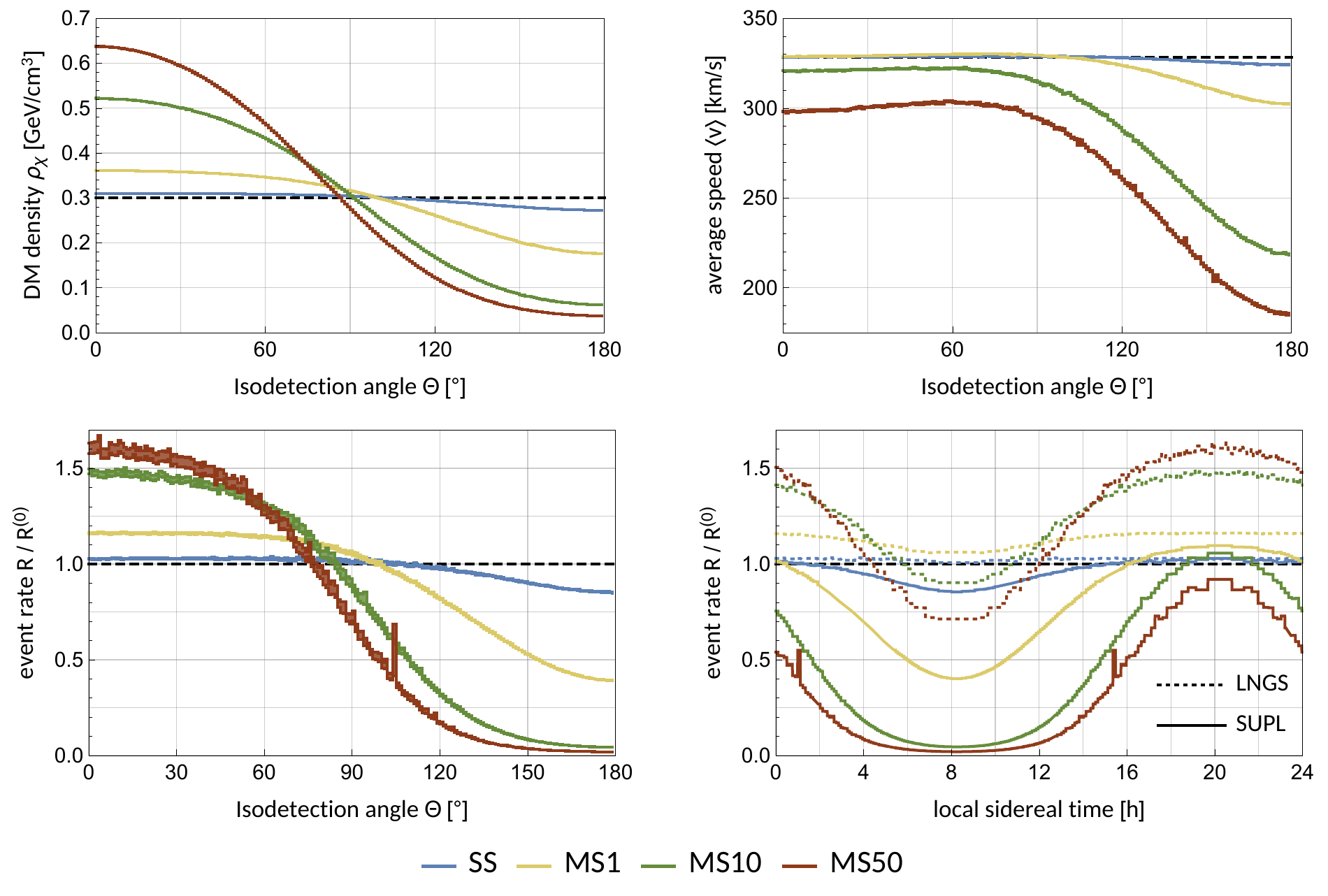}
	\caption{Results of the~MC simulations: The first three plots depict the local DM density, average speed and event rate respectively as a function of~$\Theta$. The last panel shows the corresponding diurnal modulation in the local sidereal time for experiments in the northern and southern hemisphere.}
\label{fig: MS2}
\end{figure*}

These distributions are the primary output of the~\ref{code1} simulations. They entail the local DM~density and average speed and how they evolve with~$\Theta$. The result are plotted in the first two panels of figure~\ref{fig: MS2}. Depending on the cross section the density can increase by more than a factor of~2 for locations facing the DM~wind and decrease down to below~\SI{0.05}{\GeV\per\cm\cubed} in the Earth's shadow. Similarly the deceleration of DM~particles is most severe for larger isodetection angles as expected. This results in an increase (decrease) of the event rate for direct detection experiments located at small (large) isodetection angle, shown in the third panel of figure~\ref{fig: MS2}. Again, we consider a CRESST-II type experiment as an example. Using eq~\eqref{eq: isodetection angle} in connection with eq.~\eqref{eq:vearth} and~\eqref{eq:labpos}, we can relate this result to a diurnal modulation specific to the laboratory's fixed location on Earth.

The last panel of figure~\ref{fig: MS2} depicts this daily change of the event rate over the course of one sidereal day for an experiment both at the \ac{SUPL} (\ang{37.07},\ang{142.81}E) in the southern hemisphere and at the~\ac{LNGS} (\ang{45.454} N, \ang{13.576} E) in Italy. Note that the phases only coincide, because we plot the signal rate as a function of the~\ac{LAST}. By transforming the modulation in e.g.~\ac{UT}, we can obtain the modulations' phases as well.

\begin{figure*}
\centering
\includegraphics[width=0.67\textwidth]{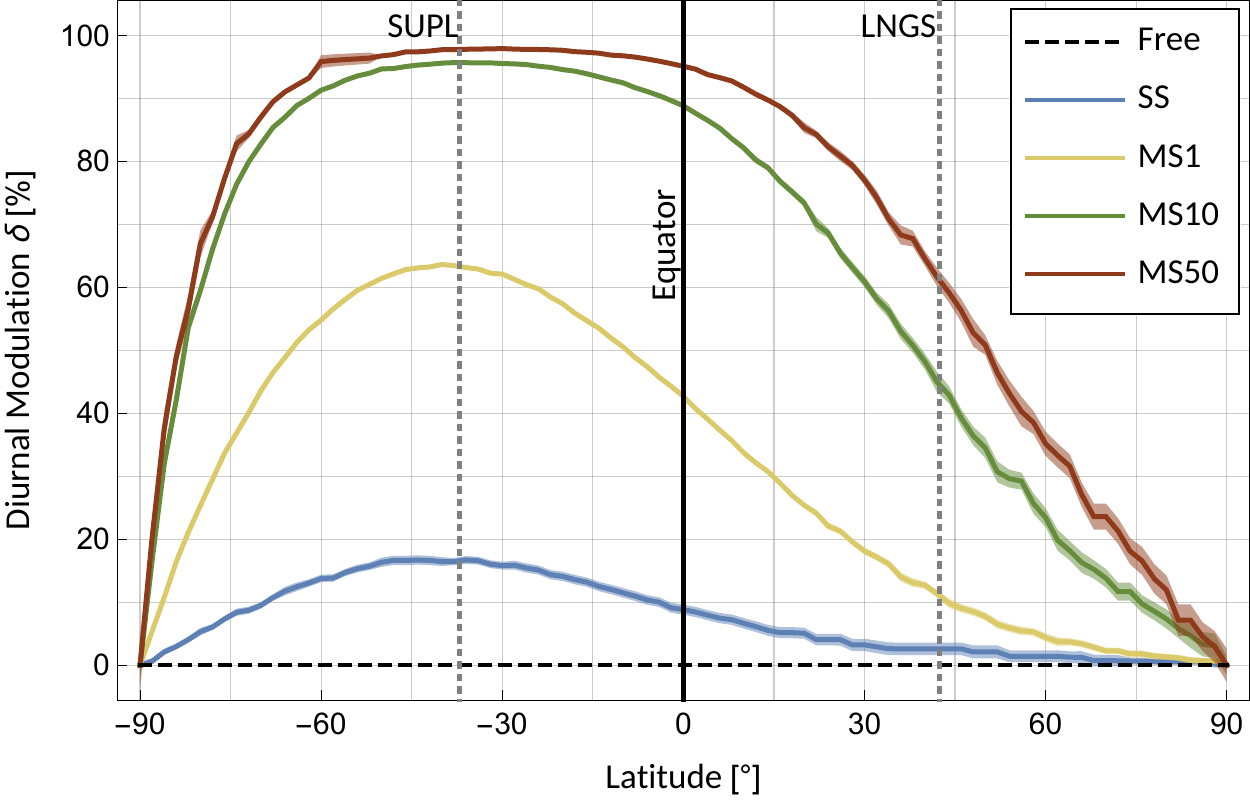}
\caption{The percentile signal modulation as a function of latitude.}
\label{fig: percentile modulation}
\end{figure*}

In the single scattering regime, experiments in the northern hemisphere are insensitive to diurnal modulations caused by Earth scatterings. However, they are no longer reserved for the southern hemisphere at larger cross sections. While it is still generally true that experiments in e.g. Australia maximize the daily modulation, we find that northern experiments can expect a sizeable effect as well, at least in the diffusion regime. To study the modulation amplitude's general dependence on the laboratory's latitude, we define the percentile signal modulation,
\begin{align}
	\delta(\Phi) =100\; \frac{R_{\rm max}-R_{\rm min}}{R_{\rm max}}\;\%\; ,
\end{align}
where $\Phi$ is the laboratory's latitude. The modulations~$\delta$ for the benchmark points are shown in figure~\ref{fig: percentile modulation}. With the exception of the poles' neighbourhood, diurnal modulations can be significant almost all over the globe, if incoming DM~particles scatter multiple times on terrestrial nuclei. For the benchmark points~`MS1',~`MS10', and~`MS50', an experiment at~\ac{LNGS} could expect to observe modulations of~10,~45, and almost~60\% respectively. The corresponding modulations at the ~\ac{SUPL} naturally exceed these values with about~18,~65 and more than~90\%.

For even higher cross sections we can expect underground scatterings to occur even in the overburden of an experiment, for example in the crust or atmosphere. Above some critical value these layers shield off not only background but also DM~particles. Terrestrial experiments are not able to probe~DM above some critical cross section. Finding this cross section for different scenarios is the main goal of the next chapter.

\section{Constraints on Strongly Interacting DM}
\label{s:simp constraints}

The diurnal modulations of the signal rate of DM~particles with moderate DM-nucleon interaction were the consequence of the shadowing or shielding effect of the Earth's bulk. The planet's mass reduces the DM~flux at the detector, especially if the DM~wind has to pass large parts of the Earth to reach the laboratory. For even higher cross sections experiments might lose sensitivity altogether. The vast majority of direct detection experiments are located deep underground, typically~$\sim$~1~km beneath the surface, in order to reduce the background such as atmospheric muons or highly energetic cosmic rays. Above a certain critical DM-nucleon cross section, the rocky overburden of the Earth crust does not just shield off the detector of this undesired background, but also the actual DM~signal it is set up to observe. The~DM not only scatters in the detector, but also on terrestrial nuclei in crust and atmosphere. These scatterings deflect and decelerate incoming particles and reduce the detectable DM~flux at the experiment. Consequently, cross section above a critical value are not probed, and only a band of cross sections can be excluded. 

Already mentioned in the original direct detection by Goodman and Witten~\cite{Goodman:1984dc}, this issue has been studied in various contexts. The central question has been, whether or not this effect reduces constraints on strongly interacting~DM in a way that an `open window' in parameter space emerges~\cite{Starkman:1990nj}. The excluded band is bounded by underground direct detection experiments from below and astrophysical constraints from above. Strong~DM-baryon interactions would have an observable impact on various astronomical observations. Strong interactions of this kind would lead to momentum transfer between the visible and dark sector and affect the~\ac{CMB}'s anisotropy~\cite{Chen:2002yh,Dvorkin:2013cea,Gluscevic:2017ywp}, affect primordial nucleosynthesis~\cite{Cyburt:2002uw}, or introduce a collisional damping effect during cosmological structure formation~\cite{Boehm:2004th}. Furthermore, collisions of DM~particles and cosmic rays would lead to the production of neutral pions and hence observable gamma-rays~\cite{Cyburt:2002uw,Mack:2012ju,Hooper:2018bfw}, or modify the cosmic ray spectrum via elastic scatterings~\cite{Cappiello:2018hsu}. Non-astrophysical constraints on strongly interacting~DM are set by the satellite experiment IMP7/8~\cite{SnowdenIfft1990,Wandelt:2000ad}, experiments on the Skylab space station~\cite{Shirk1978,Wandelt:2000ad}, the X-ray Quantum Calorimeter~(XQC) placed in a rocket~\cite{Erickcek:2007jv}, early balloon-based experiments~\cite{Rich:1987st}, and searches for new nuclear forces~\cite{Fichet:2017bng}.

During the last 30 years, several allowed regions in parameter space were found for both light and heavy DM~masses~\cite{Starkman:1990nj,Zaharijas:2004jv}. These windows have since then been closed by arguments based on DM~capture by the Earth and heat flow anomalies~\cite{Mack:2007xj}, observations from the IceCube experiment~\cite{Albuquerque:2010bt} and the Fermi Gamma-ray Space Telescope~\cite{Mack:2012ju}. Another window in the low mass region was closed more recently by a re-analysis of old data from DAMIC and XQC~\cite{Mahdawi:2017cxz} and new data from the CRESST 2017 surface run~\cite{Davis:2017noy}. Other works focused on super-heavy~DM, where an allowed window was closed by a re-analysis of CDMS-I data~\cite{Kavanagh:2017cru}.

The exact exclusion limits for direct detection experiment taking the terrestrial effects into account have so far been determined mostly based on the analytic stopping power~\cite{Starkman:1990nj,Sigurdson:2004zp,Kouvaris:2014lpa,Foot:2014osa,Kouvaris:2015laa,Davis:2017noy,Kavanagh:2017cru,Hooper:2018bfw,Armengaud:2019kfj}, but also by using \ac{MC}~simulations~\cite{Zaharijas:2004jv,Mahdawi:2017cxz,Mahdawi:2017utm,Mahdawi:2018euy}, which are best suited for this problem. The analytic approach fails to describe deflections of DM~particles and will in most cases either under- or overestimate the overburden's stopping power depending on the DM~mass and interaction type. In this section, we present a new systematic \ac{MC}~treatment of constraints on strongly interacting~DM both for DM-nucleus and DM-electron scattering experiments. The main objective of these simulations is to systematically find the critical cross section, above which any given experiment becomes blind to~DM itself. For DM-electron experiments, we do not limit our analysis to contact interactions as before and consider also electric dipole and long range interactions, mediated by ultralight dark photons. The \ac{MC}~simulations loosely resemble the ones from in the last chapter, but differ in a number of important points, as we will discuss in chapter~\ref{ss: MC of stopping}.  Versions of the \ref{code2}~code developed for this purpose were released with \ref{paper3} and \ref{paper5} and are publicly available~\cite{Emken2018a}.

\subsection{Analytic methods using the DM stopping power}
\label{ss: stopping power}

There are multiple channels for a DM~particle to lose energy while traversing a medium.
\begin{enumerate}
	\item Nuclear stopping: Elastic collisions with terrestrial nuclei.
	\item Electronic stopping: Inelastic scatterings with bound electrons, leading to ionization and excitation.
	\item Atomic stopping: Elastic and inelastic collisions with bound electrons, where the electron remains bound and the whole atom recoils. This process is also relevant for very  light DM~particles, whose kinetic energies fall below the atomic binding energy scales~$\mathcal{O}$(10)eV.
\end{enumerate}
In the cases relevant for us, the nuclear stopping dominates, which is why our simulations only include elastic~DM-nucleus interactions~\cite{Kouvaris:2014lpa}. Electronic and atomic stopping can however become relevant in certain models, where DM-quark interactions are not allowed or heavily suppressed~\cite{Kopp:2009et}. The quantitative description of the electronic stopping power is a great challenge requiring methods from condensed matter and geophysics. In particular, the computation is complicated by the electronic structure of the chemical compounds found in rocks, such as e.g. $\text{SiO}_{\text{2}}$, $\text{Al}_{\text{2}}\text{O}_{\text{3}}$, or FeO. In \ref{paper5}, we discuss the electronic and atomic stopping power in greater detail. We also derive an estimate for the DM~particles' energy loss due to ionization and atomic scatterings, potentially relevant for leptophilic models and use these estimates to set approximate constraints on leptophilic DM~models.

Naturally, the higher the average number of scatterings prior to reaching the detector, the lower the detectable DM~flux at the underground detector. We discussed in chapter~\ref{ss:free path length} that the scattering probability is given in terms of the mean free path~$\lambda$. However, the actual energy loss does not just depend on the number of interactions, but also on the relative energy loss in each interaction. This in turn depends critically on the DM~mass~$m_\chi$ relative to the target mass and the interaction type. In a contact interaction between a DM~particle and target nucleus of similar mass, the DM~particle can lose a significant fraction of its kinetic energy. If however, the same scattering is mediated by an ultralight dark photon, forward scattering is heavily favoured, and the DM~particle's deceleration is marginal. The average relative energy loss in a single scattering is given by
\begin{align}
	\left\langle \frac{E_R}{E_\chi}\right\rangle = \int\limits_{-1}^{1}\dd \cos \theta\, \frac{E_R(\cos\theta)}{E_\chi}f_\theta(\cos \theta)\, ,
\end{align}
where the scattering angle's~PDF is given in eq.~\eqref{eq: PDF scattering angle}. A~10~GeV DM~particle with a speed of~$\text{10}^{\text{-3}}$ scattering on an oxygen nucleus illustrates the importance of the interaction type. For contact interactions, it loses~48\% of its kinetic energy on average. However, if the mediator is ultralight, it loses only~$\sim$~\SI{6e-4}{\percent} in the same scattering. It will take a lot more scatterings and a much shorter mean free path in the second case to effectively attenuate the detectable DM~flux in the overburden. In conclusion, the mean free path is generally not a good measure to quantify the overburden's ability to weaken the DM~flux. Instead, the actual local quantity of interest should be the~\emph{stopping power}, the average loss of energy per distance travelled in the medium.

\paragraph{Nuclear stopping power}
As a DM~particle from the halo with kinetic energy~$E_\chi$ passes through a medium, it will lose energy due to elastic scatterings with nuclei. The average energy loss per travelled distance, the nuclear stopping power, is a local quantity and can be computed via~\cite{Starkman:1990nj,Kouvaris:2014lpa},
\begin{subequations}
\label{eq: nuclear stopping power}
\begin{align}
	\mathcal{S}_n(\mathbf{x},E_\chi) \equiv - \frac{\dd \langle E_\chi\rangle}{\dd x}&=\sum_i n_i(\mathbf{x})\sigma_i \langle E_R\rangle\\
	&=\sum_{i}n_i(\mathbf{x}) \int\limits_{0}^{E_R^{\rm max}}\dd E_R\; E_R \frac{\dd \sigma_{i}}{\dd E_R}\, . 
\end{align}
\end{subequations}
Note that the average recoil energy~$\langle E_R\rangle$ depends on~$E_\chi$, and this is a first order differential equation. The solution yields the energy (or speed) as a function of distance~$d$ travelled in the medium $\langle E_\chi\rangle(d)$ (or $\langle v_\chi\rangle(d)$) for some specified initial conditions.

In the case of \ac{SI}~interactions of light~DM, where the nuclear form factor can be neglected and the differential cross section simplifies to eq.~\eqref{eq: diff cs SI light DM}, the average recoil energy of isotropic contact interactions is simply
\begin{align}
	\langle E_R\rangle&=\frac{E_R^{\rm max}}{2} = \frac{\gamma}{2}E_\chi = \frac{2\mu_{\chi T}^2}{m_\chi m_T}E_\chi \, ,
	\intertext{and the corresponding stopping power takes the form}
	\mathcal{S}^{\rm SI}_n(\mathbf{x},E_\chi) &= \sum_i n_i(\mathbf{x}) \frac{2\mu_{\chi i}^2\sigma_i^{\rm SI}}{m_i m_{\chi}}E_\chi\equiv 2\Delta E_\chi\, .
\end{align}
Here, we cleaned up the notation and introduced the factor
\begin{align}
	\Delta\equiv \sum_i n_i(\mathbf{x}) \frac{\mu_{\chi i}^2\sigma_i^{\rm SI}}{m_i m_{\chi}} \, .
\end{align}
For a homogenous medium, inside which the density and composition remains constant, it is possible to find explicit solutions. After travelling a distance~$d$ in this medium, the average energy of particles with initial energy~$E_{\chi}^{(0)}$ decreased exponentially,
\begin{subequations}
 \label{eq: SI energy loss}
\begin{align}
	\langle E_\chi\rangle(d) &= E^{(0)}_{\chi} \exp \left[ -2\Delta d\right]\, ,\label{eq: stopping solution SI}
	 \intertext{or in terms of the average speed,}
	  \langle v_\chi\rangle(d) &= v^{(0)}_{\chi} \exp \left[ -\Delta d\right]\, .
\end{align}
\end{subequations}
We confirmed the accuracy of this expression for the average speed with \ac{MC}~simulations in~\ref{paper1}.\\[0.3cm]

In the dark photon model, the situation is complicated by the charge screening, which can not be neglected for low-mass~DM and an ultralight dark photon mediator. A closed form solution for long range interactions in analogy to eq.~\eqref{eq: SI energy loss} cannot be found. However, we can evaluate the integral in eq.~\eqref{eq: nuclear stopping power},
\begin{align}
	\mathcal{S}_n(\mathbf{x},E_\chi) = \frac{\sigma_p q_{\rm ref}^4m_\chi}{16\mu_{\chi p}^2}\frac{1}{E_\chi}\sum_i\frac{n_i(\mathbf{x})Z_i^2}{m_i}\left[ \log\left(1+x_i\right)-\frac{x_i}{1+x_i}\right]\, ,
\end{align} 
where~$x_i\equiv a_i q_{\rm max,i}^2=\frac{8\mu_{\chi i} a_i^2}{m_\chi}E_\chi$ depends on the DM~particle's kinetic energy. Assuming that the nuclear composition, i.e. the isotopes' mass fractions~$f_i$, do not change along the DM~particle's path, we can separate the dependences on~$\mathbf{x}$ and~$E_\chi$,
\begin{align}
	\mathcal{S}_n(\mathbf{x},E_\chi) =\rho(\mathbf{x}) \underbrace{\frac{\sigma_p q_{\rm ref}^4m_\chi}{16\mu_{\chi p}^2}\frac{1}{E_\chi}\sum_i\frac{f_i Z_i^2}{m_i^2}\left[ \log\left(1+x_i\right)-\frac{x_i}{1+x_i}\right]}_{\equiv f(E_\chi)}\, ,
\end{align}
factoring out the mass density~$\rho(\mathbf{x})$ as the only position-dependent quantity. For a DM~particle moving a distance~$d$ in the medium, we can then write the implicit solution as
\begin{align}
	\int\limits_{E_\chi^{(0)}}^{\langle E_{\chi}\rangle(d)}\frac{\dd E_\chi}{f(E_\chi)} = -\int \dd \mathbf{x}\, \rho(\mathbf{x})\, . \label{eq: stopping implicit solution}
\end{align}
Further steps, e.g. solving for~$\langle E_{\chi}\rangle(d)$, must be performed numerically.

Most proposed analytic methods to find the critical cross sections are indeed based on the stopping power. We will review two of these methods and discuss the reasons of their inferiority compared to a \ac{MC}~approach in most scenarios.

\paragraph{Analytic speed criterion}
Arguably the simplest method is to find the cross section for which even the fastest halo particles get decelerated that they are no longer detectable. In other words, even the highest recoil they could cause in the detector falls below the recoil threshold~$E_{\rm thr}$. For a DM~particle to be detectable means to have kinetic energy of at least
\begin{align}
	E_{\chi}^{\rm min}= \frac{E_{\rm thr}}{\gamma}=\frac{m_Tm_\chi }{4\mu_{\chi T}^2}E_{\rm thr}\, . 
\end{align}
Here, $m_T$ is the mass of the detector's lightest target nuclei. To model the path of the DM~particles towards the detector, we can assume that the particles take the direct path between Earth's surface and the detector, i.e. a straight path of length corresponding to the detector's underground depth~$d$. The fastest particles in the halo have kinetic energy of~$E_\chi^{\rm max}=\frac{1}{2}m_\chi(v_{\rm esc}+v_\oplus)^2$. If even these particles are on average slowed down to non-detectabilility, the detector is assumed to have lost sensitivity. 

As the Earth's crust can be modelled as a homogenous medium, the situation is particularly easy for \ac{SI}~contact interactions, where we have the closed solutions given in eq.~\eqref{eq: SI energy loss}. Then the critical DM-nucleon scattering cross section~$\sigma_p^{\rm SI}$ is the solution of
\begin{align}
	E_\chi^{\rm min} &= E_\chi^{\rm max} \exp \left[ -\sum_i f_i\rho \frac{2\mu_{\chi i}^2\sigma_i^{\rm SI}}{m_i^2 m_{\chi}} d\right]\, ,\\
	\Rightarrow \sigma_p^{\rm SI}  &=\left[ \sum_i f_i\rho \frac{2\mu_{\chi i}^4A_i^2}{\mu_{\chi p}^2m_i^2 m_{\chi}} d\right]^{-1}\times\log\left(\frac{E_\chi^{\rm max}}{E^{\rm min}_\chi}\right)\, . \label{eq: speed criterion}	
\end{align}
This cross section can serve as a first estimate of the critical cross section. The analytic speed criterion was first formulated in~\cite{Starkman:1990nj} and applied e.g. in~\cite{Kouvaris:2014lpa,Kouvaris:2015laa}, as well as in~\ref{paper1} and~\ref{paper4}, which also contains a method comparison. Knowing the full \ac{MC}~results, this method turns out to yield a reasonable estimate a posteriori, yet it comes with a list of deficiencies.
\begin{enumerate}
	\item The critical cross section obtained via eq.~\eqref{eq: speed criterion} depends solely on the detector's threshold and is not found by computing expected numbers of events using e.g. Poisson statistics. The experiment's details, in particular the exposure, do not enter the estimate, and the two boundaries of the exclusion limits are not on equal footing.
	\item Similarly, the knowledge of the halo DM~particles' speed distribution is not being used, except for the~PDF's hard cutoff, where the distribution is truncated. Setting~$v_{\chi}^{(0)}=v_{\rm esc}+v_\oplus$ might be conservative, but is also rather arbitrary.
	\item The incoming particles do not follow straight paths towards the detector. Deflections prolong the underground trajectories, and the average distance travelled will be larger than the underground depth. Furthermore, deceleration is not the only process, which attenuates the underground DM~flux. Particles can also get reflected back to space, which is not taken into account. However, there are some cases, where forward scattering is heavily favoured. This is typically the case for light mediators or very heavy~DM. In these cases, this problem is less severe.
	\item In the case of GeV scale~DM and contact interactions, where a single scattering causes a significant relative loss of kinetic energy, the analytic stopping equation overestimates the reduction of the detectable DM~flux, as pointed out by Mahdawi and Farrar~\cite{Mahdawi:2017cxz,Mahdawi:2017utm}. We emphasized that the stopping power describes the~\emph{average} energy loss. If the reduction of the flux occurs through only a hand full of scatterings, it fails to account for the small, but non-negligible number of rare particles, which scatter fewer times. Although their number is naturally suppressed, this is compensated by the large cross section and probability to trigger the detector, once a particle reaches the detector depth. Higher cross sections increase both the overburden's shielding as well as the event rate in the detector. This problem is most severe, if the DM~mass is close to the mass of terrestrial nuclei.
\end{enumerate}
The last two problems can not be addressed by analytic means alone\footnote{Deflections can in principle be described analytically, but this has only been done in the single-scattering regime~\cite{Kavanagh:2016pyr}.}. To solve these problems, particle transport simulations are the method of choice. However, the problems~1 and~2 can indeed be addressed by substitution of the stopping power into the computation of the DM~distribution and therefore the detection rates.
 
\paragraph{Analytic signal criterion}
The central DM~property which enters the direct detection rate in eq.~\eqref{eq: nuclear recoil spectrum} is the DM~velocity distribution. Typically, we use the halo distribution of the~\ac{SHM}, assuming that the matter surrounding the experiment has no impact on the DM~flux. It is however important to keep in mind that the correct distribution is the one~\emph{at the detector}. It is possible to include the impact of the overburden's stopping power into the local underground speed distribution $f_\chi^d(v_\chi)$ at depth~$d$, provided that we have a handle on the stopping solution~$\langle v_\chi\rangle(d)$, as e.g. for \ac{SI}~contact interactions in eq.~\eqref{eq: SI energy loss}. 

\begin{figure*}
	\centering
	\includegraphics[width=0.67\textwidth]{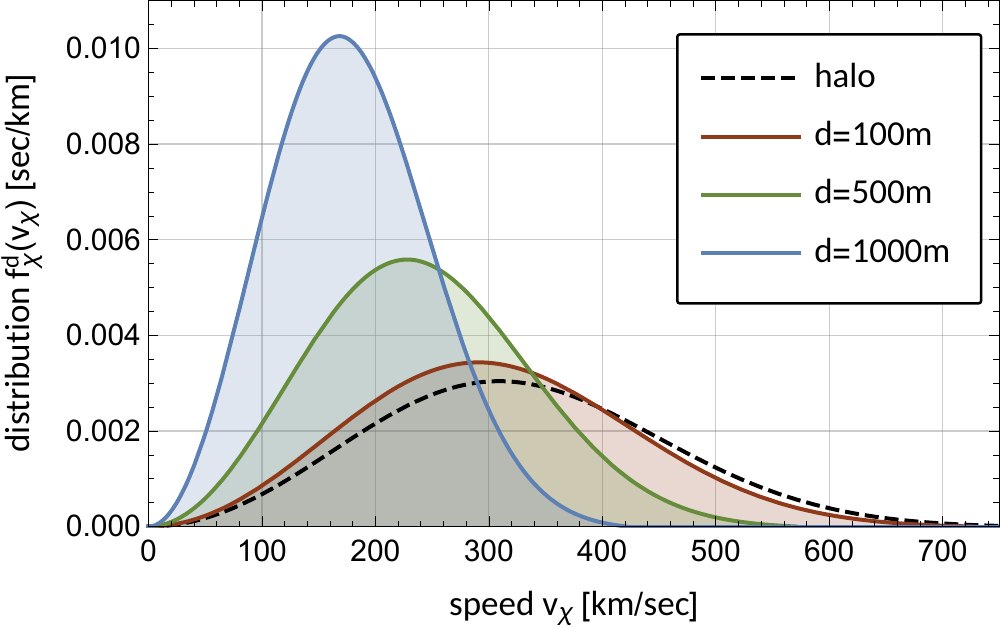}
	\caption{The underground DM~speed distribution of eq.~\eqref{eq: SI underground PDF} for~$m_\chi=$~1~GeV and~$\sigma_p^{\rm SI}=\text{10}^{\text{-30}}\SI{}{\cm\squared}$ at different depths.}
	\label{fig: underground speed distribution}
\end{figure*}

Based on our assumption that the particles move along a straight path and lose energy continuously, no particles get lost and the particle flux is conserved,
\begin{align}
	f_\chi^d(v_\chi^d) v_\chi^d \dd v_\chi^d &= f_\chi(v_\chi^{(0)})v_\chi^{(0)}\dd v_\chi^{(0)}\, .
\end{align}
For contact interactions, we can use the solutions in eq.~\eqref{eq: SI energy loss} and find
\begin{align}
f_\chi^d(v_\chi^d) &=\exp [2\Delta d]f_\chi(\exp[\Delta d]v_\chi^d)\, ,\label{eq: SI underground PDF}
\end{align}
which is shown for a few examples of stopping in the Earth crust\footnote{The Earth crust model, in particular the density and nuclear composition, can be found in app.~\ref{a:earth}.} in figure~\ref{fig: underground speed distribution}.  The local DM~density increases as the particles get slower, which is reflected by the fact that this distribution is no longer normalized. 
For the recoil spectrum, we substitute the result into eq.~\eqref{eq: nuclear recoil spectrum speed} and express it in terms of the halo distribution,
\begin{align}
	&\frac{\dd R}{\dd E_R}(d) = \frac{1}{m_T} \frac{\rho_\chi}{m_\chi}\int_{v_\chi> e^{\Delta d}v_{\rm min}(E_R)}\dd v_\chi \; v_\chi f_\chi(v_\chi) \frac{\dd \sigma^{\rm SI}_{i}}{\dd E_R}(E_R,e^{- \Delta d}v_\chi)\label{eq: dRdER d}\, .
\end{align}
This spectrum can be used to compute signal rates and counts, constraints etc., and the overburden's attenuating effect is automatically taken into account. Above a certain cross section the signal rate starts to decrease with higher interaction strengths. Beyond the critical cross section, the number of expected events falls so much that the parameter space is unconstrained. Comparing the analytic speed and signal criteria, we see that the critical cross section obtained by the first method corresponds to the cross section, for which eq.~\eqref{eq: dRdER d} predicts zero events. The critical cross section based on the speed criterion will always be a slight over-estimation when comparing to this method.

This method can be improved by not just taking straight paths between the Earth surface and detector depth, but taking the directionality into account~\cite{Davis:2017noy}. Around half of the particles approach the detector from below, not above, yet for cross sections close to the critical cross section, these will surely not contribute to the detectable DM~flux. This would lead to additional attenuation of roughly~50\%. Since the signal rate drops very sharply, as we will see later on, the effect on the value of the critical cross section is negligible. Finally, the nuclear form factor could be included to correctly compute constraints on super-heavy~DM~\cite{Albuquerque:2003ei,Kavanagh:2017cru}.\\[0.3cm]

For light mediators, the situation is not that simple, as we have to numerically solve eq.~\eqref{eq: stopping implicit solution} to obtain an incoming particle's final speed at the detector and therefore the underground speed distribution. More details can be found in~\ref{paper5}.

\subsection{MC simulations of~DM in the overburden}
\label{ss: MC of stopping}
\begin{figure*}
	\centering
	\includegraphics[width=0.67\textwidth]{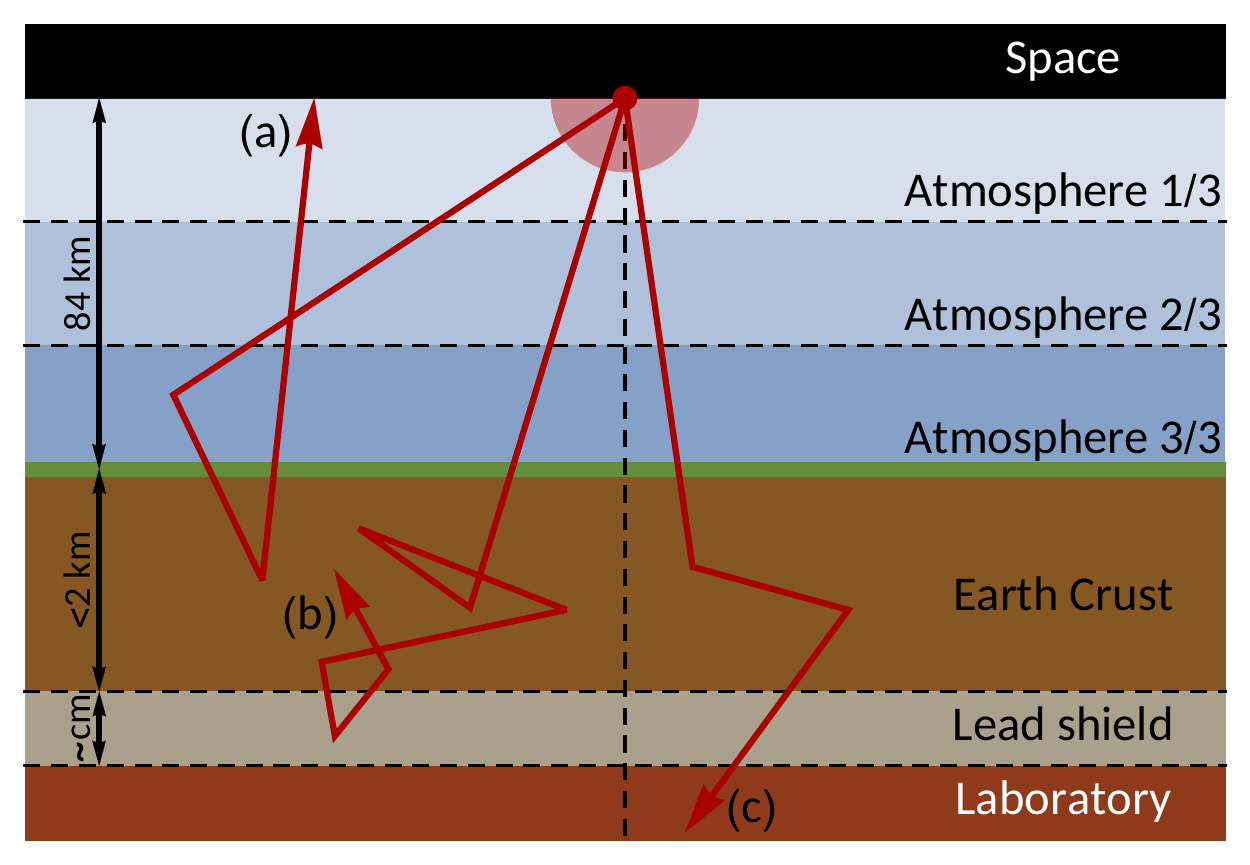}
	\caption{Sketch of the MC simulations of parallel shielding layers (not to scale).}
	\label{fig: crust sketch}
\end{figure*}

The \ac{MC}~simulation of the DM~flux attenuation due to an experiments overburden follows the principles introduced in chapter~\ref{s: MC simulations}. They track individual particles incoming from the halo, as they travel on straight paths until they collide elastically on a terrestrial nucleus, changing direction and losing energy in the process. The most important difference to the~\ref{code1} simulations of chapter~\ref{s: diurnal modulation} is the simulation volume. Instead of the whole planet, we reduce the volume to the shielding layers between the detector and space, which could include the rocky Earth crust, the atmosphere, and additional layers made of e.g. lead or copper. For cross sections close to the critical cross sections, virtually all detectable DM~particle reach the laboratory from above. The relevant scatterings occur in a~km scale volume. On these scales the Earth surface's curvature may be neglected, and we can model the overburden as a stack of parallel, planar layers as illustrated in figure~\ref{fig: crust sketch}. To check this approximation's accuracy, we can monitor the particles' horizontal displacements along their trajectories. If they stay within a few~km at the critical cross section, the inclusion of the Earth's geometry is indeed not necessary. 

The simulation code is optimized for a variable number of user-defined \emph{constant density} layers, using the algorithm of figure~\ref{fig: L algorithm}. A layer is characterized by its thickness, density, composition, and location relative to the other layers. For a layer of non-constant density, the atmosphere being the obvious example, we simply divide it into a larger number of sublayers, each of which is again of constant density. For more details on the atmospheric or the Earth crust model, we refer to app.~\ref{a:earth}.

\paragraph{Trajectory simulation}

The main objective of the simulations is to yield an accurate estimate of the underground distribution of detectable particles, which in turn can be used to compute local event rates for direct detection experiments. A DM~particle is defined as \emph{detectable}, if is able to trigger the detector with a recoil energy above its threshold. It is by definition faster than
\begin{align}
	v_{\rm min} = \sqrt{\frac{m_T (E_{\rm thr}-3\sigma_E)}{2\mu_{\chi T}^2}}\, ,\label{eq: vmin}
\end{align}
where we specify an experiment's threshold~$E_{\rm thr}$, its lightest target's mass~$m_T$, and energy resolution~$\sigma_E$. For light DM~searches the interval of interest~$[v_{\rm min},v_{\rm esc}+v_\oplus]$ typically falls into the halo distribution's high energy tail, and we would like to avoid wasting time by simulating undetectable particles. Therefore, we only simulate particles from this interval.

Before any particle is simulated, we need to specify how initial conditions are being sampled. Since our setup of stacked, parallel, planar shielding layers is invariant under horizontal translations. Therefore the underground DM~distribution solely depends on the depth, and we can pick any initial position above the overburden. For the velocity, we average over the halo distribution's anisotropy and sample the initial speed~$v_0\in[v_{\rm min},v_{\rm esc}+v_\oplus)$ via rejection sampling of the speed distribution in eq.~\eqref{eq: DM speed PDF}. This has the advantage that we do not have to specify, where and when the experiment took place. To determine the velocity's direction we define~$\alpha$ as the angle between the initial velocity~$\mathbf{v}_0$ and the vertical line. Since the initial positions are effectively distributed uniformly in space, and we send off all particles from a fixed altitude, this angle is not isotropically distributed. In order not to overestimate the number of particles with shallow incoming angle (i.e.~$\alpha$ close to~\ang{90}), we have to use the correct~PDF for this angle,~$f_\alpha(\cos\alpha)=2\cos\alpha$, with~$\cos\alpha\in [0,1)$. We transform a sample~$\xi$ of~$\mathcal{U}_{[0,1]}$ into a sample of~$\cos\alpha$ via inverse transform sampling,
\begin{align}
	\int_{0}^{\cos\alpha}\dd(\cos\alpha^\prime)f_\alpha(\cos\alpha^\prime) = \xi \quad\Rightarrow \quad\cos\alpha = \sqrt{\xi}\, .
\end{align}
This leaves us with the initial conditions
\begin{align}
	t_0=0,\, \quad \mathbf{x}_0=\begin{pmatrix}0\\0\\0\end{pmatrix}\, , \quad \mathbf{v}_0= v_0\begin{pmatrix}\sin \alpha\\0\\-\cos \alpha\end{pmatrix}=v_0\begin{pmatrix}\sqrt{1-\xi}\\0\\-\sqrt{\xi}\end{pmatrix}\, .
\end{align}
Here, we picked a particular horizontal direction, which is allowed by the simulation volume's rotational symmetry.

As illustrated in figure~\ref{fig: crust sketch}, there are three triggers terminating a trajectory simulation. The DM~particle is simulated and tracked unterground, until
\begin{itemize}
	\item[(a)] the particle gets reflected back to space,
	\item[(b)] the particle's speed falls below the threshold of eq.~\eqref{eq: vmin}, or
	\item[(c)] the particle succeeds and reaches the detector, while being detectable.
\end{itemize}
In the last case, we record the particle's speed and statistical weight. Otherwise we simply count the failed particle for the estimate of the attenuation.

\paragraph{Rare event techniques} The critical cross section is typically very high and would naively predict a large number of events in a detector. However, the critical cross section is defined as the value, where the detector loses sensitivity to~DM. To compensate signal rates of large magnitude, the underground DM~flux has to be attenuated by the same amount. Therefore, the instance~(c) of a detectable particle reaching the detector without being reflected or decelerated below threshold is extremely rare. Yet, these are the particles we are interested in. To estimate the detectable DM~flux, we need to know how many and how fast DM~particles reach the detector despite the overburden. If the success probability for a single particle is vanishingly small, the problem arises that the simulations are computationally extremely expensive, inefficient and wasteful. Rare-event simulation is a well-studied challenge to particle transport simulations, going back to von Neumann and neutron transport simulations for nuclear reactors. We implement the two most common rare event techniques, which increase the success probability in a controlled manner~\cite{Kahn1951,Bucklew:2004,Haghighat2016}. We should stress that these techniques are absolutely essential for \ac{MC}~simulations in this context to be applicable at all.

If certain statistical properties of the desired data set are known, one can introduce a bias into the simulations'~PDFs which increases the success rate and amplifies sampling favourable values, while keeping track of this bias by a statistical weighing procedure. This method is called~\ac{IS}. Its general principles and specific application for our purpose is reviewed in the app.~\ref{a:IS}. It was first applied for this purpose by Mahdawi and Farrar~\cite{Mahdawi:2017cxz,Mahdawi:2017utm}. The method yields stable and fast results for~GeV scale~DM, where a single scattering can cause significant energy losses. However, if the DM~flux gets attenuated only by hundreds or even thousands of scatterings, this method is not reliable.

Another standard rare event technique is \ac{GIS}, which was first applied to DM~simulations in~\ref{paper5}. This method supports the simulation of successful particles by identifying~`interesting' particles which get close to the detector and then splitting these particles in a number of identical copy, each of which gets tracked independently. This in turn increases the chance that one of them reaches the detector. Furthermore, particles which get less interesting by moving away from the detector have a certain chance to be terminated by what is typically called~\emph{Russian Roulette}. Compared to \ac{IS}, the splitting technique turned out to be more generally applicable and yielded stable and fast results for various cases including trajectories with large numbers of scatterings. One reason is that~\ac{GIS} leaves the simulation's underlying \ac{PDF}'s untouched. In the \ref{code2}~code, the splitting of the particle in copies of itself is realized by a recursive particle simulation function. We discuss the technical details in app.~\ref{a:GIS}.

\paragraph{Finding the critical cross section} 
The \ac{MC}~simulations provide knowledge about how the underground scatterings in the overburden redistribute and attenuate the incoming DM~particles. The generated data needs to be connected to a prediction of the direct detection event rate. This is done by estimating the local speed distribution~$f_\chi^d(v_\chi)$ at depth~$d$. Based on the speed data, we obtain an estimate of the normalized speed distribution through~\acf{KDE}, a non-parametric probability density estimation procedure, which returns a smooth function. The technique of~\ac{KDE} is reviewed in app.~\ref{a:kde}. The second necessary part is the overall attenuation factor~$a_d$, the fraction of incoming particles which passed the distance~$d$ with enough energy to cause a signal. The distribution is hence given by
\begin{align}
	\hat{f}_\chi^d(v_\chi) =  a_d\times \hat{f}^{\rm KDE}(v_\chi)\, .\label{eq: MC PDF}
\end{align}
The attenuation factor is simply the fraction of successful particles. If the total number of incoming particles is~$N_{\rm tot}$, and $N$ trajectories ended at the laboratory depth with respective weights~$w_i$, then
\begin{align}
	a_d =\frac{1}{{N_{\rm tot}}}\sum\limits_{i=1}^{N} w_i\, .
\end{align}
Here, we used that both the average \ac{IS} weight and the initial \ac{GIS} weight is equal to 1 such that~$\sum_{i=1}^{N_{\rm tot}}w_i = N_{\rm tot}$. We have to keep in mind that the total number of simulated particles~$N_{\rm sim}$ is~\emph{not} the same as the total number of incoming particles, as we only picked initial conditions from the interval of interest~$[v_{\rm min},(v_{\rm esc}+v_\oplus)]$ in order speed up computations. However, we can easily relate the two using the initial particles' speed distribution, i.e. eq.~\eqref{eq: DM speed PDF},
\begin{align}
	N_{\rm tot} = \frac{N_{\rm sim}}{\int\limits_{v_{\rm min}}^{v_{\rm esc}+v_\oplus}\dd v_\chi\; f_\chi(v_\chi)}\geq N_{\rm sim}\, .
\end{align}
The speed distribution of eq.~\eqref{eq: MC PDF} is used to compute detection signal rates with e.g. eq.~\eqref{eq: nuclear recoil spectrum speed} or~\eqref{noble} in the usual way. At this point, we are able to compute numbers of events and likelihoods for any kind of direct detection experiment, based on either nucleus or electron scatterings.

We summarize the procedure. For a point~$(m_\chi,\sigma)$ in parameter space, the simulation of DM~particle trajectories through the overburden generates speed data $\{(v_1,w_1),...,(v_N,w_N)\}$ and the attenuation factor~$a_d$. Using~KDE, we obtain a smooth estimate of the local, attenuated speed distribution, which in turn determines the recoil spectra, event rates, likelihoods, etc., depending on the experiment of interest. For a given mass, we start at the usual lower bound on the cross section and systematically increase the cross section, repeating the procedure from above, to find the point where the predicted number of signals starts to decrease. Above that value, the overburden's shielding power dominates the signal rate in the detector, and we carefully keep increasing the cross section in smaller and smaller steps. It is crucial to avoid going beyond the critical cross section, as this is typically the regime, where the necessary computation time grows dramatically. Instead, we carefully approach the critical cross section from below. Once the likelihood grows above~(1-CL), for a specified certainty level~CL, the experiment no longer sets constraints on the point in parameter space. The exact value of the critical cross section is finally obtained by interpolation of the likelihood as a function of the cross section and solving it for~(1-CL).

There are a few effects we neglect in our simulations, which affect the predicted number of events by order one.
\begin{itemize}
	\item Around half the particles approach the detector from below and get shielded, whereas we assume all particles to reach the Earth from above the experiment. The exact additional attenuation depends on the experiment's location on Earth relative to the DM~wind. As we do not specify this and average over the halo distribution's anisotropy, we neglect this attenuation of order~1/2.
	\item A few DM~particles could pass the detector's depth subsequently get reflected back up, which would increase the prediction for the event number.
	\item It was reported that the modelling of the atmosphere as a planar stack of parallel layers leads to an error in the predicted number of detectable particles at the detector of less than a factor~2 when compared to a more accurate geometric setup~\cite{Mahdawi:2018euy}.
\end{itemize}
It should be emphasized that these errors lead to order one changes in the predicted event numbers, which have only minimal effects on the value of the critical cross section. This is due to the fact that, as soon as the signal rates start to drop above some cross section, it drops extremely steeply, as e.g. shown in figure~\ref{fig: number of events}. The Earth crust turns `opaque' to~DM rapidly. However, the quantity of interest is the critical cross section, not the underground event rate. The simulations are not meant nor able to produce precise predictions of detection signal rates, which would require much more details, even if the three issues mentioned above were accounted for.
 
\subsection{Nuclear recoil experiments}
\label{ss: constraints nuclear}
In~\ref{paper4}, we focused on \ac{SI} contact interactions and constraints from nuclear recoil experiments. This is an interesting case, as the fundamental process which is probed in the detector is also the source of the signal's attenuation. Elastic nuclear recoils in the overburden and detector act as antagonizers, where the stopping effects dominates above some interaction strength. In this section, we compare the different methods of the previous chapters and derive the constraints on \ac{SI}~contact interactions for a number of direct detection experiments.

\begin{figure*}
	\centering
	\includegraphics[width=0.8\textwidth]{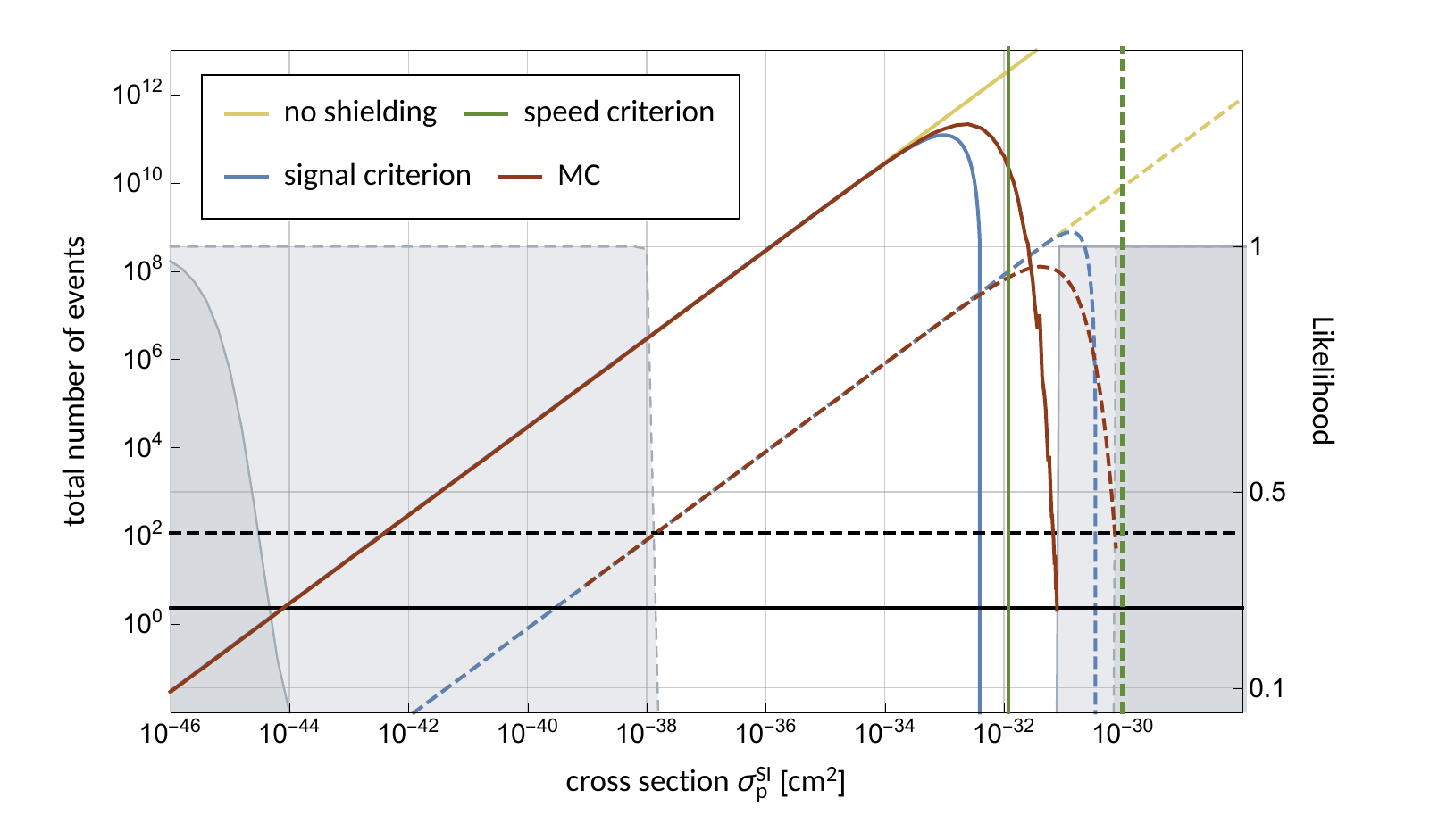}
	\caption{Comparison of the different methods and the predicted number of events for XENON1T (solid lines) and DAMIC (2011) (dashed lines) for a DM~mass of 10 GeV.}
	\label{fig: number of events}
\end{figure*}

The three methods of obtaining the critical cross section discussed in chapter~\ref{ss: stopping power} and~\ref{ss: MC of stopping} are compared for the example of DAMIC~(2011)~\cite{Barreto:2011zu} and XENON1T~\cite{Aprile:2017iyp} in figure~\ref{fig: number of events}. The plot shows the predicted number of events as a function of the~DM-proton cross section determined with the analytic stopping power in eq.~\eqref{eq: dRdER d}, as well as the more accurate \ac{MC}~simulations. It also includes the simple speed based estimate of eq.~\eqref{eq: speed criterion}, which in the case of XENON1T underestimates the critical cross section, while it slightly overestimates it for DAMIC. Without the \ac{MC}~results being available, a quality assessment of this simple criterion is not possible, but it can clearly serve as an easy-to-compute first guess on its own.

For low cross sections, the overburden has naturally no impact on the local speed distribution in the laboratory nor the predicted event numbers, since the mean free path is much longer than the underground depth. Hence, the signal numbers increase linearly with the cross section. The recoil spectrum of eq.~\eqref{eq: dRdER d} reproduces the qualitative behaviour nicely, where the signal strength above a certain value of the cross section starts to drop very steeply. However, compared to the \ac{MC}~based predictions, it overestimates the pace with which the signal number decreases and therefore underestimates the critical cross section by up to an order of magnitude. By looking for the two values of the cross sections, where the likelihood crosses~(1-CL), we obtain the excluded interval. 

For cross sections of order~100pb, we see that the shielding has no visible effect on the signal rates, which seems to be in conflict with the results of chapter~\ref{s: diurnal modulation}, where we had signal modulations of order~100\%. We mention once more that some of the assumptions, e.g. that all particles approach the detector from above, are only valid for cross sections close to the critical cross section. For intermediate values, the particles from below are additionally attenuated since they have to pass the entire planet before they pass the target material. This regime can only be studied by simulating the whole planet, as we did earlier.

The advantages of the \ac{MC}~simulation of trajectories start to emerge at this point. The analytic stopping description neglects the rare, stubborn particles, which scatter fewer times than expected and thereby overestimates the attenuation of the DM~flux. However, in other cases, where reflection, not deceleration, is the dominant process of DM~attenuation, it can also yield an underestimate. The simulations will produce a more accurate, realistic, and consistent result in all cases, which might also be more constraining.

It was claimed that the analytic stopping equation should not be applied for the derivation of the critical cross section of strongly interacting~DM~\cite{Mahdawi:2017utm}. The authors justify this statement with the discrepancy in the event numbers of many orders of magnitude. We would argue that, looking at figure~\ref{fig: number of events}, any conservative method of finding the critical cross section, such as the speed criterion or a recently proposed similar, even more conservative estimate~\cite{Hooper:2018bfw}, will yield values, for which corresponding \ac{MC}~simulations will predict huge event rates.  The resulting exclusion bands may be conservative, but are still valid.

\begin{figure*}
	\centering
	\includegraphics[width=0.75\textwidth]{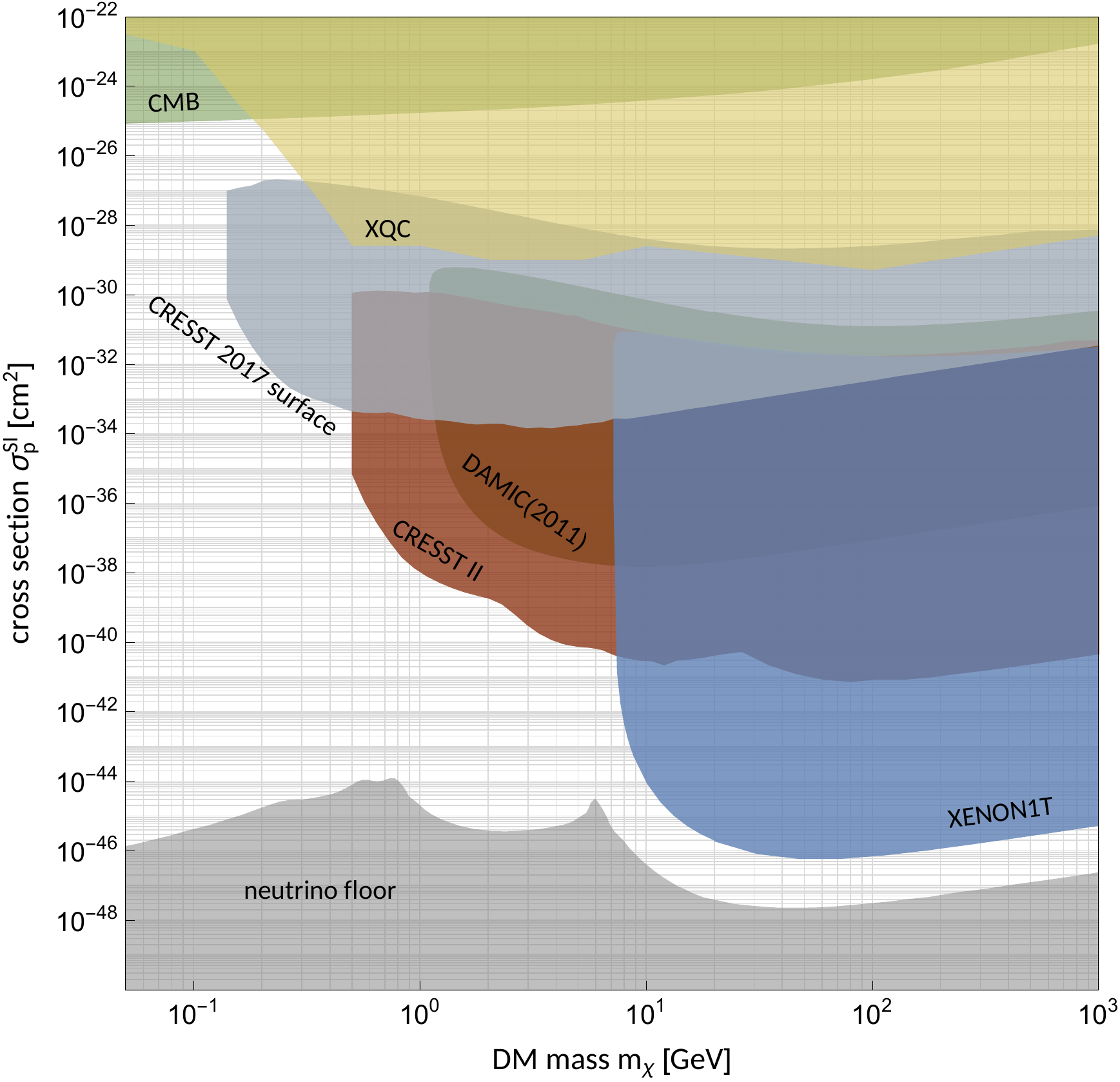}
	\caption{The constraints~(90\%~CL) by various direct detection experiments, together with limits from XQC and the CMB.}
	\label{fig: nuclear constraints}
\end{figure*}

 Using the \ref{code2} simulation code, we derive exclusion limits based on CRESST-II~\cite{Angloher:2015ewa}, the CRESST 2017 surface run~\cite{Angloher:2017sxg}, DAMIC~(2011)~\cite{Barreto:2011zu}, and XENON1T~\cite{Aprile:2017iyp}, which are plotted together with constraints from XQC~\cite{Erickcek:2007jv} and the~\ac{CMB}~\cite{Gluscevic:2017ywp} in figure~\ref{fig: nuclear constraints}. For the more specific details of the different experiments, we refer to the respective app.~\ref{a:nucleus}. We also include the neutrino floor~\cite{OHare:2016pjy}. The direct detection constraints cover DM~masses between~100~MeV and~1~TeV, and cross sections between~$\text{10}^{\text{-46}}$ and~$\text{10}^{\text{-27}}\text{cm}^{\text{2}}$ can be excluded depending on the experiment. For each DM~mass, we obtain a clearly defined, consistent exclusion interval, where both bounds are computed the same way.
 
We included constraints of the DAMIC 2011 run, despite the fact that these constraints are fully covered by the two experiments by the CRESST collaboration. These results are useful since they allow direct comparison to the results obtained with the~\textsc{DMATIS} code~\cite{Mahdawi:2017cxz}, which serves as a independent evaluation. The two results agree to a reasonable precision with deviations of around~10\%.  

Both CRESST-II and XENON1T were located deep underground at~\ac{LNGS} under 1400m of rock. They are therefore insensitive to strongly interacting~DM, with their respective constraints reaching up to $\sim\text{10}^{\text{-30}}\text{cm}^{\text{2}}$ and $\sim\text{10}^{\text{-31}}\text{cm}^{\text{2}}$ in their low mass region. On the other side, the CRESST 2017 surface run of a prototype detector developed for the~$\nu$-cleus experiment is much more powerful in this context. It was performed in a surface laboratory with only a few cm of concrete and the atmosphere as shielding. Despite its tiny exposure it is much more successful in probing high cross sections and extends the excluded intervals by around three orders of magnitude towards stronger interaction strengths. All allowed windows between underground experiments and the XQC rocket experiment can be closed this experiment by the CRESST collaboration. Only a tiny window between~\ac{CMB} constraints remains, which might could get narrowed by including constraints from cosmic rays~\cite{Cyburt:2002uw} or even closed by considerations of heat flow in the Earth~\cite{Mack:2007xj} or newer limits due to gas cloud cooling~\cite{Bhoonah:2018wmw}.

The comparison between CRESST-II and XENON1T illustrates furthermore that the exposure has only a marginal effect on the critical cross section. The exposure of XENON1T exceeds the one by CRESST-II by a factor of~$\sim$700, yet the critical cross section is only up to~$\sim$10\% higher. Compared to DAMIC and the CRESST surface run, it is clear that the underground depth is the dominant factor.  In order to probe strongly interacting~DM with a direct detection experiment, larger exposure are not an efficient strategy. Instead the detectors must be placed above ground, preferably at high altitudes, such as mountains, balloons, rockets, or possibly satellites.\\[0.3cm]

In this chapter, we systematically determined the critical cross section above which a nuclear recoil experiment loses sensitivity to~DM using \ac{MC}~simulations of trajectories in the crust and atmosphere. We presented constraints on \ac{SI}~contact interactions between DM~particles and nuclei coming from a number of experiments. The constraints reach from the interesting sub-GeV mass regime to~1~TeV. Next, we want to focus exclusively on sub-GeV~DM and experiments using~DM-electron scatterings as potential discovery process.

\subsection{Electronic recoil experiments}
\label{ss: constraints electron}
DM particles of sub-GeV mass do not have enough kinetic energy to trigger a conventional detector, which is why DM-electron scatterings were proposed as a new search channel for low masses. Nonetheless, for large cross sections, undetectable DM-nucleus scatterings can indirectly affect DM-electron scattering experiments by redistributing the underground DM~particles. DM-electron scatterings in the crust or atmosphere might in principle do the same, but the stopping due to electron scatterings tends to be a subdominant effect, as discussed in chapter~\ref{ss: stopping power}. In the dark photon model,~DM couples to both electrons and protons with a hierarchy between the respective cross sections. However, the electronic stopping of~DM is much weaker than nuclear stopping in this model, as we discussed in greater detail in~\ref{paper5} and plays a significant role only, if the model does not allow DM-quark interactions. The only truly leptophilic models, where~DM-nucleus interactions are not even generated on the loop level, include pseudo-scalar and axial vector interactions~\cite{Kopp:2009et}. In~\ref{paper5}, we also presented an analytic estimate for the constraints on strongly interacting~DM in these class of models.

A certain model dependence is unfortunately unavoidable, as we need a relation between the two cross sections, since the stopping and detection mechanisms are no longer identical. The dark photon model relates the two cross sections in a simple way, fixing the ratio to the ratio of reduced masses, see eq.~\eqref{eq:sigma ratio}. 

While the algorithmic procedure of finding the critical cross section is the same as in the previous chapter, the~\ref{code2} code~\cite{Emken2018a} has been extended in the following ways.
\begin{enumerate}
	\item More general interactions: We do not limit ourselves to contact interactions and consider light mediators and electric dipole interactions as well. This changes the scattering kinematics critically, as we discussed in chapter~\ref{ss:scattering angle}.
	\item New rare event \ac{MC}~technique: Geometric Importance Splitting was first implemented in this context. The method's details are reviewed in app.~\ref{a:GIS}.
	\item New analyses: For~DM-electron scatterings, the computation of event rates and energy spectra are more involved and require the corresponding ionization form factors. The necessary relations have been discussed in chapter~\ref{ss: direct detection electron}.
\end{enumerate}
In this chapter, we will study how scatterings in overburdens affect direct searches for light~DM. For contact, electric dipole, and long range interactions,  the detection constraints on strongly interacting~DM are determined via \ac{MC}~simulations, and the constraints' general scaling behaviours under change of exposure or underground depth is studied. We re-visit the most recent experiments setting constraints on the DM-electron scattering cross section and find the limits' extend towards strong interactions, before the crust or atmosphere shields off the detector. Namely, we analyse the data of XENON10~\cite{Angle:2011th,Essig:2017kqs}, XENON100~\cite{Aprile:2016wwo,Essig:2017kqs}, SENSEI~(2018)~\cite{Crisler:2018gci}, and SuperCDMS~(2018)~\cite{Agnese:2018col}. In addition, we present projections for future runs of DAMIC and SENSEI at different underground laboratories. Finally, we explore the prospects of a high-altitude run of a direct detection experiment. A semiconductor target could be performed on either a balloon or in Earth orbit. For such an experiment, the strong orbital signal modulation due to the Earth's shadowing effect would be of great value to distinguish a potential signal from an expectedly large background.

\paragraph{Exposure and depth scaling}

\begin{figure*}
	\centering
	\includegraphics[width=\textwidth]{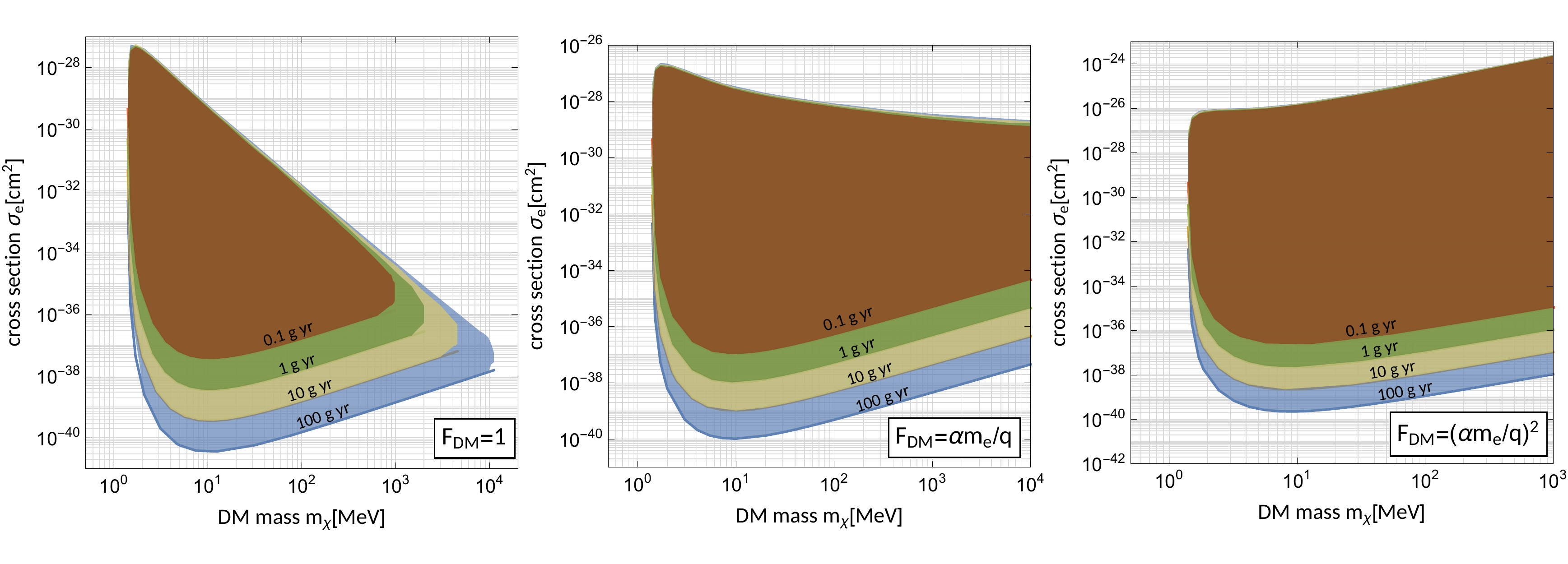}
	\caption{Exposure scaling of the constraints~(95\%~CL) for a generic semiconductor detector.}
	\label{fig: electron projections exposure}
\end{figure*}

Increasing the exposure by running experiments of larger targets for longer times allows to probe weaker DM~interactions and extends the exclusion band towards lower cross sections. In the case of nuclear recoil experiments, we previously found that such an increase has only little effect on the upper boundary of the exclusion band. The critical cross section turned out insensitive to the exposure. Here, we return to this question to check if this conclusions holds also for DM-electron scattering experiments and, more importantly, light mediators. For that, we consider a generic silicon semiconductor experiment, where we fix the ionization threshold to~2 electron-hole pairs, the underground depth to~100m, and assume the absence of background events. The exposure is varied between~0.1 and 100~gram~year. The resulting exclusions are shown in figure~\ref{fig: electron projections exposure}.

The critical cross section is again found extremely insensitive to the exposure. The exposure increase of four orders of magnitude yielded an improvement of the limits by~$\sim$60\% or less. It seems to be a generic feature, that the overburden decreases the detection signal rate very rapidly for a cross section increase beyond some critical value. The shielding layers turn effectively opaque to~DM rather suddenly, and the experimental parameters are not the dominant factors. The only advantage of larger exposures in this case concerns contact interactions, where the probed interval of DM~masses can indeed be extended. Here, larger exposures increase the sensitivity to heavier~DM.\\[0.3cm]

\begin{figure*}
	\centering
	\includegraphics[width=\textwidth]{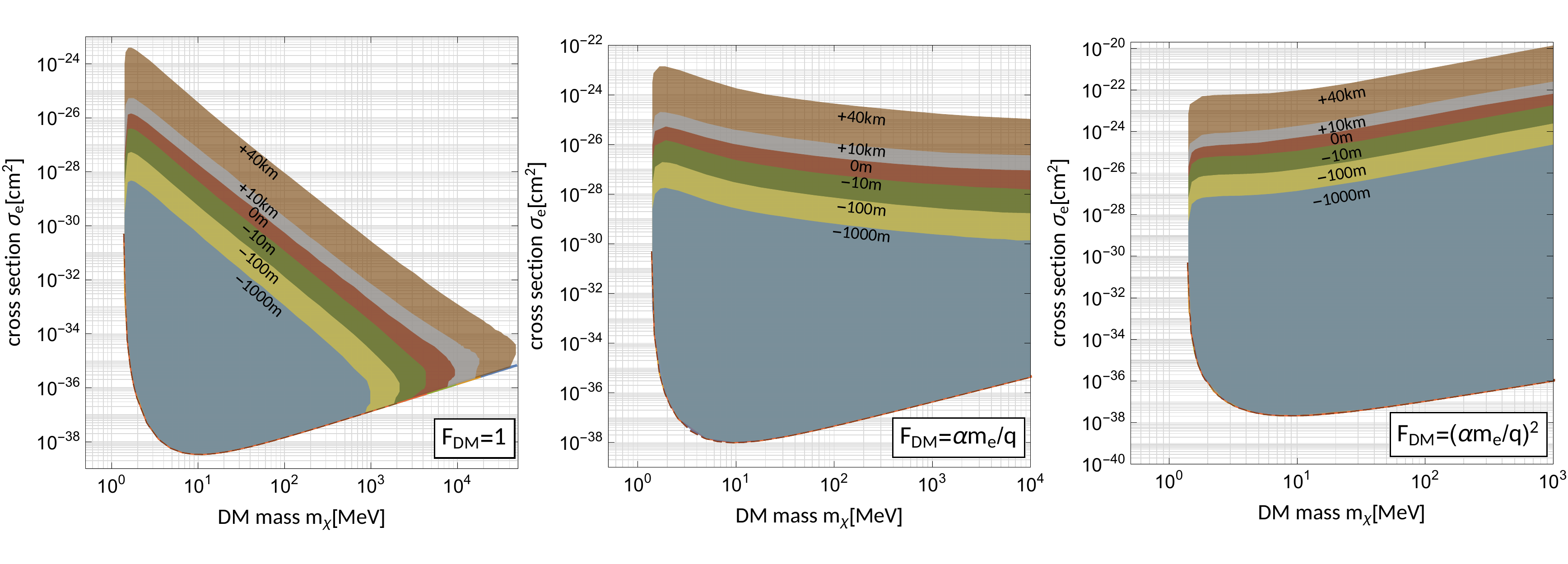}
	\caption{Underground depth scaling of the constraints~(95\%~CL) for a generic semiconductor detector.}
	\label{fig: electron projections depth}
\end{figure*}

The exact dependence on the depth should be quantified. We consider the same experiment as before, but fix the exposure to 1~gram~year. This time, we vary the underground depth between~1~km underground, to 40~km above ground, an altitude a balloon experiment might reach. There the only shielding comes from the high, very thin layers of the atmosphere and the experimental setup itself. Here, we approximate the shielding layers of the apparatus as one layer of steel~(2mm) and one of copper~(1mm). The entire atmosphere's stopping power corresponds to the stopping power of about~$\sim$5m of rock or~$\sim$1m of steel. The atmosphere can be neglected in the simulations for 100m and 1000m.

The resulting constraints~(95\%~CL) are shown in figure~\ref{fig: electron projections depth} demonstrating the fact that the critical cross section grows by about one order of magnitude for each order of magnitude the underground depth is decreased. Having a detector in a laboratory on a mountain will further improve the situation by less than one order of magnitude. To gain more sensitivity, it could be a good idea to run a small-scale balloon-borne experiment, since the exposure is not important at all. We will come back to this idea later in this chapter.

\paragraph{Constraints}

Direct detection experiments, both with noble and semiconductor targets, have been performed to search for~DM-electron interactions. In the absence of a positive result, various collaborations have published constraints on the DM-electron scattering cross section. We reanalyse their data on the basis of our \ac{MC}~simulations, determining the extend of the excluded band in parameter space. Since it is nuclear scatterings which attenuate the DM~flux but electron scatterings which get detected, we are forced to assume some model, in order to have a relation between the proton and electron cross sections. In~\ref{paper1}, we presented first estimates for the dark photon model based on a simple speed criterion. In~\ref{paper5} and this thesis, we present the constraints for the same model, but based on a full data analysis such that lower and upper boundary of the exclusion band are truly on the same footing.

\begin{figure*}
	\centering
	\includegraphics[width=\textwidth]{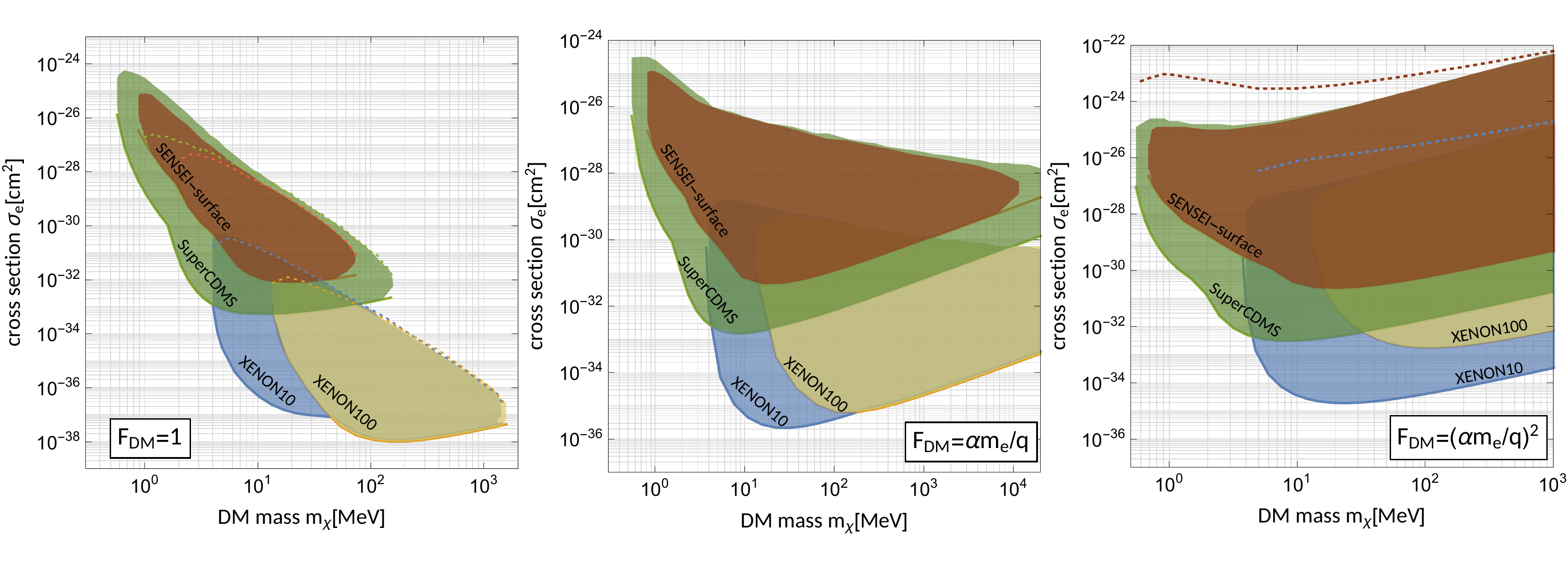}
	\caption{Direct detection constraints~(95\%~CL) by various DM-electron scattering experiments.}
	\label{fig: electron constraints}
\end{figure*}

In particular, we present updated constraints~(95\%~CL) from the S2-only data of XENON10~\cite{Angle:2011th,Essig:2017kqs} and XENON100~\cite{Aprile:2016wwo,Essig:2017kqs}, as well as the surface runs of SENSEI~(2018)~\cite{Crisler:2018gci}, and SuperCDMS~(2018)~\cite{Agnese:2018col}. The details of the respective analyses are presented in app.~\ref{a:experiments}. The constraint plot for contact, electric dipole, and long range interactions are shown in figure~\ref{fig: electron constraints}. 

The constraints for contact interactions in the left panel also include the constraints in the absence of charge screening as dotted lines~($F_{A}(q)=1$). For very light~DM, the charge screening effectively decreases the DM-proton scattering cross section, such that the overburden stopping is less effective and the critical cross section is higher. The dashed line in the right panel show an analytic estimate of the critical cross section for light mediators based on eq.~\eqref{eq: stopping implicit solution}. For low masses, the analytic approach fails to account for deflections or reflections of DM~particles and yields an overestimate. For GeV masses however, we found that the DM~particle scatter sharply in a forward direction and move on relatively straight lines continuously scattering on nuclei. This is well captured by the analytic description of nuclear stopping, and the two estimates converge around~$m_\chi\approx 1$~GeV. Above this mass, \ac{MC}~simulations become more and more impractical, because the number of scatterings increases extremely, where each individual scattering changes the DM~speed only minimally. Then, the analytic method should yield reliable results.

While the lower boundary of the exclusion band depends crucially on the experiments' details, the critical cross section is mostly determined by the overburden. The extend towards larger scattering cross sections is similar for SENSEI and SuperCDMS, since both experiments were performed on the surface. The same goes for XENON10 and XENON100 at the~\ac{LNGS} 1400m underground. For contact interactions, we observe a decrease of the critical cross section for heavier DM~masses, which is caused by the increasing ratio in eq.~\eqref{eq:sigma ratio}, $\sigma_p/\sigma_e \approx \frac{m_\chi^2}{m_e^2}$. For electric dipole interactions and ultralight mediators, this effect is counteracted by the DM~form factor. In these two cases, the momentum transfer grows with the DM~mass, while the differential cross section gets suppressed by~$\sim q^{-2}$ and~$\sim q^{-4}$ respectively. For light mediators, the suppression of large momentum transfers is strong enough to dominate over the increasing cross section ratio, and the critical cross section increases with DM~mass.

\paragraph{Projections and outlook}
\begin{figure*}
	\centering
	\includegraphics[width=\textwidth]{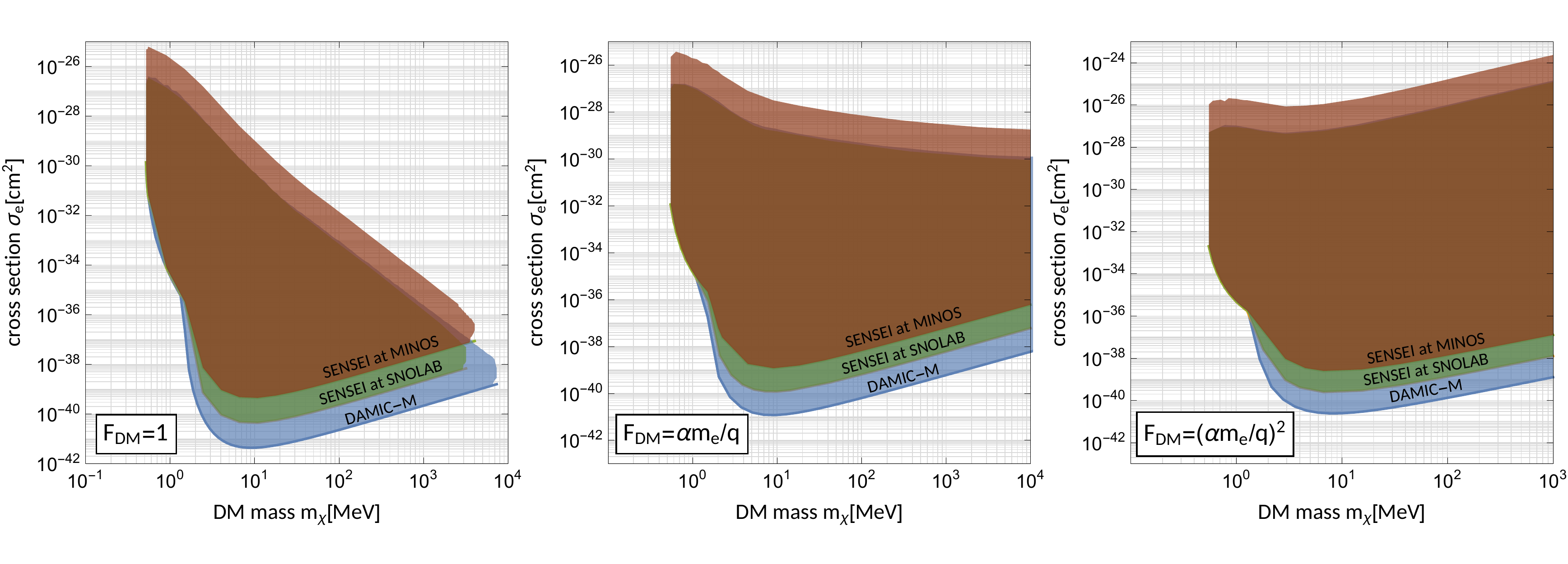}
	\caption{Projected constraints~(95\%~CL) for the planned experiments DAMIC-M and SENSEI.}
	\label{fig: electron projections damic sensei}
\end{figure*}
There are two experimental collaborations preparing a direct detection experiment using CCDs, where the silicon semiconductor acts as the target, DAMIC~\cite{Aguilar-Arevalo:2016ndq} and SENSEI~\cite{Tiffenberg:2017aac}. We derive constraint projections for these next-generation experiments and obtain their sensitivity to strongly interacting~DM.

The DAMIC-M experiment is planned to run at the~\acl{LSM} in France at a depth of 1780m and will most likely have the largest exposure. SENSEI is planned to be located at SNOLAB slightly deeper at 2km underground. However, a smaller detector might also be used at the MINOS facility at Fermilab at a relatively shallow depth of 107m. The projected exposures are 10, 100, and 1000~gram~year for SENSEI at Minos, SENSEI at SNOLAB, and DAMIC at Modane respectively. We assume the optimal ionization threshold of one electron-hole pair for all three detectors. Regarding the background, we assume to have observed $\text{10}^{\text{3}}$, $\text{10}^{\text{4}}$, and~$\text{10}^{\text{5}}$ events in the~$n_e=$1 bin. The results are shown in figure~\ref{fig: electron projections damic sensei}. Having the largest target mass, DAMIC-M would probe the smallest cross section. Being at the shallowest site, a SENSEI run at MINOS would cover the strong interaction regime the furthest.\\[0.3cm]

\begin{figure*}
\sbox\twosubbox{%
  \resizebox{\dimexpr.95\textwidth-1em}{!}{%
   \includegraphics[height=6cm]{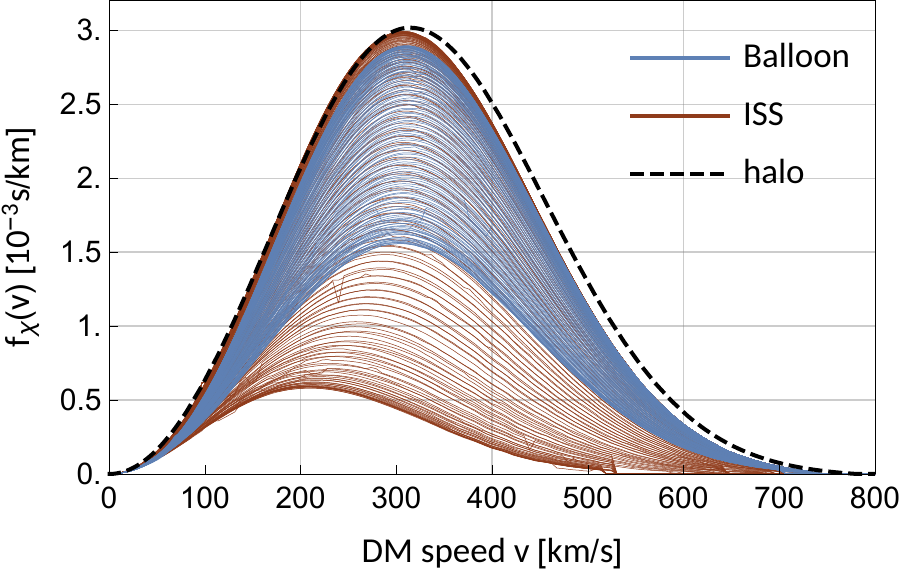}%
    \includegraphics[height=6cm]{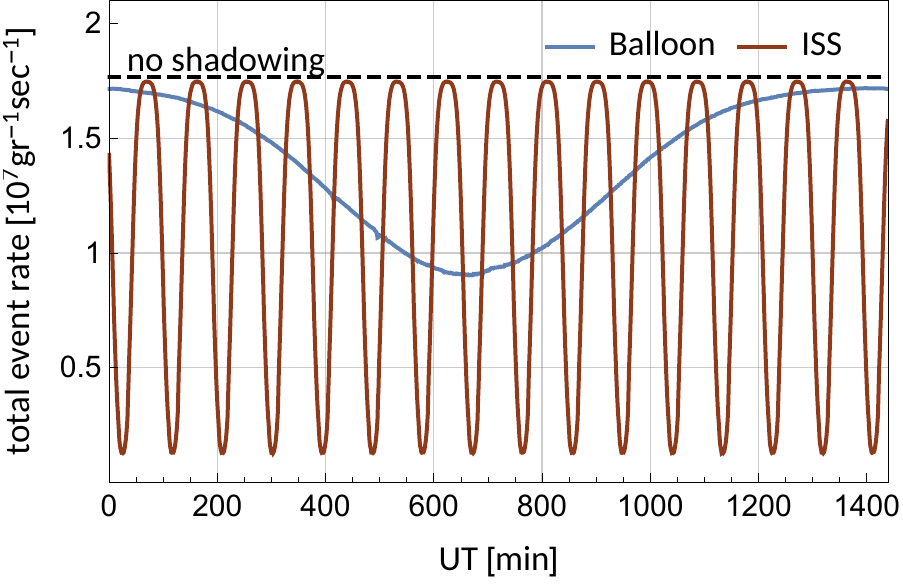}%
  }%
}
\setlength{\twosubht}{\ht\twosubbox}
   	\centering
  	\includegraphics[height=\twosubht]{plots/Shielded_PDF.pdf}%
	\quad
  	\includegraphics[height=\twosubht]{plots/Modulation}%
	\caption{Orbital modulation of the DM~speed distribution and detection rates for a balloon and a satellite experiment for~$m_\chi=\SI{100}{\MeV}$, $\sigma_e=\text{10}^{\text{-23}}\text{cm}^2$, and a light mediator~($F_{\rm DM}\sim 1/q^2$).}
	\label{fig: balloon and satellite}
\end{figure*}
To probe cross sections above the reach of terrestrial experiments, we can entertain the idea to run a direct detection experiment at great heights, either in the upper atmosphere in a balloon or in a lower Earth orbit onboard a satellite. The exposure of a balloon-borne experiment would be of order~$\sim$gram~hour, but shielded only by the upper atmospheric layers of low density. On a satellite, the situation improves further, allowing longer exposures, while the shielding is minimized to the apparatus' setup and self-shielding by the target.

There is an intermediate regime of cross sections, where the Earth blocks off the observable DM~flux entirely, while high altitude experiments are still sensitive. The DM~signal in these experiments would show a strong modulation due to the Earth's shadowing effect~\cite{Collar:1992qc,Kouvaris:2014lpa}, similarly to the diurnal modulation we studied in chapter~\ref{s: diurnal modulation}. For balloon experiments, we should expect a diurnal modulation. For a space-borne experiment, the modulation frequency would depend on the satellite's orbit. We can estimate this modulation without the need for \ac{MC}~simulation by computing the local DM~speed distribution at the experiment's location~$\mathbf{x}$~via
\begin{align}
	f(\mathbf{x},v) &= \int \dd \Omega\; v^2 f(\mathbf{v}) \times \mathfrak{S}(\mathbf{x},\mathbf{v})\, ,
	\intertext{where we filter out particles, which have to pass the planet's mass, to reach the location. This is done with a simple `shadowing function', defined as}
	\mathfrak{S}(\mathbf{x},\mathbf{v})&=\begin{cases}
		0\, ,\quad\text{if }\left|\mathbf{x}+t \mathbf{v}\right|=R_\oplus \text{ has a solution }t<0\, ,\\
		1\, ,\quad\text{otherwise.}
	\end{cases}
\end{align}
We discuss two examples of a silicon semiconductor target experiment.
\begin{enumerate}
	\item A geostationary detector running onboard a balloon, 30km over Pasadena, Ca~(34.1478${}^\circ$N, 118.1445${}^\circ$W). The time evolution of the position vector in the galactic rest frame is given in eq.~\eqref{eq:labpos}. Besides the upper 54~km of the thin, upper layers of the atmosphere, the target is assumed to be shielded by two additional layers, namely 1~mm of copper\footnote{Copper: $\rho=$~8.96~gram/$\text{cm}^3$.} and 5~mm of mylar\footnote{Mylar: $\text{C}_{\text{10}}\text{H}_{\text{8}}\text{O}_{\text{4}}$~ with~$\rho=$~1.4~gram/$\text{cm}^{\text{3}}$, (62.5\% carbon, 33.3\% oxygen, and 4.2\% hydrogen)~\cite{Mylar}.}.
	\item A small detector in low Earth orbit. We take the ISS orbit as an example, which orbits our planet at about 400km altitude with an orbital period of around 90 minutes. The orbit has an inclination of around 50${}^\circ$, and the ISS transitions between being exposed to or hidden from the DM~wind. Here, the target is assumed to be shielded by 1~mm of mylar only.
\end{enumerate}
The variation of the locally attenuated speed distribution of DM~particles is shown in the left panel of figure~\ref{fig: balloon and satellite}. The plot on the right side shows the corresponding orbital signal modulations for a DM~particle with~$m_\chi=$~100~MeV and~$\sigma_e=\text{10}^{\text{-23}}\text{cm}^{\text{2}}$, which interacts via an ultralight mediator. This parameter point is chosen as an example, which is still allowed by terrestrial experiments. Especially the population of fast DM~particles gets depleted within the Earth's shadow. 

The fractional modulation is defined as
\begin{align}
	f_{\rm mod}\equiv \frac{R_{\rm max}-R_{\rm min}}{R_{\rm min}+R_{\rm max}} = \frac{R_{\rm max}-R_{\rm min}}{2\langle R \rangle} \, ,
\end{align}
in terms of the minimum and maximum signal rates~$R_{\rm min}$ and~$R_{\rm max}$. In both cases, we find a significant modulation. For the balloon, we obtain a smaller fractional modulation with~$\sim$~30\%. The same experiment in the southern hemisphere would be more sensitive to the diurnal modulation. For the ISS orbit, we see a high frequency orbital modulation with~$f_{\rm mod}\simeq$~85\%. The satellite moves deep into the Earth's shadow where most particles, in particular the fast ones, are getting stopped. The modulation is expected to increase for lower DM~masses, as their detection relies more on the high-energy tail of the DM~distribution.

The event modulation due to the shadow effect could be used to distinguish a DM~signal from backgrounds. For experiments without major shielding, we have to expect large background event rates. We assume an experimental run with exposure~$\mathcal{E}$, observing~$B$ background events. The~5$\sigma$~discovery reach of a modulating signal due to~DM over this background can be determined by solving
\begin{align}
	\frac{f_{\rm mod}N_{\rm tot}}{\sqrt{N_{\rm tot}+B}}=5\, ,
\end{align}
for~$\sigma_e$~\cite{Essig:2011nj}. The cross section enters this equation via the total number of events $N_{\rm tot}\equiv \mathcal{E} \langle R \rangle$. For the balloon~(satellite), we assume a background of~$\text{10}^{\text{6}}$~($\text{10}^{\text{9}}$) signals, such that the 5~$\sigma$ discovery reach corresponds to $N_{\rm tot}\simeq$~16000 (182000), where we substituted~$f_{\rm mod}=$~0.31~(0.87). It should not go unmentioned that the shielding layers are still modelled as planar, and the results should be regarded as a first, but reasonable estimate. The exact geometry of the experimental apparatus would need to be implemented for more precise estimates, and the simulations would need to be extended to more complex simulation geometries.
\begin{figure*}
	\centering
	\includegraphics[width=0.75\textwidth]{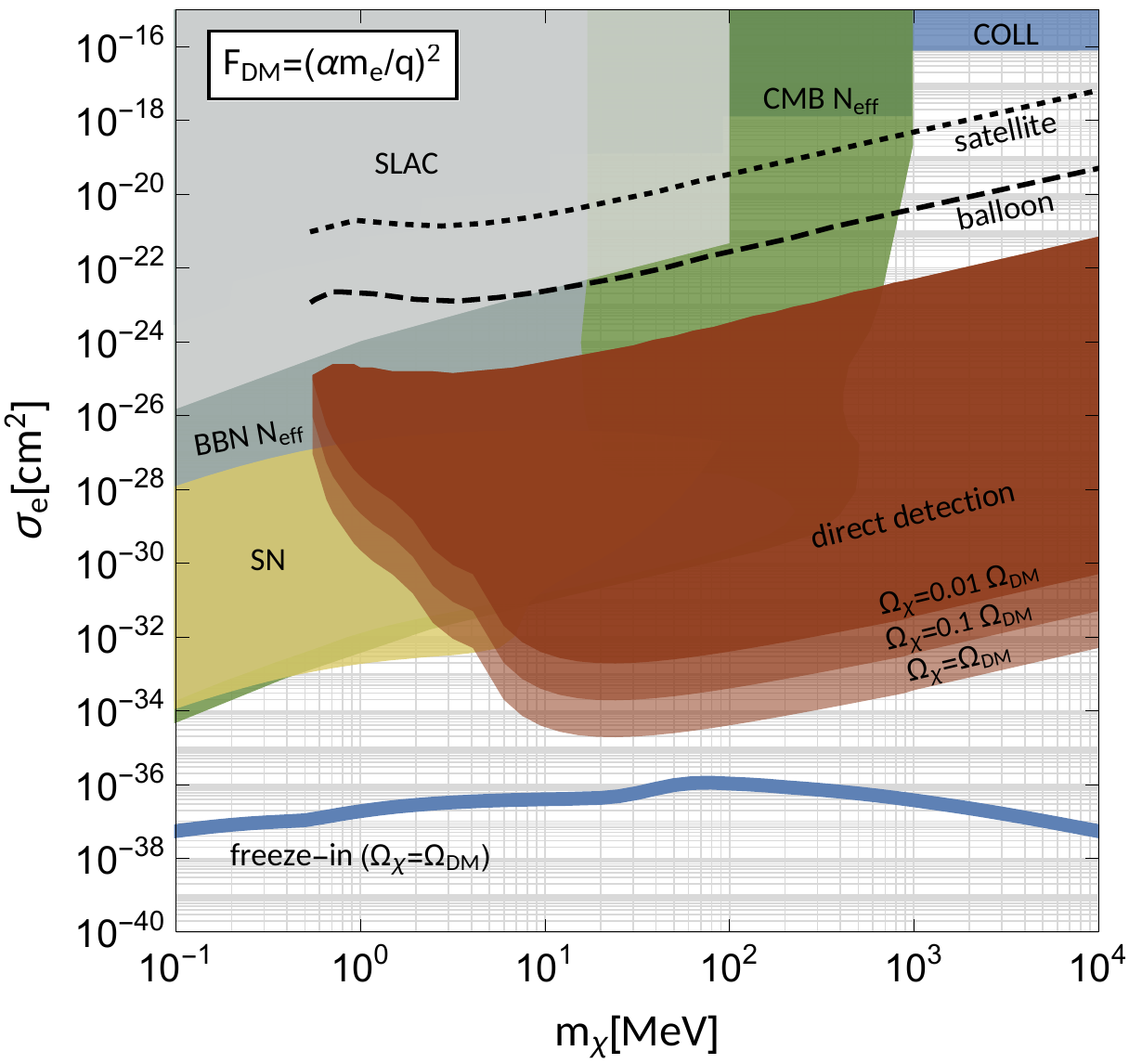}
	\caption{Constraints on~DM with an ultralight dark photon mediator.}
	\label{fig: light mediator summary}
\end{figure*}

The projected modulation discovery reaches are depicted in figure~\ref{fig: light mediator summary}. This plot also shows our results for the direct detection constraints, constraints from supernova cooling~\cite{Chang:2018rso}, cosmological bounds on~$N_{\rm eff}$ from the \ac{CMB} and \ac{BBN}~\cite{Vogel:2013raa}, and collider constraints from a search for milli-charged particles by SLAC~\cite{Davidson:1991si,Prinz:1998ua}. The blue line shows the parameter space favoured by the freeze-in production mechanism~\cite{Chu:2011be,Essig:2011nj}. 

The open window in parameter space in the top right corner is inconclusive. It might e.g. be narrowed by translating the conventional direct detection constraints from the CRESST collaboration into bounds on the electron cross section by using eq.~\eqref{eq:sigma ratio}. There are further constraints based on the requirement that~DM has decoupled at recombination~\cite{McDermott:2010pa,Xu:2018efh}. Most of these constraints assume that the strongly interacting~DM particles are the \textit{only} component of~DM, i.e.~$\Omega_\chi=\Omega_{\rm DM}$. Large parts of the parameter space open up for the scenario where a milli-charged DM~particle makes up only a fraction of the total DM~amount. This possibility has attracted a fair amount of attention in light of the 21cm anomaly observed by the EDGES collaboration~\cite{Bowman:2018yin}, see e.g.~\cite{Barkana:2018cct,Berlin:2018sjs,Mahdawi:2018euy}. In addition, some authors claimed that milli-charged~DM would get ejected from the galactic disk through its magnetic field and supernova shock waves and could therefore never be found by direct searches~\cite{McDermott:2010pa}. In a recent paper, it was claimed to the contrary that milli-charged particles from the halo continuously re-populate the disk with a highly energetic DM~flux which extends previous detection bounds significantly~\cite{Dunsky:2018mqs}. The fate of this interesting segment of parameter space is not clear at this point. If region remains allowed by the experimental bounds, a ballon- or satellite-borne experiment would be able to probe large parts of it.

\clearpage
\chapter{Solar Reflection of Dark Matter}
\label{c:sun}
Direct searches targeting low-mass~DM are complicated by the fact that light DM~particles cause smaller energy deposits in the detector. If even the largest possible recoil energy is too soft and falls below the experiment's threshold, DM is too light to be probed by that experiment. The minimum DM~mass a detector of recoil threshold~$E_R^{\rm thr}$ and target mass~$m_N$ can test is given by eq.~\eqref{eq:minimal DM mass}. Usually, the maximum speed entering this relation is set to the galactic escape velocity plus the Earth's relative motion~$v_{\rm max}=v_{\rm esc}+v_{\oplus}$. Next to the two obvious approaches to extend the experimental search to lower masses by lowering the target mass or threshold, it is also an option to consider mechanisms that might increase the maximum speed~$v_{\rm max}$. Indeed, there is a virtually halo model independent mechanism which generates a population of fast DM~particles in the solar system, provided that the DM~mass falls below the~GeV scale.

A light DM~particle may gain energy by elastic collisions on hot and highly energetic targets, reaching speeds far beyond the maximum of the~\ac{SHM}. In this chapter, we will consider the idea of DM~particles getting accelerated by the hot constituents of the Sun. This idea was first proposed in~\cite{An:2017ojc} and independently shortly after in~\ref{paper4}. Recently, cosmic rays have also been proposed as a potential DM~accelerator~\cite{Bringmann:2018cvk}.

A DM~particle entering the Sun could scatter on a nucleus in the solar core and get reflected with great speed. This mechanism is effective, if the kinetic energy of infalling~DM falls below the thermal energy of the solar targets and therefore applies only to low-mass~DM. Heavy~\acp{WIMP} most likely lose energy by scattering in the Sun and might get gravitationally captured permanently. 

To obtain a significant scattering rate in the Sun and thereby a potentially observable particle flux from~\textit{solar reflection} of~DM, the respective interaction must be strong enough. However, if the cross section is too large, the cooler outer layers of the Sun shield off the hot solar core. If the last scattering before leaving the star occurs on a colder target, the DM~particle would most likely have lost energy. The only hope to observe reflected~DM relies on the reflection spectrum extending beyond the halo particles' maximum speed. Below that value, the standard halo~DM will always dominate.

After falling into the gravitational potential of the Sun, the DM~velocities are completely dominated by the solar escape velocity. The halo's original distribution therefore has no significant effect on the final reflection spectrum, and the dependence on the assumed halo model is very weak. The obligatory existence of an additional flux of very fast DM~particles in the solar system can be used to extend the detection sensitivity of low-threshold direct detection experiments to lower masses. Depending on the experimental exposure, this might set a novel kind of constraints on sub-GeV~DM.

The general idea to look for~DM, which gained energy inside the Sun was first proposed for particle evaporation after getting gravitationally captured~\cite{Kouvaris:2015nsa}. While evaporation relies on the captured~DM being thermalized, which is only true down to a certain DM~mass, the mechanism of solar reflection does not require such assumptions.\\[0.3cm]

In chapter~\ref{s: sun analytic} we present the computation and results of~\ref{paper3}. Therein, we extend the analytic formalism by Gould~\cite{Gould:1987ir,Gould:1987ju,Gould:1991hx}. In particular, the new treatment takes the Sun's opacity fully into account, smoothly connecting the opaque and transparent regime. Finally, we find the spectrum of solar reflection via a single scattering and study the implications for direct detection. In the second part, chapter~\ref{s: sun mc}, we show first results for a \ac{MC}~treatment of solar reflection including the effect of multiple scatterings and finish by discussing the prospects of this promising approach. The results are preliminary and have not been published at this point. Throughout this chapter, we use the \ac{SSM} to model the Sun's interior. The~\ac{SSM} is introduced in app.~\ref{a:sun}.

\clearpage
\section{DM Scatterings in the Sun}
\label{s: sun analytic}
Compared to the description of underground scatterings in the Earth, we have to extend the simulation algorithms for~DM trajectories in the Sun.
\begin{enumerate}
	\item The Sun's gravitational well is much deeper than the Earth's and cannot be neglected. In fact, the DM~speed distribution inside the Sun will be dominated by the kinetic energy gained through falling into the Sun.
	\item With the target being a hot plasma, it is not a valid approximation to assume resting nuclei. In fact, this is precisely the reason why DM~particles can get accelerated by an elastic collision in the first place.
\end{enumerate}

 \begin{figure*}
 \centering
 \includegraphics[width=0.67\textwidth]{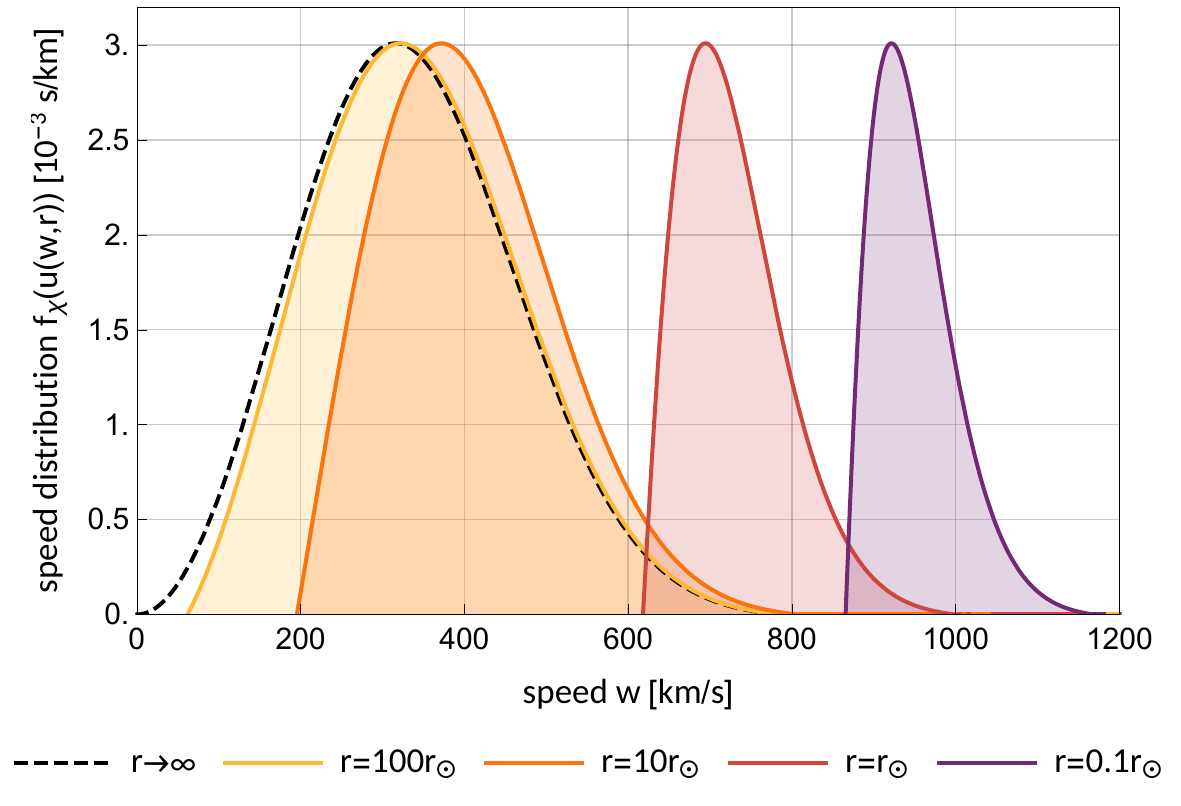}
 	\caption{Gravitational acceleration of DM close to the Sun}
 	\label{fig: sun blueshift}
 \end{figure*}

Asymptotically far away from the Sun, the DM~speed follows the halo distribution~$f_\chi(\mathbf{u})$ introduced in chapter~\ref{s: halo model}. Note that in this chapter, the asymptotic velocity (speed) is denoted by~$\mathbf{u}$ ($u$). As a halo particle approaches the Sun, it gains kinetic energy. Using energy conservation, the speed~$w$ at radius~$r$ is given by
\begin{align}
	w(u,r) &= \sqrt{u^2+v_{\rm esc}(r)^2}\, .
\end{align}
The Sun's local escape velocity~$v_{\rm esc}(r)$ is part of the~\ac{SSM}~\cite{Serenelli:2009yc} and can be found in eq.~\eqref{eq: sun escape velocity}. The direction of~$\mathbf{w}$ is unknown at this point and depends on the full~$\mathbf{r}$ and~$\mathbf{u}$ vectors. The local speed distribution of infalling~DM at radius~$r$ results from Liouville's theorem~\cite{Gould:1991hx},
\begin{align}
	f_\chi(w,r) = f_\chi(u(w,r))\, ,\quad\text{with }u(w,r) = \sqrt{w^2-v_{\rm esc}(r)^2}\, .
\end{align}
This `blue shift' of the halo distribution in the Sun's neighborhood is shown in figure~\ref{fig: sun blueshift}.

The second necessary adjustment is due to the fact that the solar medium is a thermal plasma. The nuclear velocities follow an isotropic Maxwell-Boltzmann distribution,
\begin{align}
	f_N(\mathbf{v}_N,r)\dd^3\mathbf{v}_N&= \left(\frac{\kappa}{\sqrt{\pi}}\right)^3e^{-\kappa^2 v_N^2}\dd^3 \mathbf{v}_N\, ,\quad\text{with }\kappa\equiv\sqrt{\frac{m_N}{2T_\odot(r)}}\, ,\label{eq: maxwell boltzmann}
\end{align}
 where the dependence on the radius~$r$ enters via the solar temperature~$T(r)$ given in app.~\ref{a:sun}.
 
\subsection{Scattering rate}
To describe solar reflection (or capture), we need to know the rate at which DM~particles scatter to a specified final speed~$v$. We divide the Sun into spherical shells of infinitesimal thickness~$\dd r$. The differential rate of DM~particles colliding on solar targets within a given shell can be written as a product of three factors,
\begin{align}
	\dd S = \underbrace{\dd \Gamma}_{\text{\footnotesize free DM passing rate}}\times\underbrace{\dd P_{\rm scat}}_{\text{\footnotesize scattering probability}}\times\underbrace{P_{\rm shell}}_{\text{\footnotesize probability to reach the shell}}\, . \label{eq: sun dS}
\end{align}
The first factor is the rate with which DM~particles in an infinitesimal phase space volume would pass a spherical shell, if the Sun was completely transparent. Naturally, particles might scatter already before reaching that particular shell. The real passing rate is therefore rather~$\dd \Gamma\times P_{\rm shell}$, with~$P_{\rm shell}$ being the probability to make it to the shell without colliding beforehand. Finally,~$\dd P_{\rm scat}$ is the probability to scatter while passing the spherical shell. This means that eq.~\eqref{eq: sun dS} yields the rate of the \emph{first} scattering after a DM~particle enters the Sun. 

For the computation of the first two factors, we use results obtained by Press, Spergel~\cite{Press:1985ug}, and Gould~\cite{Gould:1987ju,Gould:1987ir} in the late '80s. The third factor generalizes their framework to include the Sun's opacity. This way, we avoid that the final equations apply only to a certain regime of the scattering cross section.

\paragraph{DM passing rate inside the Sun}
We start by imagining a spherical surface of radius~$R\gg R_\odot$ with the Sun at its center. Being asymptotically far away from the Sun, the~DM follows the standard halo distribution. The differential DM~flux through this surface \emph{into} the sphere, i.e. the number of particles entering the sphere per unit area and time, is obtained to be~\cite{Press:1985ug}
\begin{align}
	\dd \Phi_\chi = \frac{1}{4}n_\chi f_\chi(u) u \dd u \dd(\cos^2\theta)\, ,\quad \text{with } 0<\theta<\frac{\pi}{2}\, .
\end{align}
The angle~$\theta$ lies between the DM~velocity and the surface normal. We transform the variable~$\cos^2\theta$ to the orbital invariant~$J^2$, the angular momentum (per mass),
\begin{subequations}
\begin{align}
	J&=u R \sin\theta\, ,\\
	\Rightarrow \dd \Phi_\chi &= \frac{1}{4}n_\chi f_\chi(u)  \dd u \frac{\dd J^2}{R^2u^2}\, ,\quad \text{with }0<J^2<R^2u^2\,\, .
\end{align} 
\end{subequations}
The total rate of particles entering the spherical volume is
\begin{align}
	\dd \Gamma = 4\pi R^2 \dd \Phi_\chi = \pi n_\chi f_\chi(u)\frac{\dd u \dd J^2}{u}\, .\label{eq: sun dGamma}
\end{align}
Particles entering a volume of radius~$R$ are not guaranteed to pass the Sun's surface or a given spherical shell of radius~$r<R_\odot$. Incoming DM~particles follow an unbound hyperbolic Kepler orbit towards the Sun, while the orbit inside the Sun is no longer Keplerian. The question whether or not a particle passes a shell of radius~$r$ is equivalent to the question whether its orbit's perihelion distance satisfies~$r_p<r$. We can use the fact that the angular momentum~$J$ is an invariant and that~$\theta=\pi/2$ holds at the perihelion. Thus, the condition for a particle of asymptotic speed~$u$ to pass a spherical surface of radius~$r$ is
\begin{align}
	J<w(u,r)r\, .
\end{align}
For example, the total rate of DM~particles entering the Sun can be computed to be
\begin{subequations}
	\label{eq: sun dm rate into sun}
\begin{align}
	\Gamma_\odot(m_\chi) &= \pi n_\chi \int_0^\infty\dd u\,\int_0^{w(u,R_\odot)^2R_\odot^2}\dd J^2\,\frac{f_\chi(u)}{u}\\
	&=\pi R_\odot^2 n_\chi \left[\langle u \rangle+v_{\rm esc}(R_\odot)^2\langle u^{-1}\rangle\right]\\
	&\approx 8\cdot 10^{29} \left(\frac{m_\chi}{\text{GeV}}\right)^{-1} \SI{}{\per\second}
\end{align}
\end{subequations}

It should be noted that we used the DM~speed distribution of eq.~\eqref{eq: DM speed PDF}, where the anisotropy of the full velocity distribution due to the Sun's velocity in the galactic rest frame is averaged out. The rate in eq.~\eqref{eq: sun dGamma} is to be understood as the average over the spherical surface.

\paragraph{Scattering probability in a spherical shell}
The scattering probability inside the Sun cannot be described in terms of eq.~\eqref{eq:dPscat Earth}, which we applied for the Earth. The reason is that the medium consists of a plasma of energetic targets instead of effectively resting nuclei. The relative velocity is not dominated by the DM~velocity, and the thermal motion of nuclei inside the Sun cannot be neglected. Instead of a mean free path~$\lambda$, the central quantity is the collision frequency or scattering rate~$\Omega$. This can be understood by considering a resting DM~particle. Using~\eqref{eq:dPscat Earth}, this particle would never scatter unless it moves relatively to the targets. In a hot medium however, one of the thermal nuclei would eventually scatter on even a resting DM~particle.

\begin{figure*}
 \centering
 \includegraphics[width=0.8\textwidth]{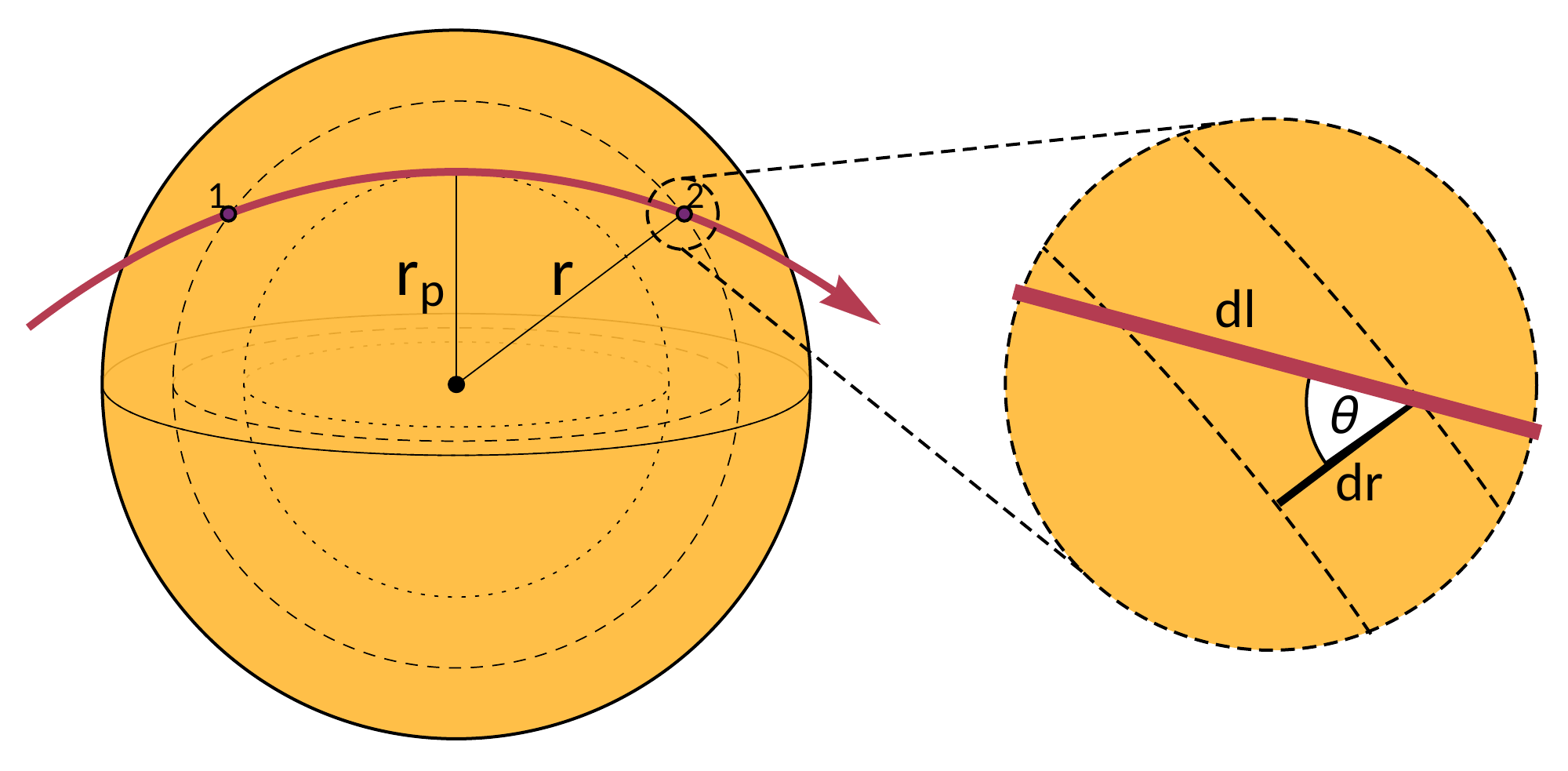}
 	\caption{Sketch of a DM trajectory crossing a solar shell of radius~$r$ twice.}
 	\label{fig: sun trajectory}
 \end{figure*}

The probability for a DM~particle of velocity~$w$ to scatter inside a spherical shell of radius~$r$ can be expressed in terms of the collision frequency~$\Omega$,
\begin{align}
	\dd P_{\rm scat} &=\underbrace{\frac{\dd l}{w}}_{\text{\footnotesize time spent in shell}}\times\underbrace{\Omega(r,w)}_{\text{\footnotesize scattering rate}}\, .\label{eq: dPscat}
\end{align}
The distance travelled inside the shell is denoted with~$\dd l$, as illustrated on the right hand side of figure~\ref{fig: sun trajectory}. The differential collision frequency on target species~$i$ with initial velocity~$\mathbf{v}_i$ is
\begin{align}
	\dd \Omega(r,w)= \sigma_i\, |\mathbf{w}-\mathbf{v}_i|\, n_i(r) f_i(v_i) \dd^3\mathbf{v}_i\, , \label{eq: dOmega} 
\end{align}
where~$f_i(\mathbf{v}_i)$ is the Maxwell-Boltzmann distribution of the target, and $\sigma_i$ is the total scattering cross section. Therefore the total scattering rate is obtained by integrating over the targets' velocities and summing over all target species with number density~$n_i$,
\begin{align}
	\Omega(r,w) = \sum_i n_i(r) \langle \sigma_i|\mathbf{w}-\mathbf{v}_i|\rangle\, .
\end{align}
The brackets~$\langle\cdot\rangle$ denotes the thermal average. If the cross section does not explicitly depend on the velocity, the collision frequency simplifies to
\begin{align}
	\Omega(r,w) &= \sum_i n_i(r) \sigma_i \langle |\mathbf{w}-\mathbf{v}_i|\rangle\, .
\end{align}
We can express the thermal average of the relative speed explicitly using eq.~\eqref{eq: maxwell boltzmann},
\begin{subequations}
\label{eq: average relative speed}
	\begin{align}
		\langle |\mathbf{w}-\mathbf{v}_i|\rangle &= \int\dd^3\mathbf{v}_i\, |\mathbf{w}-\mathbf{v}_i| f_i(\mathbf{v}_i)\\
		&=\frac{1+2\kappa^2w^2}{2\kappa^2 w}\erf{\kappa w}+\frac{1}{\sqrt{\pi}\kappa}e^{-\kappa^2w^2}\, .
	\end{align}
\end{subequations}
We note that this thermal average depends on the radius~$r$ implicitly via the temperature.

If we insist on defining a mean free path such that~$\dd P_{\rm scat} = \dd l/\lambda$, it has to be written as
\begin{align}
	\lambda^{-1}(r,w)=\frac{\Omega(r,w)}{w} = \sum_i n_i \sigma_i  \frac{\langle|\mathbf{w}-\mathbf{v}_i|\rangle}{w}\, ,\label{eq: MFP sun}
\end{align}
 provided that the DM~particle is in motion. For resting targets,~$|\mathbf{w}-\mathbf{v}_i|\approx w$, and we re-obtain eq.~\eqref{eq: MFP}. At this point, we should point out a common misconception in the literature, where the mean free path in the Sun is taken to be~$(\sum_i n_i\sigma_i)^{-1}$ (or~$(\sum_i n_i\langle\sigma_i\rangle)^{-1}$), instead of eq.~\eqref{eq: MFP sun}~\cite{Peter:2009mi,Taoso:2010tg,Lopes:2014aoa,Garani:2017jcj,Busoni:2017mhe,An:2017ojc}. This can result in a significant overestimation of the mean free path, which could critically alter the results as the mean free path enters the scattering probability in the exponent. This is especially severe in the case of electron scattering.\\[0.3cm]
 
Whether a particle gets captured or reflected by the Sun, depends not only on the total scattering rate, but also on the final speed of the scattering~$w\rightarrow v$. Starting from eq.~\eqref{eq: dOmega}, we can derive an expression for the differential scattering rate to final speed~$v$. For a derivation, we refer to the app. of~\cite{Gould:1987ju}.
\begin{subequations}
\label{eq: sun Omega}
\begin{align}
	\frac{\dd \Omega_\pm}{\dd v}(w\rightarrow v)&=\frac{2}{\sqrt{\pi}}\frac{v\dd v}{w}\sum_i\frac{\mu_+^2}{\mu}\sigma_i n_i(r)\nonumber\\
	&\qquad\times\left[ \chi(\pm\beta_-,\beta_+)e^{-\mu \kappa^2(v^2-w^2)}+\chi(\pm\alpha_-,\alpha_+)\right]\, .
\intertext{Here, we took over Gould's notation,}
\mu &\equiv \frac{m_\chi}{m_i}\, ,\quad\mu_\pm\equiv\frac{\mu\pm 1}{2}\, ,\\
\chi(a,b)&\equiv\frac{\sqrt{\pi}}{2}\left[\erf{b}-\erf{a}\right]\, ,\\
\alpha_\pm &\equiv \kappa(\mu_+v\pm\mu_-w)\, ,\quad \beta_\pm\equiv \kappa(\mu_-v\pm\mu_+w)\, .
\end{align}
\end{subequations}
The plus sign corresponds to acceleration~($v>w$), whereas the minus sign applies to deceleration~($v<w$).

\paragraph{Probability to reach a spherical shell}
The probability to travel a given path freely without interaction can be inferred from~\eqref{eq:cdf path length},
\begin{align}
	P_{\rm surv} = \exp\left( -\int_{\text{path}}\dd P_{\rm scat}\right)\, ,
\end{align} 
where the scattering probability can be found in eq.~\eqref{eq: dPscat}. We focus on a path starting at radius~$r_A$ travelling to a larger radius~$r_B>r_A$. Using the symmetry of the free DM~orbits, we can construct all relevant paths in terms of
\begin{subequations}
\begin{align}
	P_{\rm surv}(r_A,r_B)&=\exp\left[-\int_{r_A\rightarrow r_B}\frac{\dd l}{w(u,r)}\Omega(r,w(u,r))\right]\\
	&=\exp\left[-\int_{r_A}^{r_B}\dd r\,\frac{\dd l}{\dd r}\frac{\Omega(r,w(u,r))}{w(u,r)}\right]
\end{align}
\end{subequations}
Looking at the right hand side of figure~\ref{fig: sun trajectory}, we can express~$\frac{\dd l}{\dd r}$ in terms of~$J=rw\sin\theta$,
\begin{align}
	\frac{\dd l}{\dd r}=\left(1-\frac{J^2}{w^2(u,r)r^2}\right)^{-1/2}\, .
\end{align}
As mentioned previously, a DM~particle can only pass a spherical shell of radius~$r$ if~$J<w(u,r)r$. Then, the particle will pass the shell up to two times. This is illustrated on the left hand side of figure~\ref{fig: sun trajectory}. The probability to reach the shell the second time without scattering is naturally lower than for the first time. We use the orbit's symmetry around its perihelion at distance~$r_p$ such that both probabilities add up to
\begin{subequations}
\label{eq: Pshell}
\begin{align}
	P_{\rm shell}(r) &= \underbrace{\Theta\left(w(u,r_p)r_p-J\right)}_{\text{\footnotesize shell passing condition}}\nonumber\\
	&\qquad\quad\times\left[\underbrace{P_{\rm surv}(r,R_\odot)}_{\text{1st passing}}+\underbrace{P_{\rm surv}(r,R_\odot)P_{\rm surv}(r_p,r)^2}_{\text{2nd passing}}\right]\\
	&=P_{\rm surv}(r,R_\odot)\left[1+P_{\rm surv}(r_p,r)^2\right]\,\Theta\left(w(u,r_p)r_p-J\right)\, .
\end{align}
\end{subequations}
Since this is the sum of two probabilities~$0\leq P_{\rm shell}<2$.

Now all three factors in our original eq.~\eqref{eq: sun dS} are in place, and we can write down the result for the differential scattering rate inside the Sun.

\paragraph{Final result for DM scattering rate}
 The differential rate of DM~particles falling into the Sun to scatter for the first time in a spherical shell of radius~$r$ to final speed~$v$ is obtained by substituting eqs.~\eqref{eq: sun dGamma},~\eqref{eq: dPscat},~\eqref{eq: sun Omega}, and~\eqref{eq: Pshell} into eq.~\eqref{eq: sun dS},
 \begin{subequations}
 \label{eq: sun dSdrdv}
 \begin{align}
 	\frac{\dd S}{\dd r\dd v} &= \int\dd\Gamma \frac{\dd P_{\rm scat}}{\dd r\dd v}P_{\rm shell}(r)\\
 	&=\pi n_\chi \int\limits_0^\infty \dd u\mkern-18mu\int\limits_{0}^{w(u,r)^2r^2}\mkern-18mu\dd J^2\,\frac{f_\chi(u)}{u}\frac{\dd \Omega}{\dd v}(w(u,r)\rightarrow v)\left[w(u,r)^2-\frac{J^2}{r^2}\right]^{-1/2}\nonumber\\
 	&\hspace{4.5cm}\times P_{\rm surv}(r,R_\odot)\left[1+P_{\rm surv}(r_p,r)^2\right]
 \end{align}
  \end{subequations}
 This equation captures the first scatterings regardless of the cross section. It smoothly connects the opaque and transparent regime. In the latter,~$P_{\rm surv}\approx 1$, and the integral over~$J^2$ can be evaluated analytically,
 \begin{align}
 	\frac{\dd S}{\dd r\dd v}\approx 4\pi r^2 n_\chi\int\limits_0^\infty\dd u\,\frac{f_\chi(u)}{u}w(u,r)\frac{\dd \Omega}{\dd v}(w(u,r)\rightarrow v)\, ,
 \end{align}
 which reminds of Gould's capture equation~(2.8) of~\cite{Gould:1987ir}. In the opaque regime, we find
\begin{align}
 	P_{\rm surv}(r_p,r)\approx 0\, ,\;\text{and }P_{\rm surv}(r,R_\odot) \frac{\Omega(w)}{w} \frac{\dd l}{\dd r}\approx\delta(r-R_\odot)\, ,
 \end{align}
 where the left hand side of the second equation can be understood as the probability density in terms of~$r$. In other words, in the extreme opaque regime, the particle scatters immediately upon arriving at the solar surface. This can be seen from the total scattering rate, which can be evaluated to be
 \begin{align}
 	S\approx \pi R_\odot^2 \left[ \langle u\rangle+v_{\rm esc}(R_\odot)^2\langle u^{-1}\rangle\right]\, .
 \end{align}
Comparing to.~\eqref{eq: sun dm rate into sun}, this is nothing but the total rate of particles entering the Sun.

Next, we have to quantify how many of the scattered particles get captured and how many manage to escape the star's interior without scattering a second time.

\subsection{Solar reflection and capture}
A DM~particle which scatters on a solar nucleus at radius~$r$ is kicked into a new orbit with perihelion distance~$r_p^\prime$, angular momentum~$J^\prime$, and speed~$v$. If it loses kinetic energy of at least~$\frac{m_\chi}{2}u^2$, the final speed falls below the local escape velocity, and the particle gets captured. Otherwise, it gets reflected, if it survives back to the solar surface without re-scattering. The probability to get captured is simply
\begin{align}
	P_{\rm capt}(v,r)&=\Theta(v_{\rm esc}(r)-v)\, .
\end{align}
On the other side, the chance of getting reflected is
\begin{align}
	P_{\rm leave}(v,r) &=\underbrace{\Theta(v-v_{\rm esc}(r))}_{\text{\footnotesize escape condition}}\nonumber\\
	&\qquad\times\frac{1}{2}\left[\underbrace{P_{\rm surv}(r,R_\odot)}_{\text{\footnotesize short path out}}+\underbrace{P_{\rm surv}(r,R_\odot)P_{\rm surv}(r_p^\prime, r)^2}_{\text{\footnotesize long path out}}\right]\, .
\end{align}
This is an expression similar to~$P_{\rm shell}$ of eq.~\eqref{eq: Pshell}. But here, we average over the two possible directions a DM~particle could take to leave the star.

\begin{figure*}
	\centering
	\includegraphics[width=\textwidth]{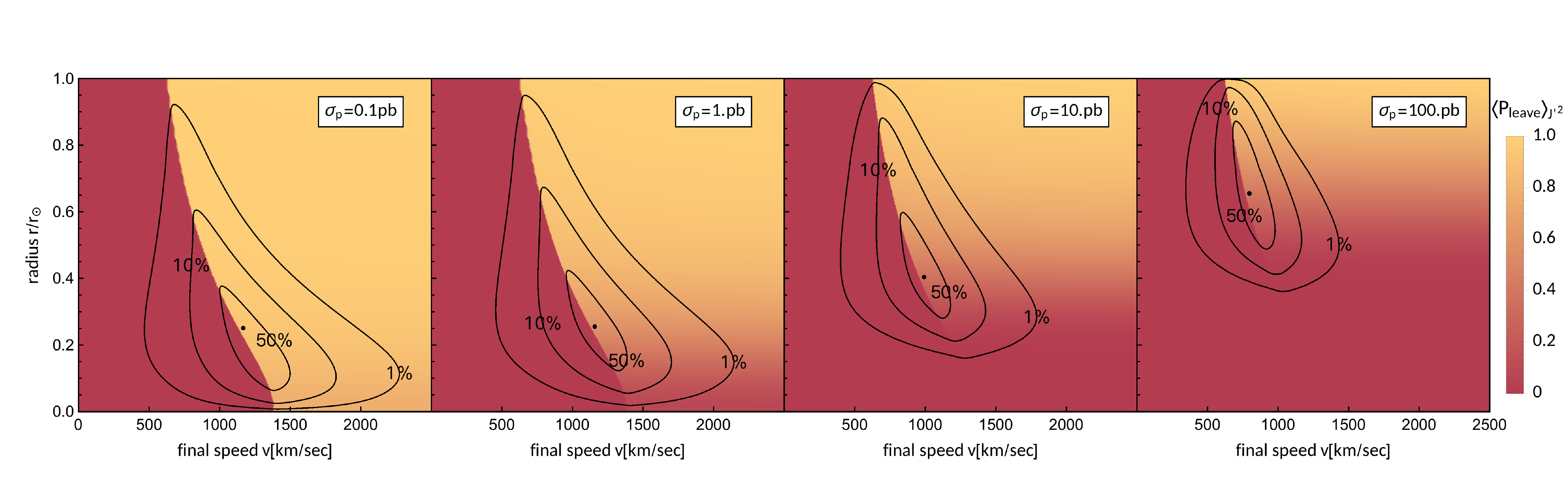}
	\caption{Average probability~$\langle P_{\rm leave}(v,r)\rangle_{J^{\prime 2}}$ to escape the Sun without re-scattering for $m_\chi=\SI{100}{\MeV}$ and increasing cross sections.}
	\label{fig: Pleave and dSdrdv}
\end{figure*}

These two probabilities can be used to filter out the captured or reflected particles out of the scattering rate. The differential capture and reflection rates are
\begin{align}
	\frac{\dd \mathcal{C}}{\dd v\dd r}&=\frac{\dd S}{\dd v\dd r}P_{\rm capt}(v,r)\, , \label{eq: capture rate}\\
	\frac{\dd \mathcal{R}}{\dd v\dd r}&=\frac{\dd S}{\dd v\dd r}\langle P_{\rm leave}(v,r)\rangle_{J^{\prime 2}}\, .\label{eq: reflection rate}
\end{align}
In the last line, we average over final angular momenta, again assuming isotropic scatterings. To get an idea about the particles getting reflected or captured, we plot ~$\langle P_{\rm leave}(v,r)\rangle_{J^{\prime 2}}$ and overlay contour lines of~$\frac{\dd S}{\dd r\dd v}$ in figure~\ref{fig: Pleave and dSdrdv}. It illustrates beautifully how an increase in the scattering cross section shifts the location of the first scattering towards the solar surface and renders the core more and more opaque. As the outer shells are cooler, the DM~particle's final velocity decreases correspondingly, and the scattering rate peaks at lower values for~$v$. About half of the particles have a good chance of leaving the Sun without re-scattering, the other half either gets captured or scatters again. Strictly speaking, there is no single scattering regime, even for very weak interaction strengths, as half the DM~particles get captured, enter a bound orbit and will eventually scatter a second time.

\begin{figure*}
	\centering
	\includegraphics[width=0.75\textwidth]{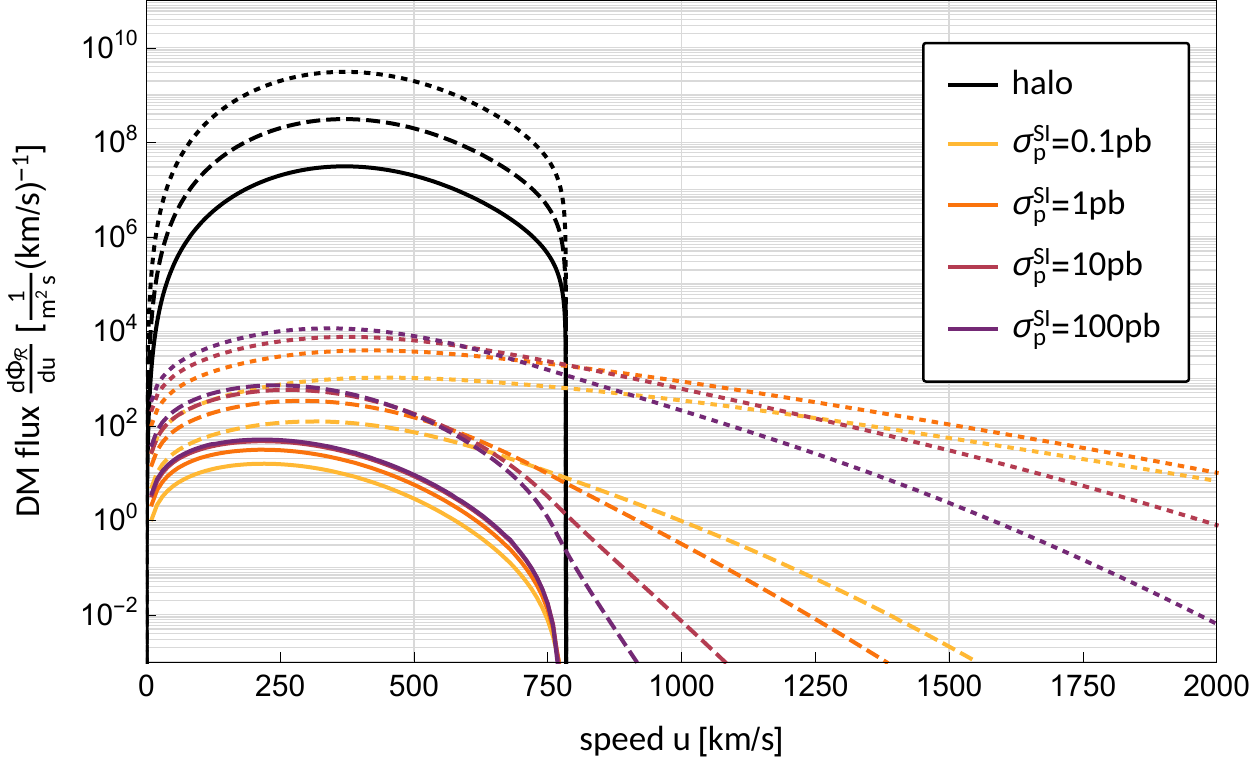}
	\caption{Differential DM~particle flux for the halo and solar reflection for DM~masses of~10~GeV~(solid), 1~GeV~(dashed), and 100~MeV~(dotted).}
	\label{fig: sun flux}
\end{figure*}

By integrating over the Sun's volume and `red-shifting' the reflected particles' speeds, we finally obtain the spectrum of reflected DM~particles on Earth, where they might pass through a detector,
\begin{align}
	\frac{\dd \mathcal{R}}{\dd u} =\int_0^{R_\odot}\dd r\,\frac{\dd \mathcal{R}}{\dd v\dd r}\left.\frac{\dd v}{\dd u}\right|_{v=\sqrt{u^2+v_{\rm esc}(r)^2}}\, .\label{eq: dRdu analytic}
\end{align}
Following from that expression, the differential particle flux is
\begin{align}
	\frac{\dd \Phi_{\mathcal{R}}}{\dd u} = \frac{1}{4\pi \ell^2}\frac{\dd \mathcal{R}}{\dd u}\, ,
\end{align}
where~$\ell=$~1~A.U. is the distance between the Sun and the Earth. This establishes a new population of potentially fast DM~particles in the solar system. Their flux through Earth is compared to the standard halo flux in figure~\ref{fig: sun flux}. For speeds below the halo distribution's cutoff, the reflection flux is naturally suppressed. However, the differential reflection flux has no speed cutoff and extends to much higher energies than present in the DM~halo, provided that the DM~mass is similar or lighter than the targets. By increasing the cross section, scattering on solar targets becomes more and more likely. The flux of slow reflected particles therefore increases as well. However, the hot core also gets shielded off more in this case, and very fast DM~particles are less likely to escape, which explains the flux's decrease with increased cross section for high energies. 

In all our computation we include the four largest targets in terms of~$n_i\sigma_i$, namely hydrogen~\isotope{H}{1}, helium~\isotope{He}{4}, oxygen~\isotope{O}{16}, and iron~\isotope{Fe}{56}. Smaller targets may be neglected. They might slightly increase the scattering rates, but also marginally shield off the hot core.

\section{Direct Detection of Reflected DM}
\begin{figure*}
\centering
\includegraphics[width=0.67\textwidth]{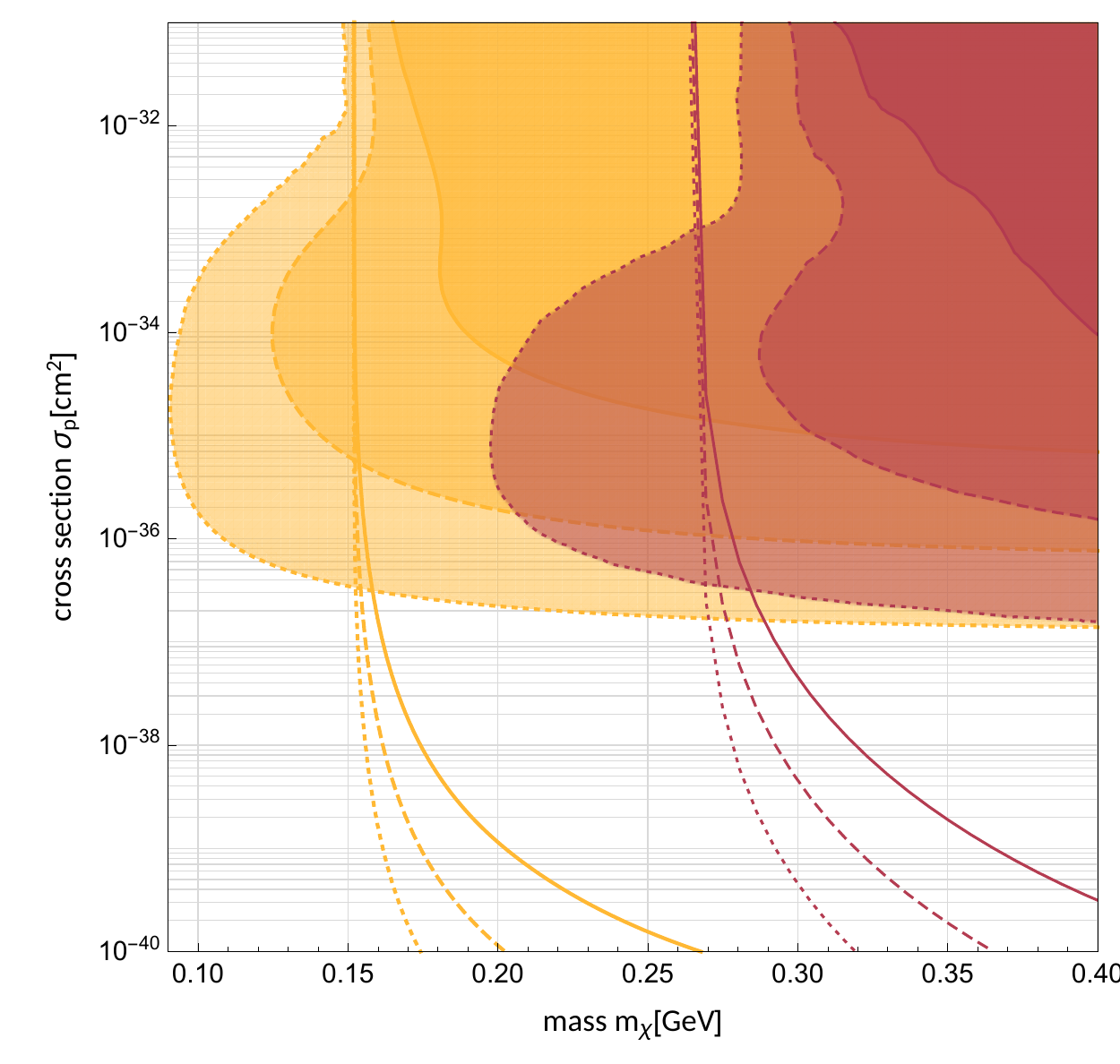}
\caption{The red (yellow) contours are projected constraints for a CRESST-III type (idealized sapphire) detector with exposures of 1,~10, and~100~ton~days (10,~100,~and 1000~kg~days) in solid, dashed, and dotted lines respectively.}
\label{fig: solar constraints}
\end{figure*}

The event rate in a direct detection experiment is proportional to the DM~flux through the detector, as covered in chapter~\ref{s: recoil spectra}. Solar reflection is the source of a new flux component, which can easily be incorporated into eq.~\eqref{eq: nuclear recoil spectrum},
\begin{align}
		\frac{\dd R}{\dd E_R} &= \frac{1}{m_N}\int\limits_{u_{\rm min}(E_R)}^{\infty}\dd u\; \left[ \underbrace{\frac{\rho_\chi}{m_\chi} u f_\oplus (u)}_{\text{\footnotesize halo DM}}+\underbrace{\frac{1}{4\pi \ell^2}\frac{\dd \mathcal{R}}{\dd u}}_{\text{\footnotesize reflected DM}}\right]\frac{\dd\sigma_N}{\dd E_R}\, . \label{eq: dRdER with reflection}
\end{align}
We present results for a detector of the CRESST-III type~\cite{Strauss:2016sxp,Angloher:2017zkf,Petricca:2017zdp}. For these projections we assume that phase 2 will have an exposure of 1 ton~day and a recoil threshold of 100eV. The app. includes a summary of the experimental setup, see chapter~\ref{ss: CRESST-III}. The constraints~(90\% CL) for solar reflected~DM are shown in figure~\ref{fig: solar constraints} as red shaded regions and compared to the corresponding constraints from halo~DM.

At this point, the reflection constraints are sub-dominant, but they differ from the halo constraints in one crucial quality. For increased exposure, lower and lower masses are being probed, whereas the halo constraints never reach below the minimum mass given by eq.~\eqref{eq:minimal DM mass}. For a detector of the CRESST-III type with exposures larger by a factor of~10 to~100 and no additional background, parameter space below the naive minimum mass becomes accessible. Above a certain exposure, the minimal probed DM~mass starts to decrease for larger exposures. These constraints are widely insensitive to the halo model and its velocity distribution, as the gravitational acceleration and subsequent scattering erase most of the initial distribution's information.

The sensitivity to reflected~DM increases significantly for lower recoil thresholds. Then, solar reflection could be used to extend the halo constraints even for smaller exposures. Figure~\ref{fig: solar constraints} also contains projections for an idealized detector made of sapphire with perfect resolution, no background, and a threshold of 20~eV. A similar detector with such a low threshold was already realized by the CRESST collaboration~\cite{Angloher:2017sxg}. For exposures above around~10~kg~days, the solar reflection flux would already be the source of observable numbers of events.

There are multiple ways to distinguish reflected sub-GeV~DM from standard halo~DM of heavier mass.
\begin{enumerate}
	\item The recoil spectrum would have a non-Maxwellian tail.
	\item Directional detectors would show the Sun as the signal's source.
	\item If scatterings in the Sun are common, they would also occur in the Earth. For a given cross section, the Sun and the Earth are similarly opaque. In chapter~\ref{s: diurnal modulation}, we already discussed in detail that this would result in a diurnal modulation of the signal.
	\item Due to the eccentricity of the Earth's orbit and the associated variation of the distance to the Sun, the reflection flux, and therefore the signal rate, will have a~$\sim$ 7\% annual modulation peaking at the perihelion around January 3rd as opposed to the standard annual modulation, which peaks in Summer.
\end{enumerate}   

The central result of this chapter is a demonstration of the existence of an additional, highly energetic DM~population in the solar system and the possibility to detect those particles in standard direct detection experiments of low threshold. For their first results of the CRESST-III experiment, a threshold below~100eV was already achieved~\cite{Petricca:2017zdp}. There are no additional assumptions underlying this result. For low thresholds and exposures above a critical value, these particles set constraints on sub-GeV masses, where ordinary~DM cannot, regardless of exposure. Furthermore, their spectrum is largely independent of the halo model. As a side product, we extended Gould's expression for the DM~capture rate in eq.~\eqref{eq: capture rate} to account for the Sun's temperature and opacity.

The results of this chapter are conservative, as the computed flux contains single scattering reflection only. Nonetheless, due to the correct treatment of the Sun's opacity, the equations apply to all masses and cross sections and smoothly connect the transparent and opaque limits. Using \ac{MC}~simulations will improve the results by accounting for multiple scatterings in the Sun, which increases the reflection flux. 

\clearpage
\section{MC Simulations of Solar Reflection}
\label{s: sun mc}
The results of the previous chapter should be considered as conservative, as they only include the first scattering between~DM and solar nuclei. Solar reflection from the second, third or 100th scattering is not captured by our formulae, and the resulting DM~flux is a conservative underestimate. In order to improve the result and include the effect of multiple scatterings inside the Sun, \ac{MC}~simulations of DM~trajectories inside the Sun are the appropriate tool. The simulations are similar to what we presented in chapter~\ref{s: diurnal modulation}. However, there are a few major adjustments to be implemented.
\begin{enumerate} 
	\item It is no longer acceptable to assume straight paths for the DM~particles trajectories in between scatterings. The effect of the Sun's gravitational potential is crucial, both for the shape of the DM~orbits an the speeds. This will involve the solution of the equations of motion with numerical methods.
	\item As discussed in chapter~\ref{s: sun analytic}, the desired effect is an acceleration of incoming DM~particles due to collisions on hot nuclei. The thermal motion of the targets can not be neglected.
	\item Due to gravitational focusing and acceleration, the generation of initial conditions in chapter~\ref{ss: initial conditions earth} has to be improved.
	\item Since DM~particles can be accelerated, it is no longer sensible to include a minimum speed cutoff~$v_{\rm min}$ into the algorithm, as done in the simulation algorithm of figure~\ref{fig: simulation algorithm}.
\end{enumerate}

This type of \ac{MC}~simulations to describe DM~trajectories inside the Sun have been performed already in the~'80s\cite{Nauenberg:1986em} to verify Gould's analytic formalism and more recently to study solar reflection of~DM via electron scatterings~\cite{An:2017ojc}. Therein, the effect of solar nuclei has been neglected. Even if the nuclei do not contribute to the reflection of accelerated DM~particle, they also decrease the DM~flux by shielding off the hot solar core, unless the considered model does not allow interactions with quarks.

\subsection{Simulating DM orbits and scatterings inside the Sun}

\paragraph{Equations of motion}
Putting aside scatterings for the moment, the DM~particles' trajectories are standard Keplerian orbits with the modification that they may penetrate the Sun's interior. The particles' motion happens in a plane perpendicular to the conserved angular momentum~$\mathbf{J}$. In terms of polar coordinates~$(r,\phi)$ of this plane, the motion is described by the following Lagrangian,
\begin{align}
	L &= \frac{1}{2}m_\chi\left( \dot{r}^2+r^2\dot{\phi}^2 \right) + \int\limits_{r}^{\infty}\dd r^\prime\;\frac{G_N m_\chi M(r^\prime)}{r^{\prime 2}}\, .\label{eq: sun lagrangian}
\end{align}
The mass-radius relation~$M(r)$ is part of the Standard Solar Model, see app.~\ref{a:sun}. The corresponding Euler-Lagrange equations are
\begin{subequations}
	\label{eq: equations of motion}
\begin{align}
	&\ddot{r}-r\dot{\phi}^2+\frac{G_NM(r)}{r^2}=0\, ,\\
	&r^2\dot{\phi}\equiv J =\text{const}\, .
\end{align}
\end{subequations}
Using the conservation of angular momentum, we will solve the equations of motion in two dimensions and translate the solutions back to~3D. Given a point in configuration space $(t,\mathbf{r},\mathbf{v})$, we can switch to the coordinate system spanned by
\begin{align}
	\hat{\mathbf{x}} &= \frac{\mathbf{r}}{\left|\mathbf{r}\right|}\, ,\quad \hat{\mathbf{z}} = \frac{\mathbf{r}\times \mathbf{v}}{\left|\mathbf{r}\times \mathbf{v}\right|} \, ,\quad\hat{\mathbf{y}} = \hat{\mathbf{z}}\times \hat{\mathbf{x}} \, ,
	\intertext{such that the polar coordinates in the orbital plane are}
	r &= \left| \mathbf{r}\right|\, , \quad \phi = 0\, ,\quad \text{and } J = \left|\mathbf{r}\times \mathbf{v}\right|\, .
\end{align}
As before, the~$\hat{\cdot}$ denotes vectors of unit length or the normalized version of a vector, i.e.~$\hat{\mathbf{V}}\equiv\frac{\mathbf{V}}{|\mathbf{V}|}$. After solving the equations of motion up to a new location given by $(r^\prime,\phi^\prime)$ we can return to our original coordinate system,
\begin{subequations}
\begin{align}
	\mathbf{r}^\prime & = r^\prime\left(\cos \phi^\prime \; \hat{\mathbf{x}} + \sin\phi^\prime\;\hat{\mathbf{y}}\right)\, ,\\
	\mathbf{v}^\prime & = \left(\dot{r}^\prime\cos \phi^\prime-\dot{\phi}^\prime r^\prime \sin \phi^\prime\right)\hat{\mathbf{x}} +\left(\dot{r}^\prime \sin\phi^\prime+r^\prime \dot{\phi}^\prime\cos\phi^\prime\right)\hat{\mathbf{y}}\, ,
\end{align}
\end{subequations}
with~$\dot{\phi}^\prime= \frac{J}{r^{\prime 2}}$. Inside the Sun, the orbits are no longer Keplerian and we need to solve the equations of motion numerically. We can write the Euler-Lagrange equations~\eqref{eq: equations of motion} as the set of first order ordinary differential equations,
\begin{align}
	\dot{r} = v\, ,\quad \dot{v} = r\dot{\phi}^2-\frac{G_N M(r)}{r^2}\, ,\quad \dot{\phi} = \frac{J}{r^2}\, , \label{eq: 1st order eom}
\end{align}
and solve them numerically with the Runge-Kutta-Fehlberg~(RK45) method~\cite{Fehlberg1969}, which we introduce in app.~\ref{a: runge kutta fehlberg}.

\paragraph{Scatterings on solar nuclei}

The probability for a DM~particle to scatter on a nucleus while travelling a distance~$L$ inside the Sun is given by eq.~\eqref{eq: dPscat}, which allowed us to define the speed dependent local mean free path~$\lambda(r,w)$ in eq.~\eqref{eq: MFP sun}.

Hence, the~\ac{CDF} of the freely travelled distance $l$ inside the Sun is identical to the corresponding expression for the Earth in eq.~\eqref{eq: MFP CDF}, but in terms of the adjusted mean free path,
\begin{subequations}
\begin{align}
	P(L) &= 1 - \exp\left(-\int\limits_0^L\frac{\dd x}{\lambda(r,w)}\right)\\
	&=1 - \exp\left(-\int\limits_0^L\dd x\,\sum_i n_i(r)\sigma_i \frac{\langle |\mathbf{w}-\mathbf{v}_i|\rangle}{w}\right)\, .
\end{align}
\end{subequations}
To find the location of the next scattering, we use inverse transform sampling and solve~$P(L)=\xi^\prime\in(0,1)$ along an orbit,
\begin{align}
	-\log(\xi)&=\int\limits_0^L\frac{\dd x}{\lambda(r,w)}\, \quad\text{with } \xi\equiv 1-\xi^\prime\, . \label{eq:log 1-xi}
\end{align}
While solving the equation of motion~\eqref{eq: 1st order eom} step by step with the~RK45 method, we add up the step's contribution to the right hand side of eq.~\eqref{eq:log 1-xi} until it exceeds $-\log(\xi)$. In other words, we continue the trajectory until
\begin{align}
	\sum_i \frac{\Delta l_i}{\lambda(r_i)} = \sum_i \sqrt{\dot{r}^2+\frac{J^2}{r^2}}\Delta t\;\lambda^{-1}(r_i) \geq -\log(\xi)\, .
\end{align}
Then the particle scatters at the current position. The identity of the target nucleus is sampled in the same way as before, using eqs.~\eqref{eq: probability target nucleus} and~\eqref{eq: sample target}, with the adjustment to use the solar mean free path. The final velocity of the DM~particle after the elastic collision can be found in eq.~\eqref{eq:vDMfinal sun}. The crucial difference to scatterings on terrestrial nuclei is the fact that the solar targets are hot. Their velocity follows an isotropic Maxwell-Boltzmann distribution, given by eq.~\eqref{eq: maxwell boltzmann}, with the temperature depending on the underground depth. This means that we not only have to sample the target's identity and the scattering angle, but also the target's velocity for each scattering. As the~\ac{CDF} of the Maxwell-Boltzmann distribution cannot simply be inverted, we can either sample the target speed via inverse transform sampling inverting the~\ac{CDF} numerically or use rejection sampling. The direction of the target's velocity can be chosen isotropically.\\[0.3cm]

With these adjustments, we can extend the previous simulation code to propagate DM~particles in the Sun, as they scatter, gain or lose energy, get captured and finally leave the Sun again as a reflected particles after having scattered any number of times. The next step concerns the initial conditions and how to sample them for the \ac{MC}~simulations.

\subsection{Initial conditions}
\label{ss: initial conditions sun}
The objective of this section is to
\begin{itemize}
	\item[(a)] generate initial conditions $(t_0,\mathbf{r}_0,\mathbf{v}_0)$ far away from the Sun, such that the particles are guaranteed to hit the Sun and are effectively distributed homogeneously in space.
	\item[(b)] to propagate the particle to a radius $r\gtrsim R_\odot$ using the analytic solution of the Kepler problem such that the DM~particle ends up in close proximity to the Sun with $(t_1,\mathbf{r}_1,\mathbf{v}_1)$ (with~$|\mathbf{r}_1|= R_i\gtrsim R_\odot$), which are the initial conditions for the numerical procedure.
\end{itemize}
Outside the Sun, the incoming, unbound DM~particles have positive total energy and follow hyperbolic Kepler orbit.

\paragraph{Hyperbolic Kepler orbits}
 Starting at $(t_0,\mathbf{r}_0,\mathbf{v}_0)$, we want to compute $(t_1,\mathbf{r}_1,\mathbf{v}_1)$ without passing by the periapsis. The orbit is characterized and fully determined by the following parameters,
\begin{subequations}
\label{eq:orbit parameter}
\begin{align}
	u^2 &= v_0^2-v_{\rm esc}^2(r_0)&\qquad &\text{(asymptotic speed)}\, ,\\
	a &= -\frac{G_NM_\odot}{u^2}&\qquad &\text{(semimajor axis)}\, ,\\
	p &= \frac{J^2}{G_NM_\odot}&\qquad &\text{(semilatus rectum)}\, ,\\
	e &= \sqrt{1-\frac{p}{a}}>1&\qquad &\text{(eccentricity)}\, , \\
	q &= a(1-e)&\qquad &\text{(periapsis)}\, , \\
	\cos \theta &=\frac{1}{e}\left(\frac{p}{r}-1\right)&\qquad &\text{(angle from periapsis)}\, , \\
	v^2&=\frac{G_NM_\odot}{p}\left(1+e^2+2e\cos\theta\right)&\qquad &\text{(speed)}\, , \\
	\nonumber\\
	\tan \phi &=\frac{e\sin \theta}{1+e\cos\theta}&\qquad &\text{(angle of velocity to the}\nonumber\\
	&&&\text{perpendicular of the} \nonumber\\
	&&&\text{radial direction)}\, , \\
	\cosh F &=\frac{e+\cos \theta}{1+e\cos\theta}&\qquad &\text{(eccentric anomaly)}\, , \\
	M &=e\sinh F-F&\qquad &\text{(mean anomaly)}\, , \\
	t-t_p &=\sqrt{\frac{(-a)^3}{G_N M_\odot}}M&\qquad &\text{(time from periapsis)}\, .
\end{align}
\end{subequations}
The second crucial ingredient is the orientation of the axes in three dimensions, such that $\theta = 0$ corresponds to the orbit's periapsis.
\begin{align}
	\hat{\mathbf{z}} = \frac{\mathbf{r}_0\times \mathbf{v}_0}{\left|\mathbf{r}_0\times \mathbf{v}_0\right|}\, ,\quad  \, \hat{\mathbf{x}} &=\cos\theta_0\;\hat{\mathbf{r}}_0+\sin\theta_0 \; \hat{\mathbf{r}}_0\times\hat{\mathbf{z}}  ,\quad\hat{\mathbf{y}} = \hat{\mathbf{z}}\times \hat{\mathbf{x}} 
\end{align}
Using eqs.~\eqref{eq:orbit parameter}, the new position and velocity can be expressed as
\begin{subequations}
\label{eq:Kepler shift}
\begin{align}
	t_1 &= t_0 +\text{sign}(r_1-r_0)\sqrt{\frac{(-a)^3}{G_N M_\odot}} (M_1-M_0)\, ,\\
	\mathbf{r}_1 &= r_1 \left( \cos\theta_1\; \hat{\mathbf{x}} +\sin\theta_1\; \hat{\mathbf{y}}  \right)\, ,\\
	\mathbf{v}_1 &=v_1\;\frac{e\sin\theta_1\;\hat{\mathbf{r}}_1+\left(1+e\cos\theta_1\right)\;\hat{\mathbf{z}}\times\hat{\mathbf{r}}_1}{\sqrt{1+e^2+2e \cos\theta_1}}\, ,\nonumber\\
	&=\sqrt{\frac{G_NM_\odot}{p}}\left(e\sin\theta_1\;\hat{\mathbf{r}}_1+\left(1+e\cos\theta_1\right)\;\hat{\mathbf{z}}\times\hat{\mathbf{r}}_1\right)\, , 
	\intertext{with}
	\theta_1 &= \text{sign}(r_1-r_0)\arccos\left[\frac{1}{e}\left(\frac{p}{r_1}-1\right)\right]\, .
\end{align}
\end{subequations}

\begin{figure*}
	\centering
	\includegraphics[width=0.5\textwidth]{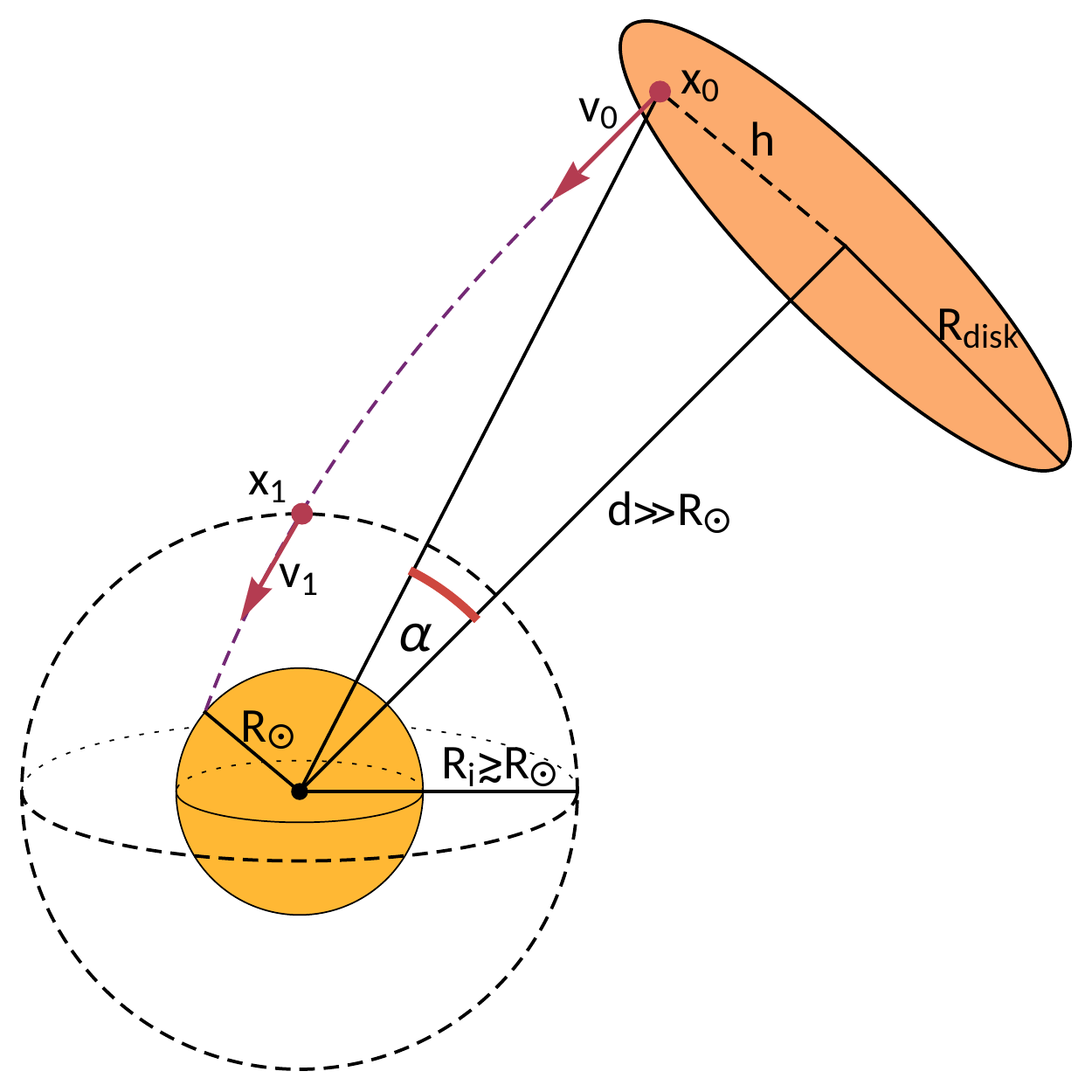}
	\caption{Sketch of the initial conditions for DM~particles entering the Sun (not to scale).}
	\label{fig: IC sun}
\end{figure*}

\paragraph{Initial position}
Let us assume, we sampled a initial velocity~$\mathbf{v}_0$ and want to sample an initial position~$\mathbf{r}_0$, such that the particle is guaranteed to penetrate the Sun's surface. The question, whether or not a given particle hits the Sun, is equivalent to the question if the orbit's periapsis is smaller than the solar radius or if the angular momentum fulfils
\begin{align}
	J &= \left|\mathbf{r}_0\times \mathbf{v}_0\right| = r_0 v_0 \sin \alpha < w(u,R_\odot)R_\odot\, ,\label{eq:hit condition}
\end{align}
where $w(u,r) = \sqrt{u^2+v^2_{\rm esc}(r)}$. Similar to the situation in chapter~\ref{ss: initial conditions earth}, all incoming particles on collision course will pass through a disk of radius $R_{\rm disk}(u) > R_\odot$ at distance $d \gg R_\odot$(such that $\frac{v_{\rm esc}^2(r_0)}{u^2}\ll 1$), oriented perpendicular to $\mathbf{u}$. Due to gravitational focussing the disk's radius is larger than~$R_\odot$ and depends on~$u$. It can be derived from eq.~\eqref{eq:hit condition} using $\sin\alpha =\frac{h}{r_0}$. The distance~$h$ to the disk's center must fulfill
\begin{align}
	h<R_{\rm disk}(u) = \sqrt{1+\frac{v_{\rm esc}^2(R_\odot)}{u^2}}R_\odot+\mathcal{O}\left(\frac{v_{\rm esc}^2(r_0)}{u^2}\right)\, .
\end{align}
This is illustrated in figure~\ref{fig: IC sun}.
Choosing a random initial position on this disk will correspond to the subset of homogeneously distributed particles on collision course with the Sun,
\begin{align}
	\mathbf{x}(t_0) = d \hat{\mathbf{z}} +\sqrt{\xi}\;R_{\rm disk}\left(\cos \varphi \;\hat{\mathbf{x}}+\sin \varphi \;\hat{\mathbf{y}} \right)\, ,\quad \text{with }\hat{\mathbf{z}} \equiv -\frac{\mathbf{u}}{u}\, .\label{eq:initial position}
\end{align}
Here, $\hat{\mathbf{x}}$ and $\hat{\mathbf{y}}$ span the disk, while~$\xi$ and~$\varphi$ are samples of the random variables~$\mathcal{U}_{[0,1]}$ and ~$\mathcal{U}_{[0,2\pi]}$ respectively. To not waste time by solving the equations of motion numerically for the infalling particle outside the Sun, we perform an analytic Kepler shift to a radius~$R\gtrsim R_\odot$ with eqs.~\eqref{eq:Kepler shift}.

This leaves us with the final recipe to sample initial conditions of DM~trajectory simulations inside the Sun.
\begin{enumerate}
	\item Sample a velocity $\mathbf{v}_0=\mathbf{u}$ far away from the Sun from the halo distribution $f_\chi(\mathbf{u})$, given in eq.~\eqref{eq:fearth}. Since we simulate the particles in the Sun's rest frame, we replace~$\mathbf{v}_\oplus$ with the solar velocity~$\mathbf{v}_\odot$ given in eq.~\eqref{eq:vsun}.
	\item Sample the initial position via eq.~\eqref{eq:initial position}.
	\item Propagate the particle analytically on its hyperbolic Kepler orbit using eqs.~\eqref{eq:Kepler shift} to a location close to the Sun.
\end{enumerate}
Finally, the resulting $(t_1,\mathbf{r}_1,\mathbf{v}_1)$ are the initial conditions for the numerical~RK45 method.

\subsection{Data collection}
\label{ss: data collection sun}
The simulation of a low-mass DM~particle continues until it leaves the Sun. As it propagates through the star's interior and scatters on the nuclei, it loses and gains energy and might get temporarily captured. Eventually, it will gain enough energy to escape the Sun's gravitational potential. If it reaches the Sun's surface without re-scattering, it counts as reflected. We analytically propagate the particle to the Earth's orbital radius with the eqs.~\eqref{eq:Kepler shift} and save the final DM~velocity as a data point. This is repeated until the data sample is sufficiently large.

\paragraph{Anisotropy and isoreflection rings}
 The DM velocity distribution in the boosted reference frame of the Sun is no longer isotropic. The anisotropy of the DM~wind might leave a trace in the reflected DM~flux. The detection signal due to these particles would consequently show a modulating behavior as the Earth orbits the Sun. This way, solar reflection could be the source of a novel annual modulation. If we want to study this effect with \ac{MC}~simulations, we have to track the directionality of the reflected particles. Again we can use the residual symmetry of the problem. The system still has a rotational symmetry around~$\mathbf{v}_\odot$. This is in principle no different from the situation in chapter~\ref{ss: data collection earth}, where we also exploited this symmetry to define isodetection rings. Again, we use the polar angle~$\Theta$ of the symmetry axis to define finite-sized rings of symmetry.

\begin{figure*}
	\centering
	\includegraphics[width=0.67\textwidth]{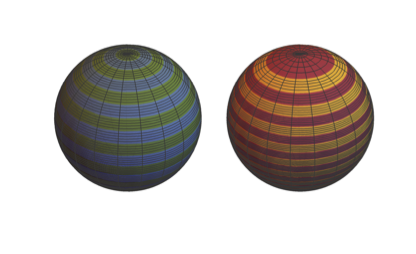}
	\caption{Comparing the equal-angle isodetection rings for the Earth simulations and the equal-area isoreflection rings for the Sun simulations.}
	\label{fig:isoreflection}
\end{figure*}

For the data collection with the terrestrial simulations, we fixed the isodetection rings' sizes by a constant angle~$\Delta\Theta$ following~\cite{Collar:1993ss,Hasenbalg:1997hs}. The area of these rings differed considerably, getting smaller towards the `poles' following eq.~\eqref{eq:ringarea}. A good angular resolution was necessary to accurately describe diurnal modulation for any terrestrial detector on the globe. Unfortunately, the data collection for the smaller rings takes much more time, as the probability for a particle to hit a small area is of course low. The varying area of the isodetection rings was an unavoidable bottleneck of the Earth simulations. For the Sun however, we do not require the same angular resolution and define the isoreflection rings to be of equal area. We call these equal-area rings \textit{isoreflection rings}. For~$N$ rings, the isoreflection rings are defined via
\begin{align}
	\Theta_{i} = \arccos \left(\cos\Theta_{i-1}-\frac{2}{N}\right)\, ,\quad \text{with }\theta_0 = 0 \text{ and } i\in\{1,...,N\}\, . 
\end{align}
 A comparison between the equal-angle isodetection rings and the equal-area isoreflection rings is shown in figure~\ref{fig:isoreflection}. The loss of angular resolution is most severe at the poles. However, the Earth never passes through these regions, and this is not a problem. 

\subsection{First results and outlook}
In this chapter of the thesis, we will present some preliminary results for the \ac{MC}~treatment of solar reflection. The simulation of DM~trajectory in the Sun can shed light on the impact of multiple scatterings. The analytic formalism presented in chapter~\ref{s: sun analytic} provides conservative results and only accounts for single scattering reflection. Similarly to the Earth simulations in chapter~\ref{ss: diurnal modulation result}, these analytic results can be used to test and verify the independent approaches.

\begin{figure*}
\centering
\includegraphics[width=\textwidth]{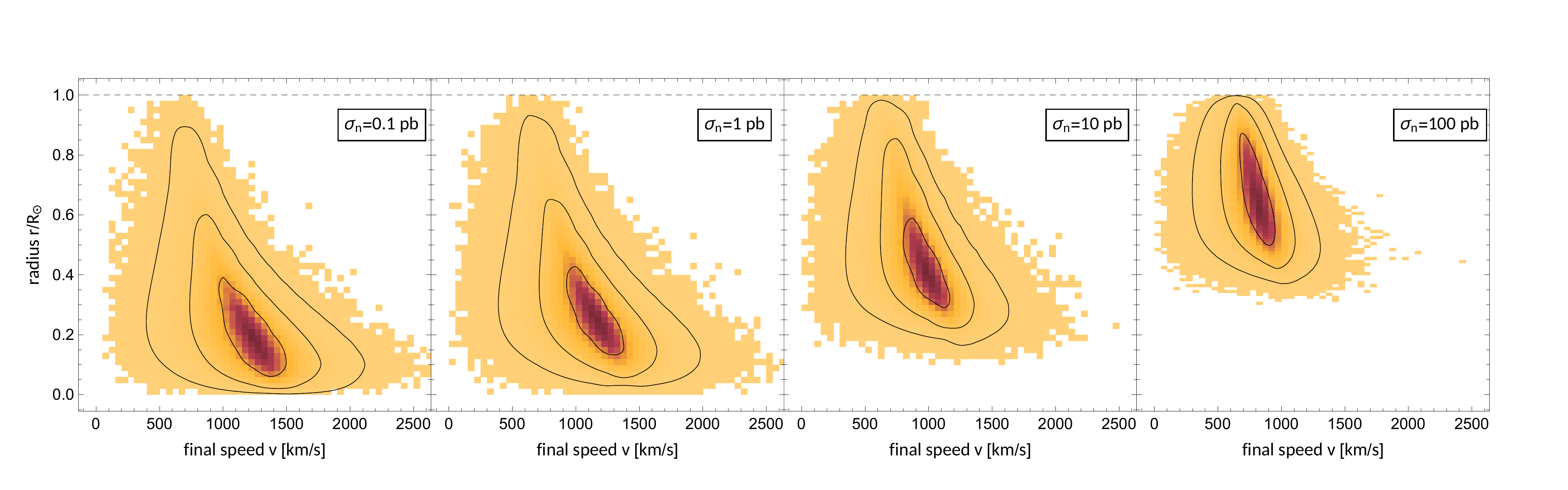}
\caption{Histogram of the radius and final DM~speed of the first scattering in the Sun. The black contours trace the analytic scattering rate~$\frac{\dd S}{\dd v \dd r}$.}
\label{fig: sun scattering rate MC}
\end{figure*}

\paragraph{Comparison of analytic and~MC result}
The number of isoreflection rings~$N$ is adjustable in the simulation code. In order to compare to the analytic results, which assumed isotropy, we can simply set it to one. Naturally, we include the same number of solar targets in the simulation. For the comparison, we are only interested in the \textit{first} scattering, as this is the one described by the analytic equations. For each simulated trajectory, we save the radius and the final speed and plot the data in a two-dimensional histogram. The differential first scattering rate in a shell of radius~$r$ to a speed~$v$ was computed previously in eq.~\eqref{eq: sun dSdrdv} and plotted as black contours in figure~\ref{fig: Pleave and dSdrdv}. In figure~\ref{fig: sun scattering rate MC}, we compare the \ac{MC}~histograms with the same contours.

We can see that the simulations yield the same location of the first scattering and the same final DM~speed, which is a first indicator for the consistency of the two results.

\paragraph{Multiple scatterings}
There is no reason why a DM~particle cannot be reflected with high energy after many collisions. In fact, since about half of the DM~particles are gravitationally captured by their first scattering, they are bound to scatter at least one more time. As discussed earlier, this statement is independent of the assumed cross section, and a `single scattering regime' does not really exist. 

\begin{figure*}
\centering
\includegraphics[width=0.67\textwidth]{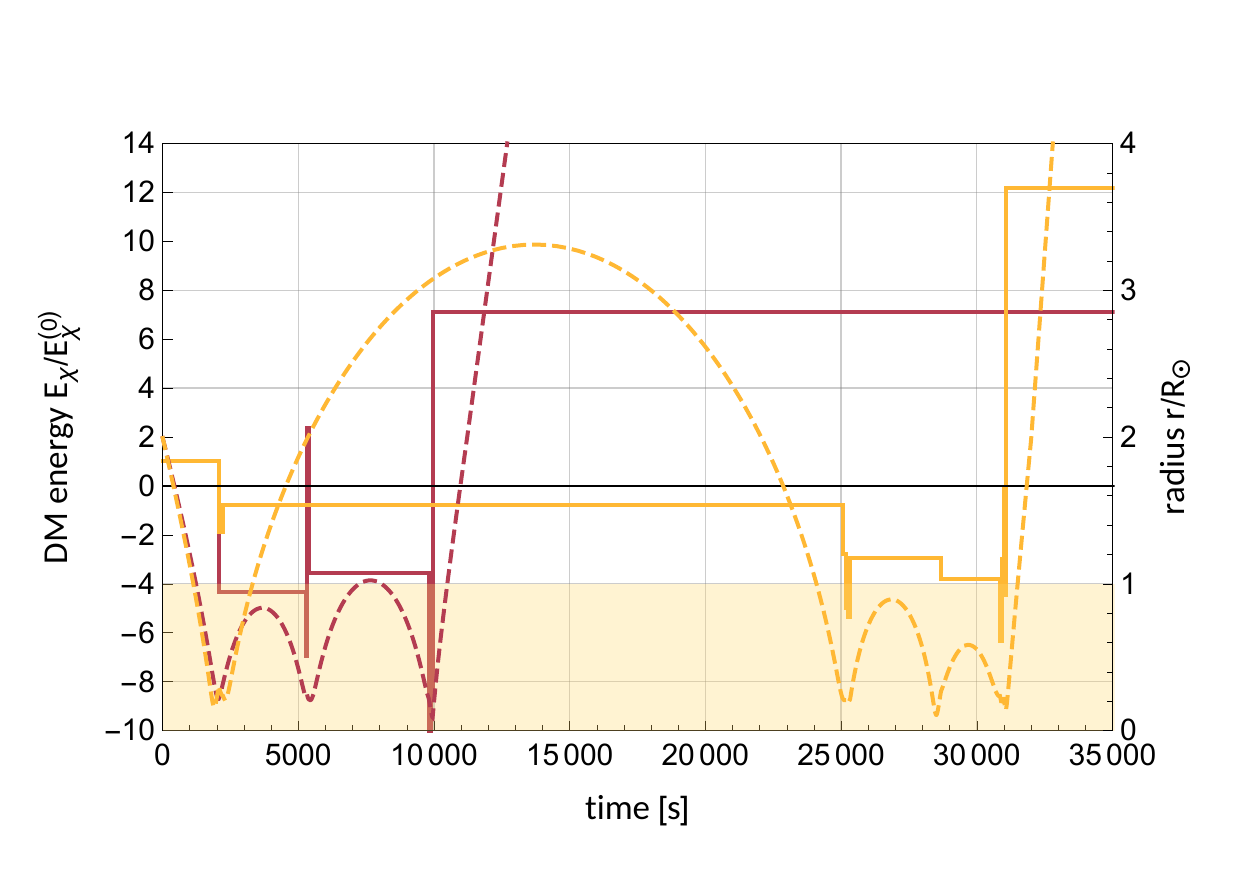}
\caption{Time evolution of the total DM~energy~(solid lines) and the radial coordinate~(dashed lines) for two example trajectories.}
\label{fig: sun mc energy and radius}
\end{figure*}

We consider two example trajectories for a DM~particle of~1~MeV and a DM-proton cross section of~1~pb. The evolution of their total energy~$E_\chi(t)=T+V$ and radial coordinate~$r(t)$ are plotted in figure~\ref{fig: sun mc energy and radius} as solid and dashed lines respectively. The red example describes a particle getting captured by the first collision. Even though the third scattering accelerate the DM~particle sufficiently to potentially escape, rendering its total energy positive, the particle does not reach the solar surface before the next scattering. Only after the seventh scattering does the particle finally leave the star with an energy increased by a factor of around 7. The yellow example shows another particle getting captured to a bound orbit partially outside the Sun. As it re-enters the hot, dense core it scatters a few more times and gets eventually reflected with more than twelve times its original energy after a total of 13 scatterings.

By counting the reflected DM~particles passing the isoreflection rings and saving their speed, we can determine the differential reflection flux. The \ac{MC}~result has to be re-scaled to yield the correct total reflection flux,
\begin{align}
	\frac{\dd \mathcal{R}}{\dd u} = \underbrace{\frac{n_{\rm refl}}{n_{\rm sim}}\;\Gamma_\odot(m_\chi)}_{\text{\footnotesize total reflection flux}} \times \underbrace{\hat{\Phi}(u)}_{\text{\footnotesize normalized flux distribution}}\, .
\end{align}
Here, we use eq.\eqref{eq: sun dm rate into sun}, the rate of halo particles entering the Sun, which can be explicitly evaluated via
\begin{subequations}
\begin{align}
	\Gamma_\odot(m_\chi) &=\pi R_\odot^2 n_\chi \left[\langle u \rangle+v_{\rm esc}(R_\odot)^2\langle u^{-1}\rangle\right]\\
	&=\frac{\rho_\chi}{m_\chi}\frac{v_0}{N_{\rm esc}}\pi R_\odot^2\nonumber\\
	&\qquad\times\Bigg[ \frac{1}{\sqrt{\pi}}\exp\left(-\frac{v_\odot^2}{v_0^2}\right)+\left(\frac{v_{\rm esc}(R_\odot)^2}{v_0v_\odot}+\frac{v_\odot}{v_0}+\frac{v_0}{2v_\odot}\right)\erf{\frac{v_\odot}{v_0}}\nonumber\\
	&\quad-\frac{2}{\sqrt{\pi}}\left(1+\frac{v_{\rm esc}(R_\odot)^2}{v_0^2}+\frac{v_{\rm gal}^2}{v_0^2}+\frac{v_\odot^2}{3v_0^2}\right)\exp\left(-\frac{v_{\rm gal}^2}{v_0^2}\right)\Bigg]
\end{align}
\end{subequations}
The galactic escape velocity is denoted as~$v_{\rm gal}$, not~$v_{\rm esc}$, to distinguish it from the solar escape velocity~$v_{\rm esc}(r)$. This expression can be used in connection with the ratio of the total number of simulated particles~$n_{\rm sim}$ and the number of reflected particles~$n_{\rm refl}$ to scale up the \ac{MC}~numbers to the realistic values. Ultimately, we will estimate the normalized flux distribution~$\hat{\phi}(u)$ using \ac{KDE} (see app.~\ref{a:kde}). However, we will start by using histograms, as this enables to study and visualize the effect of multiple scatterings. 

\begin{figure*}
\centering
\includegraphics[width=0.75\textwidth]{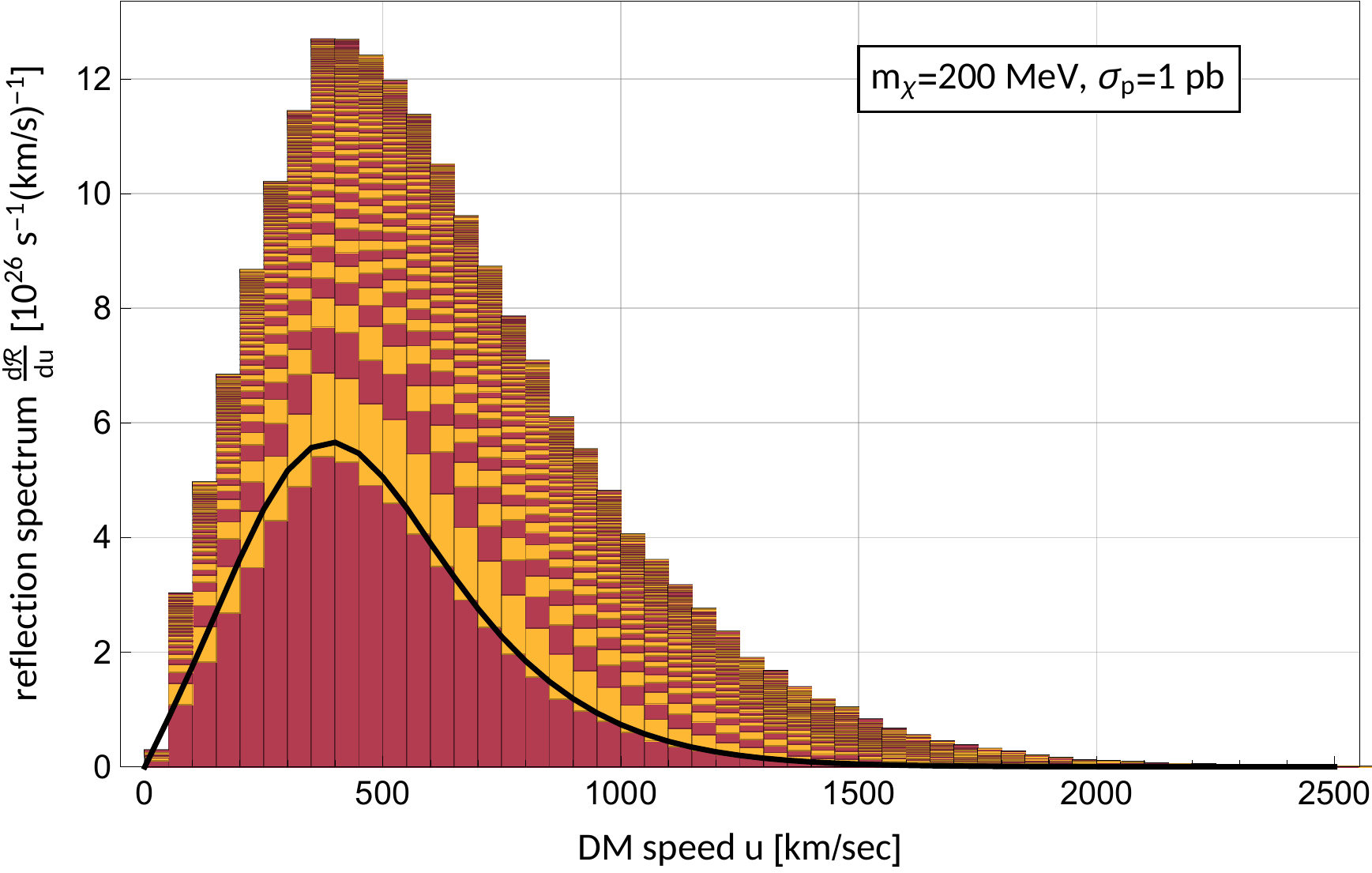}
\caption{Contribution of multiple scatterings to the solar reflection spectrum.}
\label{fig: reflection spectrum MC}
\end{figure*}

In figure~\ref{fig: reflection spectrum MC}, we show the solar reflection spectrum for $m_\chi=$~200~MeV and $\sigma_p=$~1~pb, using a histogram estimate. The bar's alternating colors indicate the contribution of reflection by one scattering, two scatterings, three scatterings, etc. The black line corresponds to our analytic result of eq.~\eqref{eq: dRdu analytic} and again shows good agreement with the single scattering fraction. It is obvious that, while single scattering reflection makes up a large fraction of the flux especially for lower speeds, the inclusion of multiple scatterings via \ac{MC}~simulations increases the reflection flux significantly. Especially the high energy tail, which is essential for DM~detection, is greatly enhanced. Whereas the individual contributions of each number of scatterings above 2 is negligible, their fractions due to large number of collisions accumulate to a sizeable flux amplification.

\paragraph{Outlook}
The results of the Sun simulations presented in this section are preliminary. The simulations have not yet been applied to derive direct detection constraints on sub-GeV~DM. An efficient procedure to scan the two-dimensional mass-cross section parameter space will be necessary, as each parameter point requires its own simulation run. Once this is set up, we can expect the resulting constraints to far exceed the analytic results in figure~\ref{fig: solar constraints}. The final hope is to find cases, where detection runs can employ solar reflection to extend their sensitivity.

 The detection of~DM accelerated by the Sun by multiple scatterings should be studied for nuclear recoil experiments, as well as DM-electron scattering experiments. Solar electrons would need to be considered as an additional target. Electrons were already considered in the context of leptophilic~DM, where the Sun's nuclear targets may be neglected~\cite{An:2017ojc}. If we want to include both target species, a model-dependent relation between the DM-nucleus and DM-electron cross sections would have to be assumed. Furthermore, the annually varying Sun-Earth distance would leave a modulation signature in the detection signal. This modulation could potentially be modified, if the anisotropy of the initial conditions due to the DM~wind leaves a trace in the reflection spectrum. \ac{MC}~simulations would allow to probe the assumption of isotropy. With the definition of the isoreflection rings, we will be able to address this question.

The mechanism of solar reflection of~DM has so far only been investigated for contact interactions, either~\ac{SI} DM-nuclear, or DM-electron collisions. The inclusion of more general interactions is one of the straight-forward extensions. It would be interesting to consider e.g. spin-dependent interactions or the general~\ac{NREFT}~operators~\cite{Fan:2010gt,Fitzpatrick:2012ix}, the relevant form factors for the solar targets have been computed in the context of DM~capture~\cite{Catena:2015uha}. Finally, implementing interactions mediated by ultralight fields into the simulations could be a promising direction.

\clearpage
\chapter{Conclusions and Outlook}
\label{c:conclusion}

Our entire knowledge about~DM rests upon its gravitational interplay with baryonic matter. The presence of large amounts of unseen matter manifests itself in astronomical observations starting in the 1930s. Back then, Fritz Zwicky found large discrepancies between the gravitational and the luminous mass of a galaxy cluster. A few years later, Horace Babcock was the first to point out a peculiar flattening of a galactic rotation curve. Furthermore,~DM revealed itself by bending the light of background sources and by dominating the formation of cosmological structures. Another crucial piece of evidence is the observation of the anisotropies of the~\ac{CMB}. Here,~DM left a unambiguous trace in the acoustic peaks of the power spectrum during the early Universe. The history and evidence of~DM was the main topic of chapter~\ref{c:DM}.

The minimal explanation for these measurements would be a new form of matter interacting with the particles of the~\ac{SM} exclusively via gravity. Gravitational interactions are unavoidable due to gravity's universality, and this could indeed be the only portal coupling to visible matter. As such,~DM would reside in a secluded sector, and it would hardly be possible to unveil its origin and properties in the foreseeable future. Nonetheless, there is the well-motivated hope that this is not the path that nature chose for our Universe. The ultimate objective of all direct detection experiments is to search for an additional interactions between the bright and dark sectors of matter by measuring the aftermath of a DM~collision in a detector. Such experiments were the focus of this thesis, and their fundamental principles were covered in chapter~\ref{c:directdetection}.

Provided that particles of the DM~halo can collide with ordinary particles, these collisions are of course not restricted to take place inside a detector's target. For high enough scattering probabilities, it is therefore illegitimate to treat the detector as an isolated object situated in the galactic halo. The surrounding medium of the Earth needs to be taken into account, since the incoming DM~particles have to pass through it to reach the detector and might scatter beforehand. The central subject of this thesis was to describe underground scatterings both in the Earth and the Sun and investigate the implications for direct detection experiments searching for low-mass, mostly sub-GeV, DM~particles.

With the exception of chapter~\ref{s: sun analytic}, the results of this thesis were obtained by the use of \ac{MC}~simulations of individual DM~particle trajectories. By tracking large numbers of particles, their statistical properties can be used to quantify the effect of multiple underground scatterings on direct DM~searches. The simulation algorithms were formulated from the ground up and implemented fully parallelized in C++. The resulting scientific codes underwent extensive testing and were released as open source code to the community together with the respective publications. Most of the results of this thesis were obtained by high-performance computations on the \textit{ABACUS~2.0} of the DeIC National HPC Center, a supercomputer with 584 nodes of 24 cores each.

\paragraph{DM in the Earth}
Concerning underground scatterings inside the Earth, we started off by simulating DM~particles on their path through our planet's core and mantle. These simulations are relevant for~DM with masses below~$m_\chi\lessapprox \mathcal{O}$(500)MeV and proton scattering cross section around~$\sigma^{\rm SI}_p\sim\mathcal{O}$(1-100)pb. For such light~DM, the constraints from direct detection rapidly grow weaker as the nuclear recoils fall below the experimental thresholds. The results can also apply to the possibility that the strongly interacting~DM makes up a subdominant component of the total amount of~DM in the Universe.

In this region of the parameter space, the probability of scattering before entering the detector is significant and must be taken into account. They distort the expected nuclear recoil spectrum, potentially increasing or decreasing the expected signal rate compared to unscattered halo particles. 

We modelled the Earth in the \ac{MC}~simulations on the basis of the~\ac{PREM} and improved the generation of initial conditions from similar works of the past. The resulting~\ref{code1} code returns a data-based estimate of the underground DM~density and velocity distribution for any location and time. For a specified direct detection experiment, we analyze the generated \ac{MC}~data and determine the time evolution of the recoil spectrum and signal rates throughout a sidereal day, quantifying both the phase and amplitude of the diurnal modulation.

For the single scattering regime, we showed that the \ac{MC}~results are in excellent agreement with previous analytic results. Therefore the simulation code passed a crucial consistency check, and we continued to study larger cross sections and the impact of multiple scatterings. We computed the diurnal modulation for a detector of CRESST-II type and a number of benchmark points. For the largest cross section, we found a sizeable modulation of almost~100\% to be expected for an experiment in the southern hemisphere.

The simulations could be extended by the considerations of more general interactions, such as long range forces or higher-order operators of the~\ac{NREFT} framework. On the side of the data analysis, it could be interesting to determine the modulation of DM-electron scattering experiments due to underground DM-nucleus scatterings.\\[0.3cm]

In the second part of chapter~\ref{c:earth}, we systematically determined how far the various direct detection constraints extend towards stronger interactions. Due to scatterings in the overburden of the experiments, which are located deep underground in most cases, the Earth crust and atmosphere can attenuate the flux of strongly interacting~DM. Ultimately, the experiment is not able to probe cross sections above a critical value. We used \ac{MC}~simulations to calculate this critical cross section for a variety of DM-nucleus and DM-electron recoil experiments. In the latter case, we shifted our focus further to models with light mediators, which modify the scattering kinematics substantially. These kind of simulations are highly relevant for direct detection experiments, as they reveal the limitations of detectors' sensitivity and might point out open parameter space above the usual constraints. We demonstrated that direct detection experiments need to be located in shallower sites, either on the surface or even at higher altitudes, if it is supposed to probe strongly interacting~DM.

Since the relevant scatterings occur in the Earth's crust and atmosphere above the laboratory, it was no longer necessary to simulate trajectories through the whole planet, and we could simplify the simulation volume to parallel planar shielding layers. The ultimate goal of the simulation is to determine the DM~distribution and density at a given underground depth. Here, the greatest challenge was the fact that close to the critical cross section, only tiny fractions of the incoming detectable DM~flux reach this depth. We solved this problem using rare-event simulation techniques such as~\ac{IS} and~\ac{GIS}. The resulting~\ref{code2} code is publicly available and can be extended to future experiments in a straight-forward fashion.

For nuclear recoil experiments, we found the that the CRESST~2017 surface run closes most open windows in parameter space, as shown in figure~\ref{fig: nuclear constraints}. The scenario of strongly interacting~DM with spin-independent contact interactions is strictly constrained. The situation is different in the presence of light mediators, which we studied for noble and semiconductor target detectors probing inelastic DM-electron collisions. As shown in figure~\ref{fig: light mediator summary}, the different bounds might leave some interesting parts of the parameter space unconstrained especially if the strongly interacting~DM particle are only a small component of the total~DM. However, the situation here is not clear at this point and requires further investigations. We presented some discovery prospects for balloon- and satellite-borne experiments targeting this region, which could become very relevant in the near future.

\paragraph{DM in the Sun}
The phenomenology of DM~particles scattering inside Sun was studied in the second main chapter of this thesis. Therein, we developed the new idea of solar reflection, when DM~particles fall into the Sun, gain kinetic energy via elastic scatterings on hot solar constituents and leave the Sun with speeds far above the maximum of the normal DM~halo population. The central consequence of solar reflection is the existence of an additional DM~population in the solar system, whose spectrum is widely insensitive to the choice of the halo model. By extending Gould's analytic framework, we established the possibility that the resulting flux of reflected DM~particles can be used to set constraints in the low-mass parameter space, where the standard halo~DM can not regardless of the experiment's exposure. Low-threshold detectors could extend their sensitivity to DM~masses below their naive minimum. In contrast to standard halo DM~constraints, the minimum testable mass due to solar reflection depends on the experiment's exposure gets lower for larger exposures. 

The constraints presented in figure~\ref{fig: solar constraints} are conservative, as they only include solar reflection by a single scattering. This is why we started to set up a \ac{MC}~description of solar reflection, which accounts for the contribution of multiple scatterings inside the Sun, where particles could also get temporarily captured gravitationally before they finally get reflected. The ground work for such simulations is presented in chapter~\ref{s: sun mc}, but there is still work to be done, before the simulations can be used to derive robust constraints.

New experiments like CRESST-III~\cite{Petricca:2017zdp} have realized recoil thresholds below 100~eV. Direct detection might be able to probe solar reflection in the near future.\\[0.3cm]

In conclusion of this thesis, we demonstrated the power of \ac{MC}~simulations to quantify the effect of multiple underground scatterings for direct searches of light~DM. The simulations of DM~particle trajectories in the Earth or Sun can be applied to answer various questions, and we applied them to study diurnal modulations of detection rates, constraints on strongly interacting~DM, and the new idea of solar reflection of highly energetic DM~particles. In these contexts, \ac{MC}~simulations are an invaluable tool to quantify or extend the sensitivity of detection experiments here on Earth. 

Future experiments will hopefully succeed to reveal a portal between the visible and dark matter sectors. For low-mass~DM with significant scattering rates, the results and tools developed in this thesis should be of great value for the discovery of~DM and further investigations of its nature.

\appendix
\clearpage
\chapter{Constants and Units}
\label{a:units}

\section{Physical Constants}
The physical parameter, couplings, and masses required for the computations of this thesis are listed in~SI units in table~\ref{tab: constants}. 
 
\begin{table}[h!]
	\centering
	\begin{tabular}{lcl}
	\hline
		Quantity	&	Symbol	& SI-Value~\cite{Tanabashi:2018oca}\\
		\hline
		\textbf{Physical constants}	&&\\
		speed of light in vacuum		&$c$	&\SI{299 792 458}{\meter\per\second} \\
		Planck constant				&$h$	&\SI{6.626070040(81)e-34}{\joule\second} \\
		electric charge				&$e$	&\SI{1.6021766208(98)e-19}{\coulomb} \\
		Newton constant				&$G_N$	&\SI{6.67408(31)e-11}{\meter\cubed\per\kilo\gram\per\second\squared} \\
		Boltzmann constant			&$k_B$	&\SI{1.38064852(79)e-23}{\joule\per\kelvin} \\
		vacuum permeability	& $\mu_0$	&\SI{4\pi e-7}{\newton\per\ampere\squared}\\
		weak mixing angle	&$\sin^2\theta_W~(\overline{\text{MS}})$	&0.231 22(4) (at $q^2=M_Z^2$)\\
		\hline
		\textbf{Masses}	&&\\
		electron mass				&$m_e$	&\SI{9.10938356(11)e-31}{\kilo\gram} \\
		proton mass					&$m_p$	&\SI{1.672621898(21)e-27}{\kilo\gram} \\
		atomic mass unit		&$m_n$	&\SI{1.660 539 040(20)e-27}{\kilo\gram} \\
		\hline
		\textbf{Derived quantities}	&&\\
		Planck constant (reduced)	&$\hbar\equiv\frac{h}{2\pi}$	&\SI{1.054571800(13)e-34}{\joule\second} \\
		vacuum permittivity	&$\epsilon_0\equiv\frac{1}{\mu_0 c^2}$ &\SI{8.854 187 817e-12}{\farad\per\meter}\\
		fine structure constant	&$\alpha\equiv\frac{e^2}{4\pi\epsilon_0\hbar c}$	&1/137.035 999 139(31) (at $q^2=0$) \\
		Bohr radius	&$a_B\equiv\frac{4\pi\epsilon_0\hbar^2}{m_e e^2}$	&\SI{0.529 177 210 67(12)e-10}{\meter} \\
		\hline
	\end{tabular}
	\caption{Physical constants and masses in SI units}
	\label{tab: constants}
\end{table}

\section{Natural Units}
Instead of~SI units, so-called \textit{natural units} are used in many fields of physics for reasons of convenience and convention. The set of natural units most common in high energy physics and used throughout this thesis, are defined by setting the speed of light in vacuum, the reduced Planck constant, and the Boltzmann constant to 1,
\begin{subequations}
	\label{eq: natural units}
	\begin{align}
		c&=1\, ,\\
		\hbar &=1 \, ,\\
		k_B&=1 \, .
	\end{align}
	\end{subequations}
Our electromagnetic units are based on the rationalized Lorentz-Heaviside units,
\begin{subequations}
\begin{align}
		\epsilon_0&=1\, ,\\
		\mu_0&=1\, ,\\
		\Rightarrow e&=\sqrt{4\pi \alpha}=\SI{0.30282212}{}\, .
	\end{align}
\end{subequations}
Consequentially, all unit dimensions can be expressed in powers of energy, and we can express physical quantities in terms of a single unit. Throughout the computations for this thesis we use~GeV as default unit. The conversion factors between~SI and natural units are found in table~\ref{tab: natural units}.

\begin{table}[h!]
	\centering
	\begin{tabular}{llcl}
		\hline
		Dimension	&	Unit	&	Conversion factor	&SI-Value\\
		\hline
		energy	&\SI{1}{\GeV}	&$\times 1$	&$=\SI{1.60218e-10}{\joule}$\\
		length	&\SI{1}{\per\GeV}	&$\times \hbar c$	&$=\SI{1.97327e-16}{\meter}$\\
		time	&\SI{1}{\per\GeV}&$\times \hbar$	&$=\SI{6.58212e-25}{\second}$\\
		mass	&\SI{1}{\GeV}	&$\times\frac{1}{c^2}$		&$=\SI{1.78266e-27}{\kilo\gram}$\\
		temperature	&\SI{1}{\GeV}	&$\times \frac{1}{k_B}$	&$=\SI{1.16045e13}{\kelvin}$\\
		Electric charge	&1	&$\times \frac{e^{\rm SI}}{e}$	&$=\SI{5.29082e-19}{\coulomb}$\\
		\hline
	\end{tabular}
	\caption{Conversion between SI and natural units}
	\label{tab: natural units}
\end{table}

\clearpage
\chapter{Astronomical requirements}
\label{a:astro}
Various astronomical details of a more technical nature are necessary for the different computations and simulations, ranging from astronomical coordinate systems to models of the Earth and the Sun. We present a review of these basics in this app. both for completeness and to serve as a reference.

\section{Constants}
\label{a: astro constants}
The relevant astronomical parameters are listed in table~\ref{tab: astrophysical quantities}.
\begin{table}[h!]
	\centering
	\begin{tabular}{lcl}
	\hline
		Quantity	&	Symbol	& SI-Value~\cite{Tanabashi:2018oca}\\
		\hline
		\textbf{Units of length}	&&\\
		light year			&ly		&\SI{9.4607e15}{\meter}	\\
		parsec				&pc		&\SI{3.08567758149e16}{\meter}	\\
		astronomical unit	&AU		&\SI{149597870700}{\meter}	\\
		\hline
		\textbf{Solar parameters}	&&\\
		Mass				&$M_\odot$	&\SI{1.98848(9)e30}{\kilo\gram}\\
		Radius				&$R_\odot$	&\SI{6.957e8}{\meter}\\
		Core temperature		&$T_\odot(0)$	&\SI{1.549e7}{\kelvin}\\
		\hline
		\textbf{Terrestrial parameters}	&&\\
		Mass				&$M_\oplus$	&\SI{5.9724(3)e24}{\kilo\gram}\\
		Radius				&$R_\oplus$	&\SI{6.371e6}{\meter}\\
		\hline
	\end{tabular}
	\caption{Astrophysical quantities}
	\label{tab: astrophysical quantities}
\end{table}

\clearpage
\section{Sidereal Time}
\label{a:siderealtime}

For a precise prediction of the diurnal signal modulation's phase, the Earth's rotation needs to be tracked precisely. The \emph{sidereal time} is the appropriate time scale as it is based on that very rotation. One sidereal day is defined as the duration of one rotation relative to vernal equinox $\Upsilon$ and is about four minutes shorter than a mean solar day~(\SI{23.9344699}{\hour})~\cite{almanac2014}. It is often given as an angle instead of a time unit. Furthermore, the \emph{\ac{LAST}} is defined as the time since the local meridian of some position on Earth passed vernal equinox $\Upsilon$ and is the primary measure for the Earth's rotational phase. In order to calculate the \ac{LAST} for any given time and location on Earth, all times will be given relative to 01.01.2000 12:00 \ac{TT}, or short `J2000.0', a commonly used reference time. The fractional number of days relative to J2000.0 for a given date $D.M.Y$ and time $h:m:s$ (\ac{UT})  is found by the following relation~\cite{McCabe:2013kea},
\begin{subequations}
\begin{align}
	n_{\text{J2000}} &= \lfloor 365.25\tilde{Y}\rfloor +\lfloor 30.61(\tilde{M}+1)\rfloor+D\nonumber\\
	&\qquad+\frac{h}{24}+\frac{m}{24\times 60}+\frac{s}{24\times 60^2} -730563.5\, ,
	\intertext{where}
	\tilde{Y} &\equiv \begin{cases}
		Y-1\quad &\text{if } M=1\text{ or }2\, ,\\
		Y\quad &\text{if }M>2\, ,
	\end{cases}
	\intertext{and }
	\tilde{M} &\equiv \begin{cases}
		M+12\quad &\text{if } M=1\text{ or }2\, ,\\
		M\quad &\text{if }M>2\, ,
		\end{cases}
\end{align}
\end{subequations}
and $\lfloor \cdot \rfloor$ is the floor function. As an example, the time of 31.01.2019 at 23:59 \ac{UT} corresponds to  $n_{\text{J2000}}=$~6970.5. It will be useful to define the \emph{epoch} as
\begin{align}
	T_{\text{J2000}}\equiv\frac{n_{\text{J2000}}}{36525}\, .
\end{align}
The first step towards the \ac{LAST} is the determination of the \ac{GMST}, expressed in seconds via the formula~\cite{almanac2014}
\begin{align}
	&\text{GMST}[\text{s}] =\nonumber\\
	&86400\times\big[0.7790\,5727\,3264+n_{\text{J2000}}\mathrm{mod}\,1+0.0027\,3781\,1911\,35448\;n_{\text{J2000}}\big]\nonumber\\
	&+\big[0.000\,967\,07+307.477\,102\,27\,T_{\text{J2000}}\nonumber\\
	&\quad+0.092\,772\,113\,T_{\text{J2000}}^2+\mathcal{O}(T_{\text{J2000}}^3)\big]\, .	\label{eq:gmst}
\end{align}
Secondly, the equation of equinoxes is added to obtain the \ac{GAST}
\begin{subequations}
\begin{align}
	\text{GAST} &= \text{GMST} + E_e(T_{\text{J2000}})\, .
	\intertext{The equation of equinoxes can be approximated as}
	E_e(T_{\text{J2000}})&\approx\Delta\psi\cos\epsilon_A + 0.000176s \sin \Omega+ 0.000004s\sin 2\Omega\, ,
	\intertext{where}
	\Delta\psi&\approx-1.1484s\,\sin \Omega - 0.0864s\, \cos 2L\, ,\\
	\Omega &= 125.0445\,5501^{\circ}-0.0529\,5376^{\circ}n_{\text{J2000}} + \mathcal{O}(T_{\text{J2000}}^2)\, ,\\
	L &= 280.47^{\circ}-0.98565^{\circ}n_{\text{J2000}} + \mathcal{O}(T_{\text{J2000}}^2)\, ,\\
	\epsilon_A &=23.4392\;79444^{\circ}-0.01301021361^{\circ}T_{\text{J2000}} + \mathcal{O}(T_{\text{J2000}}^2)	\, .
\end{align}
\end{subequations}
Finally, the \ac{LAST} for a place with latitude and longitude $(\Phi,\lambda)$ in seconds results by adding the longitude,
\begin{align}
	\text{LAST}(\lambda) = \text{GAST} + \frac{\lambda}{360^{\circ}}\SI{86400}{\second}\, .
\end{align}
Here, western longitudes are negative and \ac{LAST}$~\in$~(0,86400)s. The errors of the approximations are of the order of tens of milliseconds compared to the tables in~\cite{almanac2014}.

\section{Coordinate Systems}
\label{a:coordinates}
For the simulation of the DM~particle trajectories through the Earth, the planet's orientation has to be specified. The simulations are carried out in the galactic frame. We review all relevant coordinate systems in this chapter based on~\cite{McCabe:2013kea,almanac2014}, together with the transformations. All of the coordinate systems listed in table~\ref{tab:coordinates} are right handed and rectangular.
\begin{table*}
\centering
\begin{tabular}{lll}
\hline
	Frame	&	&	Description\\
	\hline
	galactic	&	(gal)	&	-- heliocentric\\
	&&-- $x$-axis points towards the galactic center\\
	&&-- the $z$-axis points towards the galactic north pole\\
	&&-- $x$- and $y$-axis span the galactic plane\\
	&&-- see figure~\ref{fig:galacticframe}\\
	heliocentric,	&	(hel-ecl)	& -- heliocentric\\
	ecliptic&&-- $x$-axis points towards vernal equinox $\Upsilon$\\
	&&-- $z$-axis points towards the ecliptic north pole\\
	&&-- $x$- and $y$-axis span the ecliptic plane \\
	&&--see figure~\ref{fig:heliocentriccoordinates}\\
	geocentric,	&	(geo-ecl)	& -- geocentric\\
	ecliptic&&-- $x$-axis points towards vernal equinox $\Upsilon$\\
	&&-- $z$-axis points towards the ecliptic north pole\\
	&&-- $x$- and $y$-axis span the ecliptic plane\\
	geocentric,	&	(equat)	& -- geocentric\\
	equatorial&&-- $x$-axis points towards vernal equinox $\Upsilon$\\
	&&-- $z$-axis points towards the Earth north pole\\
	&&-- $x$- and $y$-axis span the equatorial plane \\
	&&-- see figure~\ref{fig:laboratoryframe}\\
	laboratory	&	(lab)	&	-- laboratory-centric\\
	&&-- $x$-axis points towards east\\
	&&-- $z$-axis to the sky.\\
	\hline
\end{tabular}
\caption{Coordinate systems}
\label{tab:coordinates}
\end{table*}

The transformations into the galactic frame, where all computations take place, require a number of time-dependent transformation matrices.

\paragraph{(lab) $\longleftrightarrow$ (equat):} The transformation from the laboratory to the equatorial frame is done via
\begin{align}
	\mathbf{x}^{\text{(equat)}} = \mathcal{N}\mathbf{x}^{\text{(lab)}}\, ,\text{ with }
	\mathcal{N}=\begin{pmatrix}
		-\sin \phi		&-\cos \theta \cos \phi			&\sin \theta \cos \phi\\
		\cos \phi		&-\cos \theta \sin \phi			&\sin \theta \sin \phi\\
		0				&\sin \theta					&\cos \theta
	\end{pmatrix}\, ,
\end{align}
where $\theta = \frac{\pi}{2}-\Phi$, $\phi = 2\pi\frac{\text{LAST}(\Phi,\lambda)}{86400\text{s}}$, and $(\Phi,\lambda)$ are the laboratory's latitude and longitude respectively.
\paragraph{(hel-ecl)$\longleftrightarrow$(geo-ecl):} The two ecliptic frames are connected by the trivial transformation
\begin{align}
	\mathbf{x}^{\text{(geo-ecl)}}=-\mathbf{x}^{\text{(equat)}}\, .
\end{align}

\paragraph{(geo-ecl.) $\longleftrightarrow$ (equat):}
A vector $\mathbf{x}^{\text{(geo-ecl)}}$ is transformed to equatorial coordinates through the following rotation,
\begin{align}
	\mathbf{x}^{\text{(equat)}} &= \mathcal{R}\mathbf{x}^{\text{(geo-ecl)}}\, ,\text{ with } \mathcal{R}=\begin{pmatrix} 1&0&0\\0&\cos \epsilon &-\sin \epsilon \\ 0 &\sin \epsilon & \cos \epsilon \end{pmatrix}\, , 
\end{align}
where $\epsilon = 23.4393^{\circ}-0.0130^{\circ}T_{\text{J2000}}$ is the obliquity or axial tilt of the ecliptic.

\paragraph{(equat)$\longleftrightarrow$(gal):}
The equatorial frame at J2000.0 can be related to the galactic frame,
\begin{subequations}
\begin{align}
	\mathbf{x}^{\text{(gal)}} &= \mathcal{M}\mathbf{x}^{\text{(equat)}}(J2000.0)\, ,
	\intertext{with}
	\mathcal{M}_{11}&=-\sin l_{\text{CP}} \sin \alpha_{\text{GP}} - \cos l_{\text{CP}}\cos \alpha_{\text{GP}} \sin\delta_{\text{GP}} \, ,\\
	\mathcal{M}_{12}&=\sin l_{\text{CP}} \cos \alpha_{\text{GP}}- \cos l_{\text{CP}}\sin \alpha_{\text{GP}}\sin \delta_{\text{GP}} \, ,\\
	\mathcal{M}_{13}&=\cos l_{\text{CP}} \cos \delta_{\text{GP}} \, ,\\
	\mathcal{M}_{21}&=\cos l_{\text{CP}} \sin \alpha_{\text{GP}}	-\sin l_{\text{CP}} \cos \alpha_{\text{GP}} \sin\delta_{\text{GP}} \, ,\\
	\mathcal{M}_{22}&=-\cos l_{\text{CP}} \cos \alpha_{\text{GP}} -\sin l_{\text{CP}}\sin \alpha_{\text{GP}} \sin \delta_{\text{GP}} \, ,\\
	\mathcal{M}_{23}&=\sin l_{\text{CP}} \cos\delta_{\text{GP}} \, ,\\
	\mathcal{M}_{31}&=\cos \alpha_{\text{GP}} \cos\delta_{\text{GP}} \, ,\\
	\mathcal{M}_{32}&=\sin \alpha_{\text{GP}}\cos\delta_{\text{GP}} \, ,\\
	\mathcal{M}_{33}&= \sin	\delta_{\text{GP}}\, .
\end{align}
The occuring angles, namely the J2000.0 right ascension of the north galactic pole $\alpha_{\text{GP}}$, the J2000.0 declination of the north galactic pole $\delta_{\text{GP}}$ and the longitude of the north celestial pole in J2000.0 galactic coordinates $l_{\text{CP}}$, are
\begin{align}
	\alpha_{\text{GP}}=192.85948^{\circ}\, ,\quad	\delta_{\text{GP}}=27.12825^{\circ}\, ,\quad	l_{\text{CP}}=122.932^{\circ}\, .
\end{align}
\end{subequations}
The rotation in equatorial coordinates from J2000.0 to any time epoch $T_{\text{J2000}}$ is performed with the next matrix,
\begin{subequations}
\begin{align}
	\mathbf{x}^{\text{(equat)}}(T_{\text{J2000}})&=\mathcal{P}\mathbf{x}^{\text{(equat)}}(J2000.0)\, ,
	\intertext{with}
	\mathcal{P}_{11}&=\cos \zeta_A \cos \theta_A \cos z_A - \sin\zeta_A \sin z_A \, ,\\
	\mathcal{P}_{12}&=-\sin\zeta_A \cos\theta_A \cos z_A- \cos\zeta_A \sin z_A \, ,\\
	\mathcal{P}_{13}&=-\sin\theta_A \cos z_A  \, ,\\
	\mathcal{P}_{21}&=\cos\zeta_A \cos\theta_A \sin z_A+ \sin\zeta_A \cos z_A \, ,\\
	\mathcal{P}_{22}&=-\sin\zeta_A \cos\theta_A \sin z_A+\cos\zeta_A \cos z_A \, ,\\
	\mathcal{P}_{23}&=-\sin\theta_A \sin z_A \, ,\\
	\mathcal{P}_{31}&=\cos\zeta_A \sin\theta_A \, ,\\
	\mathcal{P}_{32}&=-\sin\zeta_A \sin\theta_A \, ,\\
	\mathcal{P}_{33}&=\cos\theta_A \, .
\end{align}
Here the equatorial angles were used, which are time-depending and given by
\begin{align}
	\zeta_A &=2306.083227'' T_{\text{J2000}}+0.298850'' T_{\text{J2000}}^2\, ,\\
	z_A&=2306.077181'' T_{\text{J2000}}+1.092735'' T_{\text{J2000}}^2\, ,\\
	\theta_A &=2004.191903'' T_{\text{J2000}}+0.429493'' T_{\text{J2000}}^2\, .
\end{align}
\end{subequations}

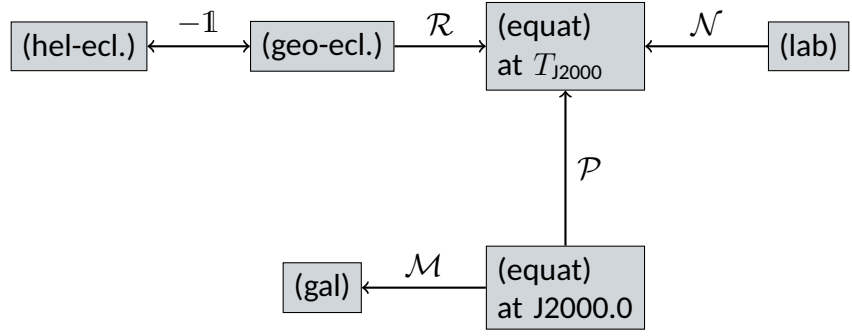
\begin{figure*}
	\centering
	\begin{tikzpicture}[scale=0.8]
			\node[draw,rectangle,fill=FlowChartBackground] (a1) at (0.0,4.0) {(hel-ecl.)};
			\node[draw,rectangle,fill=FlowChartBackground] (b1) at (4,4.0) {(geo-ecl.)};
			\node[draw,rectangle,text width=1.8cm,fill=FlowChartBackground] (c1) at (8.0,4.0) {(equat)\\ at $T_{\text{J2000}}$};
			\node[draw,rectangle,fill=FlowChartBackground] (d1) at (12.0,4.0) {(lab)};
			\node[draw,rectangle,text width=1.8cm,fill=FlowChartBackground] (c2) at (8.0,0.0) {(equat)\\ at J2000.0};
			\node[draw,rectangle,fill=FlowChartBackground] (b2) at (4,0.0) {(gal)};	
			\draw[thick,<->] (a1.east)--(b1.west)node[pos=0.5,above]{$-\mathds{1}$};
			\draw[thick,->] (b1.east) to node[pos=0.5,above]{$\mathcal{R}$}(c1.west);
			\draw[thick,->] (d1.west) to node[pos=0.5,above]{$\mathcal{N}$}(c1.east);
			\draw[thick,->] (c2.north) to node[pos=0.5,right]{$\mathcal{P}$}(c1.south);
			\draw[thick,->] (c2.west) to node[pos=0.5,above]{$\mathcal{M}$}(b2.east);
	\end{tikzpicture}
	\caption{Flowchart for transformations between the coordinate systems.}
	\label{fig: coordinate system flowchart}
\end{figure*}

\paragraph{Summary and examples}
Any transformation matrix from one frame to any other can be expressed as a product of the respective matrices. The flow chart in figure~\ref{fig: coordinate system flowchart} is a small helper to connect two frames, where the inverse rotation matrix is to be chosen if opposing the arrow's direction.
To rotate e.g. from equatorial to galactic coordinates for any given epoch $T_{\text{J2000}}$, the transformation reads
\begin{subequations}
\begin{align}
	\mathbf{x}^{\text{(gal)}}&=\mathcal{M} \mathcal{P}^{-1}\mathbf{x}^{\text{(equat)}}(T_{\text{J2000}})\, ,
	\intertext{similarly, from heliocentric, ecliptic coordinates to galactic coordinates,}
	\mathbf{x}^{\text{(gal)}}&=-\mathcal{M} \mathcal{P}^{-1}\mathcal{R}\mathbf{x}^{(\text{hel-ecl)}}\, .
\end{align}
\end{subequations}
This allows to express the axis vectors $\mathbf{e}_x=(1,0,0)^T$, $\mathbf{e}_y=(0,1,0)^T$ and $\mathbf{e}_z=(0,0,1)^T$ of the equatorial and heliocentric-ecliptic frame in terms of galactic coordinates,
\begin{subequations}
\label{eq:equat}
\begin{align}
	\mathbf{e}_{x,\text{equat}}^{\text{(gal)}} &= \mathcal{M} \mathcal{P}^{-1}\mathbf{e}_x\nonumber\\
	&=\begin{pmatrix}
		-0.0548763\\0.494109\\-0.867666
	\end{pmatrix}
	+ \begin{pmatrix}
		0.0242316\\0.002688\\-1.546 \cdot 10^{-6}
	\end{pmatrix}T_{\text{J2000}} +\mathcal{O}(T_{\text{J2000}}^2)\, ,\label{eq:equat1}\\
	\mathbf{e}_{y,\text{equat}}^{\text{(gal)}} &= \mathcal{M} \mathcal{P}^{-1}\mathbf{e}_y\nonumber\\
	&=\begin{pmatrix}
		-0.873436\\-0.444831\\-0.198076
	\end{pmatrix}
	+ \begin{pmatrix}
		-0.001227\\0.011049 \\-0.019401
	\end{pmatrix}T_{\text{J2000}} +\mathcal{O}(T_{\text{J2000}}^2)\, ,\label{eq:equat2}\\
	\mathbf{e}_{z,\text{equat}}^{\text{(gal)}} &= \mathcal{M} \mathcal{P}^{-1}\mathbf{e}_z\nonumber\\
	&=\begin{pmatrix}
		-0.483836\\0.746982\\0.455984
	\end{pmatrix}
	+ \begin{pmatrix}
		-0.000533 \\0.004801\\-0.008431
	\end{pmatrix}T_{\text{J2000}} +\mathcal{O}(T_{\text{J2000}}^2)\, .\label{eq:equat3}
\end{align}
\end{subequations}
In the same way, the axis vectors of the heliocentric-ecliptic frame can be evaluated to
\begin{subequations}
\label{eq:helecl}
\begin{align}
	\mathbf{e}_{x,\text{hel-ecl}}^{\text{(gal)}} &= -\mathcal{M}\mathcal{P}^{-1}\mathcal{R}\mathbf{e}_x\nonumber\\
	&=\begin{pmatrix}
	 0.054876\\-0.494109\\0.867666
	 \end{pmatrix} + 
	 \begin{pmatrix} -0.024232\\-0.002689\\1.546 \times 10^{-6}
	 \end{pmatrix}T_{\text{J2000}} +\mathcal{O}(T_{\text{J2000}}^2)\, ,\label{eq:helecl1}\\
	\mathbf{e}_{y,\text{hel-ecl}}^{\text{(gal)}} &= -\mathcal{M}\mathcal{P}^{-1}\mathcal{R}\mathbf{e}_y\nonumber\\
	&=\begin{pmatrix} 0.993821\\0.110992 \\0.000352 \end{pmatrix} + \begin{pmatrix} 0.001316\\-0.011851\\0.021267\end{pmatrix}T_{\text{J2000}} +\mathcal{O}(T_{\text{J2000}}^2)\, ,\label{eq:helecl2}\\
	\mathbf{e}_{z,\text{hel-ecl}}^{\text{(gal)}} &= -\mathcal{M}\mathcal{P}^{-1}\mathcal{R}\mathbf{e}_z\nonumber\\
	&=\begin{pmatrix} 0.096478\\-0.862286 \\-0.497147 \end{pmatrix} + \begin{pmatrix} 0.000227\\0.000015\\0.000018\end{pmatrix}T_{\text{J2000}} +\mathcal{O}(T_{\text{J2000}}^2)\, .\label{eq:helecl3}
\end{align}
\end{subequations}
Some of these matrices will also be used in the computation of the laboratory's velocity in the galactic frame, which we will review next.

\section{Velocity of the Sun, the Earth, and the Laboratory}
\label{a:velocity}

\begin{figure*}
\sbox\twosubbox{%
  \resizebox{\dimexpr.95\textwidth-1em}{!}{%
   \includegraphics[height=2cm]{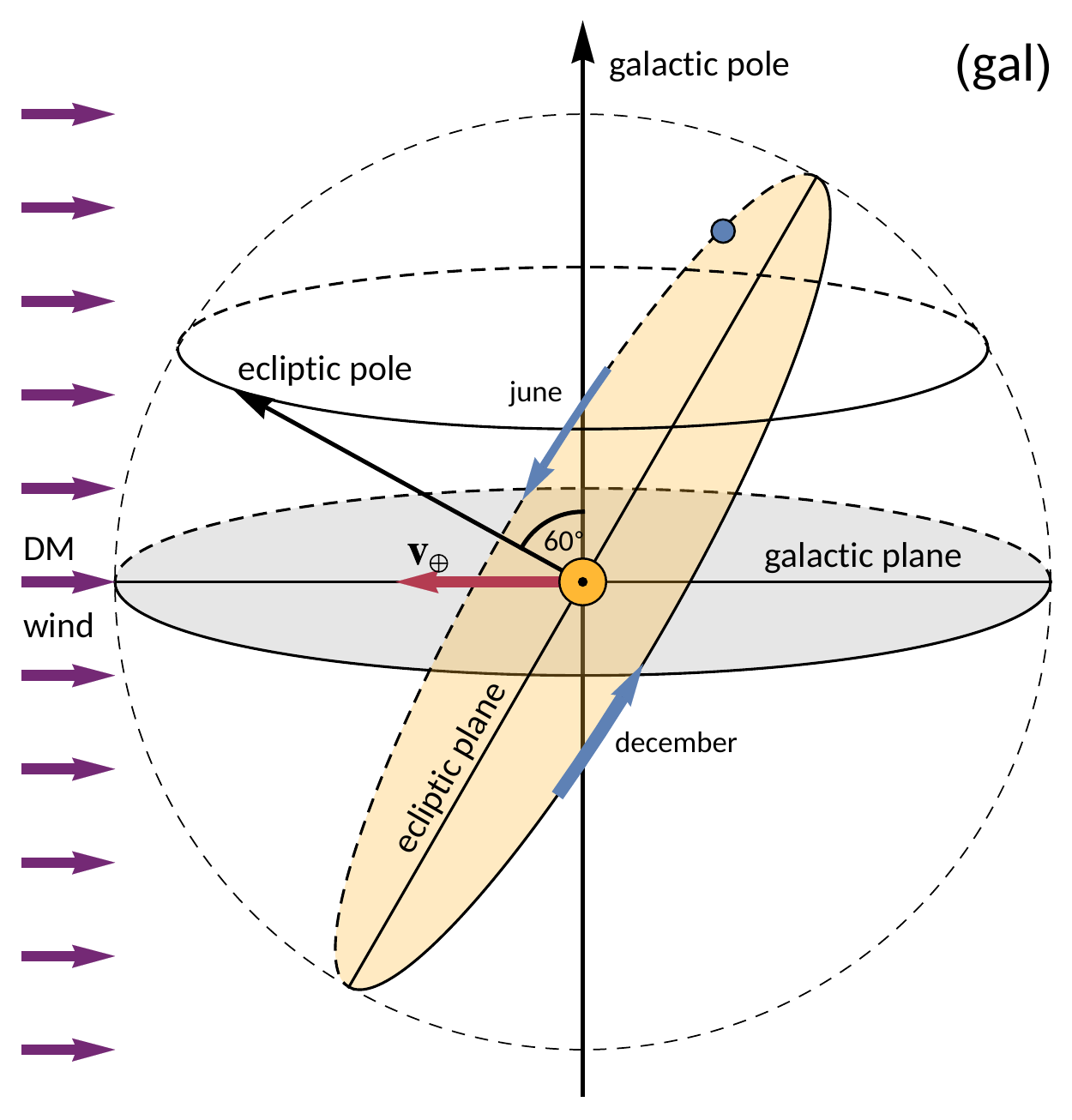}%
    \includegraphics[height=2cm]{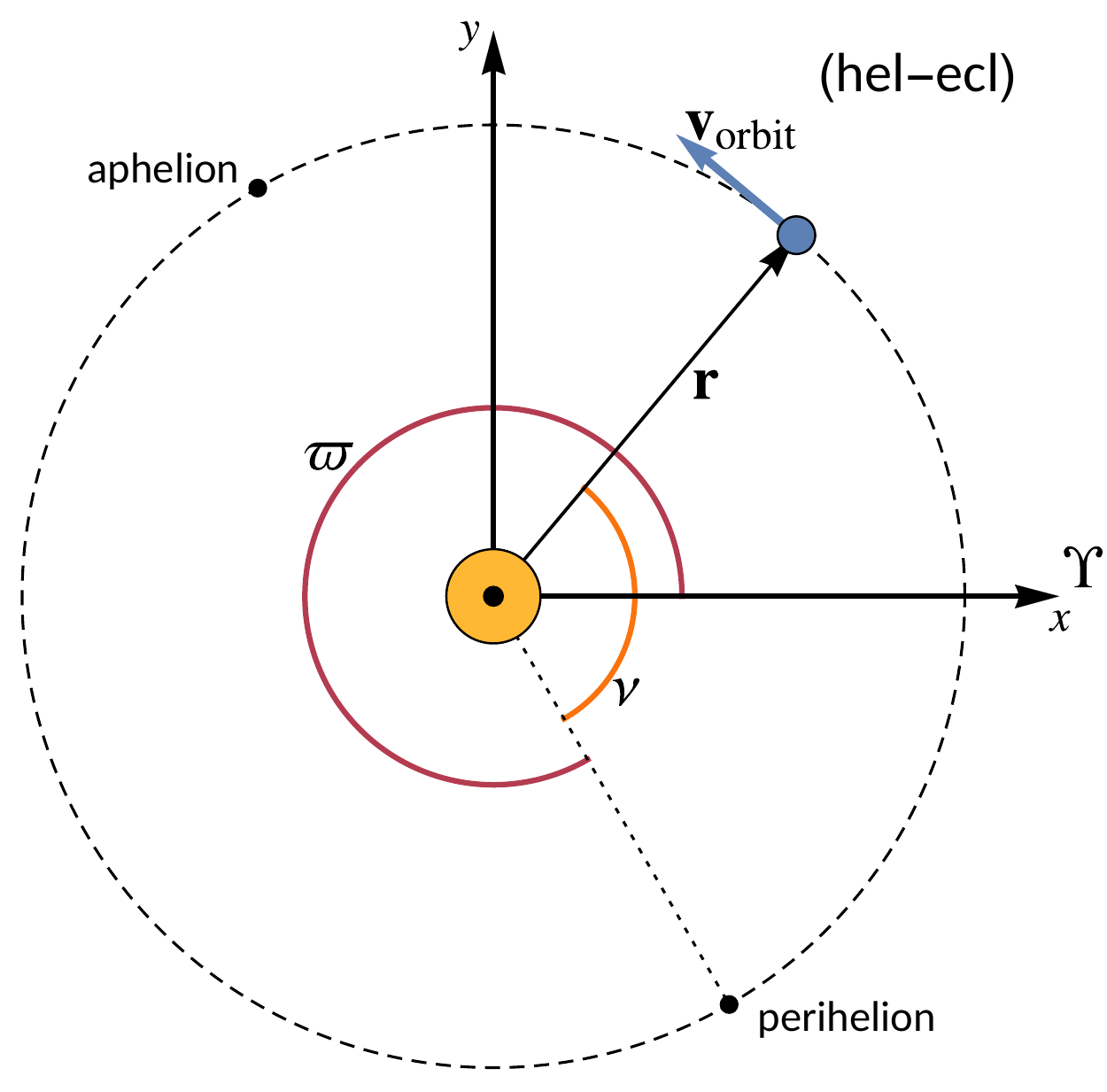}%
    \includegraphics[height=2cm]{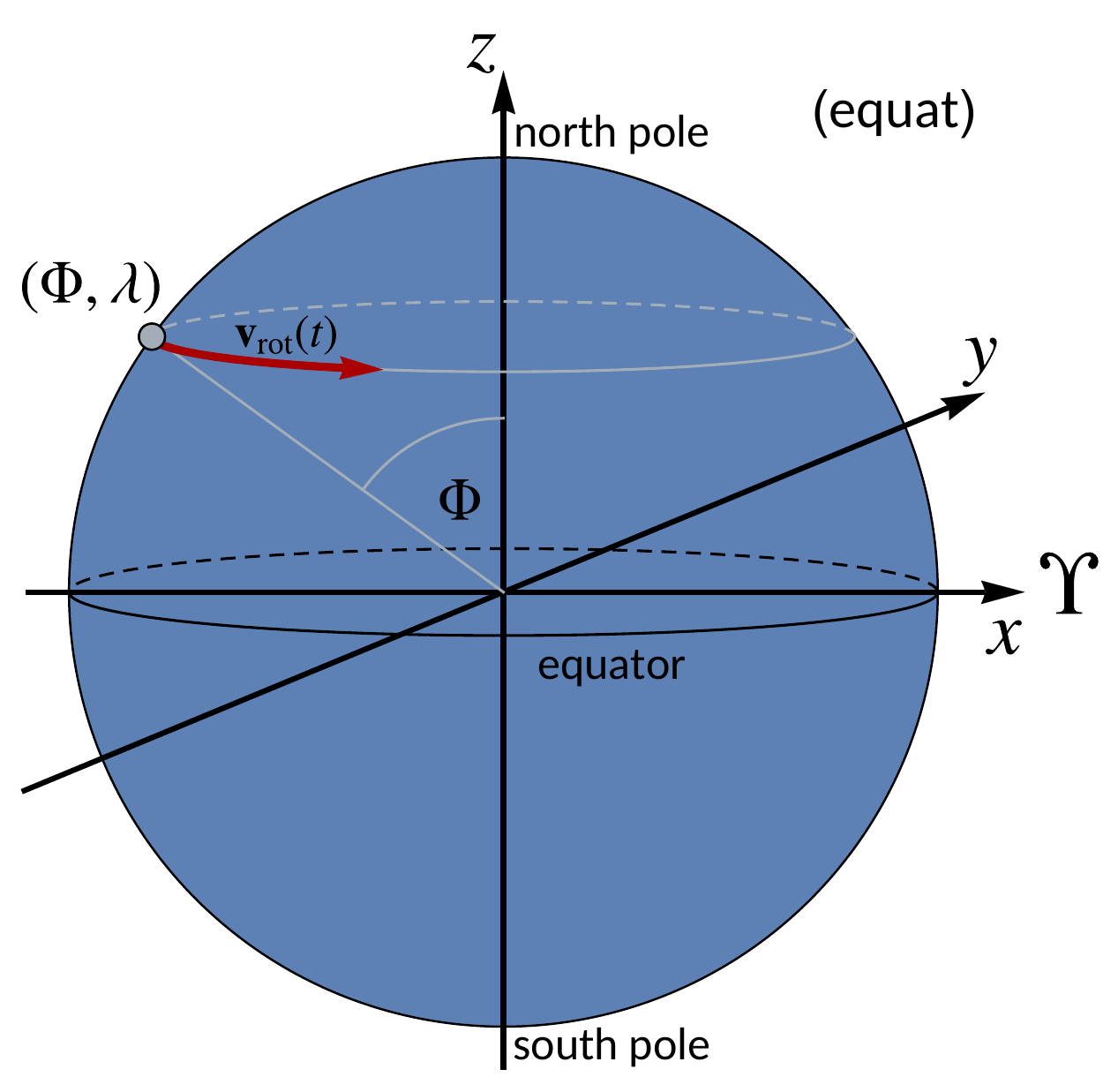}%
  }%
}
\setlength{\twosubht}{\ht\twosubbox}
   	\centering
   	\subcaptionbox{Orbital velocity in (gal)\label{fig:galacticframe}}{%
  	\includegraphics[height=\twosubht]{plots/Galactic_Frame}%
	}
	\quad
   	\subcaptionbox{Orbital velocity in (hel-ecl)\label{fig:heliocentriccoordinates}}{%
  	\includegraphics[height=\twosubht]{plots/Heliocentric_Frame}%
	}
	\quad
	\subcaptionbox{Rotational velocity in (equat)\label{fig:laboratoryframe}}{%
	  \includegraphics[height=\twosubht]{plots/Geocentric_Frame}%
	}
	\caption{The Earth's orbital velocity in the galactic and heliocentric ecliptic frame and the laboratory's rotational velocity in the equatorial frame.}
	\label{fig:coordinatesystems}
\end{figure*}

For the computation of direct detection rates one has to specify the velocity of the laboratory in the galactic frame. It is composed of the Sun's orbital velocity around the galactic center, the Sun's peculiar motion relative to neighbouring stars, the Earth's orbital velocity around the Sun, which is responsible for the annual signal modulation~\cite{Drukier:1986tm} as illustrated in figure~\ref{fig:galacticframe}, and finally the laboratory's rotational velocity around the Earth's axis. The latter is causing a similar diurnal modulation, but can be neglected in most cases.

For a long time the standard reference for the Earth's velocity in the context of direct DM searches has been the review by Smith and Lewin~\cite{Lewin:1995rx}. However, it turned out to contain an error in the first order correction related to the Earth's orbital eccentricity, which was pointed out in~\cite{Lee:2013xxa} and confirmed in~\cite{McCabe:2013kea}. This section is based on the latter.

\paragraph{The Sun}
The Sun's velocity~$\mathbf{v}_{\odot}$ in the galactic rest frame, has two components,
\begin{subequations}
\label{eq:vsun}
\begin{align}
	\mathbf{v}_{\odot} &= \mathbf{v}_r + \mathbf{v}_s\, ,
\intertext{given by the galactic rotation of the Sun,}
	\mathbf{v}_{r} &= \begin{pmatrix}0\\220\\0\end{pmatrix}\SI{}{\km\per\second}\, ,\\
\intertext{the peculiar motion of the Sun~\cite{Schoenrich:2009bx},}
	\mathbf{v}_{s} &= \begin{pmatrix}11.1\\12.2\\7.3	\end{pmatrix}\SI{}{\km\per\second}\, .
\end{align}
\end{subequations}

\paragraph{The Earth} For the Earth's velocity, we add the orbital velocity relative to the Sun, shown in figure~\ref{fig:heliocentriccoordinates}, is added,

\begin{subequations}
\label{eq:vearth}
\begin{align}
	\mathbf{v}_{\oplus}(t) &= \mathbf{v}_{\odot}  + \mathbf{v}_{\rm orbit}(t)\, ,
\intertext{with}
	\mathbf{v}_{\rm orbit}(t)&=- \langle v_{\oplus} \rangle\bigg[ \left(\sin L + e \sin (2L-\varpi)\right) \; \mathbf{e}_{x,\text{hel-ecl}}^{\text{(gal)}}\nonumber\\
	&\qquad\qquad+\left(\cos L + e \cos (2L-\varpi)\right) \; \mathbf{e}_{y,\text{hel-ecl}}^{\text{(gal)}}\bigg]\, ,
\end{align}
where the axis vectors of the heliocentric, ecliptic frame, $\mathbf{e}_{i,\text{hel-ecl}}^{\text{(gal)}}$, which are listed in eqs.~\eqref{eq:helecl}. The parameters are the mean orbital speed $\langle v_{\oplus} \rangle$, the orbit's eccentricity $e$, the mean longitude $L$, and the longitude of the perihelion $\varpi$,
\begin{align}
	\langle v_{\oplus} \rangle &= \SI{29.79}{\km\per\second} \, ,\\
	&e=\text{0.01671}\, , \\
	L&=\big[ \ang{280.460}+\ang{0.9856474}n_{\text{J2000}}\big] \text{mod}\;\ang{360}\, ,\\
	\varpi &=\big[ \ang{282.932}+\ang{0.0000471}n_{\text{J2000}}\big] \text{mod}\;\ang{360}\, .
\end{align}
\end{subequations}
The correction due to the orbit's eccentricity are mostly irrelevant for the simulation result and are only include for completeness.

\paragraph{The laboratory}
The detector's exact velocity in the galactic frame is obtained by adding the rotational velocity around the Earth axis,
\begin{align}
	\mathbf{v}_{\rm lab} = \mathbf{v}_{\oplus}(t) + \mathbf{v}_{\rm rot}(t) \, .
\end{align}
The first step is to find the spherical coordinate angles $(\theta,\phi)$ of the detector's position in the geocentric equatorial coordinate system, such that the transformation into the galactic frame is straight forward. The location of a detector on the Earth is defined via the latitude and longitude $(\Phi,\lambda)$ and the underground depth $d$ of the laboratory.

As mentioned in section~\ref{a:coordinates}, the x-axis of the equatorial coordinate system points towards the vernal equinox or March equinox. Hence, the spherical coordinate angles for a given \ac{LAST} are 
\begin{align}
\theta&= \frac{\pi}{2}-\Phi\, ,\quad	\phi(t)=\omega_{\text{rot}}\text{LAST}(\Phi,\lambda)\, .
\end{align} 
The rotation frequency is simply $\omega_{\text{rot}}=\frac{2\pi}{\SI{86400}{\second}}$. Note that these are sidereal seconds.
Finally the position vector can be transformed from the equatorial to the galactic frame,
\begin{align}
	\mathbf{x}^{\text{(gal)}}_{\text{lab}} = \mathcal{M}\mathcal{P}^{-1}\begin{pmatrix}
	(r_{\oplus}-d)\sin \theta \cos\phi\\
	(r_{\oplus}-d)\sin \theta\sin\phi\\
	(r_{\oplus}-d)\cos\theta
\end{pmatrix}\, .\label{eq:labpos}
\end{align}
Lastly, the rotational velocity follows directly,
\begin{align}
	\mathbf{v}_{\rm rot} &= \underbrace{\frac{2\pi (r_{\oplus}-d)}{T_d}}_{\equiv v_{\rm eq}}\cos \Phi\mathcal{M}\mathcal{P}^{-1}\mathbf{e}^{\text{(equat)}}_{\phi}(\mathbf{x})\nonumber\\
	&=-v_{\rm eq}\cos \Phi \bigg( \sin(\phi(t))\;\mathbf{e}_{x,\text{equat}}^{\text{(gal)}}-\cos(\phi(t))\;\mathbf{e}_{y,\text{equat}}^{\text{(gal)}} \bigg)\, .
\end{align}
The rotational speed at the equator is $v_{\rm eq}\approx \SI{0.465}{\km\per\second}$, and the axis vectors are listed in eq.~\eqref{eq:equat}. 

\section{Modelling the Earth and Sun}
\label{a:earthsun}

\subsection{The Earth}
\label{a:earth}

\begin{figure*}
\sbox\twosubbox{%
  \resizebox{\dimexpr.95\textwidth-1em}{!}{%
    \includegraphics[height=3.5cm]{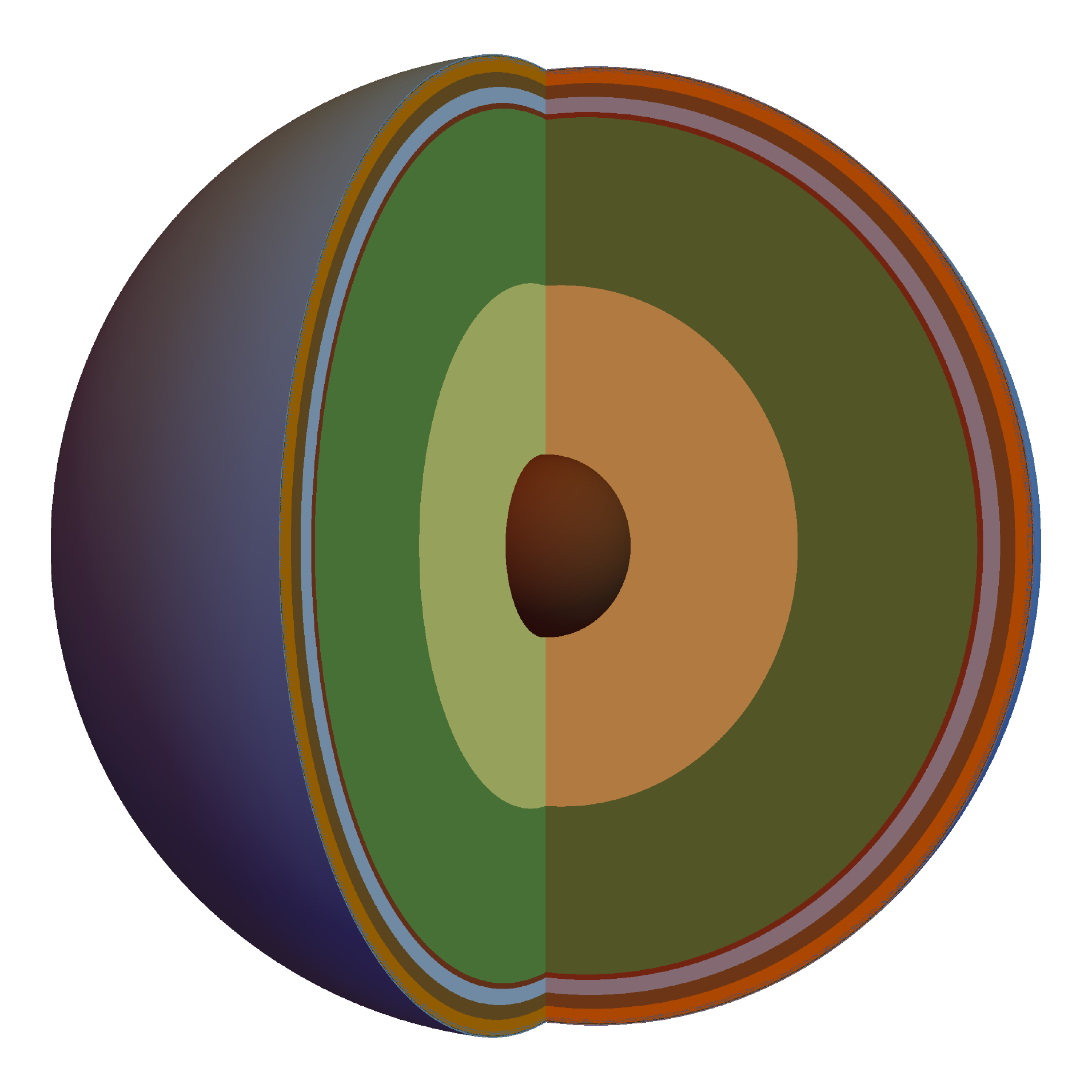}%
    \includegraphics[height=3.5cm]{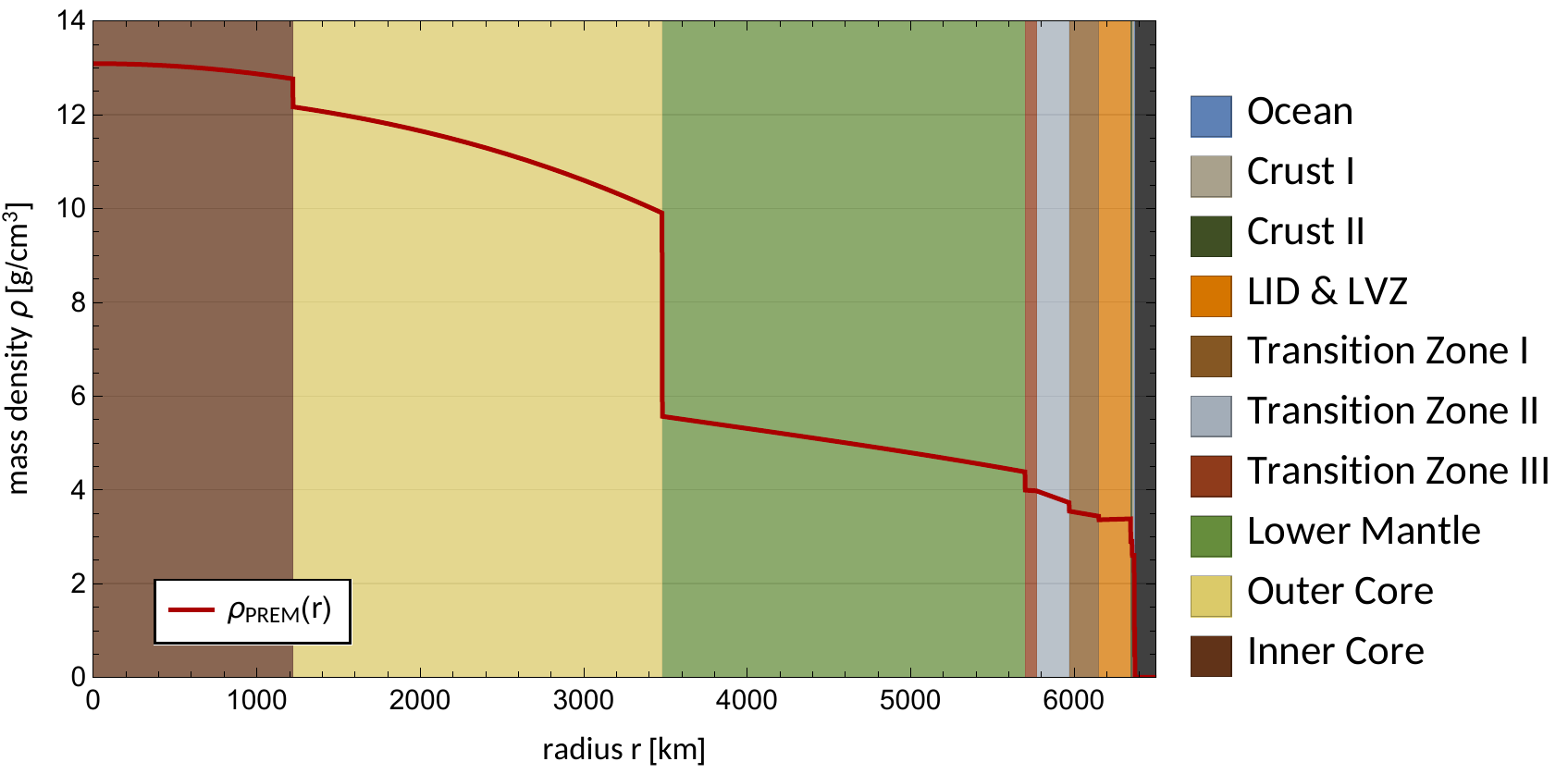}%
  }%
}
\setlength{\twosubht}{\ht\twosubbox}
   	\centering
   	\subcaptionbox{Mechanical layers.\label{fig:PREMlayers}}{%
  	\includegraphics[height=\twosubht]{plots/PREM_Layers}%
	}
	\quad
	\subcaptionbox{Mass density profile.\label{fig:PREMprofile}}{%
	  \includegraphics[height=\twosubht]{plots/PREM_Profile}%
	}
	\caption{
	The \acl{PREM}.}
	\label{fig:PREM}
\end{figure*}

\paragraph{Planetary model of the Earth} In the study of diurnal modulations, DM~particles are simulated as they traverse through the Earth and scatter on terrestrial nuclei. To model the Earth for this purpose, the planets interior structure needs to be specified, including the chemical composition and the mass density profile. We split the Earth split into two compositional layers~\cite{mcdonough2003}, namely the core with a radius of \SI{3480}{\km} and the mantle. The chemical abundances of the different nuclei are listed in table~\ref{tab:composition}.

The mass density profile of the Earth enters the computation of e.g. the mean free path of DM. A polynomial parametrization in terms of the radius~$r$ is part of the \ac{PREM}~\cite{Dziewonski:1981xy}, where the Earth's bulk is split into ten mechanical layers, as shown in figure~\ref{fig:PREMlayers}. In each of these layers the density is given by
\begin{align}
\rho_{\oplus}(r) = a_l +b_l x + c_l x^2 + d_l x^3 \, , \quad \text{where }x\equiv \frac{r}{r_{\oplus}}\, . 	\label{eq:densityprofile}
\end{align}
The ten layers are listed in table~\ref{tab:PREMdensitycoefficients}, together with their dimension and density coefficients. We plot the density in the right panel of figure~\ref{fig:PREMprofile}.

\begin{table*}
    	\centering
		\begin{tabular}{lllllll}
		\hline
		$l$		&Layer		&Depth [km]		&$a_l$ 	&$b_l$ 	&$c_l$ 		&$d_l$ 	\\
		\hline
		0	&Inner Core			&0 -- 1221.5		&13.0885	&0			&-8.8381		&0			\\
		1	&Outer Core			&1221.5 -- 3480		&12.5815	&-1.2638	&-3.6426		&-5.5281	\\
		2	&Lower Mantle		&3480 -- 5701		&7.9565		&-6.4761	&5.5283			&-3.0807	\\
		3	&TZ I	&5701 -- 5771		&5.3197		&-1.4836	&0				&0			\\
		4	&TZ II	&5771 -- 5971		&11.2494	&-8.0298	&0				&0			\\
		5	&TZ III	&5971 -- 6151	&7.1089		&-3.8045	&0				&0			\\
		6	&LVZ\& LID			&6151 -- 6346.6		&2.6910		&0.6924		&0				&0			\\
		7	&Crust I			&6346.6 -- 6356		&2.9		&0			&0				&0			\\
		8	&Crust II			&6356 -- 6368		&2.6		&0			&0				&0			\\
		9	&Ocean				&6368 -- 6371		&1.020		&0			&0				&0			\\
		10	&Space				& $>$ 6371			&0			&0			&0				&0			\\
		\hline
		\end{tabular}
		\caption{The density profile coefficients in~\SI{}{\gram\per\cm\cubed} and radii of the mechanical layers in the \acs{PREM}.}
		\label{tab:PREMdensitycoefficients}
\end{table*}

\paragraph{Earth crust and atmosphere}

\begin{table*}
	\centering
	\begin{tabular}{lllll}
	\hline
	Element				&Core	&Mantle	&Crust	&Atmosphere	\\
	&\cite{mcdonough2003} &\cite{mcdonough2003}&\cite{Rudnick20031}	&\cite{atmosphere}\\
	\hline
	\isotope{H}{1}		&0.06			&0.01			&--				&--\\
	\isotope{C}{12}		&0.2			&0.01			&--				&0.02\\
	\isotope{N}{14}		&--				&--				&--				&75.6\\
	\isotope{O}{16}		&--				&44				&46.6			&23.1				\\
	\isotope{Ne}{20}	&--				&--				&--				&0.001\\
	\isotope{Na}{23}	&--				&0.27			&2.6			&--\\
	\isotope{Mg}{24}	&--				&22.8			&2.1			&--				\\
	\isotope{Al}{27}	&--				&2.35			&8.1			&--\\
	\isotope{Si}{28}	&6				&21				&27.7			&--				\\
	\isotope{P}{31}		&0.2			&0.009			&--				&--\\
	\isotope{S}{32}		&1.9			&0.03			&--				&--\\
	\isotope{Ar}{40}	&--				&--				&--				&1.3\\
	\isotope{K}{39}	&--				&--				&2.8			&--\\
	\isotope{Ca}{40}	&--				&2.53			&3.6			&--\\
	\isotope{Cr}{52}	&0.9			&0.26			&--				&--\\
	\isotope{Mn}{55}	&0.3			&0.1			&--				&--\\
	\isotope{Fe}{56}	&85.5			&6.26			&5.0			&--				\\
	\isotope{Ni}{58}	&5.2			&0.2			&--				&--\\	
	\hline
	\textbf{Total}			&\textbf{100.26}		&\textbf{99.83}		&\textbf{98.5}	&\textbf{100}\\		
	\hline
	\end{tabular}
	\caption{Relative element abundances in wt\% of the Earth core, mantle,crust, and atmosphere.}
	\label{tab:composition}
\end{table*}

\begin{figure*}
\centering
		\includegraphics[width=0.67\textwidth]{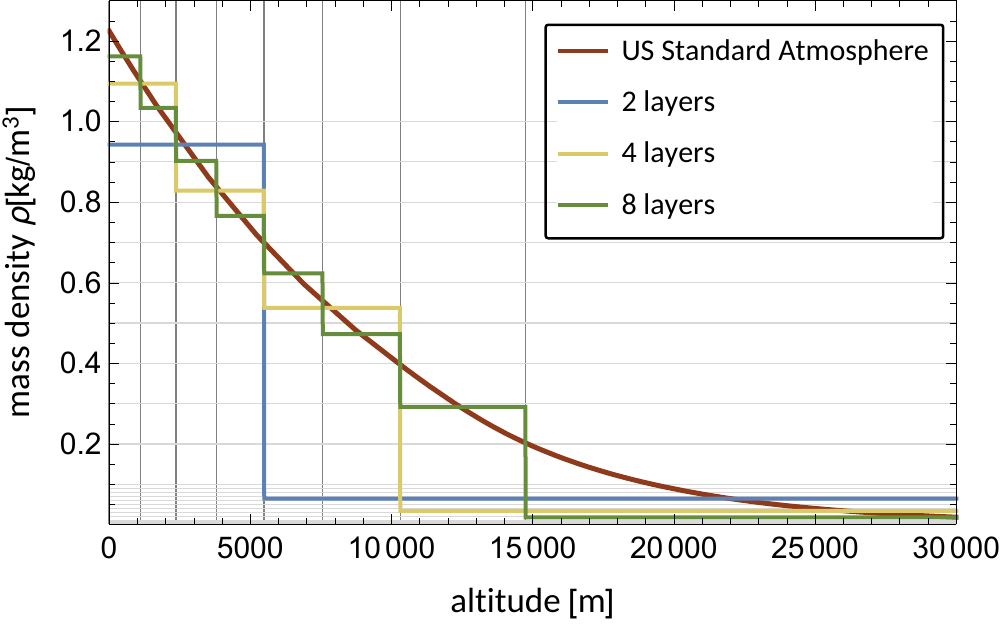}	
		\caption{Density profile of the US Standard Atmosphere and different layer approximations for the MC~simulations.}
		\label{fig:usatmosphere}
\end{figure*}

When describing the shielding effect of an experiment's overburden, the most effective DM stopping is caused by the Earth crust for underground and the atmosphere for surface laboratories respectively. In this context, where the interaction cross section is very high, there is no need to model the entire planet. The simulated geometry an be simplified to a set of parallel layers above an experiment. Hence, the Earth crust is regarded as a layer of constant mass density $\rho=\SI{2.7}{\gram\per\cm\cubed}$, consisting of nuclei, whose abundances are given in table~\ref{tab:composition}~\cite{Rudnick20031}. 

In order to account for the atmosphere's shielding effect, the US Standard Atmosphere~(1976)~\cite{atmosphere} is implemented, which extends to an altitude of 86km. The composition is also listed in table~\ref{tab:composition}. The atmospheric density decreases with altitude as shown in figure~\ref{fig:usatmosphere}. It is easier for the \ac{MC}~simulations to simulate trajectories through layers of constant density. The atmosphere is therefore split into a set of parallel layers of constant density, such that the integral $\int\rho(x)\dd x$ yields the same value for each layer. This way, each layer has a similar stopping power, as visible in figure~\ref{fig:usatmosphere} for different numbers of layers. Typically, a set of four layers is used to model the atmosphere. This discretization has been checked, the results are stable under variation of the number of layers.

\clearpage
\subsection{The Sun}
\label{a:sun}
Similarly, the Sun needs to be modelled in order to describe DM~particles as they fall into the Sun's gravitational well and possibly scatter on hot solar nuclei. The solar model used in this thesis is the \acf{SSM} AGSS09~\cite{Serenelli:2009yc}. It provides the mass-radius relation $M(r)$, the temperature $T(r)$, the mass density $\rho(r)$, and the mass fraction $f_i(r)$, and number densities $n_i(r)$ of 29 solar isotopes,
\begin{align}
	n_i(r) = \frac{f_i(r)\rho(r)}{m_i}\, ,
\end{align} 
which are required for the calculation of the local mean free path, or rather the collision frequency, inside the star. Furthermore, the local escape velocity is given by
\begin{align}
	v_{\rm esc}^2(r) =\frac{2G_NM_\odot}{R_{\odot}}\left[1+\frac{R_\odot}{M_\odot}\int\limits_{r}^{R_\odot}\dd r^\prime\;\frac{M(r^\prime)}{r^{\prime 2}}\right] \, , \label{eq: sun escape velocity}
\end{align}
which simplifies to $v_{\rm esc}^2(r) =\frac{2G_NM_\odot}{r}$ outside the Sun. 
\begin{figure*}
\sbox\twosubbox{%
  \resizebox{\dimexpr.98\textwidth-1em}{!}{%
    \includegraphics[height=3.5cm]{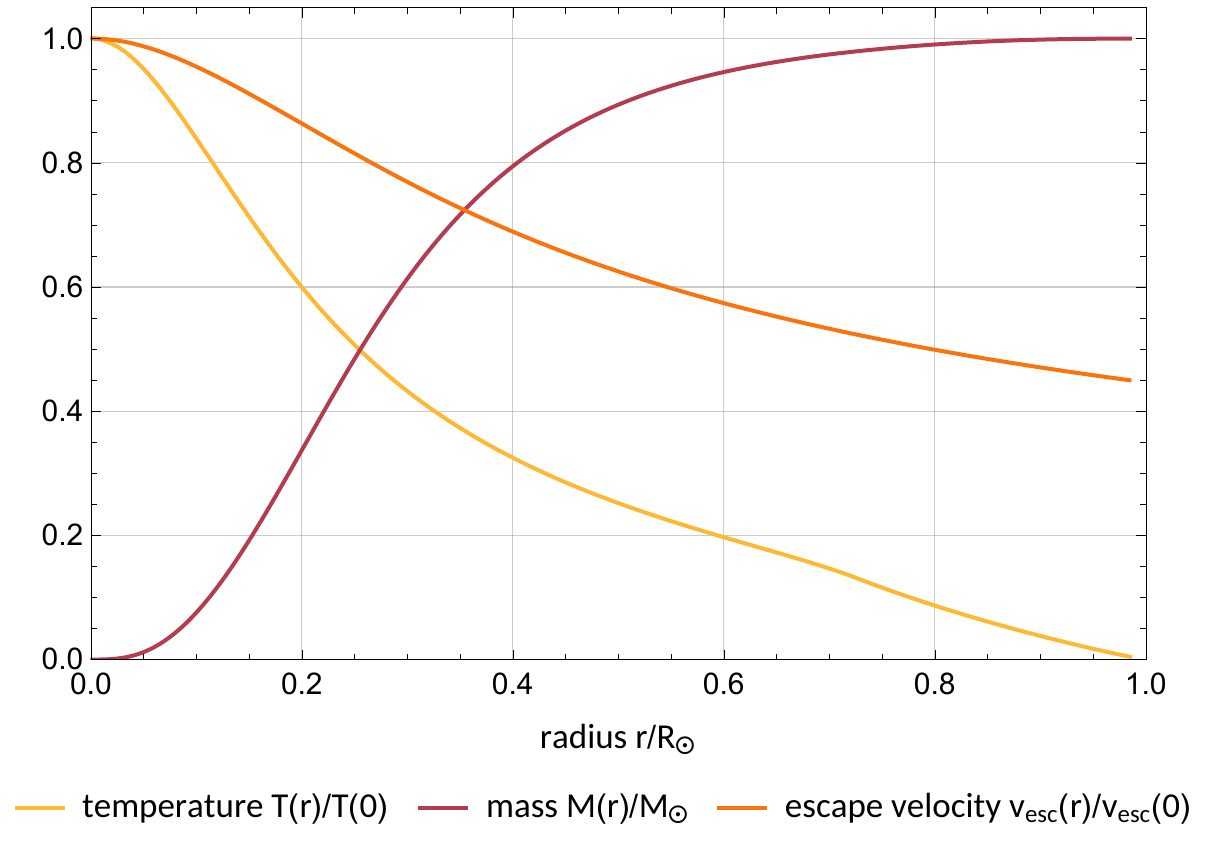}%
    \includegraphics[height=3.5cm]{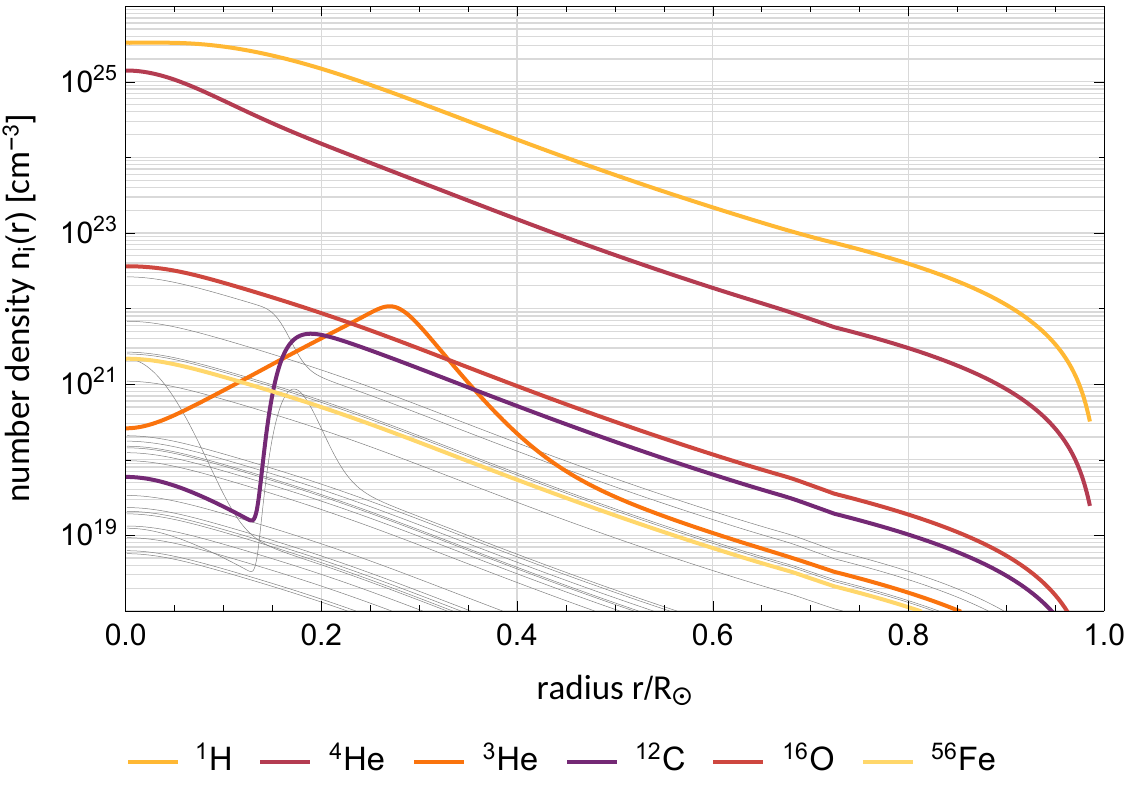}%
  }%
}
\setlength{\twosubht}{\ht\twosubbox}
   	\centering
   	\subcaptionbox{Solar mass-radius relation, temperature, and escape velocity.\label{fig:solarmodel-a}}{%
  	\includegraphics[height=\twosubht]{plots/SolarModel_1.pdf}%
	}
	\quad
	\subcaptionbox{Number density of the solar isotopes.\label{fig:solarmodel-b}}{%
	  \includegraphics[height=\twosubht]{plots/SolarModel_2.pdf}%
	}
	\caption{The solar model AGSS09.}
	\label{fig:solarmodel}
\end{figure*}

The solar mass-radius relation, temperature, and local escape velocity are shown in figure~\ref{fig:solarmodel-a}. The core escape velocity and temperature are~$v_{\rm esc}(0)\approx\SI{1385}{\km\per\second}$, and~$T(0)=\SI{1.549e7}{\kelvin}$ respectively. Furthermore figure~\ref{fig:solarmodel-b} shows the number density of the abundant isotopes.

\clearpage
\chapter{Direct Detection Experiments}
\label{a:experiments}
In chapter~\ref{c:earth} and~\ref{c:sun} of this thesis, we compute detection recoil spectra and constraints for a number of direct detection experiments, probing both DM-nucleus and DM-electron interactions. In this appendix, we summarize the detectors' details, required for various computations. In addition, the essential parameter for nuclear recoil experiments are listed in table~\ref{tab: summary nuclear}.

\section{DM-Nucleus Scattering Experiments}
\label{a:nucleus}

\subsection{CRESST-II}
\label{ss: CRESST-II}
The CRESST-II~\cite{Angloher:2015ewa} experiment was located at~\ac{LNGS} under 1400m of rock. Its phase~2 set leading constraints down to~500~MeV. The target of the `Lise' module consisted of~300~g of~$\text{CaWO}_4$ crystals. Its recoil threshold was $E_{\rm thr} =\SI{307}{\eV}$, and the energy resolution was~$\sigma_E=\SI{62}{\eV}$. With an exposure of~52.15~kg~days, the predicted signal spectrum can be computed via eq.\eqref{eq: dRdE},
\begin{equation}
\frac{\dd R}{\dd E_D} =\int_{E_R^{\rm min}}^\infty\dd E_R\,\epsilon_i(E_D)\mathrm{Gauss}(E_D|E_R,\sigma_E)\sum_{i}f_i \frac{\dd R_i}{\dd E_R} \, ,
\end{equation}
where~$i$ runs over (O, Ca, W) and $E_R^{\rm min}=E_{\rm thr}-2\sigma_E$. In the acceptance region of~$[E_{\rm thr},\SI{40}{\keV}]$ a number of 1949~events survived all cuts, and the energy data was released in~\cite{Angloher:2017zkf} together with the cut survival probability~$\epsilon_i(E_D)$ of each target nucleus. We compute the constraints using Yellin's maximum gap method~\cite{Yellin:2002xd} which, despite the fact that the collaboration used Yellin's optimum interval method, reproduces the 90\%~CL constraints on low-mass~DM to a very good accuracy. 
 
\subsection{CRESST-III}
\label{ss: CRESST-III}
The successor of~CRESST-II uses the same target at the same location. In the context of solar reflection, we regarded a CRESST-III type detector. The CRESST-III phase 2 expects an exposure of 1~ton~day according to~\cite{Angloher:2017zkf}. We assume an energy threshold of~$E_{\rm thr}=\SI{100}{\eV}$, an energy resolution of~$\sigma_E=E_{\rm thr}/5$, and efficiencies~$\epsilon(E_D)$ provided by the CRESST collaboration. The computation of the spectrum does not differ from CRESST-II.

First results of CRESST-III's phase 1 have been published in 2017~\cite{Petricca:2017zdp}. They presented new, leading exclusion limits on light~DM based on nuclear recoils down to a mass of~350~MeV. The exposure was reported as~2.39~kg~days, for which~33 events were detected in the acceptance region. We do not include results for~CRESST-III in chapter~\ref{s:simp constraints}, as the recoil data has not been published yet. Due to its similarity to CRESST-II, it will be straightforward to implement after the data release.

\subsection{CRESST surface run (2017)}
\label{ss: CSR}

The surface run of a prototype detector for the~$\nu$-cleus detector was used by the CRESST collaboration in 2017~\cite{Angloher:2017sxg}. It consisted of a small, 0.49g sapphire~($\text{Al}_{\text{2}}\text{O}_{\text{3}}$) target. Despite a tiny net exposure of~0.046~g~days, it was ideal for probing strongly interacting~DM, because it was set up on the surface in a building of the Max Planck Institute in Munich, with only 30~cm of concrete ceiling as shield. The energy threshold and resolution were remarkably low with $E_{\rm thr}\approx \SI{19.7}{\eV}$ and $\sigma_E\approx 3.74$~eV. Within the region of interest~$[E_{\rm thr},600\text{ eV}]$, 511 events were observed. The cut efficiency was set to 1 as a conservative measure. The energy data was released as ancillary files on arXiv. Apart from the different targets, the computation of the spectrum does not differ from CRESST-II or III, and the official constraints can be reproduced using Yellin's maximum gap method with only small deviations.

\subsection{DAMIC}
\label{ss: DAMIC}

The DAMIC uses CCDs of silicon as a target to probe low-mass~DM~\cite{Barreto:2011zu}. In 2011, an engineering run of a 0.5g target was used to derive first constraints. These limits constrain strongly interacting~DM more than later bigger runs, as it was located at a relatively shallow underground site at Fermilab with a underground depth of~350'($\approx\SI{107}{\meter}$). Additionally, a lead shield of 6"($\approx\SI{15}{\cm}$) thickness was installed. The main reason to include~DAMIC in our results, even though its constraints are covered by other experiments, is the fact that we can directly compare to DAMIC constraints obtained by Mahdawi and Farrar using very similar simulations~\cite{Mahdawi:2017cxz}. This is also the reason, why we follow them in their analysis.

DAMIC's exposure was reported as~0.107~kg~days, and its recoil threshold was $E_{\rm thr}=\text{0.04 keV}_{\rm ee}$. Overall, 106 events were observed in the region of interest [$E_{\rm thr}$, 2$\text{ keV}_{\rm ee}$] (in terms of nuclear recoils~[0.55,7]keV). In order to have comparable results with~\cite{Mahdawi:2017cxz}, we calculate the 90\%~CL constraints using Poisson likelihoods.

\subsection{XENON1T}
\label{ss: XENON1T}
The first results by XENON1T, which is located at the~\ac{LNGS}, were presented in May 2017~\cite{Aprile:2017iyp}. The experiment with a liquid xenon target had reported exposure of 35.6~ton~days, consistent with the background-only hypothesis. The threshold of XENON1T is~$E_{\rm thr} =\SI{5}{\keV}$, and the region of interest covers the interval~$[E_{\rm thr},\SI{40}{\keV}]$. We used a simplified recoil spectrum to compute the constraints based on eq.~\eqref{eq: nuclear recoil spectrum speed} and neglected other detector effects apart from a flat~82\% efficiency.

\begin{table*}
\centering
\begin{tabular}{llllll}
		\hline
		Experiment	&	Target	&	Shielding	& $E_{\rm thr}$ [eV]		&$\sigma_E$[eV]	&Exposure \\
		\hline
		CRESST-II	&$\text{CaWO}_4$	&d=1400m	& 307	&62	& 52.15~kg~days\\
		\cite{Angloher:2015ewa,Angloher:2017zkf}&&&&&\\
		CRESST-III	&$\text{CaWO}_4$	&d=1400m	&	100	&20	&	$\sim$1~ton~day	\\
		\cite{Angloher:2017zkf,Petricca:2017zdp}&&&&&\\
		CRESST 2017	&$\text{Al}_2\text{O}_3$	&atmosphere	&19.7	&	3.74	&0.046~g~days	\\
		 surface run	&&30cm concrete&&&\\
		 \cite{Angloher:2017sxg}&&&&&\\
		DAMIC&Si CCDs&d=106.7m&550&--&0.107~kg~days\\
			\cite{Barreto:2011zu}&&15cm of lead&&&\\
		XENON1T&Xe&d=1400m&5000&--&35.6~ton~days\\
		\cite{Aprile:2017iyp}&&&&&\\
		\hline
	\end{tabular}
	\caption{Summary of direct detection experiments based on nuclear recoils included in our analyses.}
	\label{tab: summary nuclear}
\end{table*}
\section{DM-Electron Scattering Experiments}
\label{a:electron}

\subsection{XENON10 and XENON100}
\label{ss: XENON10 and 100}
Both XENON10 and XENON100 have used the observation of~`S2-only' events for the search of light~DM~\cite{Angle:2011th,Aprile:2016wwo}. This data has been used to set constraints on DM-electron interactions~\cite{Essig:2012yx,Essig:2017kqs}. The equations necessary to describe the ionization rate of xenon atoms due to DM~particles have been reviewed in the main body of the thesis, in chapter~\ref{ss: direct detection electron}. The necessary information about the atomic shells included in the analysis is summarized in table~\ref{tab: xenon shells}. This table includes the number of secondary quanta for each shell, necessary to convert the spectrum from the primary electron's energy to the total number of ionized electrons.

\begin{table}[h!]
\centering
	\begin{tabular}{lll}
	\hline
		atomic shell		&$E_B^{nl}$ [eV]	&$n^{(2)}$\\
		\hline
		$\text{5p}^{\text{6}}$	&12.4	&0\\
		$\text{5s}^{\text{2}}$	&25.7	&0\\
		$\text{4d}^{\text{10}}$	&75.6	&4\\
		$\text{4p}^{\text{6}}$	&163.5	&6-10\\
		$\text{4s}^{\text{2}}$	&213.8	&3-15\\
		\hline
	\end{tabular}
	\caption{Binding energy~$E_B^{nl}$ and number of secondary quanta~$n^{(2)}$ of xenon's atomic shells. For the limits, the lowest number of secondary quanta is used.}
	\label{tab: xenon shells}
\end{table}

The detector specific parameters, e.g. the exposure and detector resolution, for XENON10 and XENON100 can be found in table~\ref{tab: XENON10&100}. These parameters are necessary to convert the spectra further in terms of the actual observable, the PEs, as described by eq.~\eqref{eq: PE spectrum}. Both experiments were located at the~\ac{LNGS}.

\begin{table}[h!]
\centering
	\begin{tabular}{lll}
	\hline
						&\textbf{XENON10}		&\textbf{XENON100}		\\
	\hline
	exposure			&15 kg days	&30 kg years	\\
	$\mu_{\rm PE}$		&	27		&19.7			\\
	$\sigma_{\rm PE}$	&	6.7		&6.2			\\
	$n_{PE}^{\rm min}$	&	14		&80				\\
	bin width			&27			&20				\\
	underground depth	&1400m		&1400m			\\
	\hline		
	\end{tabular}
	\caption{Parameters for the analysis of XENON10 and XENON100.}
	\label{tab: XENON10&100}
\end{table}

For XENON10, we assume a flat cut efficiency of~92\%~\cite{Angle:2011th}, which needs to be multiplied by the trigger efficiency in order to obtain the total efficiency. The trigger efficiency is given in figure~1 of~\cite{Essig:2012yx}. The corresponding acceptance and trigger efficiency for XENON100 can be found in figure~3 of~\cite{Aprile:2016wwo}.

\begin{table}
	\centering
	\begin{tabular}{llcll}
	\textbf{XENON10}			&			&\phantom{space}&\textbf{XENON100}	&			\\
	\cline{1-2}\cline{4-5}
	bin~[PE]				&events		&&bin~[PE]				&events		\\
	\cline{1-2}\cline{4-5}
	$\text{[}$14,41)			&126			&&$\text{[}$80,90)		&794			\\
	$\text{[}$41,68)			&60			&&$\text{[}$90,110)		&1218			\\
	$\text{[}$68,95)			&12			&&$\text{[}$110,130)	&924			\\
	$\text{[}$95,122)		&3			&&$\text{[}$130,150)	&776			\\
	$\text{[}$122,149)		&2			&&$\text{[}$150,170)	&669			\\
	$\text{[}$149,176)		&0			&&$\text{[}$170,190)	&630			\\
	$\text{[}$176,203)		&2			&&$\text{[}$190,210)	&528			\\
	\cline{1-2}\cline{4-5}
	\end{tabular}
	\caption{Binned data for XENON10 and XENON100.}
	\label{tab: XENON10&100 data}
\end{table}
Finally, the signal bins and event numbers for both experiments are listed in table~\ref{tab: XENON10&100 data}. For a given DM~mass, we find the upper limit on the cross section using Poisson statistics independently for each bin. The lowest of these values sets the overall limit. For XENON100 we only use the first three bins.
	
\subsection{SENSEI~(2018) and SuperCDMS~(2018)}
\label{ss: SENSEI and SuperCDMS}
The recent results by the SENSEI and SuperCDMS collaboration have a number of things in common~\cite{Crisler:2018gci,Agnese:2018col}. They use a silicon semiconductor target, presented results from a surface run in 2018 and are sensitive to single electronic excitations. The formalism necessary to compute signal spectra and numbers can be found in chapter~\ref{ss: direct detection electron}. 

SENSEI had the lower exposure with~0.07~gram$\times$456~min~$\approx$~0.02~gram~days. The observed events, taken from table~I of~\cite{Crisler:2018gci}, are also listed on the left side of table~\ref{tab: signals sensei supercdms}. The SENSEI surface run took place at the Silicon Detector Facility at Fermilab, shielded only by a few cm of concrete, aluminium, wood and cooper. These layers can safely be neglected, and the atmosphere is included as the only overburden.

\begin{table}[h!]
\centering
\begin{tabular}{lllcll}
			&\textbf{SENSEI}	&&\phantom{spa}&\textbf{SuperCDMS}	&				\\
	\cline{2-3}\cline{5-6}
	$n_e$\phantom{xxx}	&efficiency		&events		&&efficiency		&events		\\
	\cline{2-3}\cline{5-6}
	1				&0.668			&140302		&&0.88			&$\sim$53000	\\	
	2				&0.41			&4676		&&0.91			&$\sim$400	\\	
	3				&0.32			&131		&&0.91			&$\sim$74	\\	
	4				&0.27			&1			&&0.91			&$\sim$18	\\	
	5				&0.24			&0			&&0.91			&$\sim$7	\\	
	6				&--			&--				&&0.91			&$\sim$14	\\
	\cline{2-3}\cline{5-6}
\end{tabular}
	\caption{Efficiencies and observed signals in the electron bins for SENSEI~(2018) and SuperCDMS~(2018).}
	\label{tab: signals sensei supercdms}
\end{table}

\begin{table}[h!]
\centering
	\begin{tabular}{ll}
	\hline
		Element			&$f_i$~[wt\%]\\
	\hline
	\isotope{H}{1}		&0.33		\\
	\isotope{O}{16}		&52.28		\\
	\isotope{Na}{23}		&0.02		\\
	\isotope{Mg}{24}		&0.10			\\
	\isotope{Al}{27}		&0.33			\\
	\isotope{Si}{28}		&40.85			\\
	\isotope{S}{32}		&0.16			\\
	\isotope{Cl}{35}		&0.01			\\
	\isotope{K}{39}		&0.06			\\
	\isotope{Ca}{40}		&5.59			\\
	\isotope{Fe}{56}		&0.27			\\
	\hline
	\end{tabular}
	\caption{Nuclear composition of concrete in terms of the isotopes' mass fractions~$f_i$.}
	\label{tab: concrete}
\end{table}

The surface run of SuperCDMS had an exposure of~0.487~gram~days. The cut efficiency and event numbers can be found in figure~3 of~\cite{Agnese:2018col}. The event numbers, listed on the right side in table~\ref{tab: signals sensei supercdms}, were estimated from the histogram in the same figure. The SuperCDMS collaboration obtained their limits using Yellin's optimum interval method, where they removed the data more than 2$\sigma$ away from the electron peaks with a charge resolution of~$\sigma=$0.07~electron-hole pairs. We simplify the analysis and compute the constraints with simple Poisson statistics for each bin. We also take the data removal procedure into account by adding a flat efficiency factor of~$\epsilon\approx$~0.9545, corresponding to the~2$\sigma$. Regarding the shielding, we take the~60cm of concrete above the detector into account in our simulations. We model the concrete as a layer of mass density~$\rho=\SI{2.4}{\gram\per\cm\cubed}$, the composition is given in table~\ref{tab: concrete}~\cite{Piotrowski2012}.

\clearpage
\chapter{Numerical and Statistical Methods}
\label{a:numerics}
In this appendix, we review some of the methods and routines, which have been applied in the different scientific codes. The book \textit{Numerical Recipes} by Press et al. has been a great resource in this context~\cite{Press2007}.
\section{Adaptive Simpson Integration}
\label{a:integration}

In order to numerically compute the one-dimensional integral
\begin{align}
	I(f;a,b) = \int\limits_a^b\dd x\; f(x)\, ,\label{eq: integral}
\end{align}
\textit{Simpson's rule} approximates the function $f(x)$ as a second order polynomial $P(x)$ with $f(a)=P(a)$, $f(b)=P(b)$ and $f(\frac{a+b}{2})=P(\frac{a+b}{2})$ and integrates $P(x)$ analytically,
\begin{align}
	I(f;a,b)\approx S(a,b)\equiv\frac{b-a}{6}\left(f(a) + 4f\left(\frac{a+b}{2}\right)+f(b) \right)\, . \label{eq: simpson's rule}
\end{align}
Apart from a few exceptional cases, this will not be very accurate. For that reason Simpon's rule is promoted to an adaptive method~\cite{Kuncir1962}, that will compute the integral~\eqref{eq: integral} up to any tolerance $\epsilon$. If the error of $S(a,b)$ exceeds this tolerance we divide the interval $(a,b)$ into subintervals $(a,c)$ and $(c,b)$ with $c=\frac{a+b}{2}$ and apply eq.~\eqref{eq: simpson's rule} to the sub-intervals. This repeats recursively until
\begin{align}
	\left| S(a,c) + S(c,b) - S(a,b)\right| < 15 \epsilon\, ,
\end{align}
as suggested by J.N. Lyness~\cite{Lyness1969}. This method has been implemented in a recursive way, which avoids redundant evaluations of the function $f$.

\section{Interpolation with Steffen Splines}
\label{a:interpolation}
Many methods have been proposed to interpolate in between a set of number pairs $\{(x_1,y_1),...,(x_N,y_N)\}$. One particular example for one dimensional functions is the use of so-called \textit{Steffen splines}~\cite{Steffen1990}, a smooth interpolation method using third order polynomials as interpolation functions. This method guarantees continuous first derivatives and monotonic behaviour of the interpolation function and avoids relic oscillations. Local extrema can only occur directly on a data point.

\begin{figure*}
\centering
  \includegraphics[width=0.6\textwidth]{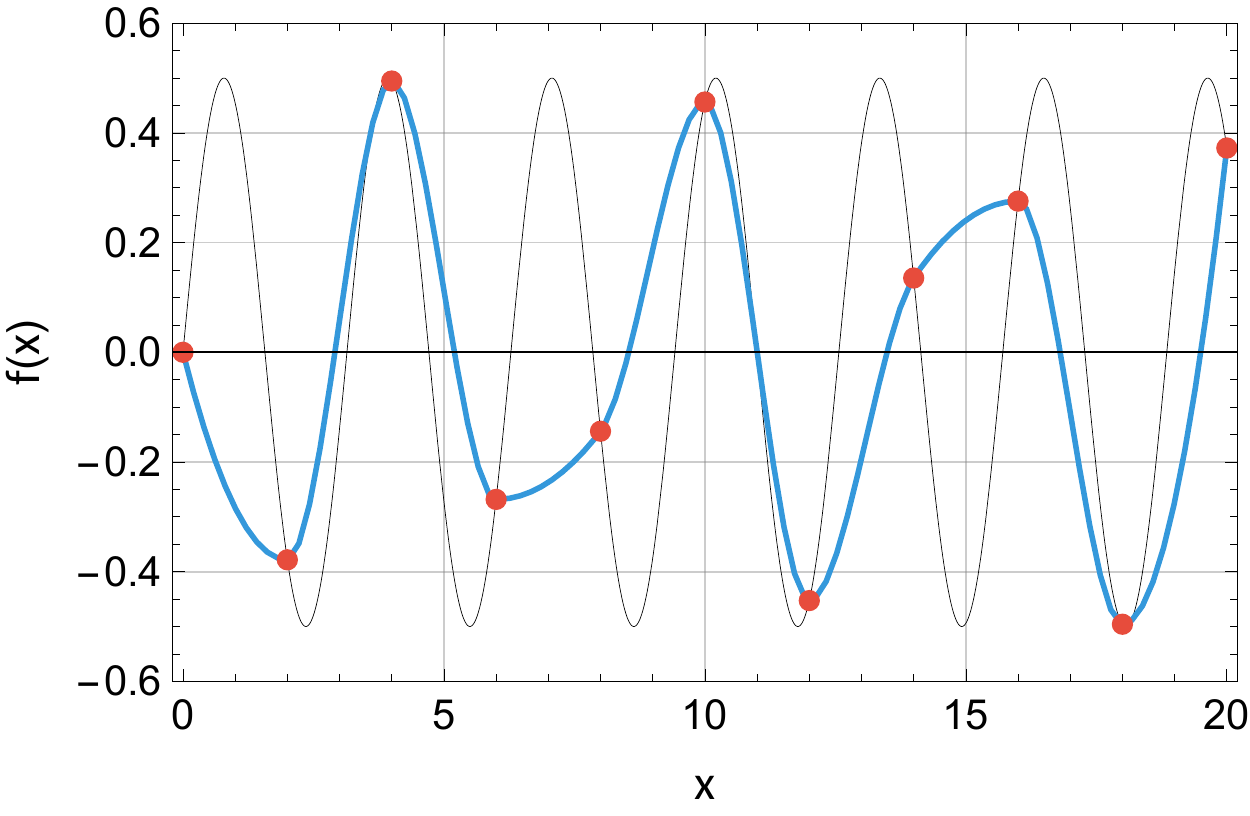}
  \caption{Example for Steffen interpolation. Ten equidistant points $(x,f(x))$ with $f(x)=\cos(x)\sin(x)$ are chosen and interpolated. Note that the extrema of the interpolation function coincide with the data points.}
  \label{fig: steffen interpolation}
\end{figure*}

In any given interval $(X_i,X_{i+1})$, the interpolation function is given by
\begin{subequations}
\begin{align}
	f_i(x) &= a_i (x-x_i)^3+b_i (x-x_i)^2+c_i (x-x_i)+d_i\, .
\intertext{The coefficients are given by}
		a_i &= \frac{y_i^\prime+ y_{i+1}^\prime-2s_i}{h_i^2}\, ,\quad  b_i = \frac{3s_i-2 y_i^\prime- y_{i+1}^\prime}{h_i}\, ,\\
		 c_i &= y_i^\prime\, ,\quad d_i = y_i\, ,
\intertext{with}
		h_i &= (x_{i+1}-x_i)\, , \quad\text{and } s_i = \frac{y_{i+1}-y_i}{h_i}\, .
\end{align}
\end{subequations}
The remaining step is to find $y_i^\prime$. For $i\in\{2,...,N-1\}$ it is given by
\begin{subequations}
\begin{align}
	y_i^\prime &= \begin{cases}
		0 &\text{if }s_{i-1}s_1\leq 0\, ,\\
		2 a \min (|s_{i-1}|,|s_{i}|)&\text{if }|p_i|>2|s_{i-1}|\text{ or } |p_i|>2|s_{i}|\, , \\
		|p_i|&\text{otherwise.}
	\end{cases}
\intertext{Here, we used}
a&=\text{sign}(s_{i-1})=\text{sign}(s_{i})\, , \quad p_i=\frac{s_{i-1}h_i+s_ih_{i-1}}{h_{i-1}+h_i}\, .
\end{align}
\end{subequations}
At the boundaries, we obtain $y^\prime$ using
\begin{subequations}
	\begin{align}
		y_1^\prime & = \begin{cases}
			0 &\text{if }p_1s_1\leq 0\, ,\\
			2s_1 &\text{if } |p_1|>2|s_1|\, ,\\
			p_1 &\text{otherwise,}
		\end{cases}
		\intertext{and}
	p_1 &= s_1 \left(1+\frac{h_1}{h_1+h_2} \right) - s_2 \left( \frac{h_1}{h_1+h_2}\right)\, .
	\end{align}
\end{subequations}
on the left and 
\begin{subequations}
	\begin{align}
		y_N^\prime & = \begin{cases}
			0 &\text{if }p_Ns_{N-1}\leq 0\, ,\\
			2s_{N-1} &\text{if } |p_N|>2|s_{N-1}|\, ,\\
			p_N &\text{otherwise,}
		\end{cases}
		\intertext{and}
	p_N &= s_{N-1} \left(1+\frac{h_{N-1}}{h_{N-1}+h_{N-2}} \right) - s_{N-2} \left( \frac{h_{N-1}}{h_{N-1}+h_{N-2}}\right)\, .
	\end{align}
\end{subequations}
on the right boundary. For further explanations and foundations of the method, we refer to the original publication. An example can be found in figure~\ref{fig: steffen interpolation}.

For the location of the argument in the tables during a function call, we implemented a combination of bisection and a hunting algorithm, which also accounts for potential correlations of subsequent function calls~\cite{Press2007}.

\section{Kernel Density Estimation}
\label{a:kde}
The \ac{MC}~simulation of DM~particles typically provides a data set of velocities or speeds, which survived e.g. an energy cut or pass a certain location on Earth. The central problem of the following data treatment is to estimate the unknown \ac{PDF}. More specifically, we assume a simulation resulting in a data set $\{x_1,...,x_N\}$ with associated weights $\{w_1,...,w_N\}$ and the underlying \ac{PDF} $f(x)$ of domain $I$, i.e. $x\in I$, is unknown. The simplest non-parametric density estimate of $f(x)$ is just the histogram of the data. Another, more sophisticated non-parametric approach is \ac{KDE}~\cite{Rosenblatt1956,Parzen1962}, which produces a continuous and smooth estimate $\hat{f}(x)$ of the true \ac{PDF},
\begin{align}
	\hat{f}_h(x) = \frac{1}{h \sum_i w_i}\sum_{i=1}^{N} w_i K\left(\frac{x-x_i}{h}\right)\, . \label{eq: kde}
\end{align}
The parameter $h$ is called the bandwidth and plays a similar role as the bin width for histograms. The function $K(x)$ is called a \emph{kernel} and satisfies
\begin{subequations}
	\begin{align}
	&K(x)\geq 0\, , \quad\text{for all } x\in I\, ,\\
	&\int_I \dd x \; K(x) = 1\, ,\\
	&\int_I \dd x \;x K(x) = 0\, .
	&\end{align}
\end{subequations}
For practical reasons, the \emph{scaled kernel} is defined as
\begin{align}
	K_h(x) &\equiv \frac{1}{h}K\left(\frac{x}{h}\right)\, ,
	\intertext{ such that}
	\hat{f}_h(x) &= \frac{1}{\sum_i w_i}\sum_{i=1}^{N} w_i K_h\left(x-x_i\right)\, .\label{eq: kde2}
\end{align}
The choice of the kernel is relatively insignificant, and a simple Gaussian kernel usually suffices,
\begin{align}
	K(x) = \frac{1}{\sqrt{2\pi}}\exp\left(\frac{-x^2}{2}\right)\, . \label{eq:gauss kernel}
\end{align}
The choice of the bandwidth $h$, being the only free parameter, is crucial, similarly to the bin width for histograms. A too large value for $h$ might smooth over interesting features of the \ac{PDF}, a value too small will resolve statistical fluctuations of the data. The bandwidth can be estimated based on the data via \textit{Silverman's rule of thumb}~\cite{silverman1986density},
\begin{align}
	h = \left(\frac{4}{3 N}\right)^{1/5}\hat\sigma\, ,
\end{align}
which takes the sample's size $N$ and standard deviation $\hat\sigma$ into account.

\begin{figure*}
	\centering
	\includegraphics[width=0.6\textwidth]{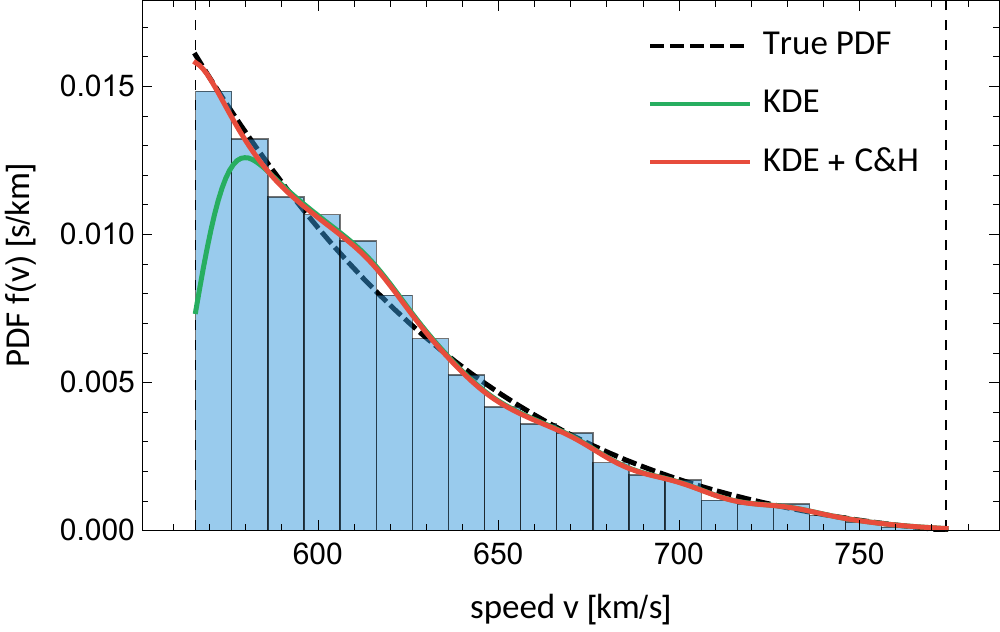}
	\caption{KDE for a bounded domain with and without the pseudo data method of Cowling and Hall (C\&H).}
	\label{fig: cowling and hall}
\end{figure*}

A well known and studied problem of \ac{KDE} occurs if the domain or support is bounded. Especially using the Gaussian kernel can be problematic and introduce significant errors since the kernel's domain is unbounded. In these cases, the \ac{KDE} estimate of eq.~\eqref{eq: kde2} underestimates the true \ac{PDF} in proximity of the domain's boundaries. The kernel does not include any knowledge of the boundary and assigns weight to the region beyond, where no data points exist. The  problem is most severe, if the true \ac{PDF} does not vanish at the domain edges. The \ac{MC}~simulations in the context of section~\ref{s:simp constraints} generate data samples with a hard speed cutoff, such that the \acp{PDF} to be estimated have a bounded domain. An example, using a data sample of 5000, generated with \ref{code2}, is shown in figure~\ref{fig: cowling and hall}, where the unmodified \ac{KDE} drops below the true \ac{PDF} close to the speed cutoff.

In order to remove this boundary bias, a plethora of methods of varying complexity has been developed and studied~\cite{Karunamuni2005}. The most simple fix is to reflect the data around the boundary. For a \ac{PDF} $f(x)$ of domain $\left[x_{\rm min},x_{\rm max}\right]$ with $f(x_{\rm min})> 0$ and a data sample $\left\{x_1,...,x_N\right\}$ with $x_{\rm min}\leq x_i \leq x_{\rm max}$, the following adjustment of eq.~\eqref{eq: kde2} can reduce the bias considerably,
\begin{align}
	\hat{f}_h(x) = \frac{1}{\sum_i w_i}\sum_{i=1}^{N} w_i \left[K_h\left(x-x_i\right)+K_h\left(x-x_i^{\rm refl}\right)\right]\, ,\label{eq: kde with reflection}
\end{align}
with the reflected pseudo data point $x_i^{\rm refl}\equiv 2x_{\rm min}-x_i$. The drawback of this easy to implement method is the fact that $\hat{f}^\prime(x_{\rm min})=0$, which introduces a new bias, unless the true \ac{PDF} shares this property. But the general idea of generating pseudo data beyond the boundary is applicable for other \ac{PDF}s as well. Cowling and Hall proposed such a pseudo data procedure~\cite{Cowling1996}. They showed that certain linear combinations of data points from within the domain into pseudo data points $(x_{(-1)},...,x_{(-m)})$ beyond the boundary can correct the flawed edge behaviour of $\hat{f}(x)$. In particular, they propose to use the modified \ac{KDE}
\begin{subequations}
\begin{align}
	\hat{f}_h(x) &=\frac{1}{\sum_i w_i}\left[\sum_{i=1}^{N} w_i K_h\left(x-x_i\right) + \sum_{i=1}^m w_{(-i)}K_h(x-x_{(-i)})\right]\, , \label{eq: kde with pseudo data}
\intertext{where one possible combination is the three-point-rule}
	x_{(-i)} &= 4 x_{\rm min} - 6 x_i + 4 x_{2i} -x_{3i}\, ,\\
	w_{(-i)} &= \frac{w_i+w_{2i}+w_{3i}}{3}\, ,\\
	 m&=\frac{N}{3}\, .
\end{align}
\end{subequations}
As opposed to eq.~\eqref{eq: kde2}, the normalization of~\eqref{eq: kde with pseudo data} on its support is no longer guaranteed, it is therefore a good idea to re-normalize the density estimate. As shown in figure~\ref{fig: cowling and hall}, the bias towards the domain boundary vanishes.

\section{The Runge-Kutta-Fehlberg Method}
\label{a: runge kutta fehlberg}
The Runge-Kutta-Fehlberg method is a member of the~Runge-Kutta~(RK) family to explicitly and adaptively solve ordinary differential equations~(ODEs) of first order~\cite{Fehlberg1969}. It requires the same number of function evaluations as RK6, but is rather a combination of RK4 and RK5, such that the error of each step can be estimated. This estimate can be used to adjust the step size adaptively.

For a 1st order ordinary differential equation with given initial conditions,
\begin{align}
	\dot{y} = f(t,y)\, ,\quad y(t_0) = y_0\, ,
\end{align}
we can find the solution via the iteration step
\begin{subequations}
\begin{align}
	y_{k+1} &= y_k + \frac{25}{216}k_1+ \frac{1408}{2565}k_3+ \frac{2197}{4101}k_4- \frac{1}{5}k_5\, ,
	\intertext{where}
	k_1 &= \Delta t\; f\left(t_k, y_k\right) \, ,\\
	k_2 &= \Delta t\; f\left(t_k + \frac{1}{4}\Delta t, y_k+\frac{1}{4}k_1\right) \, ,\\
	k_3 &= \Delta t\; f\left(t_k + \frac{3}{8}\Delta t, y_k+\frac{3}{32}k_1+\frac{9}{32}k_2\right) \, ,\\
	k_4 &= \Delta t\; f\left(t_k + \frac{12}{13}\Delta t, y_k+\frac{1932}{2197}k_1-\frac{7200}{2197}k_2+\frac{7296}{2197}k_3\right) \, ,\\
	k_5 &= \Delta t\; f\left(t_k + \Delta t, y_k+\frac{439}{216}k_1-8k_2+\frac{3680}{513}k_3-\frac{845}{4104}k_4\right) \, ,\\
	k_6 &=\Delta t\; f\bigg(t_k + \frac{\Delta t}{2},\nonumber\\
	&\qquad\qquad y_k-\frac{8}{27}k_1+2k_2-\frac{3544}{2565}k_3+\frac{1859}{4104}k_4-\frac{11}{40}k_5\bigg) \, .
\end{align}
\end{subequations}
Interestingly enough, we can use the same coefficients to get a 5th order estimate of the solution,
\begin{align}
	\tilde{y}_{k+1} = y_k + \frac{16}{135}k_1+ \frac{6656}{12825}k_3+ \frac{28561}{56430}k_4- \frac{9}{50}k_5+\frac{2}{55}k_6\, ,
\end{align}
giving us a direct measure $\left| y_{k+1}-\tilde{y}_{k+1}\right|$ of the 4th order method's error. Specifying an error tolerance~$\epsilon$, we can adjust our step size adaptively,
\begin{align}
	\Delta t_{k+1} = 0.84 \left(\frac{\epsilon}{\left| y_{k+1}-\tilde{y}_{k+1}\right|}\right)^{1/4}\Delta t_k\, .
\end{align}
The new step size is either used in the next step or to repeat the previous step, if the error exceeds the error tolerance.

\clearpage
\chapter{Rare Event Simulations}
\label{a:rareevents}
To determine direct detection constraints on strongly interacting DM with \ac{MC}~simulations, we simulate trajectories through the shielding overburden above a laboratory, where most particles fail to reach the detector. The desired data is composed of the \emph{rare events}, whenever a particle against all odds makes it to the detector, while still being energetic enough to cause a signal in the detector. These events can be so rare that millions of trajectories need to be simulated, in order to obtain a single data point. It is clear that this yields a challenge to `brute force' \ac{MC}~simulations of rare events, which are in the best case just inefficient, in the worst case would require such a tremendous amount of computing time, that they become practically unapplicable. In this app., we present and review two advanced \ac{MC}~methods for variation reduction, \acf{IS} and \acf{GIS}, which deal with this problem. Both methods are implemented in the \ref{code2} code and increase the probability of a successful simulation run, therefore reducing computational time\footnote{For a more detailed introduction to rare event simulations, we recommend~\cite{Bucklew:2004} and~\cite{Haghighat2016}.}.

\section{Importance Sampling}
\label{a:IS}
Typically, certain events in \ac{MC}~simulations are rare, if the region of interest of the involved probability densities is found in the distributions' tails. Naturally, values from these tails are rarely sampled. \acf{IS} is a standard \ac{MC}~technique and artificially increases the probability to sample these more `important' values~\cite{Kahn1951}. It introduces a bias in the simulation's \ac{PDF}s, ensuring to compensate this bias by a appropriately chosen weighting factor.

Assuming a \ac{MC}~simulation, which involves a single \ac{PDF} $f(x)$ and where a successful run is extremely unlikely. The final data set is the result of an aggressive filtering process, where most simulation runs fail and do not contribute. The statistical properties of the successful runs consequently differ considerably from the `typical' run. For example, a simulation run might be more likely to succeed, if the random variable is repeatedly sampled from the suppressed region of interest of $f(x)$. The distribution of this random variable in successful simulation differs from the distribution $f(x)$ of an average run and follows its own underlying \ac{PDF} $g(x)$ determined by the filter criteria.

The difference in the \acp{PDF} between a rare successful and an average run can be exploited. If the simulation no longer samples from the original $f(x)$, but instead of some new distribution $\hat{g}(x)$ which approximates $g(x)$, it would imitate the statistical behaviour of successful simulation runs. Consequently, the desired events occur more frequently. The modification of the \ac{PDF} introduces a bias which needs to be compensated by a data weight. Furthermore, if the biased distribution function $\hat{g}(x)$ is chosen poorly and does not approximate the true distribution of the rare events, the method becomes unstable and does no longer produce reliable results. A comparison between the unmodified and the \ac{IS} simulation for some examples should always be performed as consistency check.

To determine the weighting factor, assume a random variable $X$ with \ac{PDF} $f(x)$ and the expectation value of a quantity $Y(X)$ on a given interval $I$,
\begin{subequations}
	\begin{align}
	\langle Y\rangle_I &= \int_I\dd x \; Y(x) f(x)\, .
\intertext{This can trivially be written as}
&= \int_I \dd x\; Y(x) \frac{f(x)}{\hat{g}(x)}\hat{g}(x)\, .
\end{align}
\end{subequations}
The function $\hat{g}(x)$ can be interpreted as a new distribution function, whereas the factor $\frac{f(x)}{\hat{g}(x)}$ may be regarded as the weighting function. Importance Sampling means to sample from $\hat{g}(x)$ instead of $f(x)$ during the simulations. Under the assumption that $\hat{g}(x)$ is chosen wisely in the sense described above, the rare events of interest occur with higher probability. This way, \ac{IS} can make rare event simulation practically feasible, where `brute force' simulations fail due to finite time and resources.

This method is of special interest for the simulation of strongly interacting DM~particles in e.g. the Earth crust, where the particles can lose a significant fraction of their kinetic energy in a single scattering on a nucleus. It is remarkably powerful for GeV scale DM and contact interactions, where its application was first proposed by Mahdawi and Farrar~\cite{Mahdawi:2017cxz,Mahdawi:2017utm}. They showed how the DM~particles, which reach the detector depth, deviate in their statistical properties in this case from the average particles.
\begin{enumerate}
	\item Successful particles scatter fewer times, or in other words, propagate freely for longer distances than the mean free path.
	\item They also tend to scatter more in the forward direction, as opposed to the typical particle, whose scatterings are isotropic.
\end{enumerate}
Obviously, both properties increase the chance of reaching the detector depth before they get reflected or lose too much energy to be detectable. It also shows, how the simulation's \acp{PDF} should be modified for \ac{IS}, as will be shown in detail.

The central random variables in the DM simulations were introduced in chapter~\ref{s: MC simulations}. The first is the distance $L$ a particle propagates freely before scattering on a shielding target, with its \ac{PDF}
\begin{align}
	f_{\lambda}(x) = \frac{1}{\lambda}\exp\left(-\frac{x}{\lambda}\right)\, ,
\end{align}
where $\lambda$ is the mean free path. A reasonable \ac{IS} modification of this distribution to address point 1 is to increase the mean free path,
\begin{align}
	g_\lambda(x)= \frac{1}{(1+\delta_\lambda)\lambda}\exp\left(-\frac{x}{(1+\delta_\lambda)\lambda}\right)\, ,
\end{align}
with $\delta_\lambda>0$. If a distance $L_i$ is sampled as described in chapter~\ref{ss:MCsampling} by solving
\begin{align}
	&\int\limits_{0}^{L_i}\dd x\; g_\lambda(x) = \xi\, ,\label{eq:MCsampling}
\end{align}
the accompanying weighting factor is
\begin{align}
	w_{\lambda,i} = \frac{f_\lambda(l_i)}{g_\lambda(l_i)} =  (1+\delta_\lambda)(1-\xi)^{\delta_{\lambda}}\label{eq:weight}\, .
\end{align}
For a trajectory of $n_S$ scattering events, the total weight is
\begin{align}
	w_\lambda = \prod_{i=0}^{n_S} w_{\lambda,i}\, .
\end{align}

\begin{figure*}
\sbox\twosubbox{%
  \resizebox{\dimexpr0.98\textwidth-1em}{!}{%
    \includegraphics[height=3.5cm]{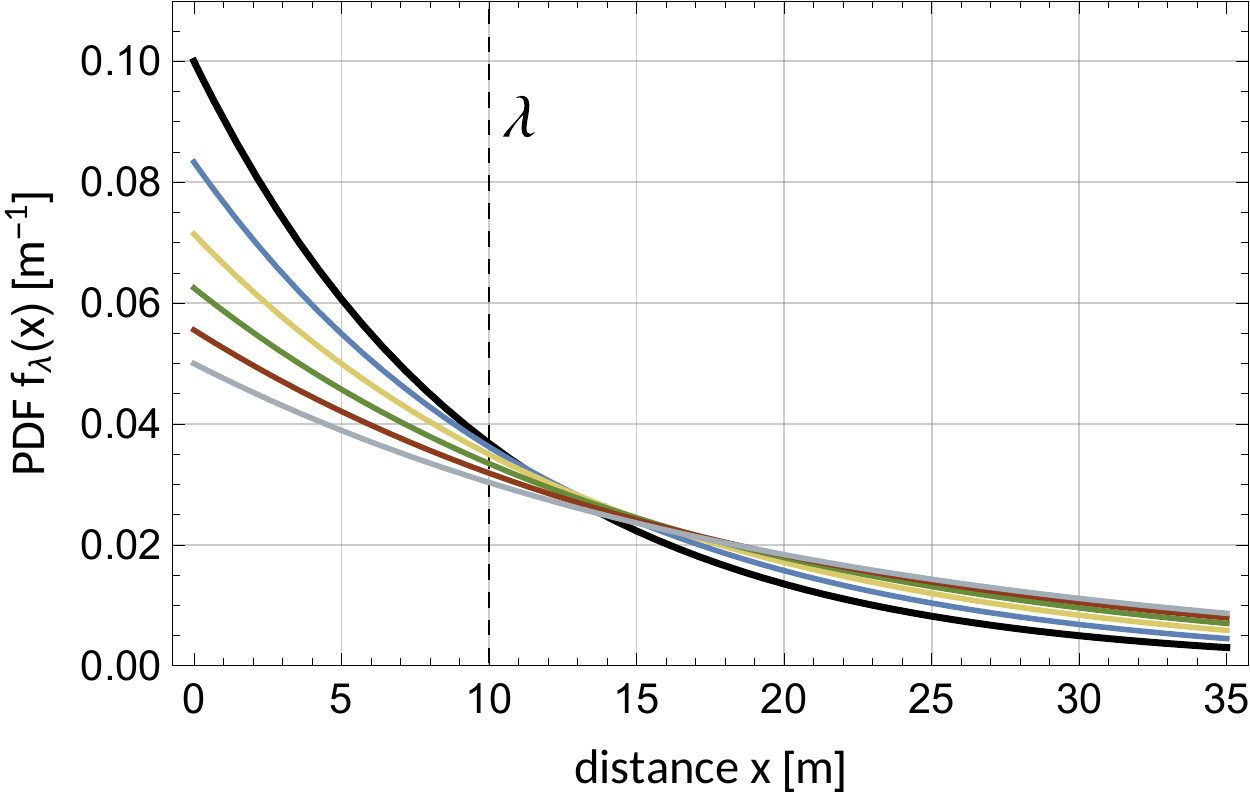}%
    \includegraphics[height=3.5cm]{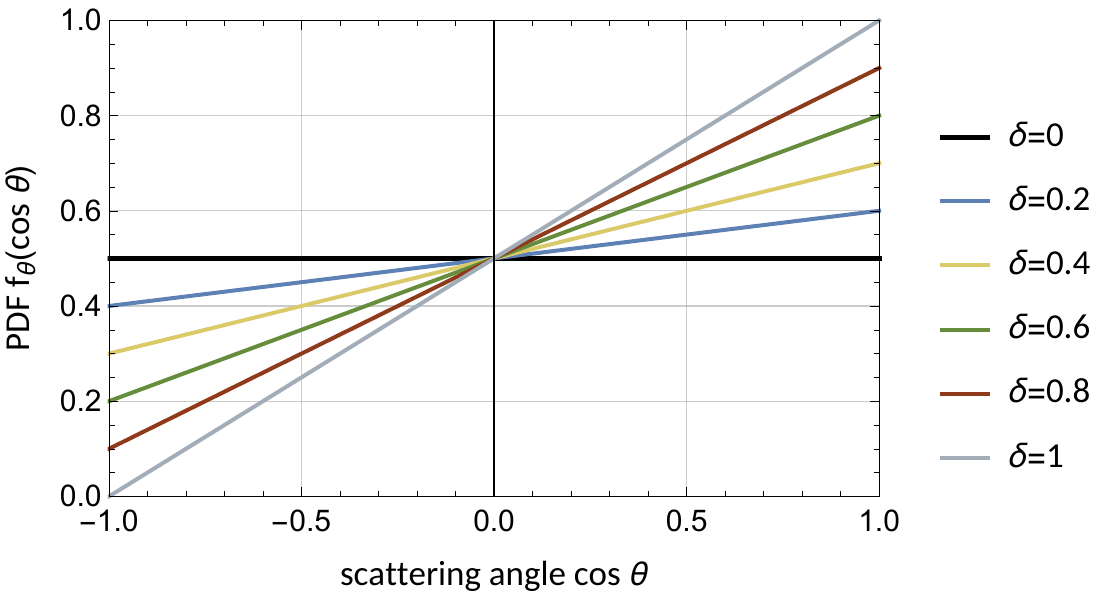}%
  }%
}
\setlength{\twosubht}{\ht\twosubbox}
   	\centering
   	\subcaptionbox{Free propagation distance.\label{fig:is lambda}}{%
  	\includegraphics[height=\twosubht]{plots/IS_Lambda.pdf}%
	}
	\quad
	\subcaptionbox{Scattering angle.\label{fig:is alpha}}{%
	  \includegraphics[height=\twosubht]{plots/IS_Alpha.pdf}%
	}

	\caption{Importance Sampling modifications of the simulation's PDFs.}
	\label{fig:importancesampling}
\end{figure*}

The random variable which can address point 2 is the scattering angle in the center of mass frame~$\theta$. As discussed in more detail in chapter~\ref{ss:scattering angle}, its distribution is isotropic for contact interactions of light~DM, as the nuclear form factor $F_N(q^2)$ can be approximated by 1. Then $\cos \theta$ is a uniform random quantity with \ac{PDF}
\begin{align}
	f_{\theta}(\cos\theta) = \frac{1}{2}\, ,
\end{align}
and domain $(-1,1)$. The \ac{IS}-modified distribution should favour forward scatterings lowering the reflection probability and the energy loss in a scattering. Both support a successful outcome of the respective trajectory. One possible implementation of this idea is the \ac{IS} biased probability density
\begin{align}
	g_\theta(\cos\theta) = \frac{1+\delta_\theta\cos\theta}{2}\, ,\quad\text{with }\delta_\theta\in[0,1]\, .\label{eq: is pdf ldm}
\end{align}
Whenever we determine a sample of $\cos \theta_i$ by inverse transform sampling,
\begin{align}
	&\int\limits_{-1}^{\cos\theta_i}\dd\cos\theta\; g_\theta(\cos\theta) = \xi\, ,\\
	\Rightarrow &\cos\theta_i = \frac{-1 + \sqrt{(1-\delta_\theta)^2+4\delta_\theta\xi}}{\delta_\theta}\, .
\end{align}
the weighting factor is
\begin{align}
	w_{\theta,i} &= \frac{f_\theta(\cos\theta_i)}{g_\theta(\cos\theta_i)} = \frac{1}{1+\delta_\theta \cos\theta_i}\nonumber\\
	& = \left[ (1-\delta_\theta)^2 +4\delta_{\theta}\xi\right]^{-1/2}\, .
\end{align}
In the non-\ac{IS} limit, isotropic scattering is re-obtained, as  $\lim\limits_{\delta_\theta\rightarrow 0}\cos\theta_i = 2\xi - 1$. The \ac{IS} modifications are shown in figure~\ref{fig:importancesampling}.

For the simulation of heavier DM~particles, the loss of coherence needs to be taken into account, and the nuclear form factor $F_N(q)$ can no longer be neglected. Instead it favours small momentum transfers and forward scatterings as we saw in section~\ref{ss:scattering angle}, with the \ac{PDF} given by eq.~\eqref{eq: scattering angle pdf helm}. Similarly to eq.~\eqref{eq: is pdf ldm}, the \ac{IS} \ac{PDF} can be defined as
\begin{align}
	g_{\theta}(\cos\theta;\, v_\chi)=f_{\theta}(\cos\theta;\, v_\chi)+\frac{\delta_\theta}{2}\cos\theta\, ,
\end{align}
in order to increase the chance of forward scattering even more. Yet it needs to be ensured that $g_\theta(\cos\theta;v_\chi) > 0$ for backwards scattering, which occurs if
\begin{align}
	\delta_\theta > 2 f_{\theta}(-1;\, v_\chi)\, .
\end{align} 

The weight of a successful trajectory with $n_S$ scatterings is given by the product
\begin{align}
	w_\theta = \prod_{i=1}^{n_S}w_{\theta,i}\, .
\end{align}
At last, the total statistical weight of a data point obtained with \ac{IS} of both involved random variables is just the product of the respective weights, $w = w_{\lambda}w_\theta$.

\section{Geometric Importance Splitting}
\label{a:GIS}

As mentioned in the previous section, \ac{IS} yields stable results, only if the simulated DM~particle can lose a large fraction of its energy in very few scatterings. In other cases, such as for very heavy DM or interactions with light mediators, this is no longer the case. Instead even successful trajectories involve hundreds or thousands of scatterings, each of which cause a tiny relative loss of energy. For these cases, we implemented an alternative \ac{MC}~method for rare event simulation, called \acf{GIS}, which goes back to John von Neumann and Stanislav Ulam and was first described by Kahn and Harris in the context of neutron transport~\cite{Kahn1951}.

\subsection{Splitting and Russian Roulette for rare event simulation}
Unlike \ac{IS}, the implementation of \ac{GIS} does not modify any the \acp{PDF}, and the sampling of the random variables is unchanged. Instead, `important' particles are being split into a number of copies, whose propagation is simulated independently. In addition, the simulation of `unimportant' particles has a chance of being terminated prematurely. This stresses the physical intuition that we do not simulate fundamental particles but packages of particles, which can be split into smaller sub-packages. The size of such a package is quantified by its statistical weight, which decreases when it gets split.

This raises the question, what it means for a particle to be `important'. The `importance' of a particle is defined by an importance function $I:\mathbb{R}^3\rightarrow \mathbb{R}$, which is the central object for simulations with \ac{GIS}. As a particle approaches the detector depth, the importance function should increase. In fact, it is a monotonously increasing function of the particle's underground depth.

The \ac{GIS} algorithm is the following. Assuming a particle of weight $w_i$ and importance $I_i$ scatters at depth $z$, its previous importance is compared to the new value $I_{i+1}\equiv I(z)$. If
\begin{align}
	\nu &\equiv \frac{I_{i+1}}{I_i} > 1\, ,
	\intertext{the particle will be split into $n$ copies of weight $w_{i+1}$, where}
n&=\begin{cases}
	\nu\, ,\quad &\text{if }\nu\in\mathbb{N}\, ,\\
	\lfloor \nu\rfloor \, ,\quad &\text{if }\nu\notin\mathbb{N} \land \xi \geq \Delta\, ,\\
	\lfloor \nu\rfloor+1\, ,\quad &\text{if }\nu\notin\mathbb{N} \land \xi<\Delta \, ,
\end{cases}\label{eq:giscases}
\intertext{and}
w_{i+1}&\equiv \frac{w_i}{n}\, .
\end{align}
Here $\Delta\equiv \nu-\lfloor \nu \rfloor$ is the non-integer part of $\nu$ and $\xi$ is a sample of~$\mathcal{U}_{[0,1]}$. The expectation value of $n$ for non-integer $\nu$ is nothing but $\nu$, as $\langle n\rangle = \Delta (\lfloor\nu\rfloor+1) + (1-\Delta)\lfloor \nu\rfloor = \nu$.
As coined by von Neumann and Ulam, the opposite action of particle splitting is called Russian Roulette. If a particle's importance is decreasing, i.e.
\begin{align}
	\nu &= \frac{I_{i+1}}{I_i} < 1\, ,
\end{align}
the particle's simulation is terminated with a probability of $p_{\rm kill}= 1-\nu$. In the case of survival, its weight gets increased via
\begin{align}
	w_{i+1} = \frac{w_i}{1-p_{\rm kill}} = \frac{w_i}{\nu} >w_i\, .
\end{align}
such that the expectation value of the new weight is just the old weight, $\langle w_{i+1}\rangle = p_{\rm kill}\cdot 0 + (1-p_{\rm kill})\cdot w_{i+1} = w_i$. Russian roulette can be regarded as a special case of eq.~\eqref{eq:giscases}.

\begin{figure*}
	\centering
	\includegraphics[width=0.67\textwidth]{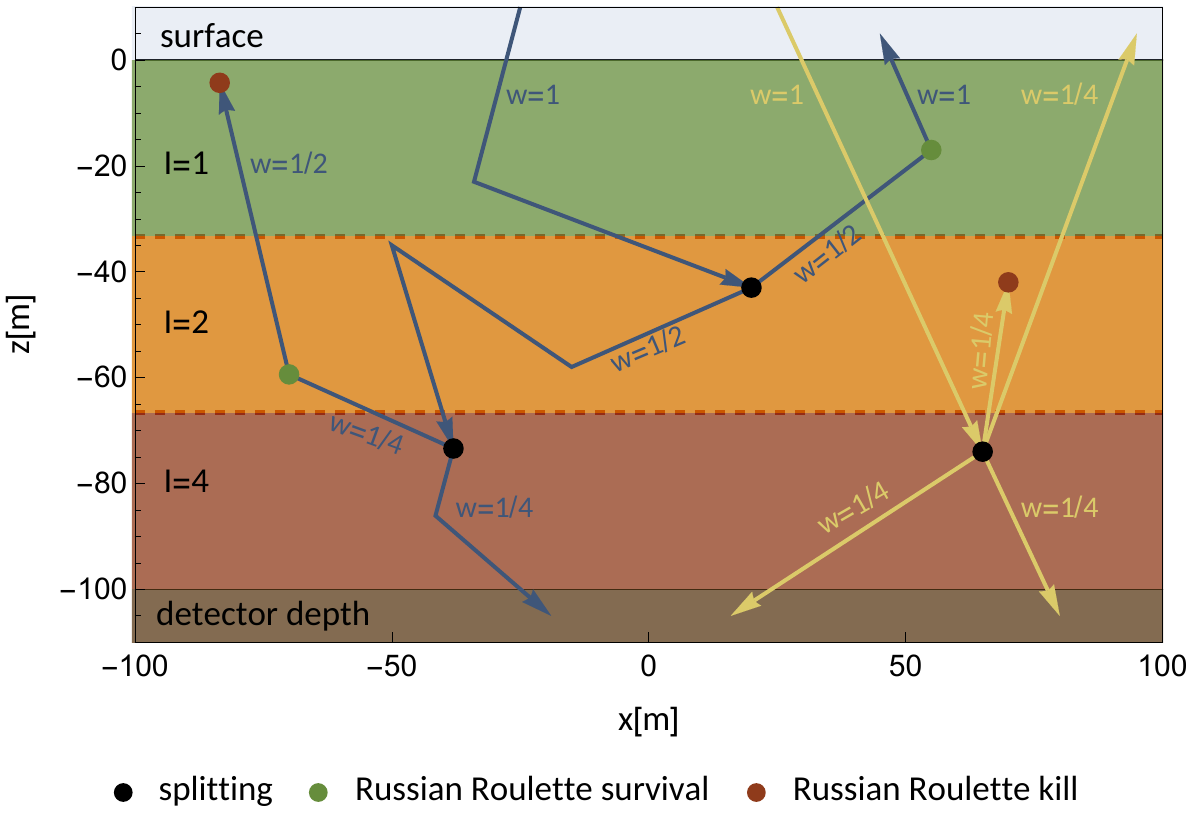}
	\caption{Illustration of two DM trajectories using~GIS with three importance domains of equal size. The particle can split into either 2 or 4 copies. The weight development along the paths is shown as well.}
	\label{fig:gis}
\end{figure*}

\subsection{The Importance Function and Adaptive GIS}
The central challenge of \ac{IS} was to find an appropriate modification of the simulation's \acp{PDF}. Similarly, the central problem of \ac{GIS} is the construction of the importance function, which determines the particle splittings.

For the purposes of this thesis, the shielding layers above a direct detection experiment at depth $d$ can simply be divided into $N_I$ importance domains, defined by a set of planar splitting surfaces of depth $0>l_1>l_2>...>l_{N_I-1}>d$. The importance of a particle changes, if it passes any of these boundaries. Assigning domain $k$ an importance of e.g. $I_k=N_{\rm splits}^{k-1}$, with $N_{\rm splits}=2,3,...$, a particle from domain $k$ reaching $k+1$ will be split into $N_{\rm splits}$ copies. However, the importances do not need to be integers.

The number of importance domains is to be determined adaptively. Choosing~$N_I$ too small will not exploit the full benefits of \ac{GIS}, choosing it too large can cause a single particle to split into a large number of copies. If these reach the detector, the data set can be highly correlated, as a large fraction of the final data originates from one particle. For example with $N_I=$10 and $N_{\rm splits}=$3, a particle which scatters in domain~9 for the first time will split into $\text{3}^\text{8}=$6561 copies.

If hard scatterings with large relative energy losses dominate the simulation, the choice $N_I \sim d / \lambda$, with $\lambda$ being the mean free path, would work well. However, it is a better idea to determine $N_I$ based on the integrated stopping power, described in chapter~\ref{ss: stopping power}, as this also works for the case of ultralight mediators,
\begin{align}
	\langle\Delta E\rangle &\equiv \int\limits_{0}^d \dd x\; S_n(m_\chi,\sigma_p,\langle v_\chi\rangle) = \sum_{l=1}^{N_{\rm layers}} t_l S_n^l(m_\chi,\sigma_p,\langle v_\chi\rangle)\, .
\end{align}
Here, $t_l$ is the thickness of the physical shielding layer $l$, and $\langle v_\chi\rangle = \int_{v_{\rm cutoff}}^{v_{\rm max}}v f(v)$ is the average initial speed of the simulated DM~particles. We can now determine $N_I$ adaptively via
\begin{align}
	N_I = \left\lceil\kappa\cdot \frac{\langle\Delta E\rangle}{\frac{m_\chi}{2}\left(\langle v_\chi \rangle^2-v_{\rm cutoff}^2\right)}\right\rceil\, ,
\end{align}
where $\kappa$ is a dimensionless parameter, which can be freely adjusted to influence the pace with which the number of importance domain is increasing.
For the location of the splitting surfaces, $l_1,...,l_{N_I-1}$, it should be ensured, that the domains are of equal integrated stopping power, i.e. $\langle\Delta E\rangle/N_I$. This way, we avoid placing a lot of the boundaries into regions of relatively weak stopping power. For example, if we simulate both the Earth crust and the atmosphere, most if not all domain boundaries should be located in the crust.

The benefit of this method is a simulation speed-up of up to two orders of magnitude. The \ac{GIS} results have been compared to non-\ac{GIS} results for different examples, to verify the method's validity. All results are invariant under (de-)activation of the \ac{GIS} method up to statistical fluctuations.

\clearpage
\bibliography{thesis.bib} 
\end{document}